\documentclass[twocolumn,english,prb,floatfix,superscriptaddress,twocolumn,amsmath,amssymb,reprint]{revtex4-1}

\usepackage{graphicx}
\usepackage{upgreek} 
\usepackage{mathrsfs}
\usepackage{bm}
\usepackage{xcolor}
\usepackage{amsmath}
\usepackage{amssymb}
\usepackage{hyperref}
\usepackage{soul}
\usepackage{xcolor}

\def\tc{$T_{\mathrm{c}}$\ }

\def\t6as{$\mathrm{(TMTSF)_{2}AsF_{6}}$\ }

\def\tmasf6sbf6{(TMTSF)$_{2}$(AsF$_{6}$)$_{(1-x)}$(SbF$_{6}$)$_x$\ }
\def\tmxns{(TMTSF)$_{2}$(ClO$_{4}$)$_{(1-x)}$(ReO$_{4}$)$_x$\ }
\def\tmc{$\mathrm{(TMTSF)_{2}ClO_{4}}$\ }
\def\tmcns{$\mathrm{(TMTSF)_{2}ClO_{4}}$}
\def\tms{$\mathrm{(TMTSF)_{2}AsF_{6(1-x)}SbF_{6x}}$\ }
\def\tmps{$\mathrm{(TMTTF)_{2}PF_{6}}$\ }
\def\tmttfsbf6{$\mathrm{(TMTTF)_{2}SbF_{6}}$\ }

\def\tmttfasf6{$\mathrm{(TMTTF)_{2}AsF_{6}}$\ }
\def\tmtsfasf6{$\mathrm{(TMTSF)_{2}AsF_{6}}$\ }
\def\tmttfbf4{$\mathrm{(TMTTF)_{2}BF_{4}}$\ }
\def\tmtsfbf4{$\mathrm{(TMTSF)_{2}BF_{4}}$\ }

\def\tmtsfreo4{$\mathrm{(TMTSF)_{2}ReO_{4}}$\ }
\def\tmno3{$\mathrm{(TMTSF)_{2}NO_{3}}$\ }
\def\tm2x{$\mathrm{(TM)_{2}X}$\ }
\def\tmttf2x{$\mathrm{(TMTTF)_{2}X}$\ }
\def\tmtsf2x{$\mathrm{(TMTSF)_{2}X}$\ }
\def\tm2xns{$\mathrm{(TM)_{2}X}$}
\def\tset2cl{$\mathrm{(TSeT)_{2}Cl}$\ }
\def\tq{TTF-TCNQ\ }

\def\tsq{$\mathrm{TSF-TCNQ}$}
\def\dsdtfq{$\mathrm{DSDTF-TCNQ}$\ }
\def\qnq{$\mathrm {Qn(TCNQ)_{2}}$\ }
\def\R{$\mathrm{ReO_{4}^{-}}$\ }
\def\C{$\mathrm{ClO_{4}^{-}}$\ }

\def\tqr{$\mathrm{TCNQ^\frac{\cdot}{}}$\ }
\def\tmr{$\mathrm{TMTSF}^{\cdot+}$}
\def\nmpq{$\mathrm{NMP^{+}(TCNQ)^\frac{\cdot}{}}$\ }
\def\f{$\mathrm{TTF}$\ }
\def\q{$\mathrm{TCNQ}$\ }
\def\nmq{$\mathrm{NMP-TCNQ}$\ }
\def\nmq{NMP-TCNQ\ }
\def\nmp{$\mathrm{NMP^{+}}$\ }
\def\pc{$P_{\mathrm{c}}$\ }

\def\hc2{$H_{\mathrm{c2}}$\ }
\def\ts{$\mathrm{TSF}$}

\def\TSF{$\mathrm{TSF}$\ }

\def\tmp6{$\mathrm{(TMTSF)_{2}PF_{6}}$\ }				
\def\tmre{$\mathrm{(TMTSF)_{2}ReO_{4}}$}
\def\tsx{$\mathrm{(TMTSF)_{2}}X$\ }
\def\tms2x{$\mathrm{(TMTSF)_{2}X}$}
\def\tm2x{$\mathrm{(TM)_{2}}X$\ }
\def\as{$\mathrm{AsF_{6}^-}$}
\def\sb{$\mathrm{SbF_{6}^-}$}
\def\pf{$\mathrm{PF_{6}^-}$}
\def\re{$\mathrm{ReO_{4}^-}$}
\def\ta{$\mathrm{TaF_{6}^-}$}
\def\scn{$\mathrm{SCN^{-}}$}
\def\br{$\mathrm{Br^-}$}
\def\cl{$\mathrm{ClO_{4}^-}$}
\def\clo4{$\mathrm{ClO_{4}}$}
\def\scn{$\mathrm{SCN^-}$}
\def\fso3{$\mathrm{FSO_{3}^-}$}
\def\no{$\mathrm{NO_{3}^-}$}
\def\4fb{$\mathrm{BF_{4}^-}$}

\def\edtasf6{$\mathrm{(EDT-TTF-CONMe_{2})_{2}AsF_{6}}$}

\def\edt2br{$\mathrm{(EDT)_{2}Br}$}
\def\tfx{$\mathrm{(TMTTF)_{2}X}$\ }
\def\tsx{$\mathrm{(TMTSF)_{2}}X$\ }

\def\ttf{$\mathrm{TTF}$\ }
\def\nmp{NMP\ }

\def\bedtttf{$\mathrm{BEDT-TTF}$\ }
\def\bedtttf2i3{$\mathrm{(BEDT-TTF)_{2}I_{3}}$\ }
\def\reo4{$\mathrm{ReO_{4}}$}
\def\bedtttfreo4{$\mathrm{(BEDT-TTF)_{2}ReO_{4}}$\ }
\def\et2i3{$\mathrm{(ET)_{2}I_{3}}$\,}
\def\et2x{$\mathrm{(ET)_{2}X}$\,}
\def\cuncnbr{$\mathrm{Cu(N(CN)_{2})Br}$\,}
\def\ket2cl{$\mathrm{\kappa-(ET)_{2}Cl}$\,}
\def\ket2xbr{$\mathrm{\kappa-(ET)_{2}Br}$\,}
\def\kappaet2cunbr{$\mathrm{\kappa-(ET)_{2}Cu(N(CN)_{2})Br}$\,}

\def\cuncncl{$\mathrm{Cu(N(CN)_{2})Cl}$\,}
\def\cuncs{$\mathrm{Cu(NCS)_{2}}$\,}
\def\betsfecl4{$\mathrm{(BETS)_{2}FeCl_{4}}$\,}
\def\bets{$\mathrm{BETS}$\,}
\def\et{$\mathrm{ET}$\,}

\def\baas{$\mathrm{Ba(Fe_{1-x}Co_{x})_2As_{2}}$}

\def\nb3sn{$\mathrm{Nb_{3}Sn}$\,}
\def\v3si{$\mathrm{V_{3}Si}$\,}
\def\nb3ge{$\mathrm{Nb_{3}Ge}$\,}

\def\v2o3{$\mathrm{V_{2}O_{3}}$}

\def\sumslashD{\mathop{\sum \kern-1.4em -\kern 0.5em}}
\def\sumslash{\mathop{\sum \kern-1.2em -\kern 0.5em}}
\def\intslash{\mathop{\int \kern-0.9em -\kern 0.5em}}
\def\intslashD{\mathop{\int \kern-1.1em -\kern 0.5em}}

\begin{document}
    
\title{
Quasi one-dimensional organic conductors: from Fr\"ohlich conductivity and Peierls insulating state
 to  magnetically-mediated superconductivity, a retrospective}

\author{Denis~Jerome} \affiliation{Laboratoire de Physique des Solides (UMR 8502), Universit\'e Paris-Saclay 91405 Orsay, France}
\author{Claude Bourbonnais} \affiliation{ Regroupement Qu\'eb\'ecois sur les Mat\'eriaux de Pointe et Institut Quantique,  D\'epartement de Physique, Universit\'e de Sherbrooke,
Sherbrooke, Qu\'ebec, Canada, J1K-2R1}
\date{\today}

\begin{abstract}
  It is indisputable that the search  for high-temperature superconductivity has stimulated the work on low-dimensional organic conductors at its beginning. Since the discovery of true metal-like conduction in molecular compounds more than 50 years ago, it appeared that the chemical composition and the quasi one-dimensional crystalline structure of these conductors  were determining factors for their physical properties; materials with incommensurate conduction band filling favoring the low-dimensional   electron-phonon diverging channel  and the establishment of the Peierls superstructure and more rarely superconductivity at low temperature, while those with commensurate band filling favor  either magnetic insulating or superconducting states depending on the intensity of the coupling between conductive chains.
In addition, the simple structures of these materials have allowed the development  of theoretical models in close cooperation with almost all  experimental findings.

Even though  these materials have not yet given rise to true high-temperature superconductivity, the wealth of their physical properties makes them  systems  of choice in the field of condensed matter physics due to their original properties and their educational qualities. Research efforts continue in this field. The present retrospective, which does not attempt to be an exhaustive review of the field,  provides a set of experimental findings alluding to the theoretical development while a forthcoming article will address in more details the theoretical aspect of low dimensional conductors and superconductors.

\end{abstract}

\maketitle

\tableofcontents

\section{Preamble}

A major objective of this overview is to address the matter of competition between insulating and superconducting phases which is perfectly illustrated by the physics of low dimensional conductors.


Furthermore, the year 2023 is particularly appropriate for writing  this overview because it is the occasion to celebrate the first  half century of existence of a scientific domain whose development during the previous decades can be considered as a particularly successful model. 

Many points will be highlighted in this article such as: the multidisciplinary character bringing together theoretical,  synthetic chemistry and experimental and theoretical physics, as well as a broad intercontinental tight cooperation which has remained very productive in the long term.

It is also to be noted that the existence of new physical properties has stimulated very actively the improvement of experimental techniques such as several aspects of physics under very high pressure, the use of high magnetic fields as well as the development of physical measurements at very low temperatures, these three topics being very often coexisting in the actual research activity.

The authors wish to make clear at the outset  that this overview is not intended to be an exhaustive review of the vast field of one-dimensional physics. Excellent books and journal articles have fulfilled this purpose and will be mentioned in the bibliography list, Sec.\ref{Bibliography} of the present article.

And last but not least, this overview is published in Comptes-Rendus Physique because Comptes Rendus are  virtuous full open access journal (diamond model) which makes them accessible to everyone even not being affiliated to a university library.  The authors feel that all products of research supported by public fundings should be accessed freely.

\section{Introduction: the early days of organic conduction}
\subsection{A long-standing need for superconductivity at high temperature}



 
Stabilizing superconductivity at higher and higher temperatures has always motivated the search for new materials  exhibiting possible new mechanisms allowing the formation of electron-electron pairs and their Bose condensation  in a state without resistivity below a critical temperature \tc\cite{Onnes11a,Onnes11}. 

\begin{figure*}
\includegraphics[width=0.75\hsize]{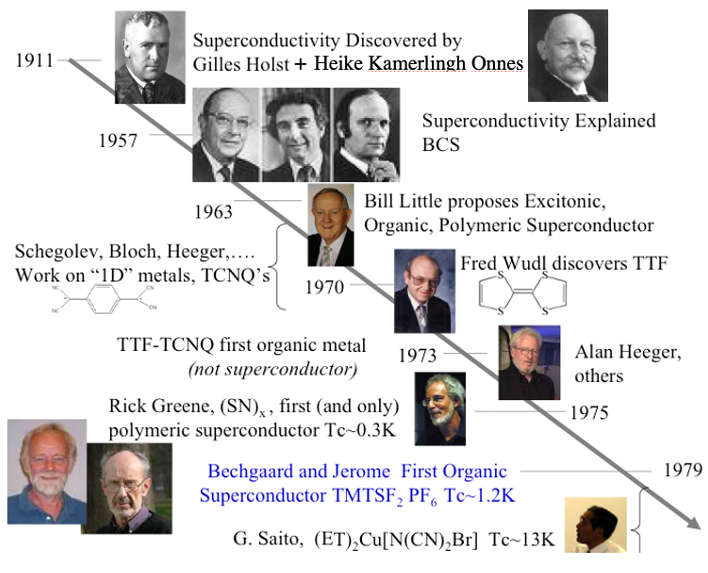}
\caption{\label{Chaikin's list} Timeline of the excitement around superconductivity until the beginning of the eighties. Slide provided by Paul Chaikin. Source: Paul Chaikin's personal website (not online anymore). Many thanks Paul for the slide ! } 
\end{figure*}

An  history   of this fascinating  discovery can be found among other articles in Ref\cite{Meijer94}  and the seminal article for the so-called   BCS theory in Ref\cite{Bardeen57,Bardeen57a}. 
An illustration for the evolution of research to make organic matter electrically conductive and ultimately superconductive has been  summarized on  Paul Chaikin's Web site, \emph{see} Fig.~\ref{Chaikin's list}.

The prerequisite for high conductivity  in  molecular compounds with the possibility of metal-like conduction is the existence in molecular crystals of stable open shell molecular species. Until the middle of the last century, molecular materials were not reputed for their high electronic conductivity.  
It is only at the beginning of the 20th century that 
the possibility of achieving electronic conduction  in molecular solids comprising open shell molecules has crossed the minds of chemists following the suggestion made  by  McCoy and Moore in their article of 1910\cite{McCoy11}. They wrote,
{\it\bf``if the electron theory of the metallic state is as fundamental
as it seems to be, there would be little reason to doubt that the
aggregate of such free radicals would be a body having metallic properties;
for such a hypothetical body would be made up of radicals which, analogous
to metallic atoms, could easily loose electrons",} and these  authors to conclude, {\it\bf "we think 
that the organic radicals in our amalgams are in the metallic state and, therefore, that it is possible to prepare composite metallic substances from
non-metallic constituent elements".}

\begin{figure} \includegraphics[width=0.8\hsize]{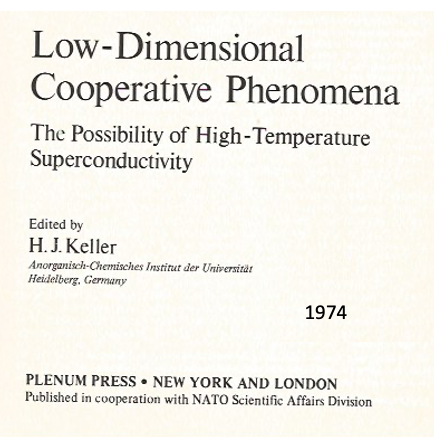} \caption{\label{Keller}  Front page of the Proceedings of the Starnberg summer school {\color{black} (September 1974)}, edited by Heimo. J. Keller, published by Plenum Press 1975\cite{Keller74}.}    \end{figure}

The first mention of the word superconductivity in organic compounds dates back to the work of Fritz London in 1935-37\cite{London35}. London hypothesized that the large susceptibility anisotropy of aromatic compounds such as benzene or anthracene molecules\cite{Pauling36} could be due to superconducting currents flowing on the loops of these unsaturated molecules under magnetic field\cite{London37,London37a}.
Even if this hypothesis does not appear to be realistic anymore after the publication of the now well established BCS theory of superconductivity\cite{Bardeen57}, the idea could still have a weak connection with the persistent currents observed in mesoscopic rings\cite{Bouchiat89}. However, it is interesting to note from the title of the Proceeedings of a summer school held in Germany in 1974 that the stimulus for this research in both chemistry and physics was the possibility of high-temperature superconductivity, \emph{see} Fig.~\ref{Keller}.

\subsection{1950, the decade of  oxidized aromatic molecules}
It is only  in the early fifties that the  concept of electronic conduction in organic solids became reality when semiconducting properties characterized by an activated conductivity had been obtained in   solids made of extended conjugated molecules containing a large number of $\pi$-electrons such as, phthalocyanine, coronene or anthracene, etc..,. It was  suggested in turns that the  current could be carried by these thermally activated $\pi$ electrons\cite{Eley51}.

In fact,  research   conducted in Japan by Akamatu and Inokuchi  did suggest in the 1950's 
that semiconductivity could be attained in organic compounds via   modest  band gaps of the order of $10^{-1} - 10^{-2}$ eV somewhat similar to  those of inorganic semiconductors\cite{Akamatu50}. Furthermore, a rather primitive application of a high pressure was able to demonstrate that the strong enhancement of conductivity observed under 8 kbar could be the result of the increase of the overlap between $\pi$ orbitals of these aromatic polycyclic compounds\cite{Inokuchi55}.
However, the first successful   attempt  to promote high conduction between open shell molecular species  came out in 1954 in complexes between  polycyclic aromatic compounds and halogens anions. The pressed-powder conductivity for the complex of perylene oxidized by bromine  reaching $10^{-3}$ to $1$ $(\Omega.{\rm cm})^{-1}$ was already considered  as fairly high at that time\cite{Akamatu54}. The conduction of this  oxidized perylene complex although due to the overlaping   $\pi$-orbitals of these polycyclic aromatic compounds turned out to be activated with an activation energy of 0.055 eV for the perylene-bromine complex. Unfortunately, this molecular salt of perylene  had the propensity to evolve over time.  
\subsection{1960, the decade of \q complexes and physicists  coming into the game}
\label{1960}
Following the above-mentioned preliminary results, 
 it is only with the synthesis of the strong electron acceptor \q (tetracyanoquinodimethane) achieved in the sixties  by Acker et-al\cite{Acker60} that organic conduction organic conduction  took off and led to the realization of what are often called ''organic metals'', although this term seems to be rather inappropriate since these new conductors, the subject of this review do not contain any metallic elements.
 
\begin{figure} \includegraphics[width=1.4\hsize]{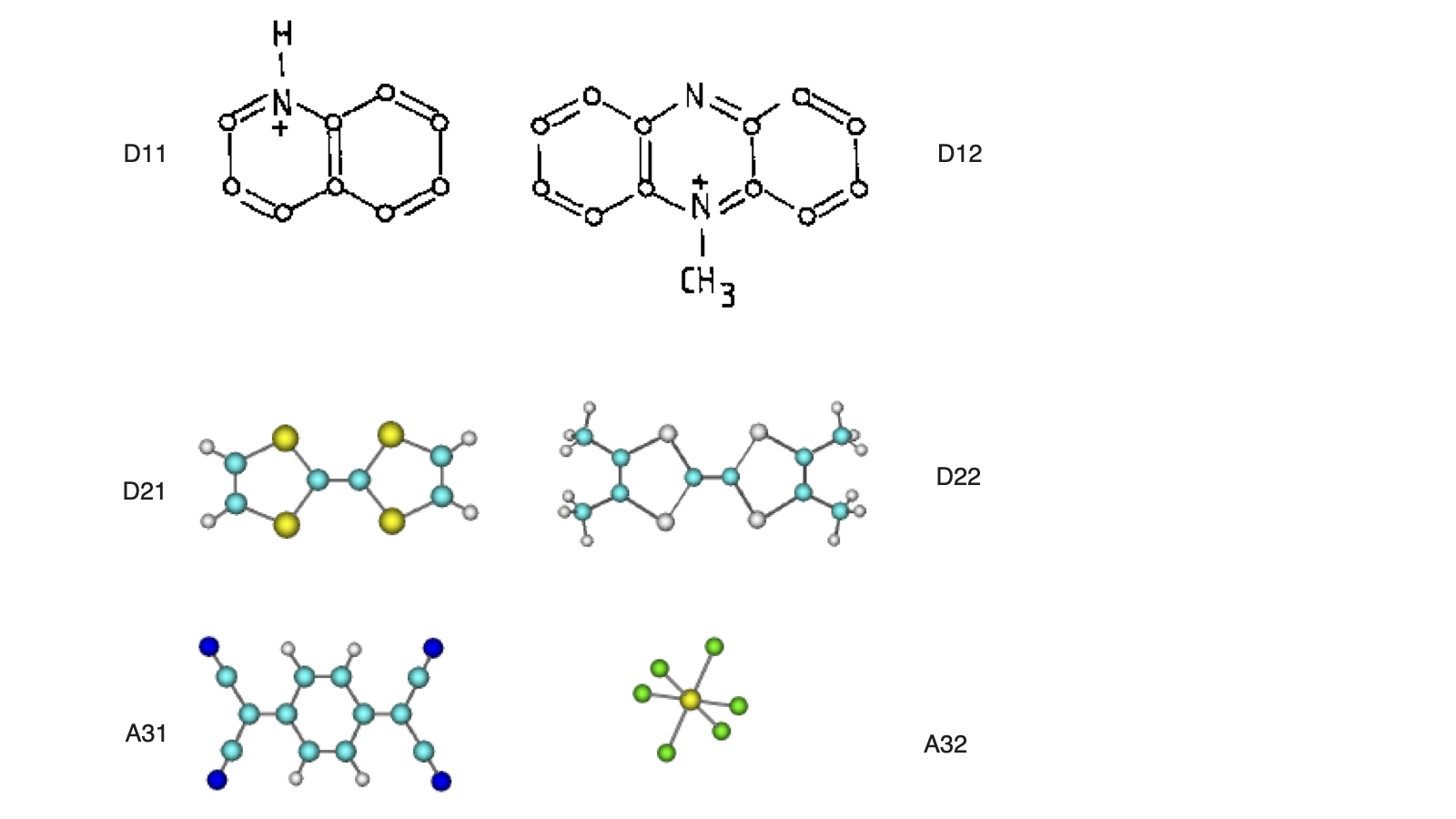} \caption{\label{molecules} Some molecular components entering the synthesis of organic conductors, donors quinolinium (D11) and N-methylquinolinium (D12), TTF (D21) and TMTTF (D22)  or the acceptors TCNQ (A31), and the monoanion \pf (A32).}    \end{figure}

 
 Most  progresses towards organic conductivity in the years 60's   have been been based on the electron acceptor \q.
 TCNQ is a large planar molecule, consisting of a quinoid cycle. It is electrically
neutral and diamagnetic with typically closed outer shells, see molecule A31 on Fig.~\ref{molecules}.  A comprehensive review of the first steps of chemistry and physics towards the synthesis of conducting organic compounds has been published in 1976 by André, Bieber and Gautier\cite{Andre76}.

The peculiarity of the TCNQ molecule is that it is an `electron-poor’ molecule
which easily accepts an extra electron to become a chemically stable open shell anion
(anion radical). TCNQ by itself is unable to conduct. The reason is that although the
molecule will easily accept electrons from its neighbours all molecules have the same
desire and finally a TCNQ crystal remains a Van der Waals solid with insulating
properties. If TCNQ is in an environment where electrons can be obtained from an electron donating
partner, however, the situation changes [4].  To mention only a few new compounds in 1960, let us first consider   simple salts involving the strong electron withdrawing  \q  molecule forming  the series of  1:1 complex  derivatives $M^{+}$\tqr  
 with complete transfer of one electron from a metallic inorganic or organic cation $M$  to \q with the formation of the anion-radical \tqr\cite{Melby62a}. These complexes are actually very poor conductors.
 
There is  also the possibility of forming complex salts comprising an organic cation  such  as quinolinium, Qn linked to two \q molecules,  one being formally neutral in addition to the  \tqr anion radical giving rise to \qnq.  
In the crystal, TCNQ molecules pile up like pancakes to form stacks.
In solids such as \qnq, one electron is injected in each pair of TCNQ molecules. The charged
counter-ion quinolinium (molecule D11 on Fig.~\ref{molecules}) thus exhibits the structure of a closed-shell (diamagnetic)
ion.
For both, 1:1 and 1:2 complexes,  the lowest unoccupied molecular orbital (LUMO) of the \q is partially filled, being half filled for the former  and one-quarter filled for the latter series.
Since the highest occupied states of the TCNQ molecule are filled on average by half
an electron only, the transfer of electrons between \q's along a given stack is
potentially possible.
The 1:2 series exhibits a much higher and very anisotropic conductivity, reaching $10^2$ $(\Omega.{\rm cm})^{-1}$ along the packing axis of single  crystals, the largest obtained in 1960 although still behaving in temperature like  a small gap semiconductor. 

\begin{figure}[h]
\includegraphics[width=1\hsize]{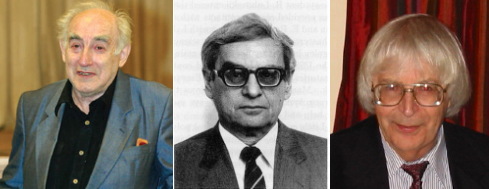}
\caption{\label{les russes} Three russian pioneers in the research on organic superconductors in former Soviet Union. From left to right, Vitaly Ginzburg (1916-2009), his major achievement is the well known theory of superconductivity together with Lev Landau. Igor Schegolev (1929-1996), who developed very successfully as early as 1970, the chemistry and the physics of organic conductors with his numerous colleagues at the  Institute of Chemical Physics of Chernogolovka. Lev Gorkov (1929-2016), he and the many researchers he has trained have contributed in a fundamental way to the development of the theory of organic conductors. As from 1992, L. Gorkov moved to   the National High Magnetic Field Laboratory at Tallahassee of which he was one of the founders.  }
\end{figure}

Regarding the series of \q simple and complexes salts, much work has been accomplished over the sixties by Igor Shchegolev, Fig.~\ref{les russes}, and his group at  Chernogolovka\cite{Shchegolev68} and an exhaustive review of the electrical and magnetic properties  of  anion radical complexes based on \q  has been published in 1972\cite{Shchegolev72}. Quite importantly, russian theoreticians at the Landau Institute of Moscow have played a major role in the early days of the development of non conventional superconductors, V. Ginzburg for his suggestion of exciton-mediated pairing\cite{Ginzburg68}  and L. Gorkov  with his students  who  were deeply involved in the physics of low dimensional conductors in close cooperation with the Orsay laboratory.


 Pursuing his inquiry onto the cation portion of the radical anion salts of \tqr and complexes, Melby\cite{Melby65} managed to synthesize the 1:1 complex N-methylphenazinium, \nmq in short, revealing an  unusually large  value of 142 ($\Omega$.cm)$^{-1} $ for the conductivity  measured on single crystals at 300K.
It is with the synthesis of 
the 1:1 salt  \nmq which presents a low resistivity at 300K  decreasing like that of a metal down to a minimum at 200K\cite{Shchegolev68,Epstein71}  that the field of organic conductors started blooming and opened to the electronic aspects of solid-state physics. 

From a developmental history of the organic conductors field, \nmq is an important compound  in that it bridges the gap between radical cation salts and charge transfer salts which will be  discussed in the following section. \nmq forms a chain-like structure in which donors and acceptors stack in uniform parallel chains. 

 There have been considerable controversies about the exact charge which is transferred  from the donor molecule \nmp  to the acceptor \q. \nmq  has first been  considered as  a radical cation salt similar to previously similar insulating salts.  As a matter of facts, early studies assumed  complete charge transfer from donor to acceptor entities\cite{Epstein71} namely \nmpq  without any  free electron left on \nmp molecules (i.e an empty LUMO band on NMP)  and a half-filled conduction band for the donor stack with an optimum $\pi$-electron interaction between  \q   leading to conduction along the \q stacks only.   However, several magnetic and NMR  as well as cristallographic experiments  concluded to an only partial charge transfer, the exact value being very ill-defined since it was found $\geq 0.8$ from NMR\cite{Devreux78} but equal to 0.42 from \q bond lengths measurements\cite{Flandrois77}.
  
  Thanks to elaborate diffuse X-ray techniques  developed previously for the study of the one dimensional  platinum chains\cite{Comes73} and successfully applied  in 1975 to one dimensional organic conductors\cite{Denoyer75} the charge transfer between \nmp and \q stacks has been measured very accurately\cite{Pouget82}, leading to an incommensurate transfer of 0.6 electron  to the \q molecule making the LUMO bands of both molecules partially charged. Altogether, there is one electron per unit cell in \nmq.

From that period  on (about 1962)  physicists became closely involved in this kind of research. In the mid-1960s they were  very motivated by the search of new superconducting compounds with \tc higher than the current 18K than   the recently discovered  (one decade earlier)   intermetallic  compounds of the
A15 structure, namely, ($\mathrm{V_{3}Si}$ or $\mathrm{Nb_{3}Sn}$)
\cite{Hardy53,Matthias54} or even 22K in $\mathrm{Nb_{3}Ge}$,\cite{Gavaler73} which  have contributed very efficiently   to the  manufacturing of superconducting wires  until now but are still constrained by the use of liquid helium for cooling. Although B. T. Matthias had been a co-discoverer of the highest \tc in 1954 this did not stop him from expressing his doubts about high temperature superconductivity, {\it \bf superconductivity at room temperature will always remain a pipedream, temperatures as high as 25-30 K are a
realistic possibility and will trigger a technological revolution.compounds,} as he wrote in 1971 in Physics Today\cite{Matthias71}. To some extent, the empirical statement of Matthias about the limitations of the A15 superconducting materials turned out to be founded. But it is only the theoretical work performed by the Friedel's school which brought a firm confirmation. It is the  hidden 1D nature of the 
A15 cubic structure which provides the  enhancement of the density of states at the Fermi
level lying close to the van-Hove singularity of the density of states near the
band edges of the 1D d-bands. Within the BCS formalism large \tc can thus be
expected. This is what is actually observed (17–23 K) but an upper limit was
found to the increase of  \tc since the large value of $N(E_{F})$ also makes the structure
unstable against a cubic to tetragonal Jahn–Teller band distortion\cite{Weger64,Labbe66}. The
theory\cite{Labbe67} showed that \tc is indeed maximized in  compounds such as $\mathrm{Nb_{3}Sn}$ or $\mathrm{V_{3}Si}$. 
In practice, the evolution of \tc in various materials  according to the year of their discovery supported the experimental findings, see Fig.~\ref{Tcyears}. However, at that time, theoretical attempts to increase \tc were still based on the phonon-mediated BCS theory and its strong coupling extension\cite{Eliashberg60}.




In order to achieve  higher \tc's  than those obtained in these years, theoreticians   became   extremely imaginative  in terms of exotic suggestions for an electron-electron pairing mechanism other than the phonon-mediated pairing proposed by Bardeen, Cooper and Schrieffer\cite{Bardeen57}. New paths were proposed. Kohn and Luttinger 
proposed a new mechanism for superconductivity still based
on the pairing idea although the attraction is no longer
phonon-mediated\cite{Kohn65}. The attraction derives from an extension
of Friedel’s density oscillations around charged impurities
 and that are present in metals due to the sharpness of the Fermi surface
\cite{Friedel58}. It is an entirely electronic  pairing mechanism  
process. However, the expected critical temperature should
be extremely small in the milli-Kelvin range or much less  depending on the symmetry of Cooper pairing, but Kohn and
Luttinger emphasized that flat Fermi surfaces and van Hove
singularities could greatly enhance the actual \tc.
\begin{figure} 
\includegraphics[width=0.55\hsize]{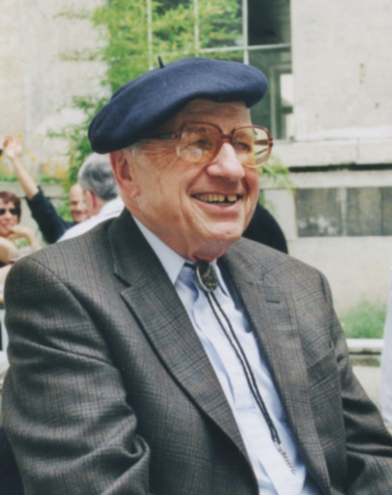}
\caption{\label{WKohn} Walter Kohn (1923-2016), Nobel laureate in Chemistry, 1998. A picture   taken during one of his numerous  visits to Paris. As Walter enjoyed coming to France, he adopted the local dress.}
\end{figure}


At about the same time, the interest for improving \tc has been strongly boostered by the announcement made in 1964 by W. A. Little that the phenomenon of superconductivity for which a satisfactory microscopic theory had been proposed only seven years earlier by Bardeen Cooper and Schrieffer\cite{Bardeen57}  could occur in hypothetical one-dimensional organic conductors at temperatures exceeding 300K\cite{Little64}.
 
\begin{figure} 
\includegraphics[width=1\hsize]{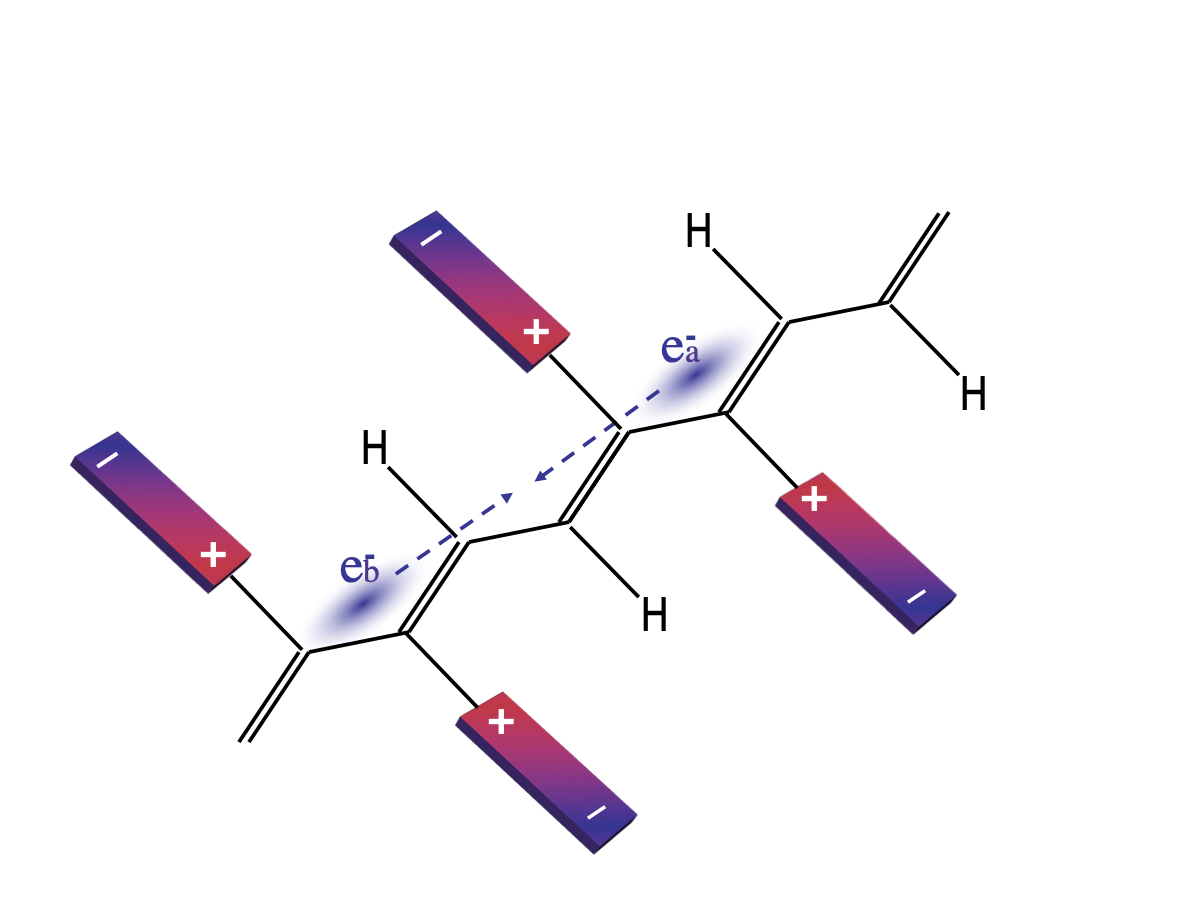}
\caption{\label{Littlemolev1} Schematic picture of Little’s suggestion. Electron $a$ and $b$ belonging to a conducting spine are bound via a virtual electronic excitation of polarizable
side groups. Electron $a$ first polarizes the side groups in its vicinity creating an electric field attracting in turn a second electron $b$, Source: Figs. 1.2 [Ref\cite{Jerome08}, p. 5].} 
\end{figure}

 Little's proposal was still made within  the pairing framework of the BCS model, but instead of the phonon-induced pairing, the attraction between electrons would come from an entirely electric process   in long conjugated polymer such as a
polyene molecule grafted by polarizable side groups\cite{Little64}, see Fig.~\ref{Littlemolev1}.

 Emboldened by the prospects of high \tc, Little published one year later in a mainstream magazine an article   under the title "Superconductivity at Room Temperature"\cite{Little65}.
\begin{figure}[h]
\includegraphics[width=0.5\hsize]{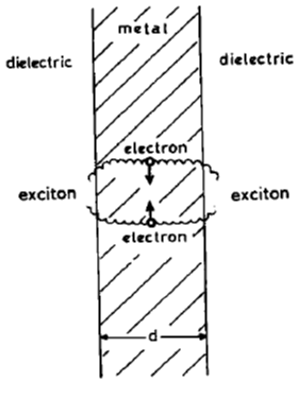}
\caption{\label{Ginzburglayer} According to Ginzburg's proposal, the attraction between the electrons in the metal layer is enhanced by the exchange of excitons
propagating mainly in the dielectric regions of the sandwich, Source: Fig. 3, [Ref\cite{Ginzburg68}, p. 370].} 
\end{figure}

At about the same time, theoreticians in Soviet Union were also very active in proposing models to go beyond the well accepted phonon mechanism for superconductivity, keeping in mind the possibility of higher \tc's. This is the case for V. L. Ginzburg and his coworkers who suggested at several occasions\cite{Ginzburg68,Ginzburg70a} that the exchange of excitons between two electrons may give rise to a net attraction and in turn to superconductivity possibly at temperatures higher than the usual phonon exchange of the BCS theory since  an electronic excitation energy about 100 or 1000 times larger than a typical  phonon energy would enter the BCS-like formula for \tc. However, they were severe technical constraints related to Ginzburg proposal: the electrons should be contained in very thin metallic  layers sandwiched between dielectrics "V. L Ginzburg was mentioning metallic layers of thickness .5 - 1 nm. Given the tremendous progresses made recently in thin film deposition, this idea may not be so unrealistic nowadays.", see Fig.~\ref{Ginzburglayer}. Given the significant progresses made recently in metallic layers preparation, the suggestion of Ginzburg deserves to be revisited.

In relation to biological substances, Ladik  and coworkers extending Little's original idea  for the possibility of superconductivity in organic polymers  to biological substances,  have suggested the possibility of superconductivity-type enhanced conductivity in some regions  of DNA polymers\cite{Ladik69}. 

All these suggestions were highly speculative  but the positive effect is that in 1970 they have awakened  the motivation for the research of new superconducting materials.

A few years after his original article,  Little succeeded  organizing in Honolulu a meeting which brought together researchers from very different fields with the same objective, namely the prospects of high temperature superconductivity,  and their cooperation proved to be decisive for the future\cite{1969LIttleconf}. Little supporting a continuous  dialogue between physicists and chemists concluded his 1969 presentation by \cite{Little70}:
 {\it \bf I would like to stress that in this field there is a desperate
need for an interdisciplinary approach. I think physicists on the whole
understand a great deal but they know very little. On the other hand the
chemist with his vast experience of reactions, compounds, conditions, etc.,
knows a great deal but . . . , well, perhaps I will not push the contrast further
but will conclude by saying that we need each other.}

What Little wrote in his conclusions at the 1969 conference has fortunately materialized from 1970 and led to new compounds with remarkable physical properties. Therefore in 1970 superconductivity became a challenge for  organic chemistry.

 But, Little's model  raised subsequently multiple questions from theorists. In particular, Bychkov {\it et al.}\cite{Byschkov66} criticized this model  regarding
its potentiality to lead to high-temperature superconductivity. As remarked by
Bychkov {\it et al.}\cite{Byschkov66}, the 1D character of the model system proposed by Little
makes it a unique problem in which there exists a built-in coupling between
superconducting and dielectric instabilities. It follows that each of these
instabilities cannot be considered separately in the mechanism proposed
by Little in one dimension. This comment turned out to be crucial as future work in the field of one dimensional conductors and superconductors has shown.  Last but not least, as shown by Landau\cite{Landau59}  and Mermin\cite{Mermin66}, fluctuations should be very efficient in a 1D conductor shifting any
long-range ordering toward very low temperature. 

Admittedly, the formidable task in synthetic chemistry did
not reached immediately  the goal fixed by Little as in the sixties the chemistry of organic conductors was barely developed, but the idea to link organic metallicity
and one-dimensionality was launched and turned out
to be a very strong stimulant for the development of organic
superconductors.

Actually,  it is with the synthesis of \qnq and \nmq  and the finding of a room temperature electrical conductivity of about $10^2$ $(\Omega.{\rm cm})^{-1}$ in these charge transfer salts that the field of organic conductors started blooming and opened to the electronic aspects of solid-state physics. From that time  on (about 1962)  physicists became closely involved in this kind of research.  

The possibility of designing and synthesizing molecules to achieve properties on the molecular scale that will lead to interesting and fundamental solid-state physics suggest an exciting future for this area of research. 

One of the main pioneers of this field has been Igor Shchegolev and his group at  Chernogolovka who published in 1972 a major review article, a  state of the art article on the electrical and magnetic properties  of  anion radical complexes based on \q\cite{Shchegolev72}.

\subsection{1970, Molecular electronics, the \ttf and  \tq decade}

A major breakthrough in chemistry arose in the 1970's when F. Wudl and his colleagues reported the synthesis of an unusually stable bis-dithiole radical cation\cite{Wudl70}. 
TTF is a planar conjugated and ``electron-rich" molecule, Fig.~\ref{molecules}, which can
easily give a charge to another species ready to accept it. 

The group of F. Wudl also established  that  remarkable electronic properties could  be attained when TTF is oxidized by  the monoanion $\mathrm{Cl}^{-}$  giving rise to a 1:1 radical cation complex with a room temperature conductivity on pressed pellets  of 0.3 $(\Omega.{\rm cm})^{-1}$  exhibiting semiconducting properties  at low temperature with an activation gap of 0.19 eV\cite{Wudl72}. 

Although the  electron donor molecule \ttf has become very popular for its ability to form intermolecular charge transfer complexes with the acceptor TCNQ leading to the celebrated charge transfer  (CT) complex \tq, we may also pay attention to an other aspect of research in which \ttf has been extensively used in the context of  organic electronic devices (semiconductors, electrochemical switches, sensors, p/n junctions, etc,...).

In this respect, TTF can be utilized as a donor in  intramolecular charge transfer materials 
in which TTF is covalently linked to an electron acceptor moiety by a variety of linking units,
sometimes giving rise to an intramolecular charge-transfer (ICT) interaction, which is most frequently manifested in the optical and electrochemical properties\cite{Bryce99}. Aviram and Ratner\cite{Aviram74} made a proposal to use the molecular  assembly donor-spacer-acceptor based on the donor \ttf to achieve molecular rectification. 

 The idea of Aviram  and Ratner has not lived up to its promise, but it has been a source of inspiration for the synthesis of a large number  intermolecular charge transfer compounds to be used for organic electronics\cite{Wudl04}.

Coming back to  intermolecular charge transfer compounds, the realization  of  non-integral charge transfer which will be extensively discussed in the following section,  has been achieved  with a lattice of TTF stacks surrounded by an halide lattice, namely the non-stoichiometric  salts such as TTF-halides$_{x}$ with $0.71 \leqslant x \leqslant 0.79$ in which the donor is partially oxidized  in a mixed valence state\cite{Laplaca75}. These materials possess a  room temperature conductivity on single crystals varying from 100 to 500 $(\Omega.{\rm cm})^{-1}$ with a small region near room temperature where the conductivity increases as temperature is lowered\cite{Chaikin80}. These TTF-(halides)$_{x}$ radical cation salts are interesting  because the non conducting halide stack produces a Coulombic potential on the conducting  \ttf stacks  at a wave length which is half the wave length defined by the 60\%  electron  filling  of the band deriving form the  HOMO level of TTF . The halide lattice although incommensurate with the \ttf latice   is thus  imposing on the one dimensional electron gas a perturbation opening gaps at values of the Fermi wave vector modulo  one reciprocal vector of the \ttf lattice, leading to a commensurate Peierls distortion \cite{Chaikin80}. It is a situation   \emph{at variance}  with the one that is observed when the Fermi wave vector is solely determined by the charge beeing  transfered from donor to acceptor  molecules, both contributing to the conduction as it is encountered in \tq\, to be introduced in the following.

Since \tq is the compound that aroused worldwide enthusiasm at the time of its discovery and which after numerous  experimental and theoretical studies led in 1980 to the synthesis of the first organic superconductors, it is useful to recall how this CT complex is formed and how high quality single crystals are obtained.

The unusual stability of the \ttf  radical cation resulting from an  oxidation  of the parent molecule has led  to to  conducting compounds with a conductivity ranging between  500 and 1000$(\Omega$.{\rm cm})$^{-1}$ at room temperature\cite{Coleman73,Ferraris73,Bright73}  but also   behaving like metals in 1973 (so-called organic metals because their resistance exhibits a positive temperature dependence down to about 60K). Coleman\emph{ et-al} published data for the conductivity of \tq culminating at 58K with provocative values larger than $10^6$ $(\Omega.{\rm cm})^{-1}$, (i.e about twice the room temperature conductivity of copper\cite{Coleman73}, which  will be commented  in more details  in the next Section about the physical properties of \tq.
 \begin{figure}[h]
\includegraphics[width=0.8\hsize]{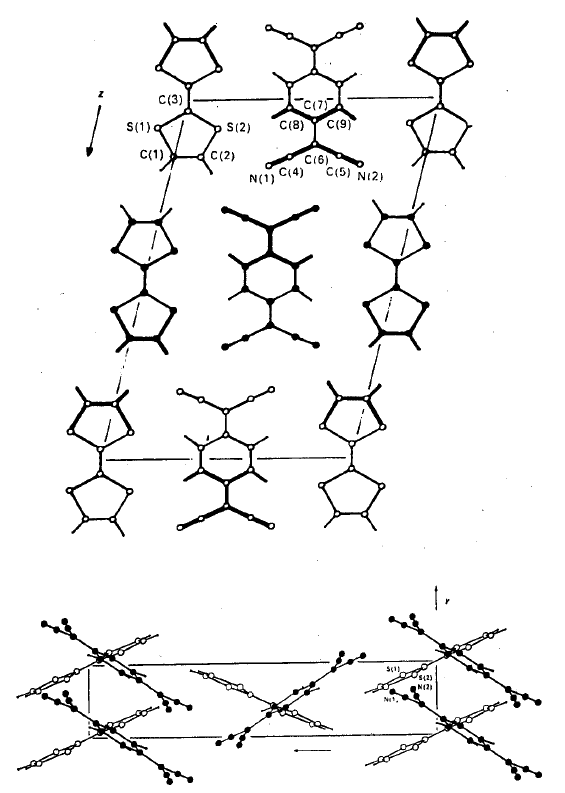}
\caption{\label{TTFTCNQvuselona}Crystal structure  of \tq viewed normal to the $ac$ plane of the crystal packing (top) and down  the [100] axis (bottom) according to Ref\cite{Kistenmacher74}. Notice the criss-cross molecular packing along the $a$-axis, Sources: Figs. 3, 4 [Ref\cite{Bechgaard80}, p. 766].}
\end{figure}

As the existence of high quality single crystals is a determining element for further physical characterization, it is important to have flavour on how they are obtained.
Neutral $\mathrm{TTF}$ and $\mathrm{TCNQ}$  
molecules combine to form a 1:1 charge transfer complex according to
a charge transfer reaction: $\mathrm{TTF+TCNQ}$ $\rightarrow$ $\mathrm{TTF}^{+\rho} \mathrm{TCNQ}^{-\rho}$, $\rho$ being less than unity,
taking advantage of the energy gain in the charge transfer reaction which reads,
$$\Delta E_{CT} = I-A -C <0$$
where $I$ and $A$ are the ionization potential of the donor and the acceptor affinity
respectively, and C includes the Coulombic (Madelung), polarization
and exchange energy contributions. In a complex such as TTF-TCNQ the amount
of charge transfer  is unknown a priori by the stoeichiometry of the complex. However, the amount of charge  transferred to the anion  can be determined by Raman spectroscopy\cite{Vanduyne86} and  even derived much  more  accurately from X-ray diffuse scattering experiments, see Sec. \ref{The TQ phase diagram}.

In order to grow crystals, highly purified TTF and TCNQ molecules are combined in the ratio 1:1 in acetonitrile solvent  and the complex precipitates from the solution\cite{Ferraris73,Anzai76}.
 In forming solids, the molecules TTF and TCNQ crystallize in segregated stacks
of TTF and TCNQ molecules, Fig.~\ref{TTFTCNQvuselona}. In the individual stacks of TTF and TCNQ
molecules the $\pi$-orbitals in the partially filled outer molecular shells overlap, so that
the discrete levels of the individual molecules are spread out to form an energy band.
The partially filled bands  thus leads to metallic-like conduction, even if the
intramolecular electron repulsion (the Hubbard $U$) is  strong compared to
that bandwidth, namely $U\gg W$.   It would be a different situation if we had the quarter-filling for which we  expect an insulator for $U$ going to infinity.
Moreover, a rather special feature of TTF-TCNQ and  related compounds, as
compared to ordinary metals is that the $\pi$-orbitals of the molecules are strongly
directional and interact mainly along the stacks, leaving only a weak interaction in the perpendicular directions. Hence, these conductors exhibit very anisotropic conduction
properties, they behave as quasi-one-dimensional conductors, although they
are of course three-dimensional crystals. The electronic properties of TTF-TCNQ will described in further details in Sec.II.

More than  thousand articles on \ttf and derivatives have been published since their discovery. The present article is  intended to provide only the very minimum amount of chemistry needed by physicists to understand the fascinating and often unseen physical properties they have revealed. 
  
 Lots of review papers have been published on this topic. A review has been published very early  in 1976 by Andr\'e and Bieber\cite{Andre76} and in addition   we refer the interested reader to  the comprehensive volume 104, issue 11, 2004 of Chemical Reviews\cite{Chemrev04}. 



The TTF molecule has been the starting point for the synthesis of many derivatives. For example,
substituting
the parent molecule with weakly donating alkyl groups, for instance methylene
($\mathrm{CH_{2}}$) groups, has led to the molecule HMTTF\cite{Greene76} and methyl groups ($\mathrm{CH_{3}}$) has
led to tetramethyltetrathiofulvalene (TMTTF) \cite{Bright74,Calas75,Brun77}. 
It is interesting to note that the charge transfer  compound TMTTF-TCNQ behaves from the point of view of magnetism and optics  in a similar way to \tq\cite{Bright74,Scott74}.  

The  TMTTF derivative turned out to be of decisive importance for the discovery of  organic superconductivity since  the four sulphur atoms of TMTTF  substituted by selenium allowed the synthesis of TMTSF\cite{Bechgaard74} (tetramethyltetraselenafulvalene), the radical cation  utilized in a salt to grow the first organic  superconducting   radical cation salts.
Furthermore, other selenium substituted molecules gave rise to new charge transfer compounds typically, \TSF (tetraselenafulvalene)\cite{Engler74,Bechgaard74} and HMTSF\cite{Bechgaard76} (hexamethylene/tetraselenafulvalene)  with interesting physical properties which will be discussed in the next Section.

As far as the acceptor side is concerned, following the precursor work of Wudl\cite{Wudl76}, a group of french chemists did show that simple radical cation salts such as \tmttf2x  could be synthesized  with X= SCN\cite{Strzelecka77} with the possibility to extend the  synthesis to an isomorphous series of simple sats  with  other anions such as \4fb, \cl and \br,  
 with a conductivity approaching 100 $(\Omega.{\rm cm})^{-1}$ at room temperature but becoming semiconducting below 200K or so\cite{Brun77,Delhaes79}. 

The  following
sections will show how these apparently minor modifications of the parent molecule
 have deeply influenced and oriented the further research of organic superconductors.

 Before moving on  to the next decade, let us emphasize that superconductivity has been a strong driving force for the development of this  field. To underline this point  E. Edelsack et-al \cite{Edelsack87} outlined during a conference in 1987 on Novel Superconductivity, one year  after the discovery of high temperature superconductivity in cuprates\cite{Bednorz86} that  {\it\bf organics are materials which forged revolutionary paths in the quest of high \tc}!.






\subsection{1980, the Bechgaard salts decade}
\label{Bechgaard salts decade}
Towards the end of the 1970's with the study of \tq the main motivation of physicists was to suppress the transition to an insulating state occurring around nitrogen temperature in order to allow eventually a superconducting state to settle. There has been a close cooperation between organic chemists specializing in organic synthesis and crystal growth, physicists both experimental and theoretical, which has been materialized by the holding of numerous multidisciplinary conferences which have led to major advances.


Substitution of the four sulfur atoms with selenium leading to  increased contacts between intermolecular chalcogen atoms, gave rise in  charge transfer compounds to a better room temperature conductivity and a lower metal-insulator transition. \textcolor{black}{Therefore, the  Copenhagen chemistry group led by Klaus Bechgaard\cite{KB}} 
 has been oriented toward selenium chemistry whose properties of some compounds will be discussed in the Sec. \ref{TSF-TCNQ}. As it was established  that  the carriers from the donor stack are dominant according to the hole-like thermopower\cite{Jacobsen78}, the chemists decided to investigate single chain compounds of TMTSF  by synthesizing radical cation salts of TMTSF with various mono-anions, somewhat analogous to the series of \tset2cl  in which a conducting phase  with $\sigma \sim 10^5$ $(\Omega.{\rm cm})^{-1}$ at helium temperature  could be attained  under  4.5 kbar\cite{Laukhin78} .

It is worth mentioning a major step which allowed the field of organic superconductivity  and all the physics surrounding it  to develop. This is the success  reached with the  growth of high quality single crystals of radical  cation  salts  using an electro-oxidation-crystallization method. Such a step allowed in turn  the development of materials enabling very detailed physical measurements concerning most of the technical possibilities  including measurements under pressure.

\begin{figure}[t]
\includegraphics[width=1\hsize]{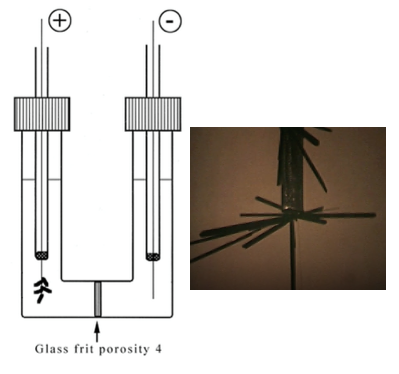}
\caption{\label{electrocrystv3} (Left) Electrochemical oxydation and crystal growth of \tmps.
A neutral TMTSF molecule dissolved in the nonaqueous tetrahydrofuran  solvent  is first electro-oxydized and precipitate at the anode (see right) with the \pf 
anion provided by the  tetrabutylammonium-hexafluorophosphate salt, ($\mathrm{NBu_{4}PF_{6}}$).(Right) A static view from the video\cite{videocrystals} showing the anode of the electrochemical cell  at Orsay from  A. Moradpour's laboratory.}
\end{figure}

\begin{figure}
\includegraphics[width=0.85\hsize]{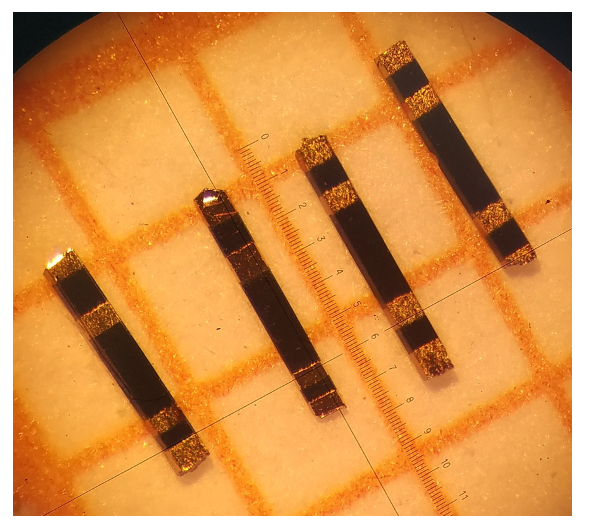}
\caption{\label{ClO4_or_KB05_x40} Crystals of  \tmc about 6 mm long viewed along their $c$-axis with evaporated gold pads for  longitudinal conductivity measurements. We thank P. Auban-Senzier for communicating this photo. }
\end{figure}

The electrochemical oxidation  growth that enabled the fabrication of high-quality single crystals is displayed  on Fig. \ref{electrocrystv3}. It has been extensively used by the Copenhagen chemistry school\cite{Bechgaard80} and reviewed by K. Bechgaard\cite{Bechgaard82} and by  P. Batail and co-workers\cite{Batail98}.
Electrocrystallization allows high-purity materials to be reproducibly obtained only if all materials and chemicals involved are properly purified beforehand. Hence, the  \pf anion for example enters in the electrocrystallization process  as a tetrabutylammonium salt which is soluble in a solvent like tetrahydrofuran, THF. 
One electron is removed from every neutral  TMTSF in the anode compartment being oxidized by  the current passing through the cell   leading in turn to the cation radical \tmr.  Then, two cation radicals combining with one  anion of the electrolyte give rise to the simple salt \tmp6 which precipitates at the anode as a non soluble molecular salt, see Fig.~\ref{electrocrystv3}. An accelerated video ($\times 20000$) shows how electrocrystallization proceeds\cite{videocrystals}, leading to centimeters long  high quality single crystals, see Fig.~\ref{ClO4_or_KB05_x40}.

\section{First metal-like organic conductors}
\subsection{Era of Peierls instability and Fr\"ohlich conduction}

\subsubsection{Introduction to the physics of \tq}
\label{Introduction to the physics}
It is   useful to present in a little more detail the physics of the compound  \tq which was historically the first charge transfer complex to exhibit  a large conductivity evolving  over a wide range of temperatures below ambient in a way similar to the conductivity of inorganic metals. In addition this organic conductor undergoes a sharp metal to insulator transition at 59K.

\tq is the prototype of the charge-transfer compounds where the metallic state is achieved first by an electron transfer from the initially filled HOMO levels of TTF to the originally empty LUMO levels of TCNQ\cite{Garito74}. Partially filled bands thus derive   from the interaction  between $\pi$ orbitals of  HOMO and LUMO levels of open shell donors and acceptors respectively forming segregated
and parallel stacks along the $b$ direction. The overlap of  molecular orbitals being
largest along the stacking direction and much weaker
between them makes the electron dispersion one dimensional, Fig.~\ref{TTFTCNQvuselona}. To a first approximation, the energy
depends only on the electron wave vector along the
$b^*$ direction in the reciprocal space. 

\begin{figure}
\includegraphics[width=1.5\hsize]{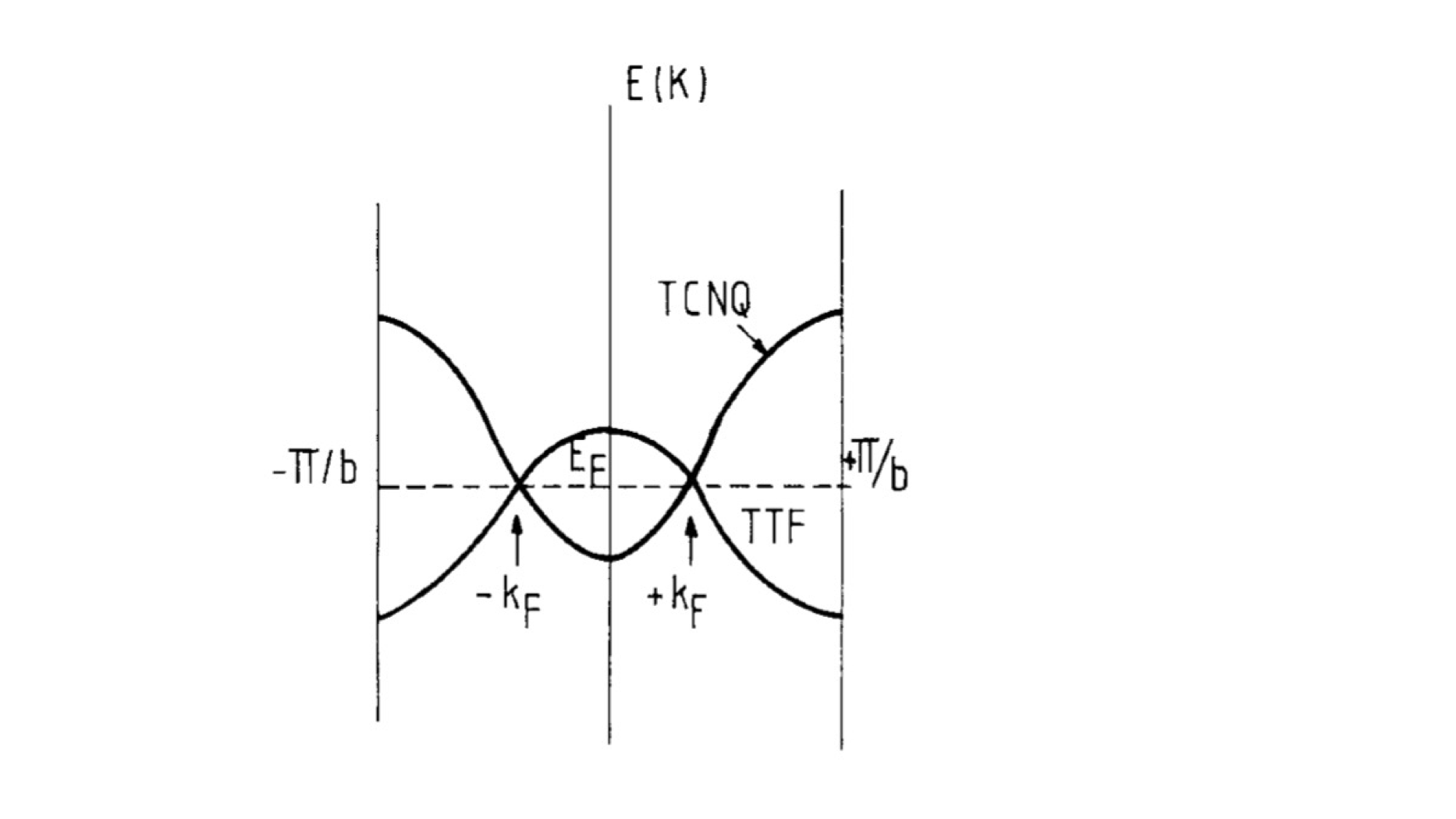}
\caption{\label{TTFQbandesinverses}A very schematic picture of the  HOMO-LUMO \tq inverted band structure,  Source: Fig. 19 left [Ref\cite{Jerome82}, p.349].}
\end{figure}

The extended H\"uckel method has been used by Berlinsky et-al\cite{Berlinsky74} to  compute the conduction band parameters of \tq  within the tight binding approximation.
The lowest electron energy on the TCNQ stacks
occurs when all molecular orbitals are in-phase (at $k= 0$ in reciprocal space). On the other hand, the band
structure of the TTF stacks is inverted with an energy maximum at the zone center, Fig.~\ref{TTFQbandesinverses}.

The  band structure calculations that have followed have confirmed that the simple tight binding approximation is relevant with bandwidths of about 0.5-0.7 and 0.4-0.5 eV for TCNQ and TTF bands respectively\cite{Berlinsky74,Shitzkovsky78}, see also \cite{Jerome82}.
\begin{figure}[h]
\includegraphics[width=0.7\hsize]{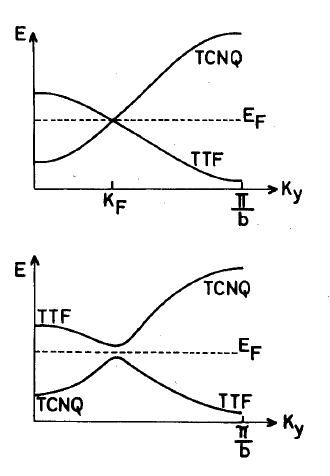}
\caption{\label{Hybridization} Energy level diagram in conducting charge-transfer compounds such as \tq; non-interacting 1-D chains (top), including coupling between 1-D chains of different nature (covalency gap), (bottom), Source: [Ref\cite{Jerome77b}, p. 353].}
\end{figure}
Such a band crossing picture ensures that both bands intersect at a single Fermi wave vector $\pm k_F$ in order
to preserve the overall neutrality\cite{Garito74}. Consequently, all states between $-\pi/b$ and $+\pi/b$ are occupied with the restriction that between -$k_F$ and +$k_F$ occupied states
belong to the TCNQ band while outside this domain
they pertain to the TTF band. The mere fact that
 charges can delocalize in TTF-TCNQ shows that
the on-site Hubbard repulsion $U$ does not overcome
the band energy $4t_{\parallel}$ gained in the band formation.
However, Coulomb repulsions still play some 
role as experiments will reveal   especially for  magnetic and structural
properties of the TTF stacks. 
\begin{figure}[h]
\includegraphics[width=0.8\hsize]{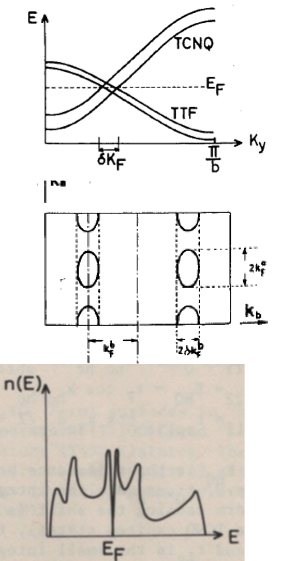}
\caption{\label{+warping} A very schematic picture of the semi-metallic Fermi surface in the low temperature
three-dimensional regime Sources:  [Ref\cite{Jerome77b}, p. 353] and Fig. 4b [Ref\cite{Soda76b}, p. 110]. The depression of the density of states at Fermi level (bottom) is the result of  the  coupling between unlike chains.  }
\end{figure}
A second crucial requirement to preserve the conducting state is the need for a uniform structure as one dimensional conducting structures are predicted to be unstable to lattice distortions due to the Peierls instability characterized by a divergent dielectric response to an external perturbation with the wave vector $2k_F$\cite{Peierls55}.

The simple band  picture in terms of non interacting neighbouring stacks is actually strongly oversimplified. Some small although non negligible transverse coupling exists between neighbouring stacks of different molecules. The bands derived from the TTF and TCNQ molecular orbitals must cross
the Fermi level where states $\psi_{Q}(k_F)$ and $\psi_{F}(k_F)$
are degenerate\cite{Jerome77b}, Fig.~\ref{TTFQbandesinverses}.

However
the existence of any finite interchain coupling between F and Q chains mixes the
states of same $k_F$ vector which become non-degenerate\cite{}. They form a bonding
state $(\psi_{Q}+ \psi_{F})/\sqrt 2$
and an antibonding state $(\psi_{Q}- \psi_{F})/\sqrt 2$
which are $2t_{\bot}$ apart in energy, the so-called covalency effect, Fig.~\ref{Hybridization}. However, in case of a coupling between like chains, the interchain coupling (Q-F) leads  to a replacement of the Q1D warped Fermi surface by small pockets  semi-metallic like\cite{Jerome82,Jerome77b} as displayed on Fig.~\ref{+warping}. The density of states at Fermi level is strongly depressed by the addition of a coupling between unlike chains (bottom). More realistic shapes of semi-metallic  Fermi surfaces at low temperature have been derived by Shitzkovsky and Weger\cite{Shitzkovsky78}.
As the Fermi energy of these electron or hole pockets is small,
of the order of $t_{\bot}$ or so, there is no hope of observing a three-dimensional Fermi
surface in a quasi-one-dimensional conductor unless the temperature is low enough
$k_BT< t_{\bot}$. In \tq, the onset of a Peierls transition around 55K precludes the observation of a 1D to 3D cross-over of the Fermi surface since $t_\bot \approx 5$meV. 

As we shall see later in Sec.~\ref{HMTSF-TCNQ-like compounds}, among charge transfer compounds undergoing a Peierls transition, the condition $k_BT_p<t_{\bot}$ required for the observation of a 1D to 3D cross-over, seems to be fulfilled only for HMTSF-TCNQ (and possibly also in TMTSF-TCNQ under pressure) when covalency effects due to strong Se--N bonds are especially large compared to the interchain coupling of \tq.

The band formation of a compound such as \tq  is thus \emph{at variance } with what has been encountered previously for radical cation salts which are single-chain conductors where the  band filling of a single band can be predicted by the stoichiometry. 

\begin{figure}
\includegraphics[width=0.6\hsize]{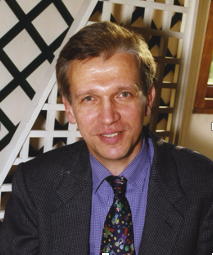}
\caption{\label{Schulz} Heinz J. Schulz at Orsay in 1998 when he was awarded the CNRS silver medal. We thank Anuradha Jagannathan for providing this photo.}
\end{figure}

\tq has  been remarkable as  a model compound for the development of  both theoretical one dimensional physics and  several aspects of experiments, that we intend to discuss briefly in the following\cite{Jerome82}. We think it is the right place in this overview to recall the central role played by Heinz. J. Schulz (1954-1998), see Fig.~\ref{Schulz}, member of the theory group in the Orsay Solid State Physics Laboratory.  Heinz was a specialist of Fermions at low dimension. He brought a very valuable collaboration to many experimentalists, in particular to one of the authors of the present article.
\subsubsection{The conductivity of \tq}
\label{The conductivity}
The year 1973 saw the announcement by two american groups, a chemistry group  led by D. O Cowan at Baltimore and the other a physics one led by A. J. Heeger\cite{Heeger} at Philadelphia  of both the synthesis and the measurement of remarkable conducting properties in \tq\cite{Ferraris73,Coleman73}. Both publications announced a conductivity  at room temperature of about 400 $(\Omega.{\rm cm})^{-1}$ along the stacking axis significantly higher than found for any other TCNQ salt such as quinolinium and   a metal-like behavior for the conductivity of \tq down to a maximum around 58 K ($\sigma_{max} \approx 10^4$ $(\Omega.{\rm cm})^{-1}$) followed by a sharp transition toward an insulating ground state. 

This report   was the subject of a post deadline invited paper at the March 1973 meeting of the American Physical Society in San Diego. According to G. Lubkin in Physics Today, the Penn report has excited great interest, but the theoretical explanation met with considerable skepticism\cite{Lubkin73}. 

Furthermore, the authors of ref.\cite{Coleman73} reported in the same article that three out of seventy measured samples     revealed a conductivity  exceeding  $\sigma>10^6$$(\Omega.{\rm cm})^{-1}$ at 60K with a slope appearing to be divergent. These same authors have suggested that this extraordinary conduction could come from superconducting fluctuations just above a Peierls-type metal-insulator transition at 58K following a one-dimensional power law divergence with the  3/2 exponent\cite{Patton71}. Moreover, this publication hypothesized that superconducting fluctuations could be at the origin of this remarkable increase in conductivity above a possibly Peierls transition occurring at 58 K by basing this argument on a phonon soft mode $\omega (2k_F)$ tending towards zero and responsible for a strong increase in the BCS-type phonons mediated  electron-electron coupling. 
\begin{figure}[h]
\includegraphics[width=0.7\hsize]{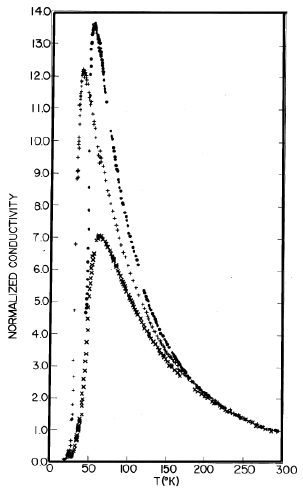}
\caption{\label{sigmaTTF}Normalized conductivity $\sigma (T)/\sigma (295K)$ of ($\bullet$)
\tq, (+) \tsq, and ($\times$) \dsdtfq, Source: Fig. 2 [Ref\cite{Etemad75}, p.742].}
\end{figure}
The conclusion of the latter paper indicated unambiguously that regarding the superconductivity mechanism their authors had in mind a standard BCS phonon type mechanism probably inspired by the physics of A15 superconductors in which van-Hove singularities were responsible for the high values of their \tc\cite{Barisic72a}. They proposed an increase of the interchain coupling in order to limit the one-dimensional Peierls divergence and thus to favor superconductivity at a temperature above the Peierls transition temperature.
The enormous conductivity peak claimed in ref.\cite{Coleman73} triggered a worldwide interest and  led to numerous verifications in various laboratories. 

Even if the claim for the observation of superconducting fluctuations at high temperature has been considered as very controversial at the time. It is clear 50 years later that it was a decisive event that contributed to strengthen a common activity between chemical synthesis, experimental and theoretical physics.

The most elaborate verification came from the Bell  Telephone and Baltimore groups\cite{Schafer74}. Without going into the details of this latter report, it turned out that the extraordinary conductivity maximum is an experimental artefact likely due to the points at which electrical contacts are made on the samples. This article even suggests recipes to guarantee meaningfull conductivity measurement  namely, the so-called unnested voltage check which should give a voltage as close as possible to zero when the current is injected through adjacent voltage and  current leads and the voltage detected on the two others leads or equivalently, a large unnested ratio. 
The controversy about a reliable value of the peak conductivity in \tq led G. A. Thomas et-al  to gather and compare experimental data from 18 different laboratories in a single article comprising 30 co-authors\cite{Thomas76},Fig.~\ref{sigmaTTF}. 

It is interesting to note that until 1974 the results of transport in molecular conductors were generally published in the form of conductivity versus temperature, when it was not the log of conduction versus inverse temperature. This habit probably originated from the history of molecular conductors, which until that time were semiconductors. It was only later, when these conductors showed a metallic conductivity, that the results were presented the way metal physicists do, in resistivity as a function of temperature.

This controversy over the interpretation of \tq conductivity measurements has shown that conductivity measurements on materials with very  anisotropic  transport properties require single crystals not only of excellent chemical purity but also of very high crystalline quality. As isotopically substituted materials often lead to crystals of better quality, the investigation of the metal-insulator transition around 53K has shown that the $^{15}$N substitution  of TCNQ molecules lead to  single crystals with a remarkably sharp transition and a peak conductivity 15 times the room temperature value\cite{Cooper78}. Crystals of the same origin used for pressure conductivity studies to be presented in the next section have confirmed values of $\sigma$(peak)/$\sigma$(300K) of between 10 and 25 with unnested ratios exceeding 20 at  $\sigma$(peak)\cite{Friend78b}. These results are also in agreement with earlier measurements\cite{Etemad75} and with an estimate based on the statistical average over the data of 18 laboratories\cite{Thomas76}. Finally, we can be confident that the longitudinal DC conductivity of \tq amounts to  400$\pm$100 $(\Omega.{\rm cm})^{-1}$ at 300K and rises up to a maximum of $\approx$ 10$^4$$(\Omega.{\rm cm})^{-1}$  at 60K, Fig.~\ref{sigmaTTF}.  Although reference\cite{Thomas76} suggested that the observed magnitude of the conductivity can be described by single particle scattering with reasonable values of mean free paths, they had left the door open to the possibility that the conduction may be enhanced by collective effects. It is this latter scenario that high pressure studies turned out to  confirm a few years later.

\subsubsection{The dilemma of transport in \tq}
\label{The dilemma of transport}
The understanding of the conduction of \tq in its metallic state has generated multiple controversies over several years. Even if a consensus about the enhancement of the longitudinal conductivity peaking at 60K has been quickly reached with a  ratio $\sigma$(peak)/$\sigma$(300K) being at most 25, there were still remaining problems regarding the behavior of the transport anisotropy both in temperature and pressure.

It is accepted that the conduction process in \tq  is coherent  and diffusive along  and perpendicularelectrocrystv3  to the chains respectively\cite{Soda77,Cooper77a}. Therefore, according to the Einstein's relation applied along the transverse $a$  direction the transverse conductivity   reads\cite{Soda77}: 
\begin{eqnarray}
\sigma_{\bot}= n _{0} D_{\bot}e^{2}/k_{B}T 
\end{eqnarray}
where $D_{\bot}$=$a^2\tau_{\bot}^{-1}$ is the diffusion constant along the $a$ direction, $\tau_{\bot}^{-1}$ being the hopping rate between neighboring chains and $n _{0}$ the number of carriers within the Fermi volume ($n _{0}$=$n(E_{F})k_{B}T$) with a degenerate   Fermi statistics, $k_{B}T \ll E_{F}$) leading to:  
\begin{eqnarray}
\sigma_{\bot}=n(E_{F})a^{2}e^{2}\tau_{\bot}^{-1}
\end{eqnarray}
On the other hand the longitudinal transport follows a Drude formulation namely, 
\begin{eqnarray}
\sigma_{\parallel} = n _{0}e^{2} \tau_{\parallel}/m^{*}
\end{eqnarray}
where $m^{*}$ is the effective mass of carriers along the conducting direction.
The conductivity anisotropy thus reads: 
\begin{eqnarray}
\label{anisotropy}
\sigma_{\parallel}/\sigma_{\bot}=(t_{\bot} / t_{\parallel})^2(a/b)^2 
\end{eqnarray}
provided  the transverse hopping rate $\tau_{\bot}^{-1}$ is related to $\tau_{\parallel}$, namely\cite{Soda77},
\begin{eqnarray}
\label{tau}
 \tau_{\bot}^{-1}= ({2\pi/
\hbar})t_{\bot}^{2}(\tau_{\parallel}/h). 
\end{eqnarray}
There is access to this three dimensional escape rate via NMR experiments according to the field dependence of the nuclear relaxation rate\cite{Soda77}. 
 \begin{figure}[h]
\includegraphics[width=0.55\hsize]{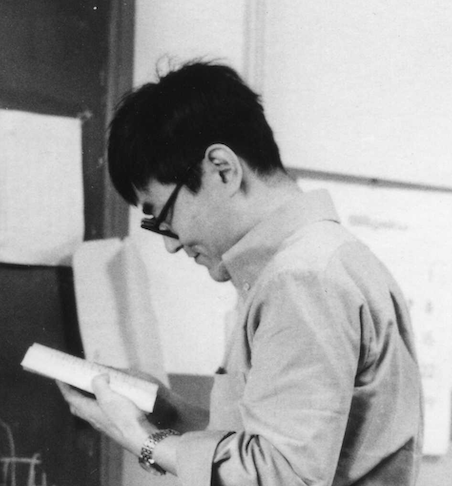}
\caption{\label{Soda} Gen Soda  was one of the lead experimentalists for the early NMR studies of one-dimensional charge transfer conductors during his stay at Orsay 1974-1976. Sadly, he died accidentally at Okazaki in 1987. }
\end{figure}
There exists a cross-over value of the electronic  resonance frequency $\omega_e$ at $\omega_e  \tau_{\bot}\approx1$  between a high field 1D regime ($1/T_1\propto H^{-1/2}$) and a low-field 3D regime ($1/T_1=\mathrm{const}$). Consequently, the field dependence of the relaxation rate allows a remarkably direct derivation of $\tau_{\bot}$ provided this time is in the limits of the NMR experiments, i.e.  $10^{-12}\mathrm{s}\leqslant\tau_{\bot}\leqslant10^{-10}\mathrm{s}$. This is the case for \tq where $ \tau_{\bot}>\tau_{\parallel}$
 \ with $ \tau_{\bot}\approx 8\times 10^{-12}\mathrm{s}$ and $\tau_{\parallel}\approx 3\times 10^{-15}\mathrm{s}$ from optical data\cite{Grant73} or the NMR derivation\cite{Soda77}. 
  However, the condition $\tau_{\parallel}$$\gtrsim\hbar/t_{\parallel}$ is actually not fulfilled strictly speaking  in \tq  at room temperature since $\hbar/ \tau_{\parallel}$ is likely to be  in-between  0.075eV and 0.2eV, with $t_{\parallel}$ being of the order of 0.1 eV.
 
 
 In short, the conduction regime which is relevant for \tq in its conducting regime  is coherent along the conducting  axis although  incoherent along the  transverse directions, namely $\hbar /t_{\parallel}<\tau_{\parallel}<\hbar /t_{\bot}$. Around and certainly above room temperature the situation  $\tau_{\parallel}<\hbar /t_{\parallel}$ is  encountered, leading in turn to a diffusive longitudinal conduction.

Equ.\ref{anisotropy} giving the anisotropy  implies that the  motion of the carriers is coherent along the chains but diffusive along transverse directions\cite{Soda77,Ong77}.   
According to the experimental data\cite{Cooper77a}  such a picture is still  valid under pressure at room temperature  
but  {\it \bf a serious problem remains} since the measurements\cite{Schafer74,Cohen74} have shown that the $b/a$ conductivity anisotropy being in the range  from 60\cite{Schafer74} to 1300\cite{Cohen74} at room temperature   is strongly temperature dependent, see for instance the conductivity data in reference\cite{Jerome79}. Actually, for the parallel component, $\sigma_{\parallel}$,  the conductivity peak ratio (CPR)   $\sigma_{\parallel}$(peak)/$\sigma_{\parallel}$(300K), lies in-between 10 and 25 depending on the measured samples in the vast majority of laboratories, whereas for the transverse conductivity $\sigma_{\bot}$, $\sigma_{\bot}$(peak)/$\sigma_{\bot}$(300K)   is no more than a factor 3 and usually sample independent. Therefore the $b/a$ conductivity anisotropy ends up being  multiplied by a factor $\sim 7$ on cooling from room temperature down to about 70K. Therefore, equ.\ref{anisotropy} cannot apply to the transport properties of \tq in the conducting temperature regime.


Understanding  the DC metallic conductivity has thus  remained
the great puzzle  in the study of TTF-TCNQ and of its derivatives. It has even generated numerous controversies in the years 1974-75. 

Bardeen\cite{Bardeen73}  and co-workers\cite{Allender74}
suggested that long living fluctuating charge density waves (CDW) of the Fr\"ohlich type\cite{Froehlich54,Patton74} could reasonably contribute to the DC longitudinal conductivity of TTF-TCNQ. But the model in question was left without  any clearcut experimental verification until  the year 1979. This is the issue which will be addressed in the next section.

 \subsubsection{The \tq phase diagram}
 \label{The TQ phase diagram}
The dilemma about the nature of the  conduction in \tq could not be solved until an extensive use of transport,  magnetism, X-rays and  neutrons diffraction techniques  very often performed under high pressure. This activity has been widely presented in numerous journal articles or contributions to book chapters. In this article we will limit ourselves to develop only a few highlights that have led to the understanding of the physics of \tq.

We thus begin with an overview of the rich  $T-P$ phase diagram before addressing the question about the conduction.
 
 \subsubsection{The \tq phase diagram in temperature}
  
\begin{figure}  
\includegraphics[width=0.8\hsize]{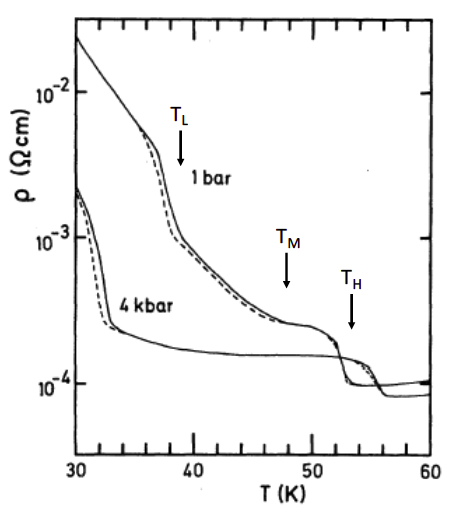} 
 \caption{\label{TQrho1baret4kbarversusT} \tq resistivity data at 1 bar and 4 kbar on cooling (dashed) and warming (solid) for a sample with CPR 25 at 1 bar, Source: Fig. 1  [Ref\cite{Friend78b} p.1048]. The temperatures of  phase transitions  are those related to the run at 1 bar.} 
 \end{figure}

 The phase diagram reveals the existence in temperature of several resistive transitions namely, two major transitions which can be seen in transport:
one at $T_H$= 54 K, where the conductivity drops by a factor of 2, and another at $T_L $=38 K\cite{Friend78b} which is the
signature of a sharp first-order transition toward an
insulating ground state, according to  the  data  1 bar data on Fig.~\ref{TQrho1baret4kbarversusT} . Furthermore, at a  temperatures $T_M$ between $T_H$
and $T_L$ the transport data reveal a sluggish and hysteretic regime related to the onset of a CDW on the TTF chain. 

The transport behavior is the consequence of a Peierls transition (a periodic lattice distortion, PLD) which has been extensively studied by X-ray diffuse scattering experiments\cite{Denoyer75,Kagoshima75} and elastic neutron scattering\cite{Comes75}. 

To summarize, between $T_H$ and $T_M$ there is $2a, 3.4b$ superstructure, and
between $T_M$ and $T_L$ the period in the $a$
direction evolves continuously and jumps discontinuously
to $4a$ at $T_L$. These are the experiments which have provided
the clear-cut signature for a Peierls instability in a
1D conductor. They have also given the first accurate
determination of the charge transferred between TTF
and TCNQ molecules. With a periodicity along the
stack axis of $\lambda=3.4b= 2b/\rho$, the charge transfer amounts to 0.59 at low temperature. Notice that the charge transfer measured at higher temperatures from diffuse X-ray experiments leads to a smaller value for the charge transfer namely, $\rho=0.54$ at K\cite{Denoyer75}. Such a difference between low and high temperature values can be understood by the thermal contraction at constant pressure  acting on the bandwidths. A temperature decrease
from 300 to  77K is essentially
equivalent as far as the structure is concerned to the effect of a 5 kbar hydrostatic pressure\cite{Jerome82}.

Furthermore, transport has shown that the Peierls transition of \tq occuring at $T_H$  is only partial and suggests that the insulator-metal transition involves only one of the chains at 54K leaving the other conducting.

This question has been settled by 
$^{13}$C Knight
shift experiments in selectively enriched samples of \tq\cite{Takahashi84a} improving previous NMR\cite{Rybaczewski76} and EPR\cite{Tomkiewicz77} studies which have shown that the loss of the spin degrees of
freedom of the TCNQ stacks is already complete
below $T_H$ with no marked spin gap on the TTF stack
between $T_H$ and $T_L$ (although a noticeable dependence
of the conductivity is observed in the same
temperature domain according to Fig.~\ref{TQrho1baret4kbarversusT}. It is only below $T_L$ that  the spin gap in the TTF stack becomes fully open. However,   there is still no common agreement in the conducting regime above $T_H$ between the three previously mentioned studies  regarding the respective contributions of the two chains to the total susceptibility.  Following the experimental results a theory for these structural phase transformations has been proposed by Bak and Emery in 1976 in terms of a Ginzburg-Landau type expansion of the free energy\cite{Bak76}.

An other way to visualize the Peierls distortion  has been achieved  via  scanning tunneling microscopy. These studies  have confirmed in a very elegant way the structural results obtained by diffraction studies\cite{Nishiguchi98} and even solved some of the pending questions\cite{Wang03}.
\begin{figure}
\includegraphics[width=0.9\hsize]{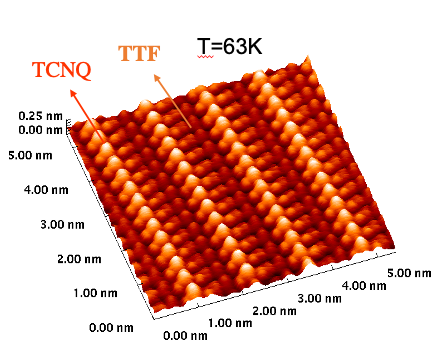}
\caption{\label{STM63K} STM image of the $a-b$ plane of \tq
taken at 63 K. The image area is $5.3\times5.3$ nm$^{2}$, Source: Fig. 1 [Ref\cite{Wang03}, p. 1].}
\end{figure}
Fig.~\ref{STM63K} displays an ultrahigh vacuum image of the $a-b$ plane of a cleaved \tq single crystal taken at 63K\cite{Wang03}.  The 1D structure of parallel chains
is clearly visible on this figure. Those containing a triplet of balls
are the TCNQ molecules, The TTF molecule appears
usually as a single ball feature in STM imaging, although
reports of doublet structures have also been made in
the literature\cite{Kato96}. The distances between
chains or between units within each chain in Fig.~\ref{STM63K}
 compare fairly well with the $a= 1.23$nm and $b=
0.38$nm lattice constants of \tq\cite{Kistenmacher74}. A salient result of this STM work is shown on Fig.~\ref{STM36KTTFQ} where the CDW is seen in direct space superimposed to the underlying TCNQ lattice at 36.5K, i.e, below the transverse lock-in occurring at $T_L$= 38K. 
\begin{figure}
\includegraphics[width=1\hsize]{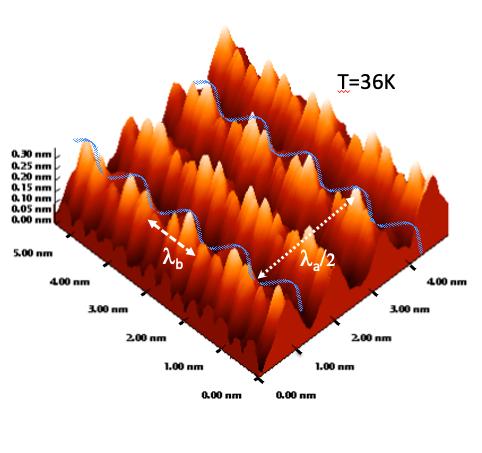}
\caption{\label{STM36KTTFQ} The  STM image of the same $a-b$ plane of \tq as Fig.~\ref{STM63K}  but taken at 36K according to Ref.\cite{Wang03}, revealing the periodic lattice distortion in blue with wave length $\lambda_{b}=3.4 b$(1.15nm) along $b$ and the transverse ordering $\lambda_{a}=4a$ along $a$. The TTF chains are not visible by STM on this figure unlike higher temperatures,Fig.~\ref{STM63K}.  This figure is adapted from the work of  J.C Girard and Zhao. Z. Wang, Source: Fig. 5.4 [Ref\cite{Wang03}, p. 156]. }
\end{figure}
These STM experiments have settled open questions related to the phase below $T_M$ when, because of the onset of a CDW on TTF chains, the transverse wave vector departs from $2a$ and increases smoothly at lower temperature. Below  $T_M$  the CDW's of 
the 2D superlattice can be described  equally well either by plane waves
with the wave vectors $q_{+} = [q_{a}(T), 2k_{F}] $
 or $q_{-} = [- q_{a}(T), 2k_{F}] $
leading to energetically equivalent
configurations\cite{Abrahams77,Bjelis76} with CDWs of fixed amplitude and a phase
varying like $q_{a}a$ along the $a$ direction. Consequently, the diffraction
pattern of the CDW state should display domains
characterized by the vectors $q_{+}$ or $q_{-}$. This is the situation which has been validated by the STM analysis between $T_M$ and $T_L$. There also
exists another possibility, namely: the superposition
of the two plane waves $q_{+}$ and $q_{-}$, which leads to a
CDW with constant phase but a modulated amplitude along the $a$ direction\cite{Abrahams77,Bjelis76}: a double-$q$ configuration.
\begin{figure}
\includegraphics[width=1\hsize]{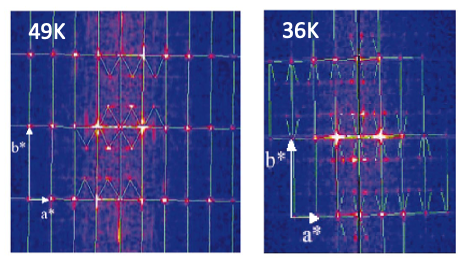}
\caption{\label{STMFourier}  Fourier transformed
pattern of \tq showing the
single-$q$ CDW in the sliding temperature
domain (left) at 49K and the
pattern  showing the
double-$q$ ($4a\times3.3b$) CDW in the
commensurate phase at 36K (right), Sources: Fig. 2, 3d [Ref\cite{Wang03}, p. 2].}
\end{figure}
The only solution which can take advantage of
the commensurability energy related to the transverse
commensurate periodicity $4a$ through the fourth
order Umklapp term in a Landau-Ginzburg expansion
is the double-$q$ configuration. Fig.~\ref{STMFourier} showing the 2D Fourier transformation of the real space data of Fig.~\ref{STM36KTTFQ} reveals even more clearly  the transverse period $4a$ and with the existence of four satellite spots around every Bragg spot, i.e, the coexistence of the two wave vectors in the same region of the real space that prevail below $T_L$. This configuration is quite different from  the Fourier transform, Fig.~\ref{STMFourier} where the wave vector  activated in the same area is the signature of a phase modulation existing in the incommensurate transverse regime  between $T_L$ and $T_M$ .

\subsubsection{Towards strong coupling}
\begin{figure}
\includegraphics[width=0.8\hsize]{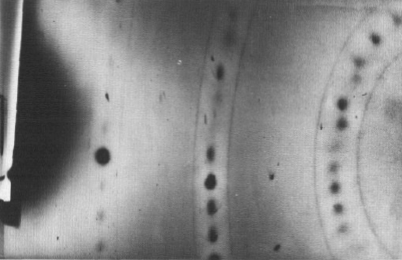}
\caption{\label{KCP} Typical diffuse X-ray scattering pattern from the 1D conductor $\mathrm{K_{2}Pt(CN)_{4}Br_{0.33}H_{2}O}$  showing the diffuse streaks on both sides of the Bragg spots, Source: Fig. 6 [Ref\cite{Comes73}, p. 573].}
\end{figure}

\begin{figure}
\includegraphics[width=0.5\hsize]{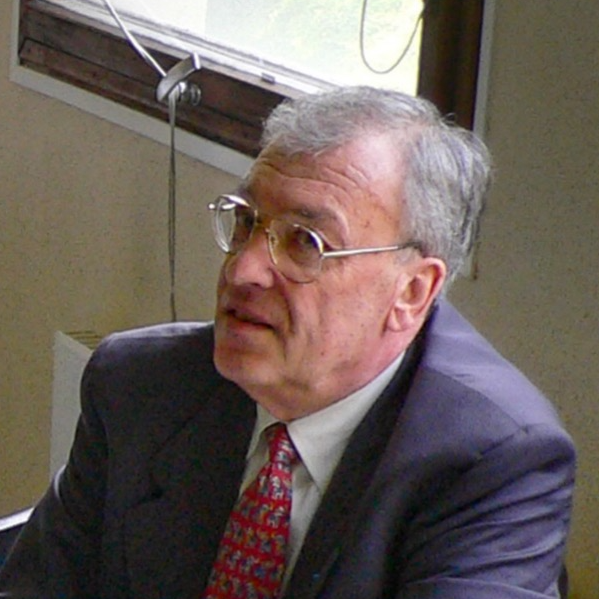}
\caption{\label{Comes} R. Comes in 2007 at the LPS Orsay,  Source: Patrick Batail.}
\end{figure}

As mentioned earlier, the metal-insulator transition occurring at 54K as a result of a new lattice periodicity $\pi/ k_F$ opening a gap at the Fermi level  in the electronic band at $\pm k_F$ is the manifestation of the Peierls instability in a one-dimensional chain.
 In the non-ordered phase above 54K, the Peierls instability is announced, by joint PLD/CDW fluctuations  being uncorrelated between neighboring chains (one-dimensionality) and whose finite-length correlation  is strongly dependent on the temperature. It is only in the very vicinity of the actual Peierls transition (a few degrees above) that 1D fluctuations on neighbouring chains become correlated with a short range lateral order which ends up in a 3D phase transition.
The physics of these lattice fluctuations is well documented by the early studies in one dimensional platinum chains $\mathrm{K_{2}Pt(CN)_{4}Br_{0.33}H_{2}O}$ where they give rise to both a giant Kohn anomaly of the acoustic phonon branch at $2k_F$ observed in inelastic neutron diffraction\cite{Shirane79} and to diffuse sheets centered at the $\pm2k_F$ wave vectors in reciprocal space on each side of the Bragg spots, see Fig.~\ref{KCP}\cite{Comes73}. R. Comes, see Fig.~\ref{Comes}, played a major role in the first experimental evidence of the Peierls transition predicted in 1955\cite{Peierls55} but only demonstrated  by diffuse X-ray scattering in the inorganic 1D conductor, $\mathrm{K_{2}Pt(CN)_{4}Br_{0.33}H_{2}O}$,  at Orsay in 1973. This result has marked the beginning of the study of one-dimensional conductors, which was rapidly extended to organic conductors.

\begin{figure}[h]
\includegraphics[width=0.85\hsize]{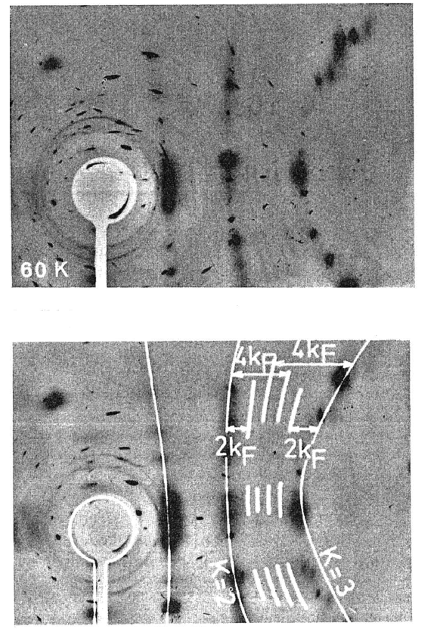}
\caption{\label{2KF4KFv1} X-ray pattern of \tq at 60K. Diffuse lines at $2k_F$ and at $4k_F$  are observed on each sides of the layers of main Bragg reflections. The sample is oriented with the $b^{\star}$ axis horizontal, Source: Fig. 1 [Ref\cite{Pouget76}, p. 437].}
\end{figure}

Similar experimental findings have been obtained in 1D organics conductors although with a much milder intensity. They are well documented by numerous publications of the Orsay X-ray group\cite{Pouget16}. However, somewhat unexpected has been the finding in diffuse X-ray patterns of a structural instability at twice the $2k_F$ wave vector in \tq, a 1D conductors which presents an incommensurate band filling\cite{Pouget76,Khanna77}. This $4k_F$ scattering is already visible  below 250K and  becomes predominant below 100K over the regular $2k_F$ Peierls scattering   occurring at 150K, Fig.~\ref{2KF4KFv1}. This scattering at wave vector $4k_F$  has been viewed as the  fingerprint on the lattice of a Wigner charge localization with a periodicity $\pi/2k_{F}$  developing into a lattice distortion at the same $4k_F$ vector because of the electron-phonon coupling\cite{Emery76}. This $4k_F$ scattering has been attributed to the TTF chain better described by a 1-D electron gas with repulsive interactions in which charge and spin fluctuations  are both gapless\cite{Voit88, Basista90} unlike the $2k_F$ related to the smaller  $e-e$ coupling TCNQ chains\cite{Pouget88}.
Without  going into much details, this phenomenon  can be understood as
the precursor of a Peierls transition in a spinless fermion gas,
where, because of  the loss of the spin degree of freedom a given $k$ state cannot be occupied by more than one fermion. This has been considered as the strong coupling limit of the Peierls transition of a Hubbard chain when $U\rightarrow\infty$\cite{Bernasconi75}. This $4k_F$ scattering will become far more dominant for compounds in which the charge transfer amounts to 0.5, i.e, the quarter-filled band situation, see Sec.~\ref{TMTSF-DMTCNQ}.


 \subsubsection{The \tq phase diagram under pressure "fluctuat nec mergitur"}
 
 Thanks to the  development of helium gas high pressure techniques up to 15 kbar\cite{Malfait69}  and teflon cells higher together with very low temperature, high pressure equipments\cite{Delplanque70}, reliable conductivity measurements on \tq single crystals  could be obtained. They have contributed to understanding both the phase diagram and the conduction process of \tq.

The transport properties of \tq have revealed a remarkably rich $T-P$ phase diagram, confirmed by magnetic, calorimetric and diffraction experiments. 
  
 It is actually the high pressure studies that have enabled us to understand transport properties of \tq. 

\begin{figure}
\includegraphics[width=0.8\hsize]{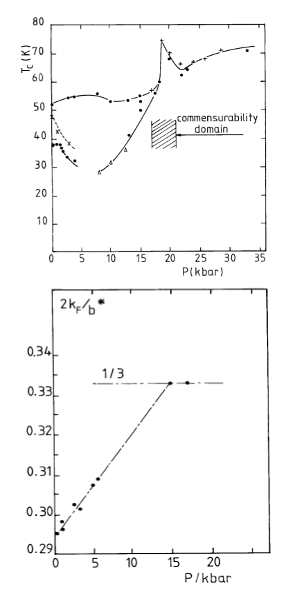}
\caption{\label{diagTQ} (Top) Pressure dependence of the phase transitions
in \tq as defined by maxima in $d lnR/dT$, the inset shows the pressure dependence of the activation energy of transport in the 3D Peierls state, Sources:  Fig. 2 [Ref\cite{Friend78b}, p. 1049]. 
(Bottom) Pressure dependence of the $2k_F$ measured by neutron elastic scattering experiments. Between $14.5$ and $17.5$ kbar, the superstructure at 35 K has the periodicity $a\times  3\times c$ , Source: Fig. 3 [Ref\cite{Megtert81}, p.876].}
\end{figure}

Somewhat surprising was  the existence  above 15 kbar of  a single transition peaking at 74K under 19 kbar, Fig.~\ref{diagTQ} \cite{Friend78b}. The width of the peaked region in the phase diagram amounts to a domain of $\sim$4 kbar in pressure units. In that pressure domain,
the phase transition also exhibits a first-order character as shown by the one Kelvin
hysteresis observed at the transition\cite{Andrieux79f,Andrieux79g}.
The peaking of the single-phase transition
temperature around 19 kbar and its first-order
character have suggested the occurrence of a (x3) commensurability ($2k_{F} = b^{*}/3$) at this pressure\cite{Friend78b} which turned out to have been  confirmed subsequently by elastic neutron scattering under pressure leading to a ($a$ x $3b$ x $c$) superstructure\cite{Megtert81}, see Fig.~\ref{diagTQ}. Such an increase in the charge transfer  under pressure should not be a great surprise since it is in line with the increase already noticed upon cooling under ambient pressure and due to the effect on the band structure of  the significant thermal contraction\cite{Jerome82}.

\begin{figure}
\includegraphics[width=0.7\hsize]{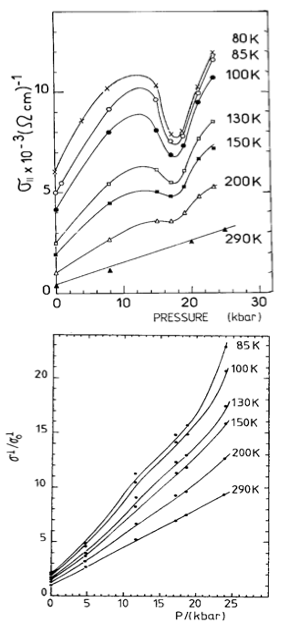}
\caption{\label{sigmaTQP} (Top) Pressure dependence of the longitudinal conductivity of \tq at various
temperatures. The ratio CPR is 25 under ambient pressure. (Bottom) Pressure dependence of the transverse conductivity of \tq at various
temperatures, Sources: Fig. 2a [Ref\cite{Andrieux79f}, p. 229] (top), Fig. 14 [Ref\cite{Jerome82a}, p. 191] (bottom).}
\end{figure}

When the longitudinal conductivity of \tq is studied under high pressure there occurs an important drop of this quantity in the pressure regime corresponding to the peak of the first order transition ($P$= 19$\pm$2 kbar)\cite{Andrieux79f,Andrieux79g}, Fig.~\ref{sigmaTQP}. The drop of
conduction in the vicinity of 19 kbar is becoming less pronounced as the temperature is increased and even practically unobservable above 250 K.


 These data imply that the longitudinal conduction contains more than one channel, actually two different channels acting in parallel though responding differently to pressure. Since the transverse component of the conductivity failed to show such a similar anomaly in the vicinity of 19 kbar, Fig.~\ref{sigmaTQP}, it was suggestive that one  channel for $\sigma_{\parallel}$ is a single particle scattering channel which follows the transverse conduction since from the previous section we know that as long as the electron scattering is a single particle process, longitudinal and transverse conductivities should exhibit similar behavior in  temperature and pressure as long as $\tau_{\bot}$ is much larger than $\tau_{\parallel}$,  a condition easily satisfied in \tq as we shall see below,   the other channel being  of a collective nature. Hence, both conduction channels acting in parallel  add in order to give  the total parallel conductivity\cite{Jerome82}, 
 \begin{eqnarray}
 \sigma_{\parallel}= \sigma_{coll}+\sigma_{\parallel,sp}
 \label{conductivities}
 \end{eqnarray}
 
 The problem now  at hands was  to determine what fraction of the total conductivity could be attributed to Fr\"ohlich fluctuating channel in the vicinity of the Peierls transition.


Actually, high-pressure optical reflectance studies\cite{Welber78}
 allow us to discard the possibility of drastic (non-monotonous) changes in the
band structure evolving slowly under pressure in particular at commensurability. 

In addition we can reject the existence of the dip as being due to an increase of the single particle scattering mechanism and/or a decrease in the
density of states at the Fermi level in the commensurability domain because of the data of transverse conductivity. Whenever the  transverse conduction is diffusive, which is verified since as long as the 1D escape time $\tau_{\bot}$ is larger than $\tau_{\parallel}^{sp} $, a  condition easily fulfilled in \tq according to the NMR results \cite{Soda77}, $\tau_{\perp}/\tau_{\parallel}^{sp} \approx 10^{3}$, $\sigma_{\parallel,sp}$  is sensitive to both the density of
states at the Fermi level and to the smearing of the one-dimensional Fermi surface by
the single particle scattering time $\tau_{\perp}/\tau_{\parallel}^{sp}$.

According to the experimental data of Fig.~\ref{sigmaTQP} displaying the pressure dependence of the transverse conductivity  no   dip in the commensurability regime is observed. Therefore, Fig. \ref{sigmaTQP}
leads to the conclusion that neither $N(E_F)$ nor $\tau_{\parallel}^{sp}$ are significantly affected 
by commensurability.
Hence, we  may conclude that the dip of longitudinal conductivity around 19 kbar in \tq can be taken as a serious indication for the existence of a significant part  of a conduction channel originating in the 
collective (fluctuating Fr\"ohlich conduction) becoming pinned in this pressure range by a low order commensurability effect.
 


\begin{figure}
\includegraphics[width=1\hsize]{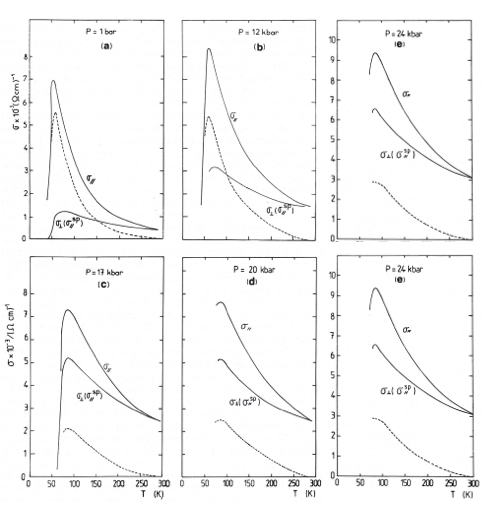}
\caption{\label{sigmaTP} Temperature dependence of the total conductivity $\sigma_{\parallel}$ and of the  $\sigma_{\bot}$ normalized to $\sigma_{\parallel}$. The dashed line corresponds to the fluctuating contribution. The room temperature value of $\sigma_{\parallel}$ is  400($\Omega.{\rm cm}) ^{-1} $. Notice, the strong depression of $ \sigma_{coll}$ in the commensurability range, Source: Fig. 16 [Ref\cite{Jerome82a}, p. 196,197].}
\end{figure}

Given the data of parallel and transverse conductivities against temperature obtained at various pressures, see Fig.~\ref{sigmaTP} including the commensurability range  we can analyze in more details how Fr\"ohlich fluctuations  behave in temperature at various pressures. 
Assuming both conductivity  channels to be additive  according to Eq.~\eqref{conductivities}, we can  extract the fluctuating contribution from the  measured total  conductivity.
To accomplish this decomposition two assumptions
are required. 

The first one is that at 300K,  $\sigma_{coll}$
 is negligibly small compared to $\sigma_{\parallel,sp}$. This assumption is justified by the monotonous increase of
 $\sigma_{\parallel}$ through the commensurability regime around room temperature. The accuracy
of the data on Fig.~\ref{sigmaTQP} shows that  $\sigma_{coll}$(300 K) is smaller than, say, $\sigma_{\parallel,sp}$/10 at 
 (19 kbar, 300 K) thus leading to $\sigma_{coll}$(300 K)$\ll$ 250 ($\Omega.{\rm cm})^{-1}$.

The second assumption of the derivation is based on the fact that in the  one-dimensional regime  $\sigma_{\bot}$ and $\sigma_{\parallel,sp}$ exhibit the same temperature dependence as already discussed above 

 The model decomposition of the conductivity\cite{Jerome82a}  is illustrated by the experimental situations of Fig.~\ref{sigmaTP} where at every pressure we have plotted the temperature dependence of $\sigma_{coll}$ normalized to the value of $\sigma_{\bot}$ at room temperature with the assumption of  zero $\sigma_{coll}$  under ambient conditions.
 
\begin{figure}
\includegraphics[width=0.9\hsize]{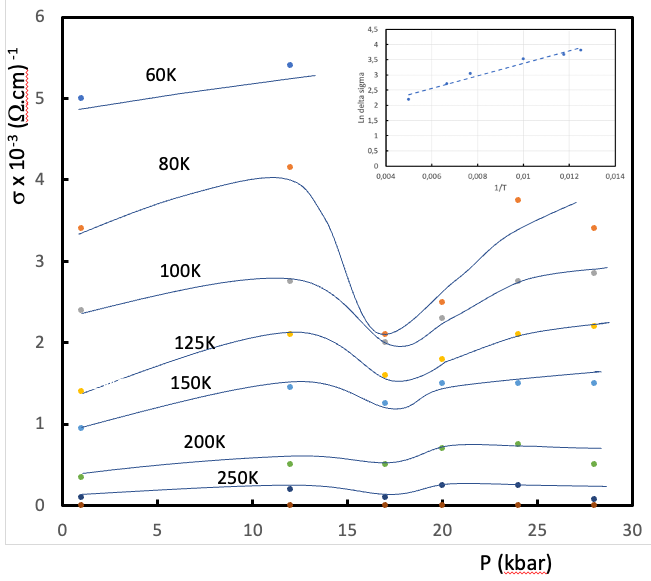}
\caption{\label{sigmaflv1} The Fr\"ohlich fluctuating contribution versus pressure at various temperatures for a \tq sample with a CPR=17 at 1 bar. The 60K contribution cannot be measured above 12 kbar since the critical temperature becomes larger than 60K in the commensurability range. The dependence of this collective contribution in pressure is small compared to its temperature dependence. (inset) Activation of the  CDW pinned  by commensurability with the activation gap of 55K.}
\end{figure}

 We may notice that except for the commensurability regime, the fluctuating conductivity is only mildly pressure dependent  and consequently the
strong pressure dependence of the total conduction which is observed especially at high temperature must be ascribed  to the single-particle contribution enhanced by strongly pressure dependent correlation effects.
\begin{figure}
\includegraphics[width=1\hsize]{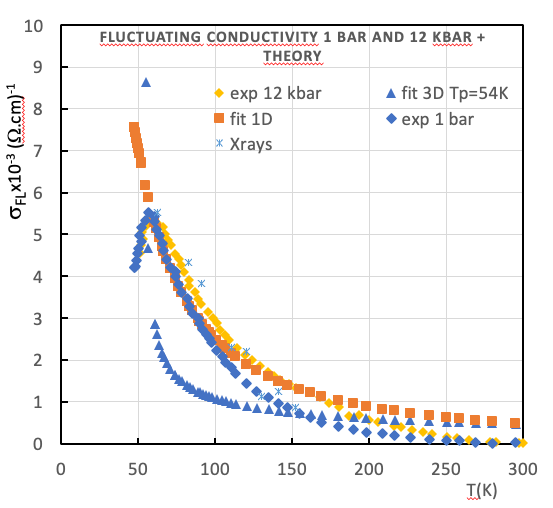}
\caption{\label{sigmafl+exp+theo} Temperature dependence of the experimental $\sigma_{coll}$ at 1 bar and 12 kbar together with the peak intensity of the $4k_F)$ scattering divided by $T$, Source: [Ref\cite{Khanna77}]  compared to the expected dependences for 1D  and 3D ($T_p$= 54K) theories. Obviously the 1D fit is much better than the 3D one above 60K. }
\end{figure}
Given the data on Fig.~\ref{sigmaTP} we are thus  able to derive $\sigma_{coll}$ versus temperature, see Fig.~\ref{sigmafl+exp+theo} 
at 1 bar and 12 kbar and also   at selected temperatures versus pressure, Fig.~\ref{sigmaflv1}.
Let us make a few remarks.

First, from the temperature dependence of $\sigma_{coll}$ on Fig.~\ref{sigmafl+exp+theo} we notice that the fluctuating contribution agrees fairly well with the divergence of a 1D law $\alpha$ $T^{-3/2}$ above 60K  at variance with a 3D behaviour $\alpha$ $T^{-1/2}$ expected below 60K in the very vicinity of the Peierls transition which would not agree with the data  above 60K. Both transport and the intensity of X-ray diffuse scattering data\cite{Khanna77}  point to one dimensional physics in \tq above 60K. Regarding the 3D Peierls fluctuations domain, diffuse X-ray scattering locate it in the range from  54K to 60K\cite{Denoyer75}. As far as the coherence length is concerned, X-ray studies give a value along the chains  of 20 nm at 55K\cite{Denoyer75} decreasing to 6 nm at 100K\cite{Khanna77} with a  transverse coherence length less than the interchain distance which is consistent with the one-dimensionality.

Second, as shown on Fig.~\ref{sigmaflv1}, the fluctuating contribution to the total longitudinal conduction is much less pressure dependent (if any) at all temperatures than the single particle part.

Third, the appreciable drop of $\sigma_{coll}$ shown on  Fig.~\ref{sigmaflv1} around 19 kbar without any effect on the transverse component as shown on Fig.~\ref{sigmaTQP}, has been interpreted  in terms of the pinning
by the commensurability potential of sufficiently long-lived CDW fluctuations. We may notice that the drop of conduction in the commensurability regime is somewhat moderate, about only half the total fluctuating contribution at 80 K, the signature of a  modest pinning potential. Hence, the temperature dependence of the relative drop of conduction enables us to estimate the strength of the pinning potential  $\Delta$ as $\sigma_{min}(T)/ \sigma_{max}\ \propto\  \exp (-\Delta/T)$. 


The inset of Fig.~\ref{sigmaflv1} shows that such an activation law is followed  with $\Delta \approx 55K$.

\begin{figure}
\includegraphics[width=0.7\hsize]{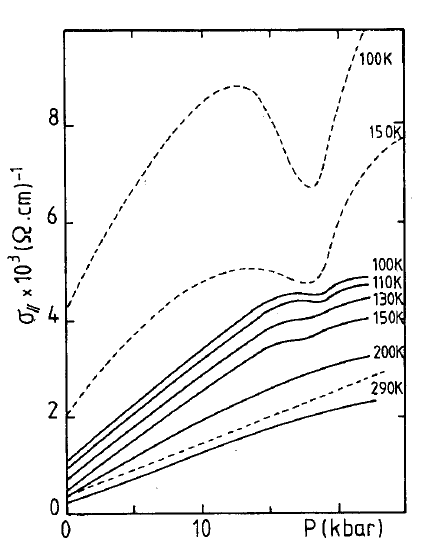}
\caption{\label{TTFQirradiation} Pressure dependence of the longitudinal conductivity at constant temperature for
irradiated \tq (continuous line), Source: Fig.2 [Ref\cite{Bouffard81}, p. 406]. The dashed lines indicate the behaviour of
pristine samples according to the data on Fig.~\ref{diagTQ}.}
\end{figure}

Additional evidence for an important contribution of CDW fluctuations to the
longitudinal conductivity comes from experiments on irradiated samples\cite{Bouffard81}
showing a suppression of the 19 kbar conduction dip, as shown in Fig.~\ref{TTFQirradiation}. Defects,
introduced by irradiation 
should 
very effectively pin the free motion of CDW
fluctuations, thus suppressing the CDW-fluctuation conductivity even in the
incommensurate situation. Correspondingly,  the drop of conductivity
 observed for irradiated samples near commensurability is much smaller that the one observed in pristine samples. Were the drop of conductivity due to changes in the single-particle scattering, one would expect
(from Mathiessen’s rule) a similar drop to appear in the irradiated samples. This is
not observed. Irradiation has shown  that an important part of the metallic conductivity
of (incommensurate, non-irradiated) \tq is due to the CDW-fluctuation
mechanism.


Detailed far-infrared measurements on \tq from 25K to 300K have also provided a strong support for a charge density wave mechanism contributing to the dc conductivity in \tq\cite{Tanner81}. 

\begin{figure}
\includegraphics[width=0.8\hsize]{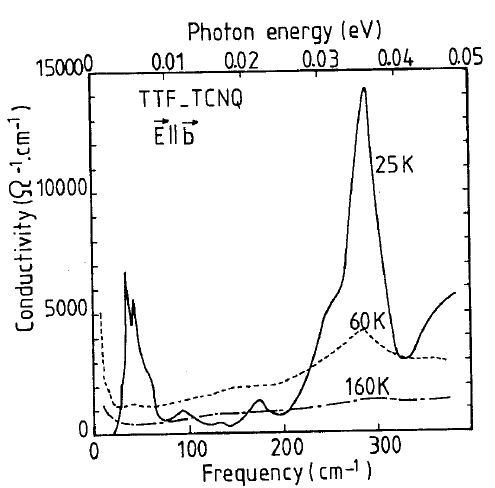}
\caption{\label{FIRTanner} Frequency-dependent $b$-axis conductivity of
\tq at four temperatures: 25, 60, and
160 K, Source: Fig. 2 [Ref\cite{Tanner81}, p. 598]. The low frequency conductivity at 60K is in very good agreement with the DC data of Fig.~\ref{sigmaTP}. }
\end{figure}

Fig.~\ref{FIRTanner} shows the $b$-axis frequency-dependent conductivity of \tq derived from the reflectance data by a Kramers-Kr\"onig analysis following
the procedure described by Tanner et-al\cite{Tanner81}. The strong peak of the conductivity
around 40 cm$^{-1}$ at 25K has been taken as the signature of the phase-mode of the
pinned CDW in the Peierls state. At 60 K, the peak of conductivity has shifted to
zero frequency. This phenomenon can be considered as the existence of a collective
mode at zero frequency: the collective mode giving rise to the sliding of finite lifetime
CDW's.

\begin{figure}
\includegraphics[width=0.8\hsize]{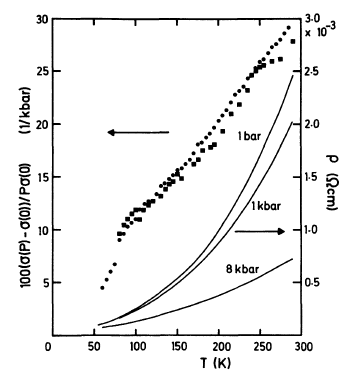}
\caption{\label{Pdependence} Temperature dependence of the pressure dependence of the conductivity of \tq, Source: Fig. 3 [Ref\cite{Friend78a}, p. 136]. }
\end{figure}

These FIR studies have also led to a better knowledge of this  conduction mode  with an effective mass about $20 m^*$ and a fluctuation lifetime at 60K of  $\tau_c= 1.6\times  10^{-12}$~s in contrast with the much shorter single particle lifetime  of  $\tau_{\parallel, sp} \approx 4.6\times  10^{-15}$~s derived at 300K from the NMR measurements\cite{Soda77} or $3.1\times 10^{-15}$~s from optical reflectance at room temperature increasing up to $6\times 10^{-15}$~s at 70K \cite{Bright74}. Let us notice that already in 1974, the authors of Reference\cite{Bright74} concluded that the dc conductivity should be carried in a collective manner.

Now it is interesting to comment on the pressure dependence of the \tq resistivity reported at various temperatures in the conducting regime, Fig.~\ref{Pdependence}. According to the data in Ref.\cite{Friend78a}, the pressure dependence of the resistivity dropping from 28 to 5\%  between room temperature and 60K. Such a temperature dependence is in very good agreement with  a collective conduction very weakly pressure dependent as shown above, becoming dominant at low temperature. Taking the collective contribution into account, we are left with a pressure coefficient for the single particle conductivity of about 25\%kbar$^{-1}$ showing hardly any temperature dependence. This is still a very large volume dependence which cannot be understood solely by the modest  volume dependence of the bare band theoretical parameters\cite{Herman77,Herman77a} of the order of 2\%kbar$^{-1}$.  High pressure measurements concluded that the large volume dependence of the conductivity can be interpreted as a result of a large electron-electron interaction\cite{Welber78}. However, given the predominance of a collective contribution to the conductivity below 100K, identified more recently, a new visit to the question of the single particle conduction would be valuable.


In short, since  we can be confident that the longitudinal conduction data of pristine \tq carry the signature for the existence of
a large conductive Fr\"ohlich fluctuating component  present anywhere but in the commensurability domain up to about 250K, the variety of CPR reported in the literature  over the past 1974-1975 years  is thus likely to be attributed (besides incorrect measurements of the conductivity)   to  differences in the amount of impurities contributing to an efficient pinning of Fr\"ohlich fluctuations in different samples of \tq. This can be as inferred from the study on irradiated samples\cite{Bouffard81} with  much milder effects on the single particle contribution. 

The decomposition of the conduction into two components has shown that the fluctuating part is hardly
sensitive to pressure\cite{Jerome82a} see, Fig.~\ref{sigmaflv1}, (besides the suppression which is related to the commensurability domain).


We have seen above that thanks to the one dimensional character of the electronic structure of \tq,  the fluctuating Fr\"ohlich mode provides a major contribution  to the longitudinal conduction whenever there exists  no correlation between CDW fluctuations of adjacent stacks.

\begin{figure}[h]
\includegraphics[width=0.6\hsize]{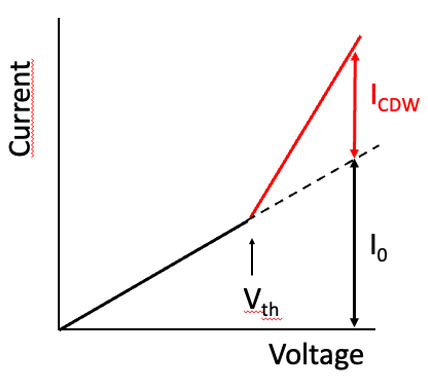}
\caption{\label{ICDW} Schematic I-V characteristics in the presence of a  collective extra contribution from the CDW condensate active above a threshold voltage (field). The conductance I/V become non-linear at Vth.  }
\end{figure}

In addition, \tq provides a brilliant and instructive illustration of the non-linear conduction mechanism observed in several inorganic reduced dimensionally trichalcogenides extensively studied in the eighties\cite{Monceau}.
As far as \tq is concerned  there exists the possibility to put the 3D-ordered CDW condensate into motion and thus observe an additional  conduction of collective origin. However this collective channel becomes active only when the applied electric field exceeds a
certain threshold field,Fig.~\ref{ICDW}, making in turns the conduction  non-linear in terms of the electric field.
The existence of nonlinear conduction is well-known from quasi
1D inorganic materials such as  $\mathrm{NbSe_{3}}$
below the Peierls transition, where the nonlinear
conduction has been interpreted as resulting from
CDW depinning\cite{Monceau}.

The mechanisms which can pin a CDW are
impurity pinning, commensurability pinning, and
Coulomb interaction between oppositely charged
chains in two-chain systems such as \tq.
The collective conduction channel 
is  observed in TTF-TCNQ below $T_H$  with $E_{T}$
reaching a minimum of 0.25 V/cm at 51.2 K in the
narrow temperature interval between $T_H$ and $T_M$ in which TTF-TCNQ
behaves as a single chain CDW\cite{Forro87}, see also Fig.~\ref{TQrho1baret4kbarversusT}.

\begin{figure}
\includegraphics[width=0.8\hsize]{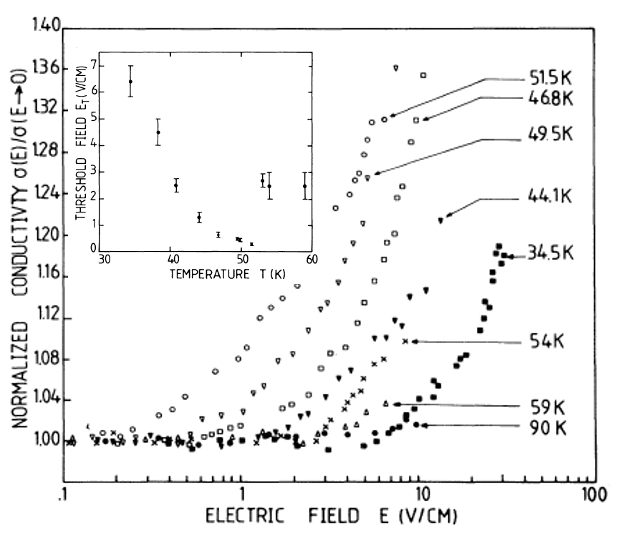}
\caption{\label{NLconductivityTQ} Normalized conductivity $\sigma(E)/\sigma(E\rightarrow0)$ vs
logE of TTF-TCNQ at 1 bar. The inset shows the temperature dependence of the threshold field, Source: Fig. 1, 2 [Ref\cite{Lacoe85}, p. 2552].  }
\end{figure}
\begin{figure}[h]
\includegraphics[width=0.8\hsize]{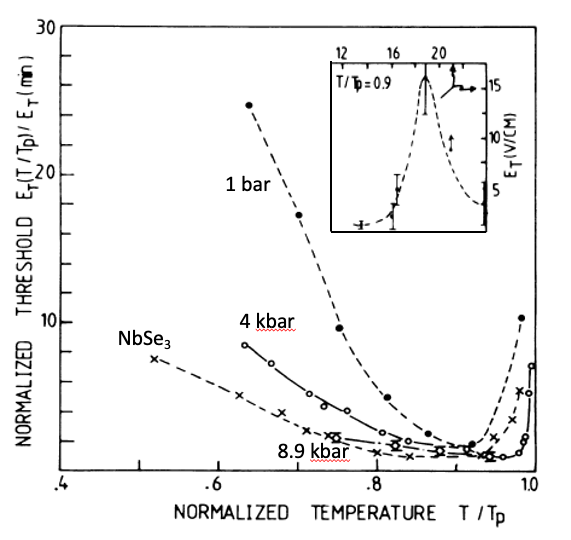}
\caption{\label{NLconductivityTQP} Normalized threshold field $E_T/E_{T} (min)$ versus the normalized Peierls temperature at different pressures,  Source: Fig. 1 [Ref\cite{Lacoe87} , p. 263], including the data for $\mathrm{NbSe_{3}}$ at 1 kbar, Source:  inset [Ref\cite{Gruner89}, p 137].  }
\end{figure}

A fast increase of $E_T$ is
observed on cooling below TM when a CDW arises on
the TTF chains, with $E_T$ becoming of the order of
1-10 V/cm\cite{Lacoe85},see Fig.~\ref{NLconductivityTQ}. This rise of $E_T$ has been understood in
terms of the sliding of joint CDW's on TTF and TCNQ
stacks moving together in the same direction with a
pinning mechanism governed by impurities\cite{Lacoe85}. According
to a Landau type theory of the CDW formation\cite{Bak76},
the enhancement of the impurity pinning mechanism
is linked to the growth of a CDW on TTF chains below $T_M$ and
the existence of a Coulomb attraction between oppositely
charged chains\cite{Lacoe85}. In the absence of any
coupling between TTF and TCNQ chains, an electric
field would lead to CDW motion in opposite directions
in the two-chain systems because of the electron
(TCNQ) and hole (TTF) character of the electronic
bands. 

However, the motion of the TTF and TCNQ CDW's in opposite directions, which provides the maximum CDW current would require a threshold field of $E_{T}\approx 5\times 10^{4}$ V/cm\cite{Lacoe85}. This cannot be responsible for the nonlinear effects
found here with $E_{T}\approx1-10 $~V/cm,Fig.~\ref{NLconductivityTQ}. The whole picture is confirmed by the behaviour of the threshold field in \tq under 4 kbar, Fig.~\ref{NLconductivityTQP}, when the range of one-chain CDW is more extended towards low temperatures  than at ambient pressure,Fig.~\ref{TQrho1baret4kbarversusT}, with a concomitant smaller increase of $E_T$ at low temperature.

\subsubsection{ \tq under higher pressure}
\label{under higher pressure}
\begin{figure}[h]
\includegraphics[width=0.85\hsize]{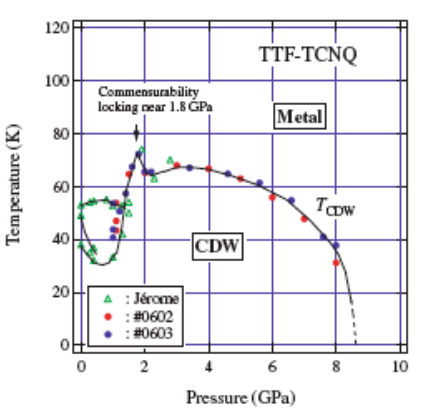}
\caption{\label{Muratav1} T–P phase diagram of \tq, Source: Fig. 4 [Ref\cite{Yasuzuka07}, p. 2]. The previous data of Ref\cite{Friend78b} are also displayed with green triangles. The authors of Ref\cite{Yasuzuka07} expect  the suppression of the Peierls transition near 9 GPa.}
\end{figure}

An investigation of the phase diagram of \tq at much higher pressures than what the Orsay had published has been performed using the cubic anvil technique up to 80 kbar\cite{Yasuzuka07},Fig.~\ref{Muratav1}. 
The motivation for such a study was the suppression of the Peierls instability as a result of the enhancement of the interchain coupling leaving an open possibility  for the stabilization of superconductivity due to the $2k_F$ phonon softening\cite{Horovitz75}. This experiment although confirming the results of the Orsay group for commensurability at 19 kbar\cite{Friend78b},  has revealed that the Peierls instability is gradually depressed under higher  pressure with a concomitant decrease of the activation energy.  Under  80 kbar ,  the resistivity
shows a metallic behaviour  down to a CDW
transition at 25K. At lower temperatures a semimetallic behaviour is observed without any sign of superconductivity at least above 2.5K. Such a failure to stabilize a superconducting state might be due to an increased gap  at the Fermi level coming from the hybridization coming from the molecular orbitals on neighbouring chains. We are not aware of any study on \tq   performed at pressures beyond 80 kbar.

 \subsection{Other charge transfer salts, selenide molecules} 
 
 \subsubsection{TSF-TCNQ}
 \label{TSF-TCNQ}
In the seventies the leading ideas governing the search for new materials likely to exhibit good metallicity and possibly
superconductivity were driven by the possibility to minimize the role of electron-electron repulsions and at the same time to
increase the electron-phonon interaction while keeping the overlap between stacks as large as possible\cite{Horovitz75}. This  suggestion not having kept its promises with the \tq family  led to other attempts and the
synthesis of new series of charge transfer compounds, for example changing the
molecular properties while retaining the same crystal structure. It was recognized that electron polarizability was important
to reduce the screened on-site e-e repulsion and that the redox potential $(\Delta E)_{1/2}$ should be minimized,
\cite{Garito74}. Hence, new charge transfer compounds with \q  have been synthesized using other
heteroatoms for the donor molecule, 
\textit{i.e.} substituting sulfur for selenium in the \f skeleton thus leading to the \ts \ molecule (($\Delta E)_{1/2}$=0.37 eV
for
\f\cite{Garito74} but only 0.32 eV for \ts\cite{Engler77}) and the synthesis of the charge transfer compound \tsq. 
This compound has the same monoclinic structure as \tq and the slight increase of the unit cell by $0.057\AA$ does not
compensate for the significant increase of $0.15\AA$ for the van der Waals radius  going from sulfur to selenium.
Consequently, the cationic bandwidth of \tsq\, is increased by 28\%  as shown by the tight binding calculation, while the \q
band is hardly affected\cite{Herman77}. It is probably the increase in the donor bandwidth with a concomitant decrease of
the e-e repulsion on the \ts \,  stacks which suppresses $4k_F$ fluctuations \textit{at variance} with  \tq where $4k_F$ scattering is
observed from 300K\cite{Pouget79}.  

Measurements of the conductivity\cite{Etemad75}, Fig.~\ref{sigmaTTF} and thermopower\cite{Chaikin76} show a single transition into a semiconducting Peierls distorted state at $T_p= 29$ K. Below this transition an incommensurate $2a\times 3.15b\times c$ superstructure
is observed \cite{Weyl76}. 
In contrast to the behaviour
of \tq no further phase changes occur down to the lowest observed
temperatures, and the electronic gap is obtained consistently from resistivity\cite{Etemad76},
susceptibility\cite{Scott78} and infrared\cite{Bates81}measurements as $E_g= 250$ K, showing that both
TSF and TCNQ stacks order at the same  Peierls transition.

Furthermore, the observation of $2k_F$ scattering  in \tsq\, already    below
230K\cite{Pouget79,Kagoshima79} prior to the Peierls transition at 29K  has led to an accurate determination of the
incommensurate charge transfer, namely $\rho = 0.63$. 

As inferred from a detailed study of the conductivity, the sliding CDW's are  also contributing to the total conduction although in a proportion much smaller than it is for \tq\cite{Thomas82}.

As far as structural precursor  effects are concerned there are significant differences between \tq and \tsq. While  precursor effects are one-dimensional  in 
\tq over almost the entire temperature domain where they are observed, the picture is different in \tsq \, since a short range 3D
coupling is observed between 29 and 50K and only a limited 2D coupling is noticed up to 100K\cite{Megtert79,Kagoshima79}.

Commensurability between CDW and the underlying lattice ($\times 3$) has been detected at $6.2$ kbar through a small peaking of
the Peierls transition and a small  concomitant  drop of the longitudinal conductivity\cite{Thomas80}. The modest amplitude of
the conductivity drop accompanying the commensurability suggests that the CDW's  in the non-ordered phase above 29K  are already
partially pinned by transverse coupling and therefore cannot contribute dominantly  to the fluctuating conduction of the
metallic domain as it is the case for
\tq\cite{Thomas82}. 

Altogether TSF-TCNQ evolves under pressure\cite{Thomas80} in a way quite similar to \tq with the single phase transition at $T_p$ increasing steadily up to 45K under 25 kbar with no sign of any suppression in this pressure range. Unfortunately, unlike \tq no higher pressure studies have been undertaken  with \tsq.

\subsubsection{HMTSF-TCNQ-like compounds}
\label{HMTSF-TCNQ-like compounds}
The goal of the effort to develop new organic metals was to find  systems which remain metallic down to low temperature a requirement   to give rise to superconductivity.

The main idea guiding the work of researchers in the 1970's for a stabilization of superconductivity in organic conductors was to increase the two-dimensional character of the electronic structure in order to suppress the Peierls instability favored by the one-dimensionality\cite{Horovitz75,Beni74}.

\begin{figure}
\includegraphics[width=0.75\hsize]{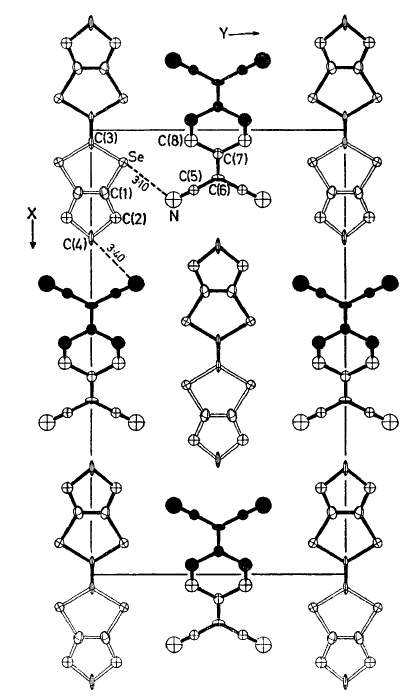}
\caption{\label{HMTSF-TCNQ} A projection of the structure of HMTSF-TCNQ onto
the plane perpendicular to the stacking axis. The chessboard pattern \emph{at variance} with the packing of TTF-TCNQ is clearly visible.  The planes of the molecules in the Figure are
tilted relative to the projection plane by 22$^{\circ}$ (HMTSF) and
34$^{\circ}$  (TCNQ).  The transverse contact Se...N  of 3.10 $\AA$ is particularly strong, much shorter than the S...N contacts in TTF-TCNQ with a criss-cross molecular packing along the $a$-axis, see Fig.~\ref{TTFTCNQvuselona}, Source: Fig. [Ref\cite{Phillips76}, p. 335]. }
\end{figure}

The attempt to increase the transverse overlap and in turn stabilize a metallic phase at low
temperature has been partly successful with the synthesis of new \q charge transfer compounds in which the structure
exhibits a chessboard-like pattern, Fig.~\ref{HMTSF-TCNQ}. This is the case for hexamethylene-donor molecules with sulfur or selenium
heteroatoms, $\mathrm{HMTTF}$ or $\mathrm{HMTSF}$ respectively\cite{Greene76,Bloch75} where metallicity is nearly
achieved at low temperature.  In $\mathrm{HMTTF-TCNQ}$ , an incommensurate CDW system with $\rho =
0.72$\cite{Megtert79}, a semimetallic character can be maintened at low temperature under pressure  above $19 kbar$ with
$\delta \rho /\delta T >0$ although a weak transition is still observed in resistivity at $30 kbar$\cite{Friend78}. It was
anticipated that the complete suppression of the distortion would require a pressure between 35 and $40 kbar$.

The selenide related compound, HMTSF-TCNQ, Fig.~\ref{HMTSF-TCNQ}, exhibits a resistance minimum prior to a phase transition at
$T_c = 24$ K\cite{Korin81} which although decreasing under pressure, never  vanishes in good quality samples\cite{Jerome82}. This
$24K$ anomaly arises from the formation of a 3D superstructure $a$ x $2.7b$$\times c$ corresponding to a charge transfer $\rho = 0.74$\cite{Pouget88}.

One of the first studies of this compound suggested that the three-dimensional order state of 24 K could be suppressed under pressure allowing the establishment above 4 kbar of a conducting state up to 100mK\cite{Cooper76b}. This study  aroused a lot of interest  since for the first time a metallic-like conductivity exceeding  $10^{4}(\Omega. cm)^{-1}$ could be stabilized at helium temperature in a charge transfer conductor member of the \tq family. A model of semi-metallic Fermi surface for the undistorted three-dimensional
Fermi surface of HMTSF-TCNQ, i.e. at T$>$ 24K has been proposed by Weger\cite{Weger76}, leading to a 3D semi-metallic Fermi surface consisting of small section cylinders or elongated
ellipsoids in the $c$-direction. However, subsequent experiments suggested that  the stabilization of the semi-metallic state previously claimed\cite{Cooper76b} could have been the result of defective samples.
But the suggestion that the coupling between neighboring donor and acceptor chains is important in determining the shape of the Fermi surface at low temperature (kT$<$ $t_{\bot}$)  proved to be important for the future.
\begin{figure}
\includegraphics[width=0.85\hsize]{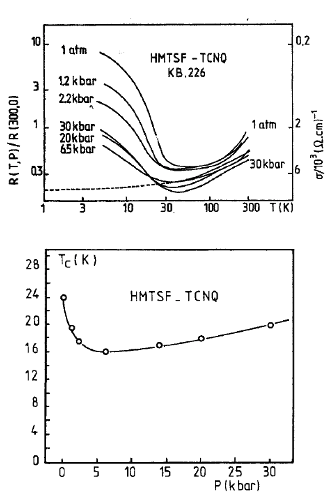}
\caption{\label{HMMiljak} (Top) HMTSF-TCNQ crystal showing no
stabilization of a semi-metallic state above 4 kbar as reported in Ref\cite{Jerome82}. The dotted line corresponds to the
stabilization of a semi-metallic state observed in a poor quality crystal. (Bottom) Pressure dependence of the Peierls phase transition of HMTSF-TCNQ. Unpublished data from M. Miljak at Orsay. }
\end{figure}
Actually, in good quality samples, the initial drop of $T_p$ below 24K observed up to 5 kbar is followed by a re-increase  up to 20K under 30 kbar as displayed on Fig.~\ref{HMMiljak},(bottom). In this picture a possible explanation for the decrease of $T_p $ with pressure in Fig.~\ref{HMMiljak} is a pressure-induced increase of the interchain transfer integrals, which increases the non-planarity of the Fermi surface so that the Peierls
instability can only affect a smaller portion of the Fermi surface and therefore occurs
at lower temperature.
\begin{figure}[h]
\includegraphics[width=0.85\hsize]{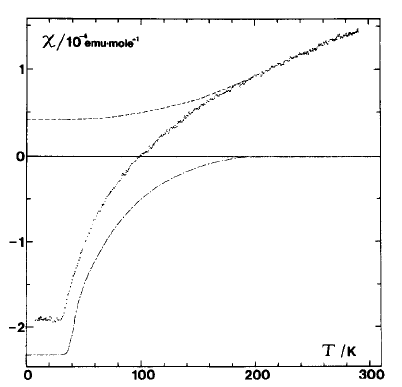}
\caption{\label{Diamag} Temperature dependence of the HMTSF-TCNQ susceptibility measured by NMR. The Pauli contribution is  the dashed line and Landau-Peierls, the dashed dotted line, Source: Fig. 3 [Ref\cite{Soda76b}, p. 108]. }
\end{figure}
It was concluded that in 
$\mathrm{HMTSF-TCNQ}$, due to the important deviation from planarity  of the Fermi surface, the Peierls transition can
only partly destroy it. There are plenty of experimental results supporting the picture of a semi-metallic Fermi surface at low temperature in HMTSF-TCNQ\cite{Jerome77b}. A striking confirmation was provided by the measurement of the Hall effect\cite{Cooper76d} providing a very large Hall constant at low temperature related to the semi-metallic behaviour with approximately 1/500-1/1000 large mobility ($\mu\approx 4\times10^{4}$ cm$^{2}$/Vs) carrier per formula unit compared to the small Hall constant measured at room temperature, signature of the 1D metallic nature.  Another confirmation was provided by the measurement of the magnetic susceptibility found to become diamagnetic below 100K\cite{Soda76b} 	and attributed to a large Landau-Peierls orbital diamagnetism associated with very small cross-sectional areas of the Fermi surface. The LP diamagnetic susceptibility being proportional to $m_0/m^{*}$ becomes appreciable at low temperature because of the small effective masses of the carriers in the semi-metallic pockets as measured by 
Shubnikov-de Haas oscillations  describing adequately the semi-metallic band structure\cite{Miljak78}.  
\begin{figure}[h]
\includegraphics[width=0.85\hsize]{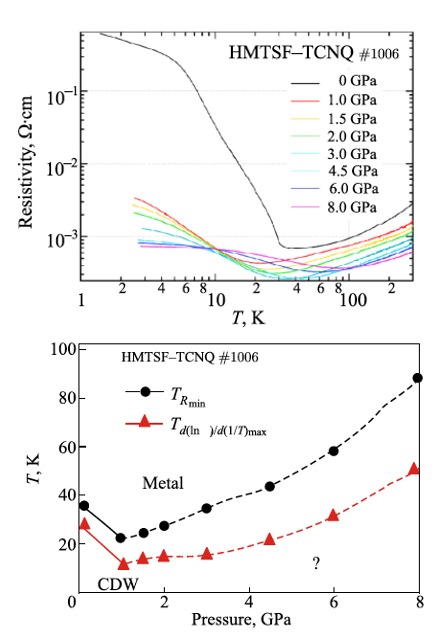}
\caption{\label{HMTSF_MUrata} \textcolor{black}{(Top) Recent temperature dependence of the resistivity under pressure in cubic anvil pressure device. According to the zero pessure data, the activation energy of the resistivity amounts to $\approx$ 35 K. (Bottom) temperature-pressure  phase diagram according to Ref\cite{Murata14}, Source: Fig. 3 and 4 [Ref\cite{Murata14}, p. 372]. It is in fair agreement with the data on Fig.\ref{HMMiljak} bottom.}}
\end{figure}

It is usually agreed that a cross-over from the 1D (planar) to a 3D (ellipsoidal semi-metallic ) Fermi surface occurs around 100K in this compound\cite{Soda76b}. In addition, the conductivity does not vary significantly through the cross-over regime since both the density of carriers and the Fermi wave-vector are changing in about the same proportion\cite{Miljak78}.

If the transition  is to be suppressed by pressure, this would require much higher pressures \textcolor{black}{
as Murata \textit{et al} recent results\cite{Murata14} imply that if it occurs, this would arise  above 8 Gpa or so, \emph{see} Fig.~\ref{HMTSF_MUrata}. The resistivity data displayed on Fig.~\ref{HMTSF_MUrata}  reveal  a ratio between 5 and 24K under ambient pressure much larger than the one reported in the early studies in the 1976's. Resistivity data would thus lead to an activation energy of the order of 35K, in line with transition temperature around 24K. This is likely the result of an improved sample purity minimizing  the role of impurity conduction in the low temperature phase. In addition, the possibility of a field-induced charge density wave state (topic not covered in the present article, \emph{see} Ref.\cite{Lebed09}  for instance,  have been claimed when the component of the magnetic field along the least conducting direction exceeds 10 T (under 1.1 GPa) accompanied  by the observation of a quantized Hall effect and angular magnetoresistance oscillations\cite{Murata14,Murata14a}. These remain interesting  possibilities which will require further confirmations.}


\subsubsection{TMTSF-DMTCNQ}
\label{TMTSF-DMTCNQ}
 The study of the two-chain charge transfer compounds went on with a system where both donor and acceptor 
molecules had been methylated namely, $\mathrm{TMTSF-DMTCNQ}$, so-called  (TM-DM), where the donor is the tetramethyl selenide
derivative of TTF and the acceptor . The outcome of this study has been truly  decisive for the quest of organic superconductivity
. This 1-D conductor undergoes a Peierls transition at 42 K detected by conductivity\cite{Jacobsen78} and
magnetic\cite{Tomkiewicz78} measurements
 where unlike $\mathrm{TTF-TCNQ}$ a distortion occurs simultaneously on
both chains \cite{Pouget81}. Several other results have triggered the attention. 
X-ray experiments had shown that the charge transfer is only $\rho =0.5$ namely,
a quarter-filled band situation \cite{Pouget81} for both acceptor and donor bands. Such a band filling leads in turn to a high order commensurate CDW with a periodicity of $4a$.  $a$ is the direction of best conductivity.  Transport and thermopower
data have emphasized the dominant role played by the TMTSF chain in the mechanism
driving the Peierls transition and also in its contribution to the conduction at
high temperature \cite{Tomkiewicz78}.  

\begin{figure}
\includegraphics[width=1.05\hsize]{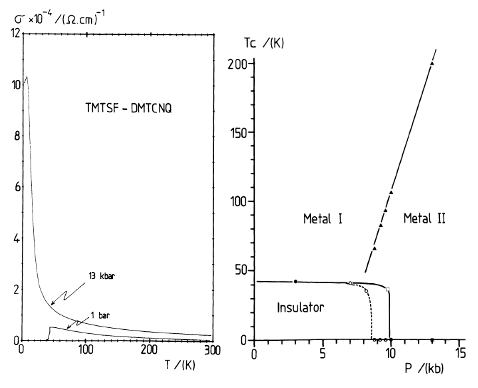}
\caption{\label{TMDM} Temperature dependence of $\sigma_{\parallel}$ in TMTSF-DMTCNQ,
at  atmospheric pressure and  under13 kbar (left).  Phase diagram of TMTSF-DMTCNQ, under pressure (right), Sources: Fig. 2, 3 [Ref\cite{Andrieux79b}, p. 1200].  }
\end{figure}

In addition, according to the pressure study displayed on  Fig. \ref{TMDM},  the commensurate state is remarkably stable under pressure since the Peierls transition stays at practically the same
temperature of 42K up to 8 kbar\cite{Andrieux79b}. 

The really new and unexpected finding has been the sharp  suppression of 
 the Peierls transition  under a pressure of about 9 kbar and the conductivity remaining metal-like down to low temperature,
reaching
$10^{5} (\Omega.{\rm cm})^{-1}$  under 10 kbar at liquid helium temperature 
\cite{Andrieux79b}, Fig.~\ref{TMDM}. 

The stabilization in TM-DM of the conducting state at low temperature under pressure, see Fig.~\ref{TMDM}, is apparently related   to a
pressure-driven  S-shape anomaly which is visible  on the temperature dependence of the resistivity.  


The very high value of the low temperature conductivity above 10 kbar has stimulated more investigations under high magnetic field in order to better characterize this high conduction state.
\begin{figure}[h]
\includegraphics[width=0.7\hsize]{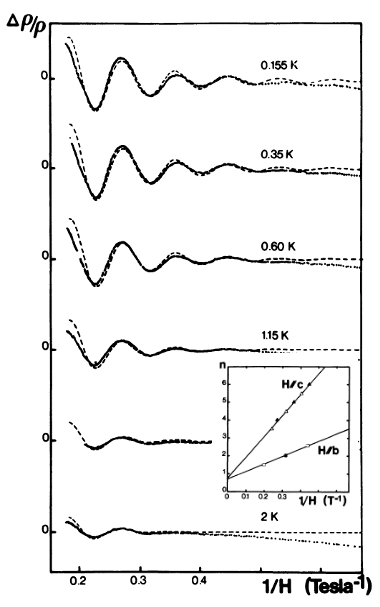}
\caption{\label{TMDMdHSh}Oscillatory magnetoresistivity of TMTSF-DMTCNQ under 12 kbar (H//$c$) versus 1/H for different
temperatures, Sources: Fig. 2, 4 [Ref\cite{Bouffard83}, p. 288, 289], the inset shows successive integers and half-integers corresponding to maxima and minima of the oscillations versus 1/H for both transverse directions. In TM-DM, the stacking axis is usually called $a$ while $b$ and $c$ are the transverse axes.}
\end{figure}

 Magnetoresistivity experiments performed  on high purity TMTSF-DMTCNQ single crystals reveal well formed Shubnikov de Haas  oscillations below 3K for the longitudinal resistance  in fields up to 7 T aligned along both transverse directions\cite{Bouffard83}, Fig.~\ref{TMDMdHSh} . 
When these magnetoresistance oscillations are analyzed in terms of the celebrated Shubnikov de-Haas theory\cite{Roth66}, the characteristics of the Fermi surface at low temperature can be derived, leading to ellipsoidal Fermi surfaces with cross sections  $1.16\times10^{17}$~m$^{-2}$ and $2.6\times 10^{17}$ m$^{-2}$ in  the $(a-b)$ and $(a-c)$ planes respectively. Furthermore, the cyclotron mass and the Dingle temperature  derived from the temperature dependence of the amplitude at fixed field and the field dependence at constant temperatures respectively amount to $\mathrm{m_c=0.56 m_0}$ and $\mathrm{T_d \approx 0.8 K}$.
If we assume parabolic bands and use the above mentionned
ShdH parameters  and figure\ref{TMDMdHSh} we obtain the Fermi energy of a pocket namely, $E_F\approx 50$K. Moreover,
the volume enclosed within such small pockets is about $3\times 10^{- 4}$ times smaller than the
volume enclosed by the original high temperature 1-D Fermi surface. Thus, the density of carriers
measured by SdH effects  in the highly conducting phase at low temperature is about $10 ^{-4}$ per formula-unit,
i.e. $3\times 10^{18}$ cm$^{ - 3}$ (assuming two pockets of such carriers per Brillouin zone, a semi-metallic situation). Given the data   from quantum oscillations, an upper limit  value for the single particle low temperature conductivity should be around 2-3$\times10^{3}$$(\Omega.{\rm cm})^{-1}$. 

The semi-metallic structure under pressure has also been confirmed by a Landau-Peierls contribution to the susceptibility occurring below 30K\cite{Hardebusch79}, contrasting with the 1D to 3D cross-over of HMTSF-TCNQ developing below $\approx$200K, see Fig. \ref{Diamag} . A conclusion  from the susceptibility study was that the TM-DM transverse coupling between unlike chains 
should be smaller than that of HMTSF-TCNQ under ambient pressure with a consequent lowering  of the 1D to 3D cross-over temperature.  

As discussed in more details in Ref.\cite{Bouffard83}, the semi-metallic picture of TMTSF-DMTCNQ at low temperature under pressure cannot explain the remarkably high value  of the conductivity under pressure which amounts to more than $\approx10^{5}(\Omega.{\rm cm})^{-1}$ as shown on Fig.~\ref{TMDM}.

The semi-metallic picture which can explain the low temperature state, quantum oscillations, Landau-Peierls diamagnetism of HMTSF-TCNQ\cite{Soda76b} and 
$\sigma \approx 10^{3}-10^{4} (\Omega.{\rm cm})^{-1}$ thus  fails to provide a reasonable interpretation for  the extraordinary large
value of the conductivity of TMTSF-DMTCNQ which is ten times larger at helium temperature.

 \begin{figure}[h]
\includegraphics[width=1\hsize]{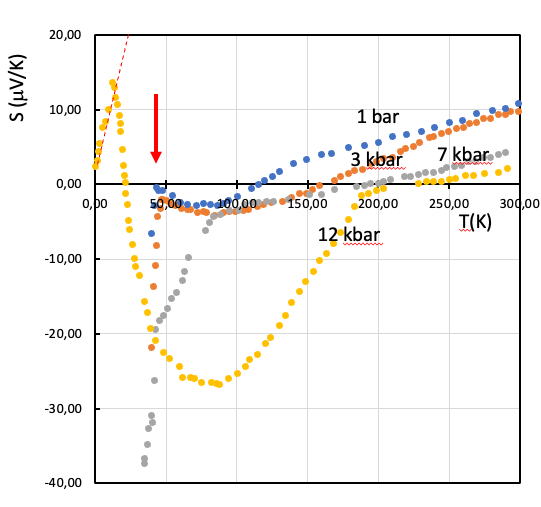}
\caption{\label{TEP_TMDM} Temperature dependence of the absolute thermoelectric power of TM-DM at 1 bar (data from \cite{Jacobsen78}) and under pressure, Source: Fig. 5 [Ref\cite{Andrieux79b}, p. 1202].   Notice the semi-metallic behaviour below 25K characterized by the large slope of S(T). The arrow marks the temperature of the Peierls transition which is nearly unchanged between 1 bar and 8 kbar. }
\end{figure}

It is now quite instructive to look how the  thermopower of TM-DM behaves in temperature and pressure, see Fig.~\ref{TEP_TMDM}. These data  displayed on this figure show together ambient\cite{Jacobsen78} and high pressure\cite{Andrieux79b} measurements. At low temperature ($T<10$K),
the inset of Fig.~\ref{TEP_TMDM} the TEP is positive and quite
linear in temperature, extrapolating close to the origin. This behaviour is indicative of a 3D semi-metallic ground state dominated by small hole pockets. Assuming  3-D semi-metallic pockets, the Seebeck coefficient reads\cite{Chaikin80},
\begin{eqnarray}
S=\frac
{\pi^2}
{2}
\frac
{k}
{e}
\frac
{kT}
{E_F}
\label{Seebeck}
\end{eqnarray}
Therefore, the low temperature experimental  data of TEP in Fig.~\ref{TEP_TMDM} would lead to the  estimate  ${E_F}\approx$ 300K according to Eq.~\ref{Seebeck}. This value is admittedly much larger than what can be derived from the data of the dHSh measurements\cite{Bouffard83}, with a crude estimate of parabolic bands\cite{Bouffard83} namely, ${E_F}\approx$ 30K. However, we can reconcile both values, as it has been proposed in Ref.\cite{Bouffard83}, assuming the existence of a channel of collective conduction acting in parallel with the single particle channel leading to a total conduction of $\sigma=\sigma_{coll}+\sigma_{sp}$. Thus, with a two-fluid model the thermopower becomes,
\begin{eqnarray}
S=\frac
{\sigma_{coll}S_{coll}+\sigma_{sp}S_{sp}}
{\sigma_{coll}+\sigma_{sp}}
\end{eqnarray}
 Since no heat can be carried by a current of collective origin, ${\sigma_{coll}S}$ is expected to be negligible and the measured thermopower becomes,
 \begin{eqnarray}
S=\frac
{\sigma_{sp}S_{sp}}
{\sigma_{coll}}
\end{eqnarray}
 
 Consequently, the Seebeck coefficient related to the small semi-metallic pockets would thus becomes ($\sigma_{coll}/\sigma_{sp}$) $\times$ the measured value of $S$, leading in turn to a reduction for the value of  the actual Fermi energy of the pockets, in  better agreement with the estimate of $E_{F}\approx$ 30K from dHSh data since $\sigma_{coll}/\sigma_{sp}$ is $>1$ at 50K.

 In order to explain the intriguing question of extra conduction in TMTSF-DMTCNQ, an other conduction mechanism should necessarily  be active in parallel with the single particle channel.  
 Since the existence of a collective conductivity seems to be the only way to reconcile fermiology and thermopower data in TM-DM under 12 kbar,  the salient question is about the actual nature of this collective contribution.
 
 At the beginning of the eighties a suggestion has been made 
 for the huge conductivity of order $10^5(\Omega.{\rm cm})^{-1}$ in TM-DM which is observed at low temperature under pressure. The interpretation of this conductivity in terms of a single-particle
conduction mechanism would lead to a huge mean free path of about 3000 $\AA$ at helium temperature. This possibility seemed to be
rather hard to accept at that time. Instead, an interpretation based on superconducting fluctuations induced paraconductivity was proposed\cite{Schulz81a}. We shall see in the following Sections that very large mean free paths have been measured in materials such as \tmp6   and they are even required to explain the stability of the 3D superconducting state which is very sensitive to the presence of non-magnetic defects. Therefore, the suggestion of a superconducting origin for the collective conductivity of TM-DM  should  be nowaday taken with a grain of salt.
\begin{figure}[h]
\includegraphics[width=0.8\hsize]{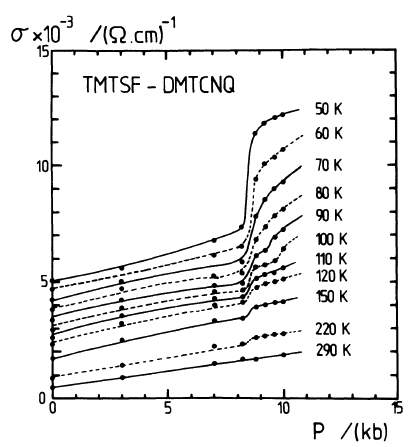}
\caption{\label{sigmaTMDMunderP}  Pressure dependence of $\sigma_{\parallel}$ in TMTSF-DMTCNQ above
42 K. $\sigma_{\parallel}$(300K) = 500 ($\Omega.{\rm cm})^{-1}$, Source: Fig. 1 [Ref\cite{Andrieux79b}, p. 1200]. }
\end{figure} 
An other possibility can also be considered. 
According to  the behaviour of the pressure dependence of the conductivity for different temperatures performed in a helium gas pressure apparatus, see Fig.~\ref{sigmaTMDMunderP}, the suppression of the Peierls insulating state is accompanied by a clear increase of the conductivity bearing some resemblance with \tq  where a dip in the $\sigma (P)$ curves is observed in the commensurability  regime, (Fig.~\ref{sigmaTQP}).  Consequently, the way the conductivity of TM-DM  behaves is  strongly suggestive for the existence of a Fr\"ohlich conducting channel becoming active outside a broad commensurability domain, i.e. above 8 kbar.

\begin{figure}
\includegraphics[width=1\hsize]{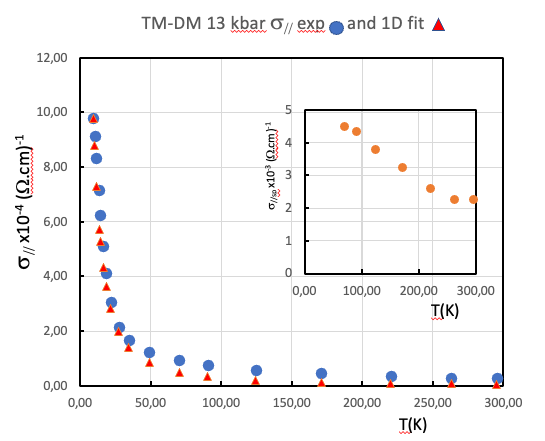}
\caption{\label{TMDM13kbfit1D} One-dimensional fit in $T^{-3/2}$ for the experimental conductivity of TM-DM at 13 kbar taken from Ref\cite{Andrieux79b} normalized at 10K, dots are experimental data while triangles are the theoretical 1D fit. The inset shows the  single particle contribution derived by subtracting the 1D collective contribution from the experiment.}
\end{figure}

Following the data analysis used for \tq, we have displayed on Fig.~\ref{TMDM13kbfit1D}  the  fit for one-dimensional  Fr\"ohlich fluctuations contributing to the longitudinal conduction  normalized to the experimental value of the conductivity under 13 kbar at 10K according to Ref.\cite{Andrieux79b}. In addition, subtracting the theory from the measured conductivity, we can derive the single particle contribution plotted in the inset of this figure. As we can see in the inset of Fig.~\ref{TMDM13kbfit1D}, $\sigma_{\parallel, sp}$ is dominant over the Fr\"ohlich  contribution  at 300K and reaches $\approx 5\times10^{3}$ ($\Omega.{\rm cm})^{-1}$ at low temperature, both channels contributing about equally around 100K. Such a behaviour is indeed reminiscent of \tq, see Fig.~\ref{sigmaTP}.
It is worth paying more attention to the very broad width of the commensurability regime in pressure   as displayed on Fig.~\ref{sigmaTMDMunderP} which is at least of 9 kbar. While at $P>9 $kbar, we have seen previously that the conduction is dominated by the Fr\"ohlich  mode of incommensurate fluctuations, we may suggest that below 9 kbar we are in presence of a state which is not  commensurate strictly speaking. Instead, in the conducting state above the metal-insulator transition at 42K commensurate and incommensurate  regions (discommensurations) could coexist. It is plausible that the proportion of incommensurate volume is evolving under pressure, reaching 100\% for incommensurate volume  only above 9 kbar. Thus, the observed S-shape anomaly in resistivity occurring at an increasing temperature with pressure, see Fig.~\ref{TMDM}, would mark the limit in pressure between a low pressure regime with a mixture of commensurate and in commensurate domains and a high pressure regime   regime in which the CDW is uniformly  incommensurate. The steep slope of this transition, with a 5 kbar difference between ambient temperature and 50K, is in a good agreement with the change in bandwidth due to the large thermal contraction   for \tq already mentioned above. 

To be noted that doubts had been expressed about the possibility of Fr\"ohlich conduction in this material under pressure\cite{Friedel82} arguing that long coherence lengths at low temperature should lead to a three-dimensional ordering of the one-dimensional fluctuations and consequently to an insulating ground  state. The argument expressed in the reference\cite{Friedel82} is understandable although the increase of conductivity noticed out of the commensurability domain on Fig.~\ref{sigmaTMDMunderP} and   subsequent dHSh results\cite{Bouffard83} seem to support the existence of a collective contribution to the conduction at least down 30-20K where a cross-over between 1D and semi-metallic electronic structure could take place.

In conclusion for TM-DM, a  transition 1D-3D is observed around 30K in the incommensurate state of this compound above 9 kbar, suppressing the Peierls ground state. In this  two-chain material, the closed FS  is thus a result  of interstack
hybridization with a value for the transverse overlap lying in-between that of \tq and HMTSF-TCNQ. Such an hybridization being responsible for a significant  lowering of the density of states at Fermi level\cite{Shitzkovsky78}  may in turn prevent the establishment of superconductivity although the Peierls transition is suppressed under pressure.



\subsubsection{Beyond TMTSF-DMTCNQ}
TM-DM is  a very  interesting system which should have deserved more work
 but  since all these phenomena were new and unexpected the effort was quickly put on a simpler
structure built on a single organic stack  comprising the novel and lucky  TMTSF molecule together with  an inorganic
monoanion  which was able to achieve the situation of a  quarter-filled band which prevails in TMTSF-DMTCNQ, at low pressure. 
Such a structure was already known from the
early and extensive  work of the Montpellier chemistry group who  synthesized and studied the series of isostructural
$\mathrm{(TMTTF)_{2}X}$ organic salts \cite{Brun77,Galigne78} where TMTTF is the sulfur
analog of TMTSF and X is a monoanion such as
$\mathrm{ClO_{4}^{-}}$,$\mathrm{BF_{4}^{-}}$ or $\mathrm{SCN^{-}}$  etc,.. All these compounds turned into strong insulators at low temperature  under atmospheric pressure.
It is the reason why they did not attract much attention until the day  when they
have been  revisited by Klaus Bechgaard who, taking advantage of   the results obtained  in  TM-DM, had the  intuition to synthesize similar radical cation salts but with the TMTSF molecule. It is the synthesis of this  $\mathrm{(TMTSF)_{2}X}$ series which has marked the beginning   of a new era in the physics and the chemistry of one dimensional conductors.

\subsection{Concluding the charge transfer era}
 Proposals of theorists in the 1960's, in search of new materials and new mechanisms likely to lead to the stabilization of superconductivity at temperatures higher than the 20-23K range of the time gave rise to a remarkably productive effort to synthesize new molecular crystals possessing electronic conduction properties unknown at the time.
 
 In 1973, the first molecular crystals  with electronic properties similar to those of metals in a wide range of temperatures appeared. These organic conductors whose protoptype is the TTF-TCNQ charge transfer complex, have led to numerous physical studies (crystallography, electrical and magnetic measurements, NMR, calorimetry, etc.) often under extreme temperature and pressure conditions which have stimulated numerous theories in the six years that followed. 
 
 What made this area of research so popular was on the one hand the hope that these systems could  eventually lead to high \tc superconductors and  on the other hand the  possibility to synthesize and grow  single crystals of large size and very high  purity to perform  sharp  physical investigations contributing to a better understanding   of the one-dimensional world.

\begin{itemize}

\item The main property of the \tq family materials is a crystal structure formed by parallel segregated columns of donors and acceptors along the $b$ stacking  axis.

 Consequently, the metal-like conduction is made possible by the overlap along the $b$ axis of the $\pi$ orbitals from HOMO and LUMO levels of the open shell donors and acceptors respectively.   The lowest electron energy on the TCNQ stacks occurs when all molecular orbitals are in-phase (at $k=0$ in reciprocal space) while the situation is inverted  for TTF stacks, leading for this stack to an energy maximum at the zone center.
Such a band crossing picture ensures that both bands
intersect at a single Fermi wave vector $\pm$$k_F$ in order to preserve the overall neutrality. Consequently, all
states between $-\pi/b$ and $+\pi/b$ are occupied with the restriction that between $-k_F$ and $+k_F$ occupied states belong to the TCNQ band while outside this domain
they pertain to the TTF band. In addition, the fact that the charges can delocalize in \tq shows that
the on-site Hubbard repulsion $U$ does not overcome the band energy $4t_{\parallel}$ gained in the band formation.

The feature of inverted bands opens many possibilities such as an incommensurate   charge transfer  becoming commensurate under pressure as it is the case for \tq in a narrow pressure window around 19 kbar.

\item Although the one dimensional crystal structure and  intermolecular interactions within each stacks  priviledge a band formation  with  planar Fermi surfaces at  $\pm$$k_F$,   non negligible although much smaller couplings exist as well along lateral directions.  Molecular overlaps between like-chains give rise to a warping of the 1D Fermi surface but overlaps between unlike-chains are destructive for the density of states at Fermi level and tend affect the shape of the surface leading at low temperature to electrons and holes semi-metallic pockets, namely a 1D to 3D cross-over for the Fermi surface. 

\item Whenever the temperature is larger than the transverse couplings, the Fermi surface looks  1D  and as such is likely to undergo a transition which lowers the energy of the electrons by a lattice modulation of such a wave length that it produces energy gaps at the Fermi level with a $2k_F$ periodic lattice modulation  and charge density wave, the so-called Peierls ground state which is the leading divergence in these compounds. Such a phenomenon has been thoroughly  studied in several members of the \tq series including the charge transfer complexes including conductors comprising various selenide fulvalene donors, TSF-TCNQ, TMTSF-DMTCNQ or HMTSF-TCNQ.

\item In addition, the chain structure has enabled to reveal a major feature of one-dimensional physics which is the existence   at high temperature of PLD-CDW fluctuations, those which condense into the long-ranged Peierls  ground state when the 3D character becomes dominant at lower temperature.
 These precursors known as  Fr\"ohlich precursors  are well established through their manifestation on X-ray scattering and on the conductivity  with a collective contribution. The latter contribution could be evidenced in \tq thanks to the possibility of making under pressure the charge transfer  commensurate with the  lattice, thus pinning the conductivity of collective origin. We believe a similar collective mechanism is at work in TMTSF-DMTCNQ under pressure.
 
\item Electron-electron repulsions though less important in this charge transfer series than in the radical cation  series of   organic superconductors  are visible through the tendency to prevent two electrons with opposite spins with a lattice modulation  to occupy the same $k$ state ending up  to fluctuating lattice modulations with a wave vector   $4k_F$. The situation is thus a superimposition of lattice distortion and strong electronic correlations (namely, the observability of  $2k_F$ and $4k_F$ X-ray diffuse scattering which is the signature 1D physics).

\item Finally, the stabilization of superconductivity turned out to be hopeless in these charge transfer conductors, possibly because of the negative role played by the interactions between unlike chains on the density of states at Fermi level.

\item However, the turning point  to superconductivity may have been guided by the commensurate quarter-filled conductor TMTSF-DMTCNQ in which the conduction band becomes actually half-filled with spinless Fermions once  the manifestations of strong electron repulsions observed experimentally are taken into account. 
\end{itemize}


\section{Organic conductors becoming superconductors: The Bechgaard salts} 
\subsection{Experimental evidences for organic superconductivity}

\subsubsection{Introduction}
 Based on metallurgical considerations, some authors at the end of the
1970's were considering 25–30 K as an upper limit for the superconducting \tc\cite{Matthias71} but for others
the saturation of superconducting critical   temperatures slightly above 20 K  was certainly a strong stimulus for the search of  both new superconducting materials and new pairing mechanisms in the framework of the BCS theory. 

\begin{figure}[h]
\includegraphics[width=0.95\hsize]{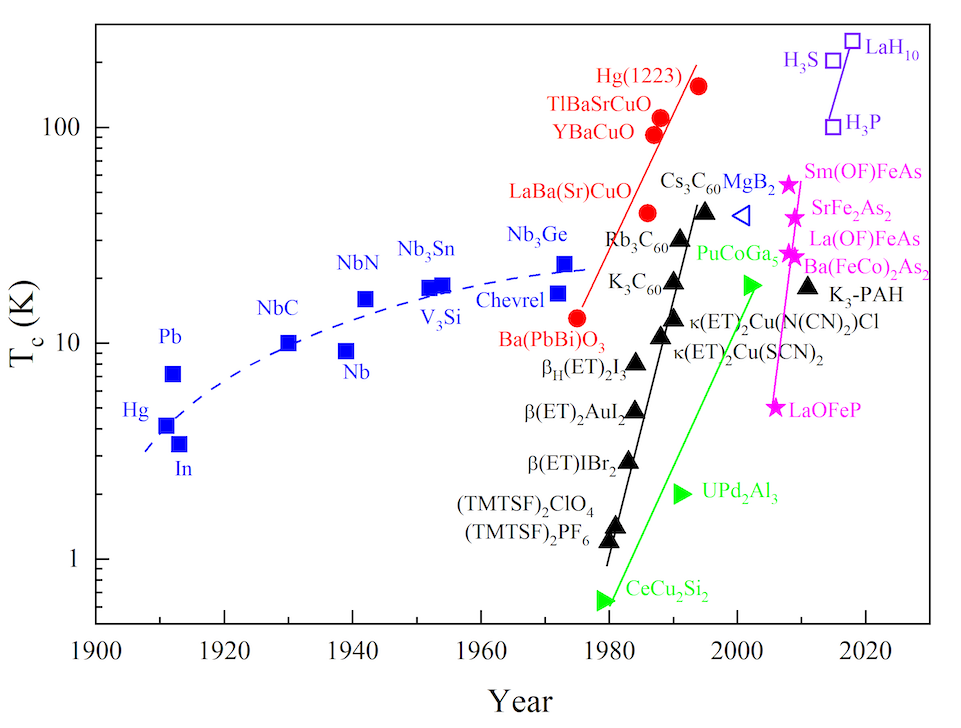}
\caption{\label{Tcyears} Evolution of \tc in various families of superconductors over the
years. Notice that superconductivity in hydrides requires  pressures in the megabar regime, Source: Fig. 1 [Ref\cite{Jerome12}, p. 633].} 
\end{figure}

As can be seen in Fig. \ref{Tcyears}, the period starting in the mid-1970's has been very fruitful in terms of progress in the field of superconductivity. The figure presents some superconducting materials and their critical temperature according to the year of their discovery. Obviously, they don't  represent at a given time, materials with the maximum \tc. 
It is interesting to note that most published  figures displaying the  \tc's of the various superconducting materials fail to mention  organic Q1D's compounds, which are nevertheless of major educational interest.
\begin{figure*}[t]
\includegraphics[width=0.8\hsize]{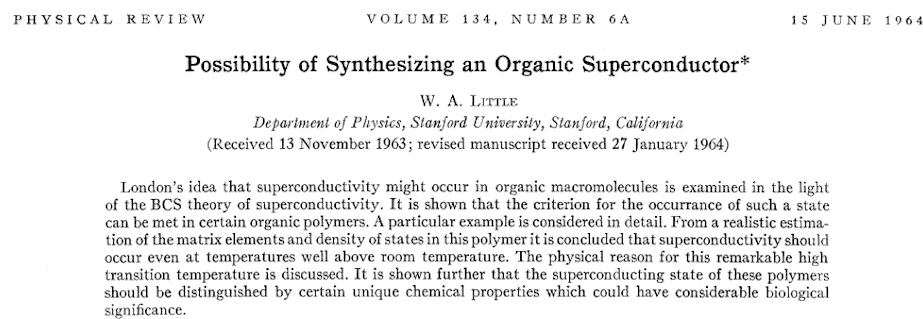}
\caption{\label{Little1964} Abstract of the seminal article published by W. A. Little in 1964.  }
\end{figure*} 

In addition to the development of new materials, extreme conditions have often been required to stabilize superconductivity,  as  illustrated with  the recent discovery of superconductivity near room temperature in hydrides   requiring extremely  high pressure conditions in  the megabar  domain\cite{Drozdov15}. 

Such is also the case for the  superconductivity in organic matter, which is the {\it raison d'être}  of this overview, since its discovery has required besides the skill of  a talented chemical engineering,    the joint use of hydrostatic pressures of the order of 10 kbar together with low temperatures $\mathrm{^{3}He-^{4}He}$ dilution refrigerators below the boiling temperature of helium  not so developed in the 1970's. Fortunately, chemical synthesis has rapidly made new organic superconductors stable under ambient pressure. The growth of very good quality single crystals  has also allowed the achievement of delicate experiments whose results could be confronted with the theory of quasi 1D conductors.

Designing new materials for superconductivity at high temperature  is the proposal of the  seminal paper of Bill Little  in 1964 (\emph{see} Fig.~\ref{Little1964}) in which the author suggested the possibility to use a virtual electronic excitation of polarizable molecules grafted on a one-dimensional polymer such as polyacetylene    for electron-mediated pairing instead of the  electron-phonon mechanism of the celebrated BCS theory\cite{Bardeen57}. The idea of Little was indeed deeply rooted in the extension of the isotope effect proposed by BCS since we could expect  an enhancement for \tc of the order of $(M/m_e)^{1/2}$ where  $m_e$ is the small electron mass compared to a much larger ionic mass $M$. It is also a virtual excitonic excitation of electronic origin which was  suggested in the same years by Ginzburg in layered structures\cite{Ginzburg68}.

\begin{figure}[h]
\includegraphics[width=0.9\hsize]{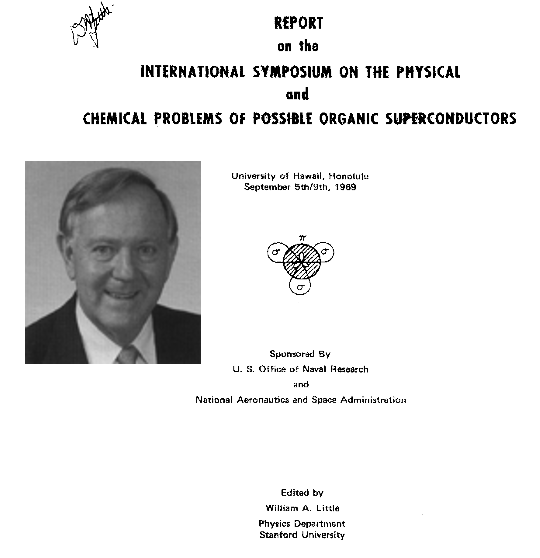}
\caption{\label{Honolulu} Masthead of the Proceedings of the Conference on Organic Superconductors organized by W. A. Little and  held at Hawaii in 1969 with the signature of the organizer.  }
\end{figure}
The conference held in Hawaii in 1969 brought together a large number of top scientists in physics and chemistry, both theoreticians and experimentalists, and was certainly the trigger for a worldwide research effort with the goal of synthesizing organic conductors undergoing a transition towards superconductivity, \emph{see} Fig.~\ref{Honolulu}. 
It took another decade to stabilize superconductivity in an organic molecular conductor. 

This has been an essential decade (from 1965 to 1974)) for the development of organic synthesis and the growth of new materials marked in particular by the discovery of the first conductors belonging to the class of inverted band charge transfer compounds, the so-called \tq series. It was also the decade during which physicists managed to develop with much profit new cristallographic X-ray and neutron diffraction  techniques and extend transport, magnetic and NMR measurements  at high pressures and down to very low temperatures. Last but not least, the development of this field has  only been possible thanks to a remarkable and close cooperation between synthetic chemists, experimental and theoretical physicists enabling the discovery of long predicted behaviours in one-dimensional conductors. An ultimate goal  was to avoid the appearance of  a metal to insulator transition so characteristic of one-dimensional conductors, as we have seen previously for compounds undergoing a Peierls instability.

Hoping to be able to stabilize superconductivity, the research was directed towards the increase of the interchain coupling either by the application of a large hydrostatic pressure or by short $\mathrm{S-N}$ or $\mathrm{Se-N}$ contacts between donor and acceptor pairs. As shown in Sec.~\ref{under higher pressure}, high pressure although responsible for the discovery of numerous basic phenomena related to 1D physics, has failed to stabilize superconductivity in the \tq family of charge transfer conductors. 

However an important step has been accomplished at the end of the 70's by the Copenhagen group led by Klaus Bechgaard, as highlighted in Sec.~\ref{Bechgaard salts decade}, with the development of the synthesis of tetraselenafulvalene molecules together with the electrochemical growth of very high quality  radical cations single crystals.

First, the selenide charge transfer compound TM-DM, (TMTSF-DMTCNQ)\cite{Andersen78} enabled the suppression of the metal-insulator under a pressure in excess of 9 kbar keeping a conductivity of the order of $10^5$ ($\Omega.{\rm cm})^{-1}$ at helium temperature even if the reason for the existence of this low temperature metallic (or semi-metallic) state is still not fully clarified\cite{Andrieux79c}.  

Second, emphasis has been put on the molecule TMTSF since for the first time in the experimental investigation of the \tq series the band filling of a charge transfer conductor, TM-DM, was found commensurate with 1/4 band filling under ambient pressure.

The new and exciting properties of TM-DM have provided a strong motivation to synthesize stoichiometric one chain  radical cations compounds with  TMTSF  emphasizing the role of the donor molecule   in a commensurate band filling environment.  

Such radical cations salts were already known in 1979 with the ($\mathrm{TSeT)_{2}Cl} $\ conducting salt known to retain a metallic state down to 24K\cite{Schegolev79} and remaining metallic under pressure down to helium temperature\cite{Laukhin80}. 

Fortunately, the Montpellier chemistry group\cite{Galigne78} had synthesized in 1978 a series of
isomorphous radical cationic conductors based on TMTTF (the sulfur analog of the TMTSF molecule) with an inorganic
mono-anion, namely, \tms2x,  with structures even simpler than those of the two stacks charge transfer compounds.  These materials were all poor conductors  under ambient conditions becoming semiconductors at low temperature\cite{Delhaes79} but they provided the motivation for the synthesis of a similar series based on the TMTSF molecule and they were inspiring for the Copenhagen group  expert in selenium chemistry. 
\begin{figure}[h]
\includegraphics[width=1\hsize]{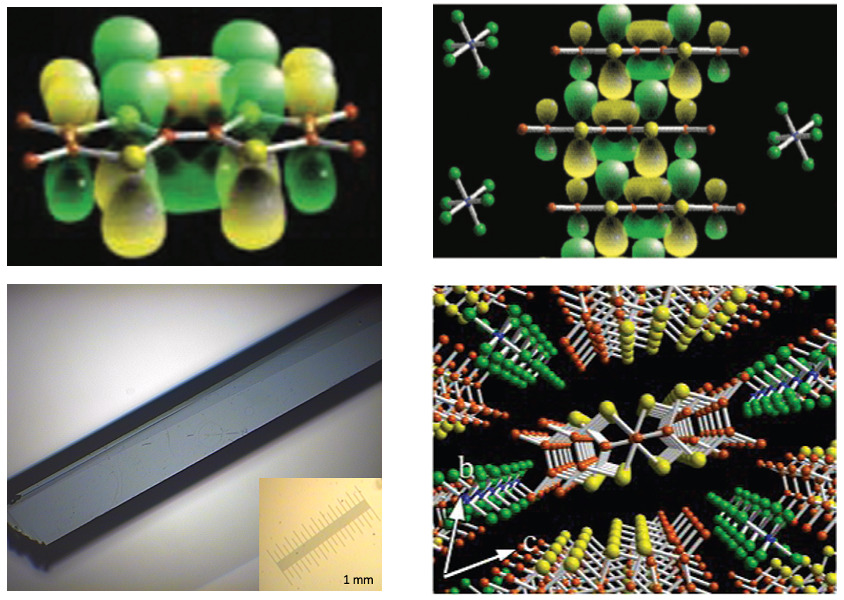}
\caption{\label{TMPF6} (Top left), The TMTSF molecule. (Top right) Side view of \tmc along the $b$ axis showing ordered  \cl anions. (Bottom left) A \tmc single crystal viewed along the $c^\star$ axis. (Bottom right) Stacking of TMTSF molecules along the $a$ axis, Sources: Fig. 5.7, 5.8, 5.9 [Ref\cite{Jerome10}, p.~160]. We thank  P. Auban-Senzier from the Orsay lab for the photograph of the crystal. }
\end{figure} 

This group  succeeded in 1979 the
synthesis of a new series of  radical cationic conducting salts all based on the  TMTSF  molecule\cite{Bechgaard80}   namely, \tms2x   where X are inorganic mono-anions with various symmetries  such as  centrosymetrical \pf, \as, tetrahedral \4fb or triangular \no anions  with a  uniform cationic stacking.  

\begin{figure}[h]
\includegraphics[width=0.7\hsize]{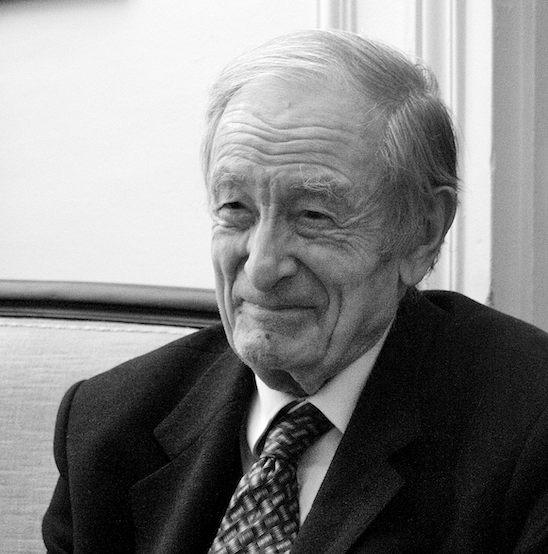}
\caption{\label{JF}  Jacques Friedel (1921-2014). An enthusiastic supporter of low dimensional physics which he used to teach in his graduate course at the University of Orsay. We thank Jean-François Dars  who kindly agreed to give us one of his pictures of Professor Friedel.
}
\end{figure} 
The success of the Orsay group is largely due to the open-mindedness of Professor J. Friedel, Fig.~\ref{JF}, one of the founders of the solid state physics laboratory, who despite his commitment to the defense of physics at the national and international level   has always been a strong supporter for the  field of low-dimensional conductors.

\subsubsection{Basic evidences for superconductivity}
\begin{figure}[h]
\includegraphics[width=0.85\hsize]{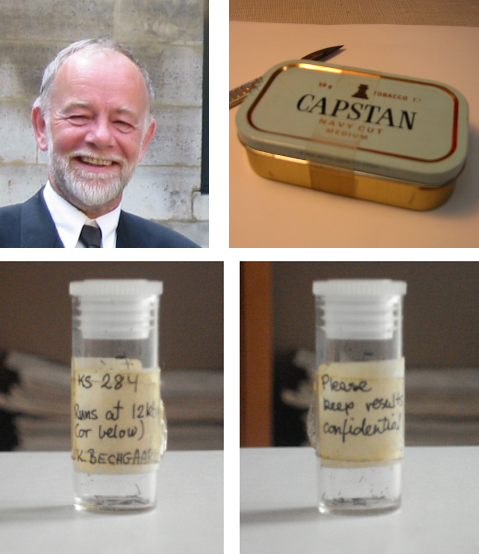}
\caption{\label{sampletubes} Klaus Bechgaard (1945-2017) used to ship his samples to Orsay in these famous tobacco boxes. It is interesting to read on one side of the tube KS 284 "\emph{runs at 12 kbar or below}" and on the other "\emph{keep results confidentia}l".  Great premonition! }
\end{figure} 
The compound with X= \pf \, caught much attention 
since the conductivity reaches the value of about
$10^{5}(\Omega. {\rm cm})^{-1}$ at 12K  before the onset of an insulating ground state at this temperature  with still a strong temperature dependence. In addition, very  large and nice-looking crystals good for  transport properties measurements could be grown via the electrochemical route cited in Sec.~\ref{Bechgaard salts decade}.

For historical reasons,  Fig.~\ref{sampletubes}  shows the tube sent by K. Bechgaard which contained the samples from batch KS 284 that were synthesized  in Copenhagen. It is instructive to read K. Bechgaard's comment on the tube, \emph{runs at 12 kbar or below} on one side and \emph{please keep result confidential} on the back side. The authors of Refs.\cite{Bechgaard80,Bechgaard80c} also noticed that the temperature of the metal to insulator transition is apparently not enhanced by commensurability since the period of the presumably Peierls distortion  amounts to two crystallographic units according to the stoichiometry. All these reasons were strong motivations for further investigations of this compounds.
\begin{figure}[h]
\includegraphics[width=1\hsize]{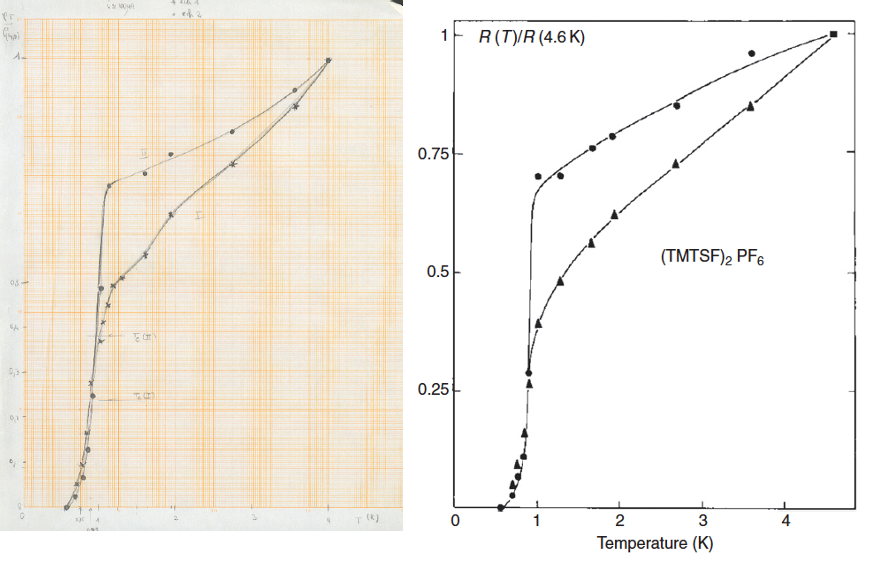}
\caption{\label{Firstsupra} (Left) \emph{Fac simile }of the lab book of the doctoral student, Alain Mazaud, displaying the first observation of superconductivity from the resistance   of two samples of  \tmp6 under 9 kbar at the end of November 1979. This time was before the common use of computerized data acquisition. (Right) Sources: Fig. 3 [Ref\cite{Jerome80}, p.96]j. }
\end{figure} 

The behaviour of   transport properties together with the absence of any static lattice distortion\cite{Pouget81,Pouget97} concomitant with the metal-insulator transition, were new features in a field still dominated until then by the Peierls phenomelogy and this stimulated  further investigations  under pressure suppressing the insulating state at liquid helium temperature under a pressure of about 9kbar. The finding in a first step of a very small and \emph{still non-saturating} resistivity at 1.3 K.The strong temperature dependence of the resistance below 4.2K observed in Fig.\ref{Firstsupra} was by no means an experimental artefact. It has been explained by a strong single particle scattering against spin fluctuations, see Sec.\ref{conductingstate}. Since a finite and  temperature independent  resistivity is usually observed in conventional metals at very low temperature, such an experimental finding    was a strong enough motivation for the Orsay low temperature group with our colleague Michel Ribault  to trigger further  studies and develop a new equipment adapted to  measurements  under hydrostatic  pressure in a specially designed dilution refrigerator down to much lower temperatures.

Rather quickly by the end of November 1979 a narrow transition of the \tmp6  longitudinal resistance  to a non-measurable  value was observed at 0.9K under 9 kbar as shown on Fig.~\ref{Firstsupra}. As this zero resistance state was reproducible with different samples and   easily
suppressed by a moderate magnetic field transverse to the most conducting direction, superconductivity was announced in an article published in February 1980\cite{Jerome80}. 

\begin{figure}[h]
\includegraphics[width=0.9\hsize]{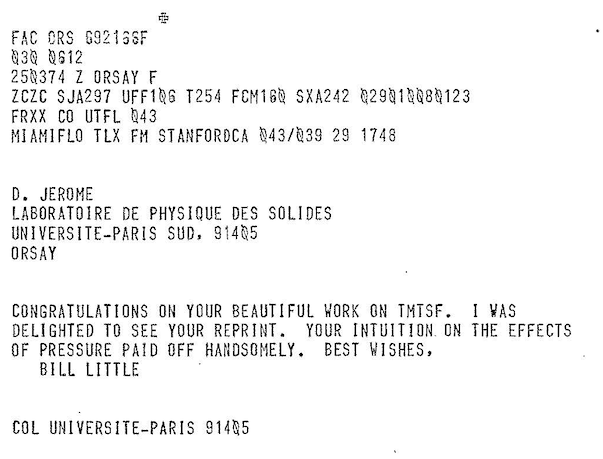}
\caption{\label{TelexLittle} Telex received by D. Jerome from W. A. Little, early 1980.  }
\end{figure}

This result has aroused much interest in the scientific world, as evidenced by numerous reactions, the first being from our colleague Bill Little who since 1964 has been supporting this field  enthusiastically, \emph{see} Fig.~\ref{TelexLittle}.
The first international event after the publication of these results took place in August 1980 at Helsing\"or, Denmark, with the International Conference on  Low Dimensional Synthetic Metals, which  which brought together 135 active researchers in this field. In this regard, it is interesting to mention the Conference summary  given by W. A Little  published in the conference Proceedings. It is illuminating to have a look at the words of Bill Little in the foreword of these Proceedings. \emph{"This announcement generated a certain euphoria, a sense of relief and represents to many a psychological barrier successfully surmounted. It is a great encouragement to the synthetic chemist and experimentalists in the field. One should note that this was not the first Conference on organic superconductors . Some eleven years ago an International Conference on possible Organic Superconductors was held in Hawaii for which I was responsible. In the publication of the proceedings of that meeting a farsighted journal editor truncated the title simply to the "Conference on Organic Superconductors" and it was thus with a sense of relief that I received the news of the properties of \tmp6}".

The finding of superconductivity was also reinforced by the observation of a large diamagnetic contribution below 0.9K\cite{Jerome80} disappearing under a transverse magnetic field of about 1 kOe at 0.1K applied along the c$^\star$ axis. The results of AC susceptibility  suggested a strong expulsion of the magnetic flux from the sample and type II superconductivity but it is the DC Meissner experiment which brought the proof of a  bulk superconductivity\cite{Andres80}. 

What is so special with \tmp6, the prototype of the so-called Bechgaard salts, unlike previously investigated
\tq, is the magnetic origin of the ambient pressure insulating state which contrasts with the Peierls-like ground states discovered until then in \tq.

Magnetic susceptibility measurements\cite{Scott80} indicate that the metal-insulator transition is  remarkably low in view of the commensurate Fermi wave vector provided data showing that the low temperature semiconducting phase is not simply Peierls in nature. The Bell Telephone research group made the suggestion  that a SDW modulation could be responsible for the disappearance of the metallic state below 12K in order to explain the non-linear electronic properties of the insulating state\cite{Walsh80}.

\begin{figure}[h]
\includegraphics[width=1\hsize]{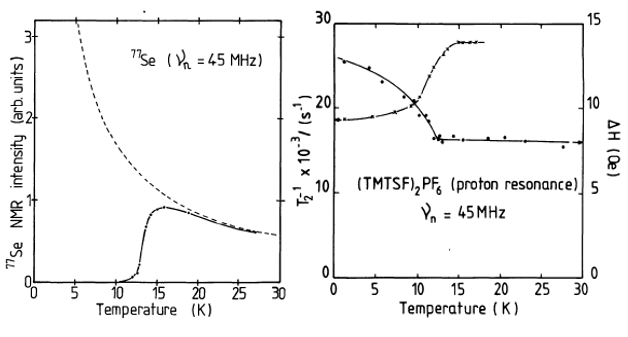}
\caption{\label{SDWNMR}  (Left) The $^{77}$Se (I=l/2) NMR signal versus temperature in \tmp6.The signal vanishing observed below 12K, instead of a Curie law temperature dependence (dashed curve), is due to the inhomogeneous broadening by internal magnetic fields in the  state. (Right) The proton NMR signal in \tmp6 revealing an inhomogeneous line broadening and a concomitant  decrease of the  homogeneous line width (1/$T_2$) below 12 K, Source: Fig. 1, 3 [Ref\cite{Andrieux81}, p. 88].  }
\end{figure}

However, it is the $^{77}$Se  NMR study that subsequently validated this hypothesis on a microscopic scale\cite{Andrieux81}, \emph{see} Fig.~\ref{SDWNMR}. The ground state of \tmp6 turned out to be a spin-density-wave state\cite{Mortensen81}, similar to the predictions of Slater Hartree-Fock theory when applied in lower dimensions\cite{Slater51}, and akin to the approaches by Lomer\cite{Lomer62} and Overhauser\cite{Overhauser62} for metals.

The onset of itinerant antiferromagnetism opens a gap at Fermi level, and since the Fermi surface is nearly planar, the gap develops over the entire surface leading in turn to a semi-metallic ground state.  These experimental findings had a decisive influence on the theoretical development of the superconducting pairing in these quasi 1D organic materials.

The magnetic origin of the insulating ground state of \tmp6 was thus the first experimental hint for the prominent role played by correlations  as induced by repulsive interactions {\color{black} and that the electron-phonon interaction is relatively weak } in these organic conductors\cite{Andrieux81}. 
The competition between a charge density-wave superstructure  and superconductivity had already
been observed, for instance, in the transition metal dichalcogenides layer crystals under pressure such as 2H–NbSe$_2$\cite{Friend79,Molinie74}. However, in the latter situation superconductivity preexisted in the low-pressure metallic phase with a Peierls distortion due to the nesting of some regions of the
Fermi surface\cite{Wilson75}, decreasing the density of states at the
Fermi level. As noticed by Friedel\cite{Friedel75}, the effect of pressure
on the \tc of 2H–NbSe$_2$ compounds is likely to result in  an
increase of $N(E_{F})$ by reducing the gaps on some parts on
the Fermi surface leading to a concomitant enhancement of \tc. It is a situation at variance with what is encountered
in \tmp6 since the whole single  component Fermi surface is involved in the transition and the SDW phase is truly insulating. However, the analogy with dichalcogenides may make sense close to the border between SDW and  superconducting phases when $T_{SDW}$ is already strongly depressed.   The reasons why superconductivity is optimized at the border with the SDW state are likely to be quite different from what has been suggested for layer compounds, as it will be further discussed in Section.~\ref{Genericphase}.

Because of its past experience in the physics of dichalcogenides  the Orsay high pressure group was not discouraged by the finding of an insulating ground state of \tmp6 stable under ambient pressure and quickly  undertook an investigation of this compound  under pressure which gave rise to the discovery of the first organic supertconductor\cite{Jerome80}. 

\begin{figure}[t]
\includegraphics[width=0.75\hsize]{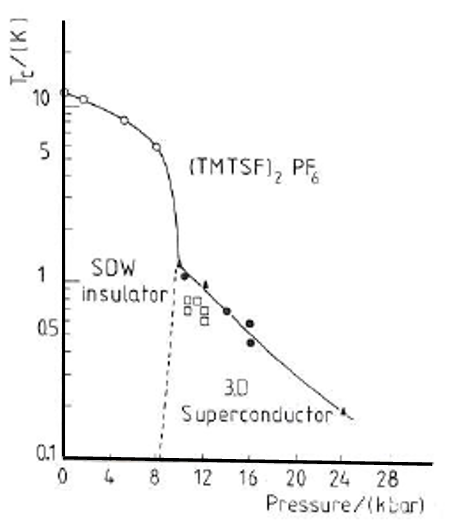}
\caption{\label{TPPhasediagram} The initial version of the low temperature phase diagram of \tmp6.
The various experimental points on the superconducting transition line
have been compiled in the literature, (triangles, Orsay\cite{Jerome82,Schulz81}, dots, Bell
Labs\cite{Andres80} and open squares, IBM\cite{Greene80}). A closer investigation of the
 phase diagram for \tmtsfasf6\cite{Brusetti82a} reveals a 
reentrance phenomenon near the triple point at \pc$\approx$8.5kbar,
\tc$\approx$1.2 K, Source: Fig. 78 [Ref\cite{Jerome82b}, p. 166].  }
\end{figure}

Shortly after the announcement of superconductivity, the IBM group performed a high pressure experiment revealing  a pressure dependence for \tc of about 0.1K per kbar  and also a coexistence of superconductivity with SDW in the low pressure regime\cite{Greene80}. Subsequently, the pressure dependence of the superconducting phase of \tmp6 has been extended up to 24kbar where \tc amounts to 0.2~K\cite{Schulz81a} and a  $T-P$ phase diagram  could be proposed for \tmp6\cite{Jerome82}
, \emph{see} Fig.~\ref{TPPhasediagram}.  The phase diagram thus obtained under pressure figured as the first clear example of a superconducting dome emerging on the brink of a parent SDW state.

Thanks to developments of chemistry, a generic diagram for one-dimensional conductors mixing high pressure and  chemistry including both \tmtsf2x and the isostructural series \tmttf2x
was proposed\cite{Jerome91}. 

This is how different ordered states coming from one-dimensional quantum liquids evolve towards superconductivity and ultimately to a Fermi liquid behavior could be established on solid experimental grounds through the whole generic diagram. This is to be covered by the following section.

\subsection{A generic \tm2x phase  diagram}
\label{Genericphase}
In the mid-1980's first, the isostructural family comprising the
sulfur  molecule TMTTF with the same series of monoanions was synthesized and second, high pressure techniques enabling access to pressures above 30 kbar under quasi-hydrostatic conditions at low temperature  have been developed\cite{Jaccard01}.
 
 It was realized that
\tmttf2x and \tms2x salts both belong to a unique
class of materials with the possibility of evolving from one to the other by means of pressure forming in turn  a generic \tm2x phase diagram\cite{Jerome91}
sketched on Fig.~\ref{Generic}. The investigation of this series as a whole under pressure has thus been extended to  centro-symmetrical anions such as \as\cite{Brusetti82a,Itoi07}, \sb\cite{Itoi08}, \ta\cite{Parkin81a}, or \br\cite{Parkin82a,Balicas94} in the case radical cation TMTTF, and also   with non-symmetrical anions \re\cite{Tomic89a} and \cl\cite{Parkin81a}. 

\tmps, although one of 
the most insulating compound of the phase diagram on Fig.~\ref{Generic}, is of particular interest  not only because it can be
made superconducting at low temperatures under a pressure
of 45 kbar,\cite{Wilhelm01,Adachi00} but also because its position in the diagram allows, thanks to the adequate adjustment of pressure, to explore not only all the ordered  phases at low temperature but the  also to provide experimental evidences for the existence  of 1D quantum  liquids as precursors of the phase transitions. These govern essentially the non ordered state in the left hand side of the phase diagram and evolve on the right  towards a 3D, though  anomalous,  metal above superconductivity under pressure.

\subsubsection{1D quantum liquid physics in a nutshell}
\label{1DPhysics}

\begin{figure}[t]
\centerline{\includegraphics[width=1\hsize]{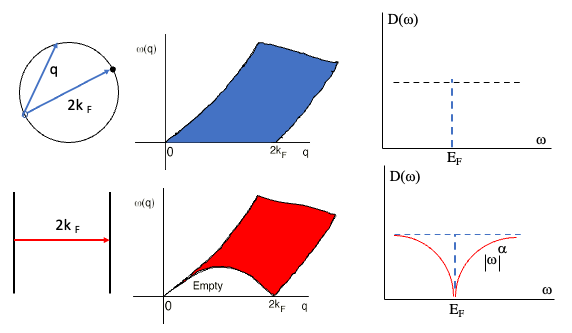}}
\caption{Basic differences between Fermi and Luttinger liquids. (Top) Electrons-holes excitations are possible in a FL for all vectors from 0 to $2k_F$ leading to a constant density of states at Fermi level. (Bottom) In a LL liquid  the only excitations are those with wave vector 0 or  $2k_F$  and the density of quasi particle states follows a power law around the Fermi energy.}
\label{1Dexcitations}
\end{figure}

\begin{figure}[h]
\centerline{\includegraphics[width=1.05\hsize]{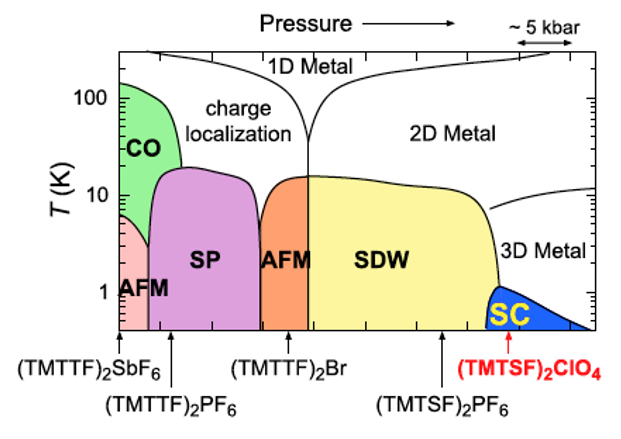}}
\caption{\label{Generic}A generic diagram for the \tm2x family. The horizontal tics correspond to a 5 kbar interval. All colored phases are long-range ordered. The origin of the pressure scale is set for the compound \tmttfsbf6 according to its ordered state at ambient pressure. Notice that all compounds marked on this figure but \tmc  exhibit superconductivity provided a high enough pressure is applied, $\mathrm{(TMTTF)_{2}PF_{6}}$\cite{Jaccard01}, $\mathrm{(TMTTF)_{2}Br}$\cite{Parkin82a} and $\mathrm{(TMTSeF)_{2}PF_{6}}$\cite{Jerome80}.  The highest pressure investigated for superconductivity in this series of organic salts is for the compound  $\mathrm{(TMTTF)_{2}BF_{4}}$ which requires a pressure of 33 kbar for a maximum \tc at 1.5K\cite{Ruetschi09}
}
\end{figure}

Given the place that one-dimensional physics holds in the non ordered states on the left of the  (TM)$_2X$ phase  diagram of Fig.~\ref{Generic}, it is helpful to recall  some basic  features of quantum physics of the interacting electron gas in 1D \cite{Emery79,Solyom79,Schulz95,Voit95,Giamarchi04}.  To do this different approaches has been extensively developed over the years,  like the fermion many-body technique,  the boson representation and of course numerical techniques. In this brief overview, we will focus on the boson representation. First, at  low enough energy the electron spectrum $\epsilon_k$ can be considered as essentially linear with respect to the  1D Fermi points  $\pm k_F$, namely ${\epsilon_k\approx v_F(|k|-k_F)}$, where $v_F$ is the Fermi velocity. In these conditions the  hamiltonian of electrons with spin  can be  rigorously transposed into a  bosonic one  associated to long wavelength spin ($\sigma$) and charge $(\rho)$ density excitations, each with a linear acoustic dispersion relation,  $\omega_{\sigma,\rho}=u_{\sigma,\rho}|q|$, where $u_{\sigma,\rho}$ is the velocity for each sound mode.

In the model,  interactions are given by the set of different scattering processes of electrons  that can be projected on the two Fermi points. The above bosonic representation remains valid so that  the low energy properties of the electron system become entirely governed by collective sound like  excitations that can be described as vibrating strings for spin and charge degrees of freedom. Both strings are  decoupled, a feature  known as spin-charge separation, and each of them  is characterized solely by two parameters, a velocity $u_{\sigma}(u_{\rho})$ and a stiffness constant $K_{\sigma}(K_{\rho})$, both  being in general non universal, that is, dependent on interactions. These excitations  are the true eigenstates of the system and have actually no available phase space to decay.  Therefore in contrast to the situation found in a Fermi liquid in higher dimension, fermion quasi-particles are absent at 
the Fermi level for a 1D system of interacting electrons at zero temperature
(\emph{see} Fig.~\ref{1Dexcitations}). This is illustrated by the   power law decay in energy predicted of  the fermion density of states at the Fermi level,
\begin{equation}
N(\omega) \propto |\omega|^{\alpha}
\label{DOS}
\end{equation}
with $\alpha$ related to $K_\rho$,
\begin{equation}
\alpha = {1\over 4}(K_\rho+1/K_\rho -2).
\label{alpha}
\end{equation}
The electron gas model for 1D conductors  with repulsive Coulomb scattering matrix elements with  only small momentum transfer, noted as $g_2$ and $g_4$ couplings in the g-ology jargon of the model, corresponds  to the exactly solvable  Tomonaga-Luttinger (TL)  model  with purely harmonic string excitations in both spin and charge sectors, with $1>K_\rho>0$ and $K_\sigma=1$ (due to rotational symmetry of the spins). The TL model is generic of the so-called Luttinger liquid phenomenology where  in addition to the Fermi level density of states (\ref{DOS}),  the $2k_F$ density-wave and superconducting susceptibilities develop power law behaviors in temperature with exponents that are non universal namely, dependent on interactions. This is the case of the antiferromagnetic or SDW susceptibility  which, aside from logarithmic corrections, reads
\begin{equation}
\chi_{\rm SDW}(2k_F,T) \propto T^{-\gamma_{SDW}}
\label{SDWKi}
\end{equation}
with the exponent
\begin{equation}
\gamma_{\rm SDW} = 1 -K_\rho,
\label{SDWexp}
\end{equation}
 which is congruent with a singular growth of SDW correlations at $T\to0$ for repulsive interactions.

One can go beyond the TL model by including other possibilities of scattering process. The backward scattering, noted $g_1$,  is one of these. For this process, particles near the Fermi points  exchange  momentum of the order of $2k_F$   introducing non linearity in the string excitations in  the spin sector.  For repulsive $g_1>0$, however, this non linearity turns out to be irrelevant, in the sense that as the energy distance from the Fermi level goes to zero this process becomes vanishingly small in amplitude. One then recovers an effective TL model in the low energy fixed point limit. Although less pertinent to (TM)$_2$X materials for which interactions are expected to be dominantly repulsive, it is worth mentioning that the outcome would have been completely different  in the attractive case ($g_1<0$), where $g_1$  then becomes relevant and evolves toward strong attractive  coupling at low energy with the formation of a gap for spin excitations alone. The presence of a gap in one of the two sectors   defines a different quantum liquid known as a Luther-Emery liquid\cite{Emery79}.

Another important extension to the TL model applies for systems like (TM)$_2$X. These materials actually  consist of  weakly dimerized stacks that  introduces a small dimerization gap $\Delta_D$  in the middle of the three-quarter filled band in terms of holes (or quarter-filled in terms of electrons) which is fixed by the 2:1 stoichiometry (\emph{see} Table~\ref{dimer})\cite{Emery82,Barisic81,Penc94b}. The spectrum then splits into an empty upper band and  a half-filled  lower band having one hole per  unit  cell with one  dimer  of TM molecules\cite{Grant82}. This underlying lattice effect leads to half-filling umklapp scattering processes, noted as $g_3$,  which  can transfer two holes from one Fermi point to the other. This is made possible {\it via} the inclusion  of the reciprocal lattice vector $G$ in the momentum conservation of electrons at the Fermi points the latter  taking the special value  $G=4k_F(=2\pi)$ at half-filling. Following the example of backward scattering, the presence of $g_3$ introduces non linearity in the string excitations but this time for the charge whose impact is of major importance. As   a relevant coupling, its amplitude grows at lower energy or temperature favoring the formation of a  gap $\Delta_\rho$ in the charge excitations known as a Mott  gap. The spin part, as we have seen, remains gapless.  The 1D electron system then becomes at low energy a  Mott insulating  Luther-Emery liquid  with the concomitant downward renormalization of $K^*_\rho\to 0$ to its minimum value in the low energy or temperature limit. According  to (\ref{SDWexp})   this yields a maximum power law singularity for the rotationally invariant SDW susceptibility ($ \chi_{\rm SDW} \sim T^{-1}$).  
For weakly dimerized systems, ${\Delta_D\ll E_F}$, the bare amplitude $g_3 \approx g_1 \Delta_D/E_F$ is weak and as a result the Mott insulating gap regime is pushed back at   lower energy values ($\Delta_\rho \ll E_F$).   

The observation of an insulating charge localised state in the intermediate temperature range on  the  left hand side   of the diagram of Figure~\ref{Generic}  for more dimerized sulfur based compounds\cite{Coulon82} has been early interpreted as the consequence of one-dimensional physics coming from electron-electron umklapp scattering at half-filling\cite{Emery82,Caron86,Brazo85} (\emph{see} Sec.~\ref{model Quasi-1D} for more experimental consequences).

The actual electronic state preceding the charge gaped one in (TM)$_2$X materials shows a   metallic resistivity at high temperature with $d\rho/dT>0$\cite{Coulon82}, rather than $d\rho/dT<0$ predicted for umklapp momentum dissipation in the weakly dimerized case at half-filling\cite{Giamarchi91,Shahbazi15} - {\color{black} neglecting here all other sources of inelastic scattering such as phonons}. This  has stimulated another route of interpretation for the origin of the charge gap \cite{Giamarchi97}. Neglecting dimerization, the band of the TM organic stack is actually 1/4-filled in terms  of electrons. The corresponding  lattice commensurability   then yields a different kind of Umklapp scattering noted $g_{1\over 4}$, which is also responsible for non linearity in the excitations of the charge string alone. In terms of the on-site Hubbard repulsion parameter $U$ alone, for instance, one finds a bare amplitude of the form $g_{1\over 4} \approx U(U/E_F)^2$. This corresponds to a  three-particle interaction linked to the higher order fourfold commensurability with  the undimerized lattice. As such, it is an irrelevant coupling. However,  the strength of interactions as a whole  can reverse the situation and make $g_{1\over 4}$ a relevant coupling. This indeed occurs with the proviso that $K_\rho$ falls below the critical value  $ K_{\rho,c} =1/4$, a condition that  can be reached when the  long-range part of the Coulomb interaction becomes sufficiently strong. In such a case an insulating charge gap $\Delta_\rho$ is present and is usually referred to as a charge ordered or a Wigner  state\cite{Mila93}, whereas the spin degrees of freedom remain gapless.

\begin{figure}[h]
\centerline{\includegraphics[width=0.65\hsize]{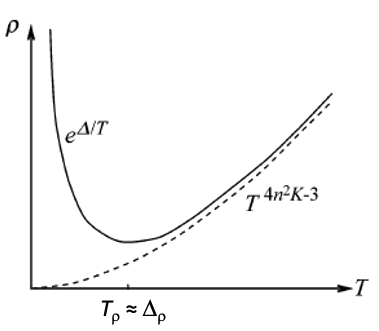}}
\caption{\label{Mottgap} The expected behaviour for the dc transport of a charged gap  insulator. At high temperature the power law dependence provides access to the Luttinger parameter $K_{\rho}$, Source: Fig. 16 [Ref\cite{Giamarchi04a}, p. 5047].
}
\end{figure}

Since  the stacks are weakly dimerized in practice one may expect that both Umklapp terms are  present and interfering with one another\cite{Tsuchiizu01}.  This may reflect in electrical transport. In the gapless  1D regime for $T>\Delta_{\rho}$, the resistivity coming from electronic Umklapp dissipation considered above has been shown to vary according to the power law\cite{Giamarchi91,Giamarchi04a}
\begin{equation}
\rho(T) \sim T^\theta
\label{rhoTh}
\end{equation}
    with the exponent
\begin{equation}
\theta=4n^{2}K_{\rho}-3,
\label{theta}
\end{equation}
 where $1/n$ is the filling of the band. At quarter filling $n=2$ and one gets a metallic regime when $K_\rho$ is not too small, with $\rho(T)\sim T^{16K_\rho-3}$, while at half-filling, $n=1$ and  $\rho(T)\sim T^{4K_\rho-3} $, which is most likely showing non metallic resistivity in lowering the temperature. Therefore if one restricts the mechanisms of momentum dissipation in transport to electronic Umklapp alone, the metallic behavior seen  in sulfur based (TM)$_2$X compounds  above the charge localisation in Fig.~\ref{Generic}  is most likely attributable to quarter-filling Umklapp scattering (See Fig.~\ref{Mottgap}).\cite{Giamarchi97} A crossover  to the second  Mott precursor regime at half-filling  then appears as   an interesting possibility for these weakly dimerized materials \cite{Tsuchiizu01}.


The generic phase diagram for the \tm2x family  displayed on Fig.~\ref{Generic} is actually based on  the sulfur compound \tmttfsbf6 taken for the origin of the pressure scale although \tmps, \tmp6 and \tmc have been the subject of most high pressure measurements because they are the samples that have focused the efforts of chemists and also because these samples can be grown as  high quality single crystals of fairly large size for physical measurements.  
Notice that \tmc  is the only compound in this family to exhibit superconductivity under ambient pressure which explain why some very delicate experimental investigations reported in this article have only been performed on this latter compound so far. In order to to appreciate the wealth of this  generic diagram and establish a link with the theory of quasi one dimensional conductors we shall base the presentation mainly on two compounds, \tmps and \tmp6 which have been thoroughly investigated using various experimental approaches. Thanks to the high pressure work they are able to exhibit practically properties of low dimensional physics.
\begin{figure*}[t]
\includegraphics[width=0.9\hsize]{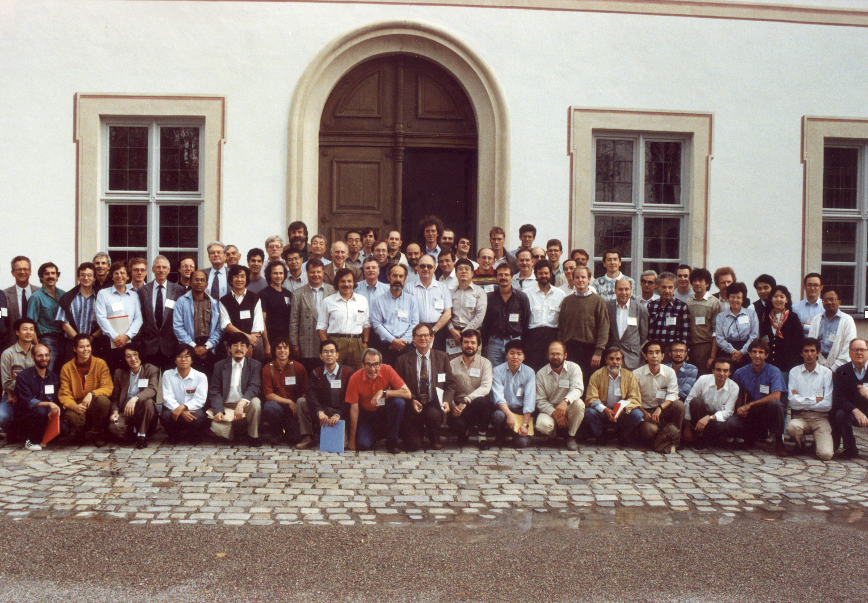}
\caption{\label{IRSEE91}  A large part of the Low Dimensional Organic Conductors family at the Gordon Conference (also first International Symposium on Crystalline Organic Metals, ISCOM meeting)  organized in Germany at Irsee in 1991 by P.M. Chaikin and D. Jerome. Participants may recognize themselves 32 years ago. Unfortunately some of them have already left us. }
\end{figure*} 

\subsubsection{\tmps: A model Quasi-1D conductor}
\label{model Quasi-1D}
 Because of their particular crystal structures, such \tm2x materials can be considered   at first glance  as protoptypes for 1D physics\cite{Bourbonnais98}.
  

The development of this field has aroused enthusiasm in the 80's and its progress is due to the close and efficient collaboration between various scientific fields ranging from synthetic chemistry to the elaborate theory of electrons in low-dimensional solids, through numerous experimental techniques. Fig.~\ref{IRSEE91}  allows us to recall some of the actors in the 1990's.

\begin{figure}[h]
\centerline{\includegraphics[width=1\hsize]{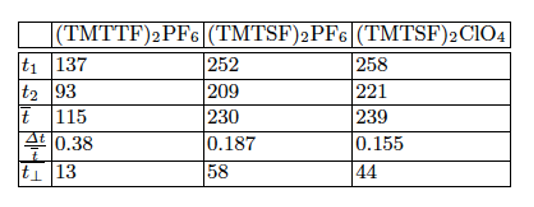}}
\caption{Calculated band parameters for three representative members of the
\tm2x series according to the room temperature crystallographic data in reference\cite{Ducasse86}. The bond dimerization is shown in the line  $\Delta \bar t/{\bar t}$. Energies are in meV.}
\label{dimer}
\end{figure}

However, an important peculiarity of \tm2x materials makes them different from the usual picture of TL conductors. Unlike incommensurate two-stacks \tq materials, $\mathrm{(TM)_{2}X}$ conductors exhibit a band filing which is commensurate with the underlying 1D lattice due to  the 2:1 stoichiometry  imposing half a carrier (hole) per TM molecule. Consequently, assuming  a uniform spacing of the  molecules along the stacking axis, the unit cell contains  1/2 carrier, \textit{i.e.} the conduction band becomes quarter-filled in terms of electrons.  
However,
non-uniformity of the intermolecular spacing  had been noticed in the early structural studies of \tfx crystals\cite{Ducasse86}. It is  due to the periodicity of the anion packing being twice the periodicity of the molecular packing. This non-uniformity is at the origin of  a dimerization of the intrastack overlap between molecules  which is important  in the sulfur series 
but  still present although less developed in some members of the  
$\mathrm{(TMTSF)_{2}X}$ series as shown on Table~\ref{dimer}. 

Hence, taking into account the symmetry imposed by the anion lattice, the conduction band may also be considered as effectively half-filled at sufficiently low energy.
\textcolor{black}{As discussed in Sec.~\ref{1DPhysics}, a dramatic consequence of this commensurate situation for $\mathrm{(TM)_{2}X}$ materials is the existence of two localization channels due to electron-electron Umklapp scatterings. The first  at half-filling with momentum transfer $G=4k_F$ to the lattice and to which the two particles scattering coupling constant $g_{3} \approx U(\Delta_D/E_F$) is associated. The second at quarter filling where $G=8k_F$ and the umklapp coupling  $g_{1/4} \approx U(U/E_F)^2$ corresponding to three-particle coupling \cite{Emery82,Giamarchi97}. This localization is a typical outcome of 1D physics in the presence of repulsive interactions and leads to  a charge gap $\Delta_{\rho}$ in the charge excitation spectrum although no ordering is expected at any finite temperature for a purely 1D system.  The spin sector remains gapless on account of the separation between spin and charge degrees of freedom in  1D conductors. Fortunately, most features of the 1D localization and the physical properties of this electron gas can be studied by transport, optical and NMR experiments on the materials. }

The system \tmps  also known under the name of  a Fabre salt since it has  been first synthesized in the Montpellier chemistry laboratory\cite{Fabre04}, can be considered as the workhorse of  Q1D conductors as far as \textcolor{black}{1D physics is} concerned because of the existence of high quality single crystals enabling  in-depth studies under very high pressure either with the piston-cylinder  or with diamond anvils techniques.

\paragraph{\textcolor{black}{Nuclear magnetic resonance  and 1D nature of spin correlations.}}

\begin{figure}[h]
\centerline{\includegraphics[width=1\hsize]{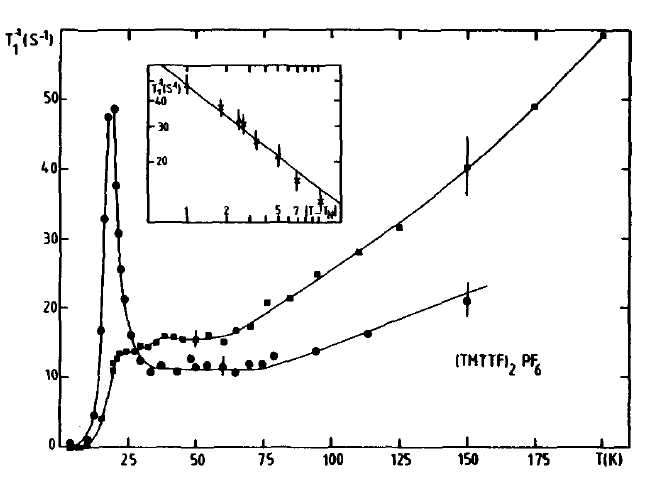}}
\caption{$^{13}$C NMR spin-lattice relaxation rate $T_1^{-1}$, as a function of temperature  for (TMTTF)$_2$PF$_6$ at $P=1$bar (full squares) and $P=13$kbar (full circles). Inset: antiferromagnetic critical behavior of $T_1^{-1} $ near $T_N$ at $P=13$kbar, Source: Fig. 1 [Ref\cite{Creuzet87a}, p. 291]. }
\label{T1Creuzet}
\end{figure}

It is instructive at this point to discuss the nature of correlations  in the low pressure domain of Fig.~\ref{Generic} for a model system like  (TMTTF)$_2$PF$_6$. An extensively used tool  to this end was NMR. In Fig.~\ref{T1Creuzet}, we show the results of Creuzet {\it et al.,}\cite{Creuzet87a} for the temperature dependent $^{13}$C nuclear spin-lattice relaxation rate $T_1^{-1}$ of (TMTTF)$_2$PF$_6$ at  $P= 1$bar and 13kbar. For both pressures an activated resistivity temperature profile confirms the existence of a charge localization gap $\Delta_\rho$ over a large temperature domain, as indicated  in Fig.~\ref{Generic}. For the  single crystals used in these NMR experiments, each  $^{13}$C of enriched TMTTF molecules has a nuclear spin $I=1/2$ that is locally  coupled to the spins of conduction electrons {\it via} an hyperfine coupling  $\bar{A}$. The coupling is responsible of the nuclear spin-lattice relaxation governed by the Moriya expression
\begin{equation}
T_1^{-1} = |\bar{A}|^2 T \int d^dq\, {\chi^{\prime\prime}(\bm{q},\omega)\over \omega},
\end{equation}
which is sensitive to the contribution of low frequency spin correlations coming from all wave vectors $\bm{q}$ through the imaginary part of the dynamic magnetic susceptibility $\chi^{\prime\prime}(\bm{q},\omega)$.

For an interacting electron system in $d=1$ spatial  dimension,  $T_1^{-1}$   takes the form
 \begin{equation}
T_1^{-1} \simeq C_{q\sim 2k_F} T^{1-\gamma_{SDW}} +\  C_{q\sim 0} T \chi^2_\sigma(T),
\label{T11D}
\end{equation}
which superimposes as a function of temperature the  power law singularity of $2k_F$ SDW correlations Eq.~\ref{SDWKi} and the long-wavelength spin correlations of the uniform static spin susceptibility $\chi_\sigma(T)$. In the  presence of $\Delta_\rho$, e.g., for a Mott Luther-Emery liquid, the charge stiffness evolves  at low energy or temperature towards its minimum value $K_\rho^*\to 0$. This  leads to the power law exponent $\gamma_{\rm SDW} =1$, which according to Eq.~\ref{T11D}, is  responsible for a constant, temperature independent, SDW contribution  to  the relaxation rate. As for the uniform part,   the regular enhancement of susceptibility $ \chi_\sigma(T)$ by interactions will give rise to a superlinear temperature dependent contribution to $T_1^{-1}$, which grows in importance as temperature is raised to ultimately become the dominant part of the relaxation rate\cite{Bourbon89}. This temperature sequence for $T_1^{-1}$  is clearly visible in the data of Fig.~\ref{T1Creuzet} at both pressures where $\Delta_\rho$ is present. This constitutes an experimental confirmation of the 1D character of spin correlations and the actual vanishing value of $K^*_\rho$ in a sizable part of the non ordered phase of the Fabre salts\cite{Wzietek93}.
\begin{figure}[h]
\includegraphics[width=0.5\hsize]{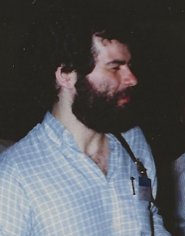}
\caption{\label{Creuzet} François Creuzet  (1957-2020) made major contributions to the understanding of the physics of Fabre and Bechgaard salts for  his PhD  at Orsay  in the eighties using NMR in close cooperation with the Sherbrooke group. He became Deputy Director, Scientific director at the Saint Gobain Company in 2016.}\end{figure} 

It is worth concluding this paragraph by noting that the NMR results of Fig.~\ref{T1Creuzet} also give precise information about the nature of long-range ordering in (TMTTF)$_2$PF$_6$ in the low pressure range of the phase diagram. At ambient pressure for instance, $T_1^{-1}$ undergoes an exponential fall below $T_{\rm SP}\simeq 20$K of the form $T_1^{-1} \sim e^{-\Delta_s/T}$ due to  the coupling of spins to phonons and the onset of a spin gap $\Delta_s$ compatible with  3D spin-Peierls order\cite{Creuzet87a,Pouget82a}(SP in Fig.~\ref{Generic}). One also notes in Fig.~\ref{T1Creuzet}  the presence of 1D precursors to the SP transition, in the form of a gradual decrease or a spin pseudogap occuring well above  $T_{\rm SP}$. Finally at $P=13$kbar, $T_1^{-1}$ presents a singularity at $T_N\simeq 19$K, signaling the crossover from 1D to 3D magnetic long-range order. The critical temperature profile  $T_1^{-1}\sim |T-T_N|^{-1/2}$ shown in the inset of Fig.~\ref{T1Creuzet} near the transition is the signature of an antiferromagnetic 3D Néel state promoted by superexchange coupling between spins of neighboring chains\cite{Creuzet87a} (AFM in Fig.~\ref{Generic}).

\paragraph{Transport.}

The investigation of  \tmps transport  properties under several pressures covering the whole phase displayed on Fig.~\ref{Generic} has offered a remarkable illustration for the evolution from a  \textcolor{black}{1D quantum liquid of different form (Luther-Emery Mott, charge or Wigner ordered insulators,  LL)  to a higher dimensional metallic state 
} thanks to the control under pressure  of the strength of both the electron-electron Coulombic repulsion and the interchain kinetic  coupling.
 \begin{figure}[h]
\centerline{\includegraphics[width=1.1\hsize]{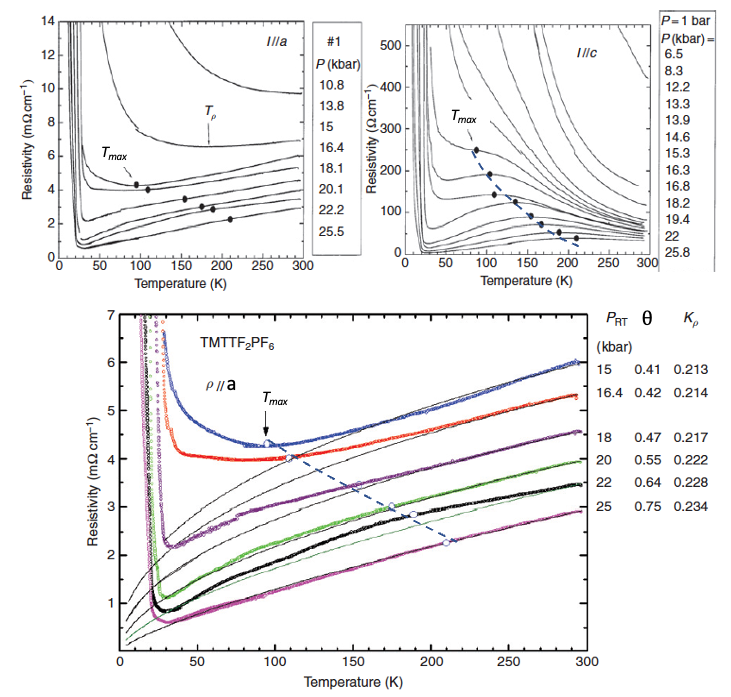}}
\caption{\tmps, longitudinal (left) and transverse (right) resistances versus temperature at different pressures. The data for $\rho_{a}$ shown in the bottom figure provide the pressure dependence of the Luttinger coefficient $K_{\rho}$ derived from  fitting  the experimental data  to $T^{\theta}$ with $\theta = 16K_{\rho}-3$, i.e, \textcolor{black}{($n$= 2)} \emph{vide infra}. Notice that the decrease in compressibility under pressure makes the constant volume correction less significant for the  temperatures dependences measured at high pressures which are \emph{at variance} with the case of \tmp6 presented on Fig.~\ref{rhoPF6constantvolume}, Sources: Fig. 1 [Ref\cite{Auban04}, p. 42] (top), Fig. 5.14 [Ref\cite{Jerome10}, p. 166] (bottom). }
\label{TransparaetperpPF6}
\end{figure}

 What makes the data displayed  on Fig.~\ref{TransparaetperpPF6}  particularly interesting is the difference noticed between the temperature dependences of  longitudinal and transverse transport along the least conducting $c$ direction\cite{Auban04}. 
  \begin{figure}[h]
\centerline{\includegraphics[width=1\hsize]{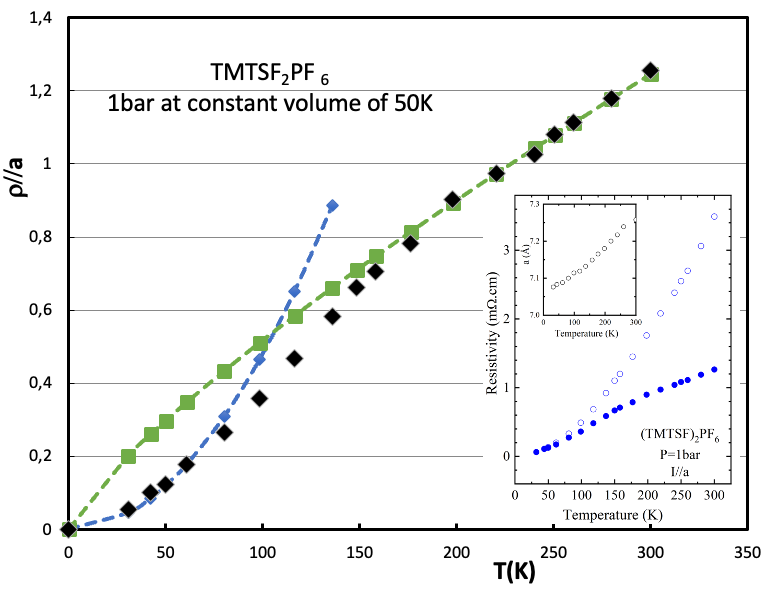}}
\caption{ Temperature of the \tmp6 longitudinal  resistance after making the constant volume transformation (black losanges). The fit to $T^{0.8}$ are the green squares while blue losanges are the $T^{2}$  low $T$ behaviour. The inset shows the strong volume dilatation of $a$ lattice parameter and the actual temperature dependence of $\rho_a$ as measured under 1 bar (open circles) plus the conversion to the constant volume $T$-dependence (full blue dots). }
\label{rhoPF6constantvolume}
\end{figure}
At ambient pressure longitudinal and transverse  components of the   \tmps transport  exhibit a strong insulating character. However, at increasing pressure, \textcolor{black}{the longitudinal resistivity $\rho_a(T)$} begins to show a weak metallic temperature dependence below room temperature down to $T_{\rho}$ where according to  Fig.~\ref{TransparaetperpPF6} a broad minimum takes place ($T_{\rho}$=180K under 13.8 kbar), and  below which the resistance becomes activated. $T_{\rho}$ is strongly depressed by pressure and tends to vanish around 15 kbar    giving way to the appearance of a sharp metal-insulator transition around 20K. 

\begin{figure}[h]
\centerline{\includegraphics[width=0.8\hsize]{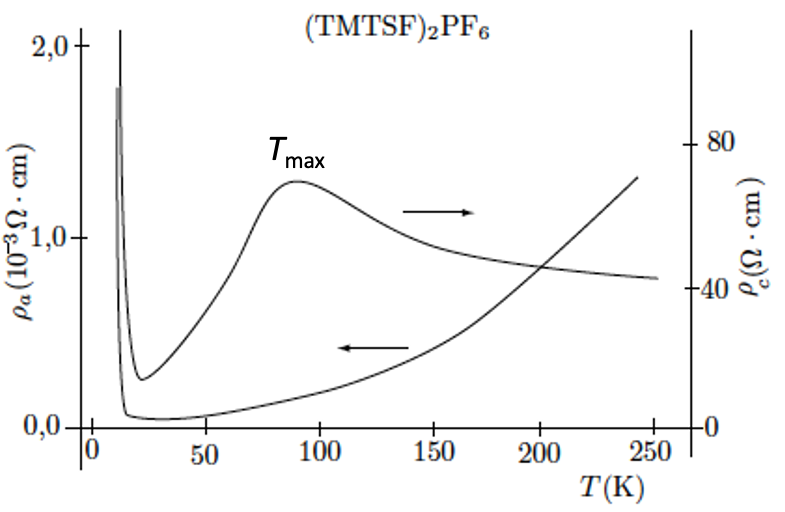}}
\caption{ Temperature dependence of transport along $a$ and $c$ for \tmp6 providing a simple illustration of the non-metallic behaviour of $\rho_c$ above $T_{max}$, Source: Fig. 31 [Ref\cite{Jerome04}, p. 5584]. }
\label{longandtrans}
\end{figure}

On the other hand, ${\rho_{c}(T)}$  remaining insulating (d${\rho_{c}}$/$dT< 0$) below room temperature from 1 bar up to about 13 kbar, develops a maximum at higher pressures  at the temperature $T_{\rm max }$ (=100K at 15 kbar) which keeps increasing  at higher pressures. The upper right part of Fig.~\ref{TransparaetperpPF6} shows how $T_{\rm max }$ moves up to room temperature under 25 kbar. Under that latter pressure, both ${\rho_{c}}$ and ${\rho_{a}}$ undergo the sharp metal-insulator transition in the temperature regime of 20K.  Furthermore,  it is important to realize that the phase diagram of  \tmps above 25 kbar becomes quite similar to the $T-P$ diagram of \tmp6 above ambient pressure. Notice that the ground state of  \tmps becomes superconducting around 45 kbar\cite{Jaccard01}, but this is not shown on this figure.  Consequently,  one may consider that there exists a pressure shift of about 35 kbar between the phase diagrams of \tmps and (TMTSF)$_2$PF$_6$.
The different behaviours of transport   are made crystal-clear with the example of  \tmp6 under ambient pressure on Fig.~\ref{longandtrans}.

\begin{figure}[h]
\centerline{\includegraphics[width=1.1\hsize]{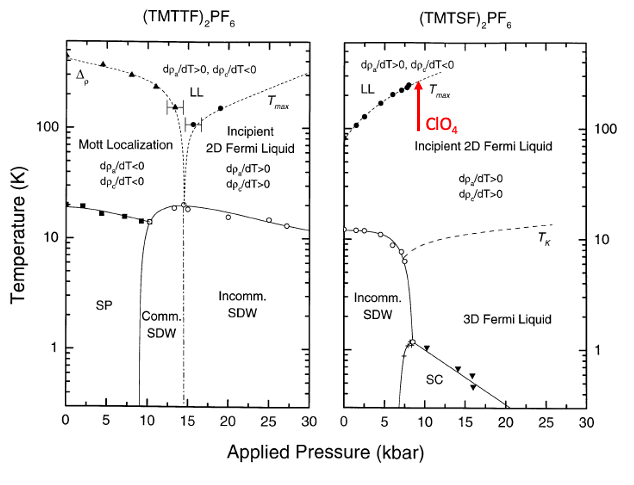}}
\caption{\label{Trho}($T-P$) phase diagram  of \tmps (left panel) and \tmp6 (right panel). 
The $T_{max}$ versus $P$ line (full circles) setting a border between a Luttinger regime (LL) and a transient phase towards a Fermi liquid is shown for both compounds. The $T_{K}$ line (right panel) marks the restoration of a 3D Fermi liquid at low temperature with $\rho_c$ becoming coherent. SP stands for the Spin-Peierls phase. Commensurate (Comm) and Incommensurate (Incomm) SDW states have been singled out. Also displayed in red, the location of \tmc at ambient pressure  for which the cross-over between Luttinger-like and Fermi-like physics occurs around 300K, Source: Fig. 2 [Ref\cite{Moser98}, p. 42].} 
\end{figure}

{All 
experimental results relating to \tmps and \tmp6  are gathered on Fig.~\ref{Trho} which can be considered as the model diagram for a commensurate quasi one dimensional conductor which  constitute the basic experimental facts to be compared with the theory  proposed for the deconfinement of a one-dimensional Mott insulator in the presence of interchain kinetic coupling\cite{Giamarchi04a,Bourbonnais88}. In the Mott-insulator phase of \tmps, the finite interchain coupling $t_\perp\approx 200K$\cite{Jerome82,Grant83,Ducasse86}  is small compared to a Mott gap of $\approx 500$K\cite{Jerome04}, a value which is such that the interchain hopping is suppressed by the insulating nature of the 1D phase. 
Consequently, the interchain coupling becomes no longer relevant in these sulfur compounds at ambient pressure\cite{Schwartz98}. The situation evolves quickly under pressure as both the Mott gap is sharply depressed while the interchain coupling increases driving the compound toward a situation where the two quantities are interplaying. From the experimental view point, Fig.~\ref{Trho} provides a good illustration for this evolution from the 1D Mott insulator to a higher dimensionality metal at a pressure overpassing 15 kbar. A measure of the gap extracted from the optical conductivity shows that at this pressure the gap is roughly of the order of magnitude of the interchain coupling\cite{Vescoli98,Degiorgi06}  These observations are supporting the existence of a deconfinement transition which has been treated theoretically very satisfactorily using a generalization of the dynamic mean field theory\cite{Biermann01}. The $T_{\rm max}$ 
line  shown on the \tmp6 panel of the general diagram, Fig.~\ref{Trho} marks the temperature where motion along  $c$ becomes coherent\cite{Moser98}. This temperature coincides also with the recovery of a $T_{1}^{-1}$ {\color{black} Korringa, though strongly renormalized, behaviour } for  spin fluctuations observed either in $\mathrm{Se-ClO_4}$ or in $\mathrm{Se-PF_6}$ under 8 kbar, \emph{see} Ref.\cite{Creuzet87b} and Sec.~\ref{precursors}.}

 In spite of the charge gap $\Delta_{\rho}$ a metal-like behaviour of the  longitudinal resistance can still be obtained at $T$ larger than  $\Delta_{\rho}$. \textcolor{black}{ According to Eqs.\ref{rhoTh} and\ref{theta} the resistance is  displaying  a power law $\rho_{a} (T) \approx T^{4n^{2}K_\rho-3}$ for  1D interacting electrons in the presence of umklapp scattering.}
 
 Experimental studies  have revealed such a metallic-like  behaviour for the  resistance either in  \tmp6 at ambient pressure\cite{Jerome82} and also  in \tmps under pressure above 15 kbar\cite{Auban04,Degiorgi06}. 
 
 Since transport properties of these molecular crystals  are   known to be strongly volume dependent\cite{Jerome82},  before any comparison between experimental data and a theoretical model it is required to derive the proper experimental temperature dependence of the resistivity  (to be compared with the theory) under a constant volume, taking into account both the thermal dilatation at constant pressure (inset of Fig.~\ref{rhoPF6constantvolume} and the volume dependence of transport at constant temperature, \emph{see} Ref\cite{Gallois87}. Then we derive the equivalent pressure $P_{i}(T )$ which is needed at each temperature to recover the reference volume of 50K. The procedure for the derivation of the constant volume temperature dependence   is detailed in the following reference\cite{constantvolume}. 

Although the constant volume transformation is crucial for \tmp6 under ambient pressure as discussed above and on Fig.~\ref{rhoPF6constantvolume} it becomes  of less importance for \tmps since for this latter compound a pressure of 15 kbar is already requested for the observation of a  metallic behavior for $\rho_{a} $ and the lattice expansion in temperature is known to be much  depressed under pressure\cite{Gallois87}. 
 
 What has been found for $\rho_{a} (T)$ in \tmps  above 15 kbar is a sublinear exponent of the order of $\theta = 0.7$  (\emph{see} Fig.~\ref{TransparaetperpPF6})  whereas  in  \tmp6  the   fit  of  $\rho_{a} (T)$ above 100K  at the  constant volume of 50K on Fig~.\ref{rhoPF6constantvolume}\cite{Auban04,Degiorgi06} would rather provide the result  $\theta = 0.8$. 

Taking into account the power law for longitudinal transport found in the metallic regime of Fig.~\ref{Trho} namely $\theta \approx 0.7-0.8$,   the assumption of a relevant a half-filling of the  band  (commensurability unity)  would lead to the Luttinger parameter $K_{\rho}$  very close to unity (corresponding to very weak repulsive interactions) since \textcolor{black}{ $\theta = 4n^2K_{\rho}-3$}, $n$ being the order of commensurabilility ($n$=1 for a half filled band. This  weak coupling ($K_{\rho}\approx 1$) can be discarded since it would be 
hard to reconcile with the exchange enhancement of the spin susceptibility\cite{Bourbonnais99}. 
 
 On the other hand, when quarter-filled Umklapp scattering is dominant at high temperature $\theta = 16K_{\rho}-3$ and consequently  $K_{\rho} = 0.23$ according to the  data of \tmp6 from Fig.~\ref{rhoPF6constantvolume}. This is also a reasonable value for the Luttinger parameter of \tmps at very high pressures of 25 kbar or so, Fig.~\ref{TransparaetperpPF6}.
Additional arguments have been given in Ref\cite{Auban99} in favour of the 1D quarter-filled band scenario  in the high temperature metallic region in both series of \tm2x compounds.

Regarding the ordered phases at low temperature in the phase diagram Fig.~\ref{Generic},  what happens for \tmps below the 1D Mott localization line is a phase transition towards a long range ordered insulating phase  observed in the charge localized temperature domain, \emph{see} the left part of Fig.~\ref{Generic}. The phase at low temperature has been  ascribed, according to NMR data, to the onset of a charge  disproportionation between molecules \textcolor{black}{of each dimer} on the molecular chains\cite{Chow00} or charge ordered (CO) state. Since the charge of this low temperature phase 
is no longer uniform, manifestations of  ferroelectricity can be expected as shown by a signature in dielectric measurements\cite{Nad00,Monceau01}. The stability of this CO state (often called a Wigner state) is a direct consequence of the long range nature of the Coulomb repulsion which, in terms of the extended Hubbard model, amounts to  finite on-site {\color{black} $U$ repulsion along with first and second nearest-neighbors Coulomb  $V$ terms}. Another example of the charge ordering in 1D conductors will be given in Sec.~\ref{1/4}.

\begin{figure}[h]
  \includegraphics[width=1\hsize]{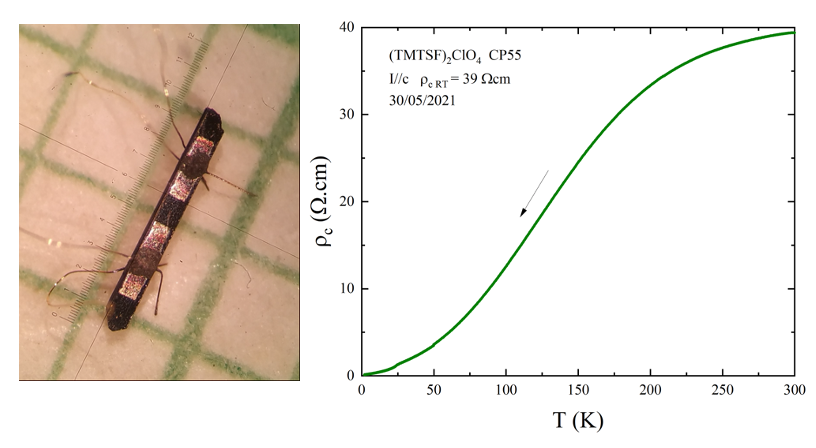} 
  \caption{\label{Rhoc}This picture  shows how samples with  $c$ axis perpendicular to the figure are prepared for the measurement of $\rho_c$. Gold pads are evaporated on both equipotential $(a-b)$ planes for current and voltage leads along $c$. Usually measurements along $c$ are free from any cracks during cooling unlike  $a$  axis data for which severe cracks are often observed under ambient pressure. The tiny kink on $\rho_{c}(T)$ at 24K is due to the \cl anion ordering transition.  We thank P. Auban-Senzier for the communication of this picture. }
  \end{figure}

We turn now to the study of transport along the least conducting direction which has contributed significantly to better understand the diagram in Fig.~\ref{Trho} and in particular the evolution from Luttinger to \textcolor{black}{higher dimensional metallic} physics under pressure. Let us notice that measurements along the $c$ axis, \emph{see} Fig.~\ref{Rhoc},  are also very reliable  as no cracks develop  on cooling unlike those along the $a$ direction (at least at ambient pressure).

 Transverse transport requires the tunneling of Fermions   between neighbouring chains ({\it at variance}\, with the longitudinal transport which is related to  1D collective modes)
 and therefore transverse transport along the least conducting direction  probes the physics of the $a-b$ planes namely, the amount of quasi-particles (QP) weight existing close to Fermi level  of the weakly interacting Luttinger chains. In the case of a Landau-Fermi \textcolor{black}{ state, these QP are} characterized by an in-chain
(or in-plane) lifetime $\tau$, there exists a proportionality relation between  single particle contribution to conduction of $\sigma_\perp$ and  $\sigma_\|$  which
has been well established for \tq
using the Fermi Golden rule for incoherent transverse
transport, \textcolor{black}{as discussed above in Sec.~\ref{The dilemma of transport}}.

\begin{figure}[h]
\centerline{\includegraphics[width=0.8\hsize]{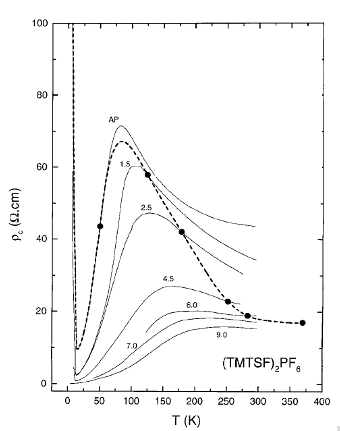}}
\caption{\label{TMTSFconstvolume} $c$-axis resistivity of \tmp6 under various hydrostatic pressures up to 9 kbar showing the procedure followed     to extract  the temperature dependence at the volume of 50 K under ambient pressure. The bold points and the dashed curve represent the $T$-dependence under constant lattice parameters and in-turn $\rho_c$ amounts to 17 instead of 42  ($\Omega.{\rm cm}$)  at 300K, Source: Fig. 4 [Ref\cite{Moser98}, p. 45]. } 
\end{figure}

\tm2x compounds behave quite differently with very often opposite $T$-dependence for the $a$ and $c$ components of transport, an activated  character for the  transverse transport at high temperature and a metal like behaviour for $\rho_a$ (\emph{see} Figs~\ref{TransparaetperpPF6} and \ref{TMTSFconstvolume}). These features  can  be interpreted as the signature of a remnant in-chain  non
Fermi-Landau behaviour \cite{Moser98}. 

\begin{figure}[t]
\includegraphics[width=1.1\hsize]{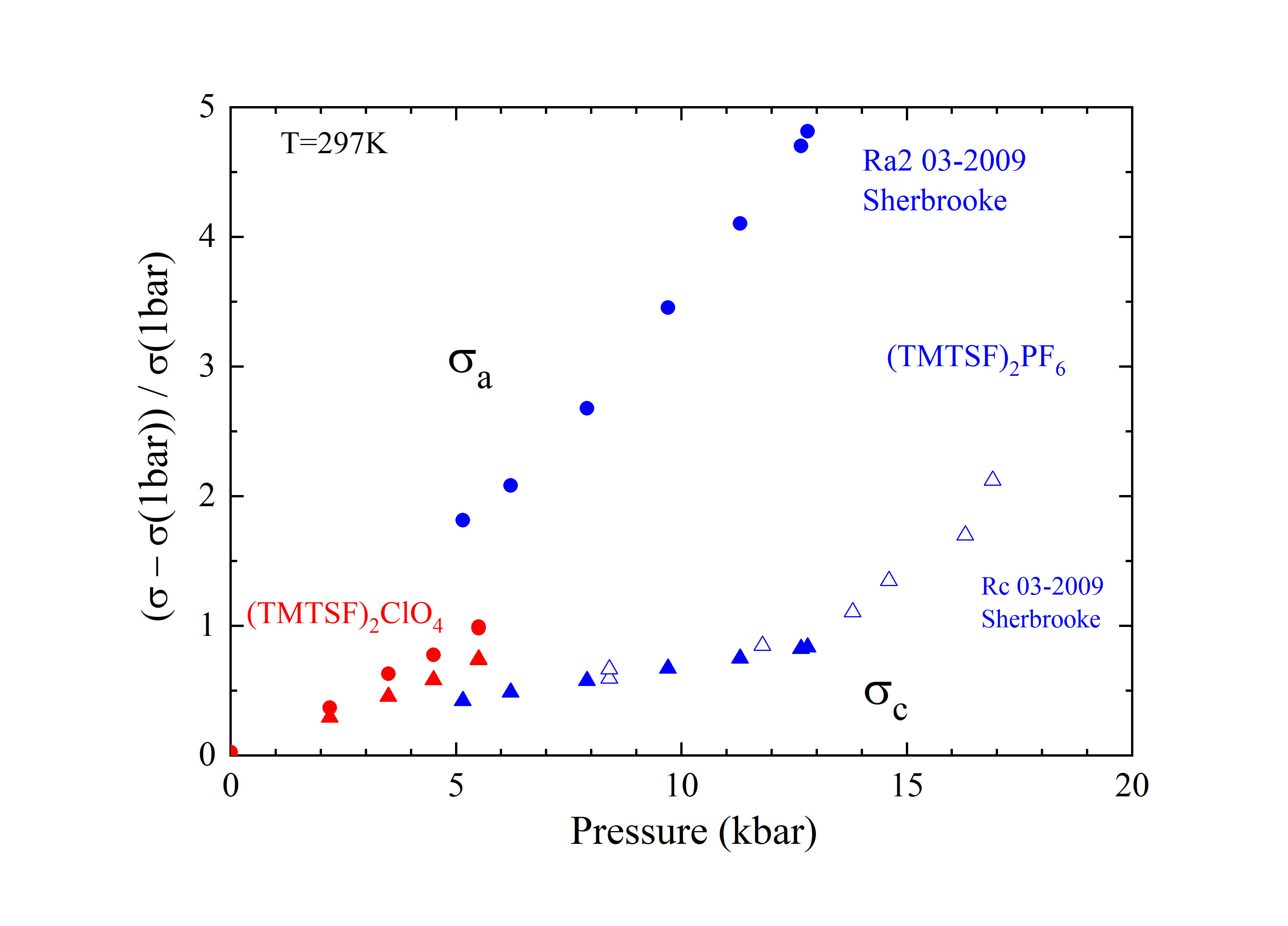}
\caption{\label{sigmaClO4PF6} Pressure dependence of conductivities at 300K along $a$ and $c$ for \tmp6 in blue and \tmc in red. According to Fig.~\ref{Trho}, the pressure regime for these \tmp6 data  is almost entirely located in the ``Luttinger" regime   since $P_{\rm max}$ corresponding  to the cross-over is about 10 kbar. On the other hand \tmc is already at ambient pressure  in the incipient 2D Fermi liquid phase in which the concept of single particle  relaxation time develops. The upturn of  $\sigma_c$ in {$\mathrm{PF_{6}}$} around 14 kbar might be attributed to  the passing through the cross-over under pressure with $\sigma_c$ evolving from an activated mode to a metal-like behaviour at high pressure.  It is noticeable that transport is much more  volume dependent in the ``Luttinger" domain  than it is in the 2D incipient Fermi liquid phase. We thank P. Auban-Senzier for the communication of these unpublished data.
}
\end{figure}

Below a temperature $T_{\rm max}(P)$  the $c$-axis transport switches from an insulating at high temperature 
to a metal-like temperature dependence which has been attributed to the manifestation of a cross-over between two regimes 
; a high temperature 1D regime
with reduced QP weight at  Fermi energy (possibly a LL liquid) and another regime in which the 
weight acquires  higher dimensional metallic  (HDM) features with decreasing temperature.
The strong pressure dependence of $T_{\rm max} $ on Fig.~\ref{TransparaetperpPF6} rules out a simple relation
$T_{\rm max}(P)\propto t_b$ since the pressure coefficient of the
bare transverse coupling should be the order of what is expected  for the rather weak volume dependence of the  kinetic couplings  
namely, 2\%/kbar. 
 The crossover is also quite visible comparing the pressure dependence of $\sigma_a$ and  $\sigma_c$ at constant temperature, \emph{see} Fig.~\ref{sigmaClO4PF6} for the case of (TMTSF)$_2$PF$_6$. Following the data on Fig.~\ref{TMTSFconstvolume} the cross-over temperature is thus increased by a factor of order 3 under 10 kbar which is definitely inconsistent with a non-interacting electrons model.

Consequently, the strong
pressure dependence of $T_{\rm max}(P)$ is very suggestive of some kind of renormalization of the transverse coupling  $t_{\perp}$, 
as pointed out in Ref.\cite{Bourbonnais84} based on the NMR data, with intrachain interactions  renormalizing downward the interchain hopping between the weakly coupled Luttinger chains  as a result of the decay of the density of states (\ref{DOS}), {\color{black} which reduces the probability for  coherent interchain quasi-particle  tunneling}.  The renormalized cross-over thus   becomes,
\begin{eqnarray}
\label{tperp}
t_{b}^{\star}\sim W(\frac{t_{\perp}}{W})^{1/(1-\alpha)},
\end{eqnarray}
where $\alpha$ is given by the exponent (Eq.~\ref{DOS}) entering in the QP density of states.  It has been argued following the optical data of  \tm2x salts\cite{Schwartz98},  that given a value for $K_{\rho}$ of the order of 0.23, which means $\alpha =0.64$. The renormalization of the transverse coupling according to Eq.~\ref{tperp} would lead to a  cross over temperature  between the 1D regime and a higher dimensional coherent regime for single particle motion \cite{Schwartz98} much too severe, below 10K or so, \emph{at variance} with transport data which locate it around 100K or higher. However, the above expression for $t_{b}^{\star}$ assumes that the exponent $\alpha$ is fixed at all temperature up to the high-energy cutoff  $W$, which  is of the order of the bandwidth along the chains. The actual $\alpha$ is likely to evolve as a function of energy in the presence of Umklapp scattering, its   averaged  - lower - value over the whole energy interval would give a more realistic scale for the crossover.  

According to the transport data of \tmp6  displayed on Fig.~\ref{TMTSFconstvolume}, the renormalisation which cannot exceed   a factor  about 3 at ambient pressure in \tmp6 	is practically suppressed under a pressure of 10 kbar. 

Hence, we may conclude that Eq.~\ref{tperp} is exaggerating the renormalisation of the transverse coupling due to  intrachain Coulombic interactions as well as its pressure dependence. {\color{black} A reexamination of this question would be useful in regard to the value of $\alpha$ extracted from $\rho_c$}.

Below the deconfinement temperature $T_{\rm max}$,   charge excitations lose  their 1D character and resemble \textcolor{black}{in certains respects} to   what is expected in a Fermi liquid with a quadratic temperature dependence for the longitudinal resistivity observed in \tfx compounds under very high pressure\cite{Ruetschi09} or even at ambient pressure in \tmp6\cite{Tomic91} although electron excitations of this apparent ``Fermi liquid'' retain  a low energy gap in the far infra-red spectrum in which  most of the oscillator strength is carried by states above the gap coexisting  with a very narrow and intense zero frequency peak in the conductivity,\cite{Cao96,Schwartz98} as we shall detail in Sec.\ref{opticalresponse}.

 Such a picture does not necessarily imply that the transport along
the $c$-direction must  become coherent below the cross-over as the $c$-axis transport may  remain
incoherent with a progressive establishment of a Fermi liquid like state  in  $a-b$ planes at temperatures much below
$T_{\rm max}$ and the emergence of a weak Drude peak with an edge along $c^{\star}$\cite{Henderson99}. As a result, the coupling $t_c$ along $c$ is of the order of 1 meV.

The question of the evolution between 1D physics and the higher dimensionality conductor has stimulated numerous theoretical treatments.
A first attempt has been made following  the  treatment of the the $c$ axis conductivity between adjacent confined Luttinger liquids in underdoped cuprate superconductors\cite{Clarke95}, with the conclusion \emph{``it would be of great interest to examine the conductivity along the weakest hopping direction in the quasi 1D and 2D organic conductors"}.



The interplane transport can
only proceed via the hopping of quasi-particles.
For
such a situation to occur a \emph{real} particle has to escape the Luttinger liquid by recombining its charge and spin
components.
The \emph{real} particle can then hop onto a neighboring
stack, contributing to $\sigma_c$  and then decay into the Luttinger
liquid again. The transverse interplane conductivity has
been derived theoretically when the physics of electrons in
chains is governed solely by a 1-D Luttinger regime and when conduction can be approximated in the tunneling approximation by\cite{Moser98}
: \begin{eqnarray}
\label{tunneling}
\sigma_c (T)\propto  t^2_{c}(\frac{T}{v_c})^{2\alpha}
\end{eqnarray}
where the exponent $\alpha$ enters the density of states near the Fermi energy according to Eq.\ref{DOS}. 

Following this model, $\rho_c$ should behave
like,
\begin{eqnarray} 
\label{alpha}
\rho_{c}(T)\propto T^{-2\alpha}
\end{eqnarray} 
 in the Luttinger liquid domain with $\alpha$ positive and ranging from infinity in a Mott insulator to
zero in a non-interacting Fermi liquid. However, fitting the high temperature regime of  the constant volume transverse transport on Fig.~\ref{TMTSFconstvolume} by Eq.~\ref{alpha} would lead to $\alpha$= 0.7\cite{Moser98} and in turn to $K_\rho \simeq 0.22$, from
solving the second order equation for $K_\rho$. Such a value would  be too large to agree   with the findings of longitudinal transport and optical conductivity at high frequency to be presented below. This mismatch between transport data calls for an improvement   of the theory for transverse resistivity, \emph{see} Sec.~\ref{transverse}. 
 In the case of a Landau-Fermi liquid in which electron states are characterized by an in-chain (or in-plane) life time $\tau$, Eq.~\ref{tunneling} recovers the proportionality between $\sigma_\|$ and $\sigma_\perp$  has  mentioned above and which is fairly well followed for \tmc at room temperature, \emph{see} Fig.~\ref{sigmaClO4PF6}. However, these conditions are not fulfilled    for \tmp6 which   according to the diagram of Fig.~\ref{Trho} is located in the ``Luttinger liquid" domain.

Fig.~\ref{sigmaClO4PF6} shows that in this phase the  pressure dependence of the conductivity is at least 25 \%/kbar for the longitudinal component while it is only $\approx$ 10-15\%/kbar along the least conducting direction \textcolor{black}{ that still shows an  activated temperature profile}. 

We are now facing a problem dealing with the constant volume dependences of longitudinal and transverse transport since the consensus on $K_{\rho}$= 0.23  giving $\alpha= 0.64$ 
 cannot  afford for  the amplitude of the temperature dependence of $\rho_c$ between 300 and 100K. Notice that a $K_{\rho}$= 0.20  would raise $\alpha$ up to 0.8.

What is learned from longitudinal transport and the attempt to explain the transverse transport starting from the 1D Luttinger model is that
the behaviour of the electronic transport of \tm2x at low temperature although at first sight looking perfectly classical, Fermi liquid type with  $T^2$ law  components in these strongly anisotropic conductors may  be nothing more than a ``Canada Dry" effect. The physics of the conducting phase is actually far more complex and subtle as we shall see, looking at crucial data brought by the examination of  the frequency dependence of the metallic conductivity regime in the far infrared range. 

Concluding this section, Fig.~\ref{Trho} provides  quite a general experimental picture based on transport for the generic diagram of \tm2x  for the physics going on at high temperature. Next, we intend to see how these data can be connected to the studies of the optical conductivity and will force us to reconsider some of the conclusions reached from longitudinal transport.

\subsubsection{Optical response in the \tm2x series}
\label{opticalresponse}
Measurements of optical properties of \tmp6 began very early after the discovery of superconductivity in this compound, first with reflectance studies for light polarized along the stacking axis\cite{Bechgaard80}  providing a plasma edge around 10.000 ${\rm cm}^{-1}$ and a bandwidth of approximately 1 eV, significantly larger than for TTF-TCNQ. 
\begin{figure}[t]
\includegraphics[width=0.8\hsize]{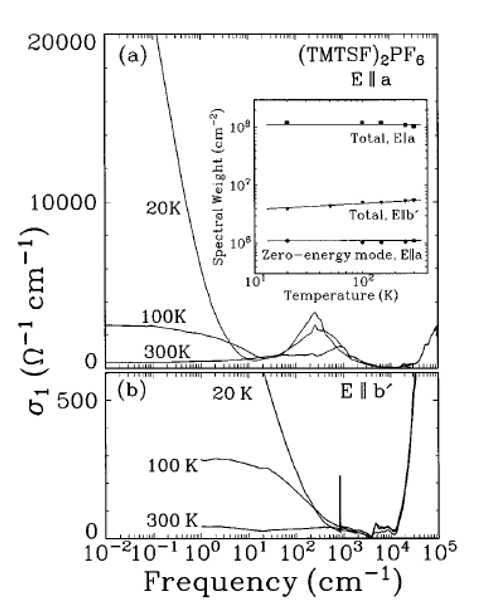}
\caption{\label{Dressel} Optical conductivity of \tmp6 obtained after a Kramers Kr\"onig analysis of the reflectance in both $\parallel$ (a) and $\perp$ (b) polarization directions. The inset shows the spectral weight of both modes with the zero energy mode carrying about 1\% of the total spectral weight, Source: Fig. 2 [Ref\cite{Dressel96}, p. 399]. }
\end{figure} 

Subsequent measurements performed  by the danish group\cite{Jacobsen81} from far infrared  to visible  pointed out the growth below 100K  of an infrared reflectance edge for electric field perpendicular ($\parallel$b') to the stacking axis resembling the plasma edge of a Drude metal. It was thus suggested that the \tmp6 compound evolves from a 1D conductor at high temperature to a 2D conductor at low temperature ($\approx 100$K). A few years later, a new investigation of the \tmp6 reflectance covering an extremely broad frequency range led after a Kramers Kr\"onig analysis to the determination of the optical conductivity for parallel and perpendicular polarizations\cite{Dressel96}.

Optical studies have been finalized in an article covering the electrodynamic response of several members of the \tm2x series  all showing a similar behaviour\cite{Schwartz98}, all compounds exhibiting a behaviour departing drastically from the response of a simple Drude  metal. An overview of the electrodynamics of Fabre and Bechgaard salts can be found in Ref.\cite{Dressel12}

\begin{figure}[h]
\includegraphics[width=0.8\hsize]{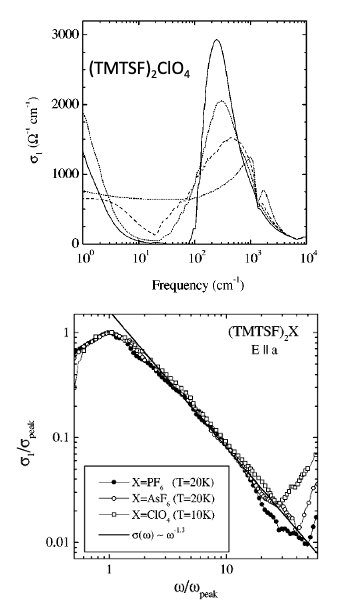}
\caption{\label{optics} The conductivity of \tmc for E$\mid\mid a$ at 300, 200, 100, and 10K. The lower part  displays the power law frequency dependence of the conductivities for several members of the \tm2x series in the  low temperature metallic  regime such as $\sigma_{\parallel} (\omega)^{- 1.3}$, after Eq.~\ref{power}, Sources: Figures 3, 4 [Ref\cite{Schwartz98}, p. 1264, 1266].
}
\end{figure}

From these optical studies it turned out that the metallic state of \tmp6 or \tmc above the establishment of long	range order (SDW or SC) is of great interest since it exhibits characteristic features of both the correlations-gapped Mott insulator around $E_{gap} = 200~{\rm cm}^{-1}$, ($\sim 275$K)  together with the high energy excitations of a 1D Luttinger liquid\cite{Vescoli98} with a conductivity revealing a power law in frequency $\sigma (\omega) \propto \omega^{-1.3}$, \emph{see} Fig.~\ref{optics}. This is the power law expected for the AC conductivity   of a Luttinger liquid\cite{Giamarchi97} taking into account the quarter-filled band  Umklapp scattering   namely,
\begin{equation}
\label{power}
\sigma_{\parallel} (\omega) \propto \omega^{16K_{\rho}-5}
\end{equation}
at $\omega > E_{gap}$, leading in turn to $K_{\rho}$ = 0.23 and $\alpha$ = 0.64 according to the experimental data on  Fig.~\ref{optics}.

\subsubsection{Back to the transverse transport}
\label{transverse}

The optical conductivity data let believe that the assumption of the high temperature phase as a set of 1D Luttinger chains is likely too naive.  As a result, the question of the transverse transport  
has been reconsidered including precursor effects revealed by  high frequency optical investigations\cite{Schwartz98,Vescoli98}which  provide remnants of    the Mott localization. Using a DMFT approach treatment of the transverse coupling establishing the link with the low temperature Fermi liquid state, a derivation for $\rho_{c}(T)$ valid all the way from the Luttinger liquid to the Fermi regime has been obtained namely,
\begin{eqnarray}
\label{Georges}
\rho_{c}(T) \approx T^{1-2\alpha} 
\end{eqnarray}

Unfortunately, Eq~.\ref{Georges} does not represent much improvement as compared to the treatment leading to Eq.~\ref{alpha} since a value $\alpha= 0.64$ 
 in agreement with $K_{\rho}$= 0.23, would lead to a constant volume  temperature dependence of the resistivity too small compared to the experiments, \emph{see} Fig. \ref{TMTSFconstvolume}.




Hence, the authors of Ref.\cite{Georges00} have made the remark that the underlying Mott physics cannot be neglected and propose that this effect is the one responsible for the significant  increase of $\rho_c$ on cooling. The phenomenological fit by an activated behaviour, 
\begin{eqnarray}
\rho_{c} (T) \propto T^{(1-2\alpha)} \mathrm{exp}{(\Delta/T)} 
\end{eqnarray}
provides the activated behaviour in Fig.~\ref{TMTSFconstvolume} using the  values of the optical Mott gap for $\Delta$. 

The authors of Ref.~\cite{Georges00} concluded that ``\emph{incoherent tunneling between LL chains in a purely metallic regime is insufficient and that the physics of
Mott localization plays an important role in the temperature
range $150 {\rm K} <T < 300 {\rm K}$}".

\vspace{1cm}

\subsubsection{
Chemistry coming to rescue the quarter-filled scenario}
\label{1/4}

 \begin{figure}[h]
\includegraphics[width=0.9\hsize]{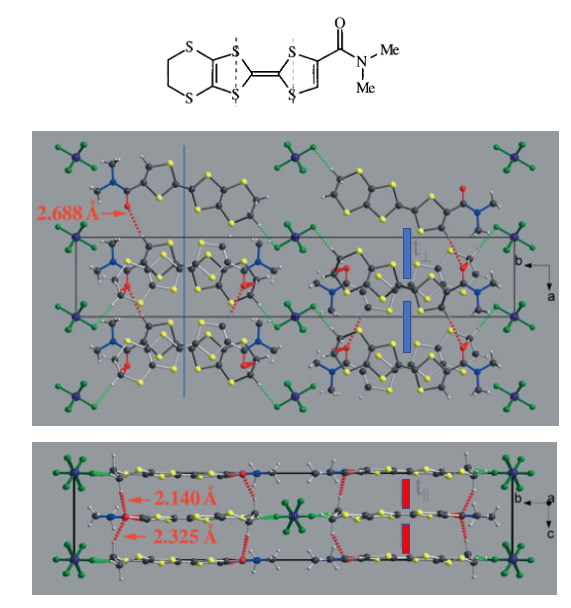}
\caption{\label{EDTv2} (Top) The non-symmetrical  radical cation molecule $\mathrm{EDT-TTF-CONMe_{2}}$. (Middle) Projection onto (110) of the low temperature monoclinic form of $\mathrm{(EDT-TTF-CONMe_{2})_{2}AsF_{6}}$. The vertical thin line is the trace of the glide plane at $3b/4$. Notice the criss-cross stacking along the $c$ axis. (Bottom) Parallel uniform stacks in the low temperature monoclinic form. The red bars show  the uniform kinetic coupling along the $c$ direction, Source: Fig. 1, 2 [ Ref\cite{Heuze03}, p. 1252].}
\end{figure} 
So far, the details of the $T/P $ phase diagram for the TMTTF compounds described
previously clearly involve a competition between half and quarter-filled Umklapp
scatterings. The dimerization in the g-ology approach to the physics of 1D
conductors\cite{Emery82} results in a localization of charges on the bonds between
adjacent molecules due to half-filled Umklapp scattering\cite{Giamarchi04a}. The ground
state is referred to as a dimer-Mott insulator. In  actual   \tm2x materials,
spin-Peierls (SP) or antiferromagnetic (AF) orderings occur at low temperature\cite{Wzietek93,Creuzet87a,Chow98a}
. On the other hand, the next near neighbor Coulomb interaction
competing with the on-site interaction is known to lead to charge ordering\cite{Mila93}
 which tends to favor an AF ground state, but here the exchange integrals,
and consequently the AF wave vector, are controlled by the CO (charge order) order
parameter. Although there already  exists a strong suspicion as discussed in Sec.~\ref{model Quasi-1D} in favour of the dominant role played by the quarter-filled Umklapp scattering in the generic diagram of the \tm2x series, it is clear that a system in which the competition between half and quarter filled Umklapp is no longer relevant is most welcomed.

The opportunity to study the progression from insulator-metal
in a structurally quarter-filled quasi-1D  compound (i.e., without dimerization) has been given by chemistry 
 with the crystallization of 2:1 salts from the-non symmetrical   molecule, $\mathrm{(EDT-TTF-CONMe_{2})_{2}X}$\cite{Heuze03}
 for
which there is no center of symmetry between the molecules along the stack,
(\emph{at variance} with the \tm2x series and their inherent dimers) as a  result of a glide plane symmetry  and where
extensive refinement of X-ray synchrotron data indicate a uniform distribution
of the molecular units, \emph{see} Fig.~\ref{EDTv2}, yielding a band quarter-filled with holes\cite{Zorina09}.
\begin{figure}[h]
\includegraphics[width=0.9\hsize]{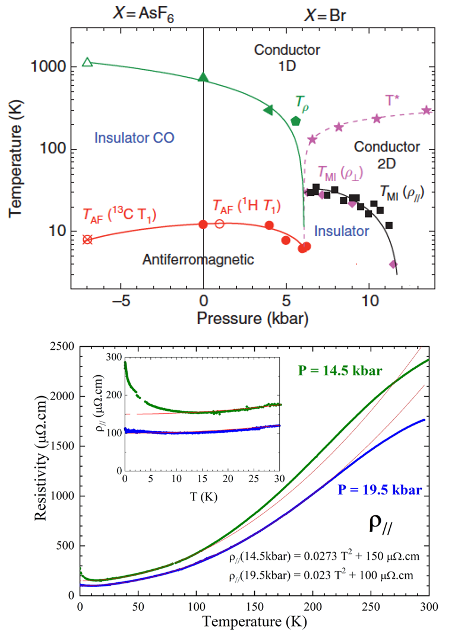}
\caption{\label{EDT2Xdiagv1} Generic ($P/T$) phase diagram for the quarter-filled compounds $\mathrm{(EDT-TTF-CONMe_{2})_{2}X}$, Source: Fig. 1 [Ref\cite{Auban09a}, p. 2].  Open and close symbols refer to $\mathrm{(EDT)_{2}AsF_{6}}$ and $\mathrm{(EDT)_{2}Br}$, respectively. The origin of pressure is taken for $\mathrm{(EDT)_{2}Br}$. High-temperature triangles provide a determination of $T_{\rm CO}$ from the $P$-dependence of the longitudinal conductivity at room temperature assuming a BCS relation between the gap and the transition temperature. $T_{CO}$ is also deduced from the kink on the $P$-dependence of the conductivity at room temperature (triangle at 4 kbar). The minimum of the longitudinal resistivity ($T_{\rho}$) in the 1D regime is reported when it is observed below room temperature (triangle at 4 kbar, losange at 5.6 kbar)).  Circles (crossed (x)) indicate the antiferromagnetic ground state observed by $H$ and $^{13}$C NMR\cite{Zorina09,Auban09a}. In all respects, the high temperature 1D regime of this diagram is similar to the generic diagram of the \tm2x series displayed in Fig.~\ref{Trho}. The insulating phase suppressed around 12 kbar is non-magnetically ordered unlike the SDW phase of the \tm2x diagram. The lower part   shows the $T$ dependence of $\rho_a$ without any sign of superconductivity above 70 mK. Notice the absence of the $T$-linear contribution to the resistivity at low temperature next to superconductivity due to AF fluctuations and  commonly observed in the \tm2x series, \emph{see} Sec.\ref{precursors}.
}
\end{figure}

 The $T/P$ phase diagram shown in Fig.~\ref{EDT2Xdiagv1}, drawn from a combination
of transport and NMR measurements on two members of this family, is simpler than for the TMTTF
series because of the absence of dimerized stacks\cite{Jerome10}. At ambient pressure
and $T$ = 300K, both compounds in the EDT series are in a  CO (Wigner) insulating state and
cooling results below about 12K in antiferromagnetically ordered ground states. The application
of pressure suppresses the CO phase in favour of a quasi-1D conductor
that, upon cooling, undergoes a transition toward which is likely to be a
Peierls state since no sign of magnetic ordering could be detected in NMR
experiments. Still higher pressures suppress the Peierls transition, presumably
as the nesting condition responsible for this Peierls transition is weakened.
Similar properties were observed for both compounds provided the
pressure is shifted for the $\mathrm{(EDT)_{2}AsF_{6}}$ material by 7 kbar relative to $\mathrm{(EDT)_{2}Br}$ and for this reason the phase diagram as presented in Fig.~\ref{EDT2Xdiagv1}  applies to both
compounds, $\mathrm{(EDT)_{2}Br}$ and $\mathrm{(EDT)_{2}AsF_{6}}$, \emph{albeit }with the appropriate shift in
pressure. 

The lesson to be draw from these compounds born intrinsically  quarter-filled 1D  organic conductors is the relevance of  1/4 Umklapp processes to induce electronic localization in highly correlated 1D conductors. 

In addition, one should not forget that in spite of serious attempts\cite{Auban09a}, no sign of superconductivity could be detected above 70 mK under high pressure in the generic diagram of Fig.~\ref{EDT2Xdiagv1}. This let us suggest that \emph{the existence  of a SDW ground state may be a prerequisite  for the stabilization of superconductivity in its proximity}. 

So, can this help answer the question raised by T.Giamarchi in his review \cite{Giamarchi04a}: \emph{" ... the
fact that these compounds, $\mathrm{(EDT)_{2}X}$, are indeed insulators and
with a structure similar to the Bechgaard salts
strongly confirms the interpretation that the dominant
mechanism is also in these systems the quarter-filling of the band. It would be of course very
interesting to investigate the phase diagram and the
physical properties under pressure of these compounds.
Since they share the same basis microscopic $\mathrm{(EDT)_{2}AsF_{6}}$ features, it is crucial to assert whether these quarter
filled systems also exhibit superconductivity under pressure as in the Bechgaard salts"}.
 
We feel that, although there is now little doubt about the predominance of the quarter-filled picture on the physics of high temperature phases,  the respective role of half-filling versus quarter-filling for the existence of superconductivity in these Q1D compounds  remains an important open question.   Why  the metallic phase of $\mathrm{(EDT)_{2}X}$ compounds under pressure does not exhibit superconductivity is a challenge, which is probably waiting for additional investigations from physicists and possibly new materials from chemists. 
Let us suggest that the dimerization which is present in the \tm2x series   may be a hidden prerequisite for the existence of superconductivity in these 2:1 conducting salts.

\subsection{Superconductivity and symmetrical versus non-symmetrical anions}
\label{non-symmetrical anions}
By the end of 1980, the Copenhagen group in the attempt  to stabilize the superconducting state under ambient pressure succeeded with the evidence  from resistance measurements  of  a superconducting transition in \tmc beginning around temperatures between 1.2K and 1.4K depending on the  samples\cite{Bechgaard81,Bechgaard81a}. The rationale behind changing the anion $X$ from octahedral \pf to tetrahedral \cl is that the latter is smaller, which might have the same effect on the conducting TMTSF stacks as pressure has in (TMTSF)$_2$PF$_6$. The idea of a synthesis using  \cl anions turned out to be justified since structural analysis shows that the two compounds \tmp6 and \tmc are isostructural with a unit cell volume for \tmc smaller by 2.8\% compared to \tmp6.\emph{"We wish to recall that it is important to be aware that the preparation of salts with {\cl} anions requires the handling of unstable and possibly explosive perchlorates. This has led to the death of a chemical engineer, Mrs Maldy, at the Orsay laboratory. We should not forget her".}
 In the triclinic structure
of \tmp6, \pf resides in a center of inversion\cite{Thorup81}. In the case of non-centrosymmetrical anions such as \cl or \re , the tetrahedral anion  located at inversion center of the structure but lacking the inversion symmetry will introduce an ordering expected  at low temperature for the entropy minimization and consequently 
change the periodicity of the lattice. It can also, by preventing perfect ordering after fast cooling, introduce  non-magnetic disorder in an otherwise  superconducting compound. This possibility will be  a determining factor in characterizing the nature of the mechanism of organic superconductivity as we shall \emph{see} in Section~\ref{Cooling rate-controlled superconductivity}.

\begin{figure}[h]
\includegraphics[width=0.7\hsize]{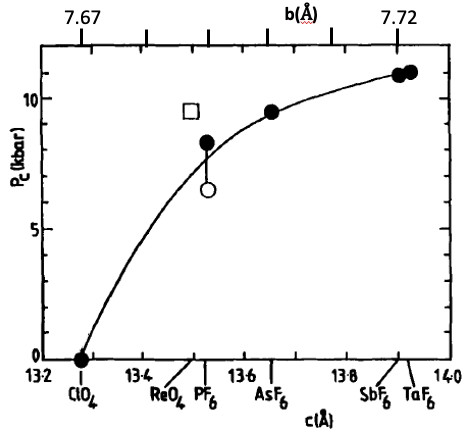}
\caption{\label{ParkinPc} A correlation between the critical pressure for the stabilization of a SC ground state and transverse lattice parameters of several superconducting  members of the \tm2x series,  Source: Fig. 9 [Ref\cite{Parkin81a}, p. 5316]. The critical pressure correlates   better with the $c$  rather than the $b$ parameters as shown on this diagram, since   by  $b$= 7.71\AA\ for \tmp6\cite{Thorup81} is actually similar to $b$= 7.72\AA\  in $\mathrm{(TMTTF)_{2}SbF_{6}}$\cite{Ducasse85}.}
\end{figure}

The investigation of superconductivity in the entire \tm2x family\cite{Parkin81a} has shown that there exists a critical pressure, above which the superconducting phase is stabilized with respect to the insulating phase stable at low
pressures and temperatures. For all compounds belonging to  the \tm2x series but (TMTSF)$_2$ClO$_4$, the critical pressure ranges between $ \approx$  8 kbar and 12 kbar. Furthermore, as implied by the extensive high pressure work\cite{Parkin81a}, the critical pressure correlates  with the separation between the stacks of TMTSF molecules at ambient pressure and room temperature, this separation being related crystallographically to the size of the different anions, Fig. \ref{ParkinPc}.

The possibility of observing  superconductivity in \tmc under ambient pressure  with a \tc around 1.2K\cite{Bechgaard81} has made sophisticated experiments much more accessible than for the case of  centrosymetrical anions. 
\begin{figure}[h]
\includegraphics[width=0.7\hsize]{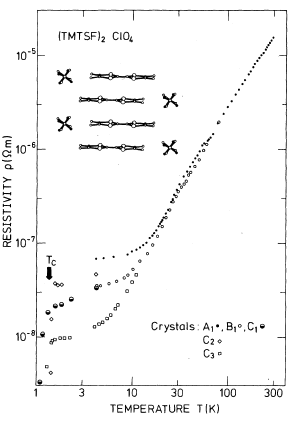}
\caption{\label{supraClO4} Temperature dependence of the resitivity for \tmc. Results for different samples are normalized  to the same room temperature value of 1.45 $\times 10^{-3} \Omega$.cm, Source: Fig.1 [Ref\cite{Bechgaard81}, p.853]. Notice that from the log plot, the resistivity is still temperature dependent at liquid helium temperature, very much like \tmp6 using a $T$-linear plot on Fig.~\ref{Firstsupra}. }
\end{figure} 
In addition, 
as far as \tmc is concerned, ambient pressure may even  be slightly above the critical pressure since it has been shown that  a SDW phase can be stabilized under an uniaxial tension\cite{Kowada07}. 

These experimental findings have had a major influence on the development of the theory of superconductivity of these quasi 1D conductors to be discussed later for which the transverse coupling is the major parameter for determining the stability of the SC state.

Although most of the physics of organic conductors
is governed by the organic molecules, the anions, the
presence of which is essential for electric neutrality,
may in some case suppress the stability of the
conducting phase. As a matter of fact, the possibility
for \tm2x compounds having non-centrosymmetrical
anions to undergo a structural phase transition can
modify the band structure and the topology of the
Fermi surface. 

The role of anions is particularly manifest
when it is compared to compounds with spherical
anions such as \pf, \as, and so forth, for which the
absence of alteration of the Fermi surface via anion
ordering entails, for example, the stabilization of spin
density-wave long-range order in zero magnetic field
at ambient pressure. The anion potential produced
by spherical anions  leads to a modulation of the charge along the
organic stack with the same periodicity as the dimerization. It may independently contribute to the
half-filled character of the band and then enhance
the strength of the electron-electron interaction at
low temperature\cite{Bruinsma83,Emery83}.

On the other hand, anions such as \cl, \re, \no, \scn,
and \fso3 have two equivalent orientations corresponding
to short and long contacts between the Se
 atoms of the TMTSF molecule and a peripheral electronegative
atom of the anion. 
\begin{figure}[h]
\includegraphics[width=0.6\hsize]{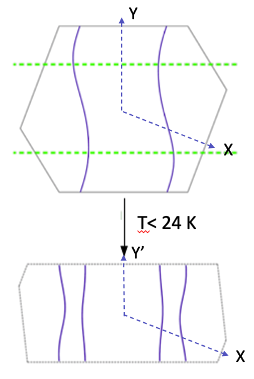}
\caption{\label{anionordering} At the \cl ordering temperature of 24 K, new Brillouin zone boundaries are formed at $\pm \pi/2b$ (top) leading to a folding of the Fermi surface (bottom). }
\end{figure} 
Consider the case of (TMTSF)$_2$ClO$_4$, a compound from which most of the characteristics of the organic superconducting state have been obtained due to the stability of this phase under ambient pressure. The anion lattice orders at 24 K, leading
to a superstructure of the Se-O contacts with a wave
vector $q= (0, 1/2, 0)$\cite{Pouget83}, \emph{see} Fig.~\ref{anionordering}.  The periodic potential
thus created,  connects two Fermi points along the transverse  $b$
direction and opens a gap with a new zone boundary which doubles the unit cell
along that direction. The folding of the Fermi surface
that results introduces two warped Fermi surfaces
near $\pm k_F$ providing an only limited disturbance to the conductivity with a slight drop of the resistivity at  24K related to the decrease of the elastic scattering.

Anion lattice superstructure has thus
important consequences on the one-particle spectrum
and in particular the nesting properties of the Fermi
surface. This plays an important role in the efficiency
of electron-electron interactions at low temperature
and in the nature of the ground states as remarked by Bruinsma and Emery\cite{Bruinsma83}. Anion ordering also controls the stability of the superconducting phase in \tmc below 1.2 K at ambient pressure when the
long range orientation is defective as already mentioned. 
The proper ordering of the \cl anions at wave vector $q= (0, 1/2, 0)$ in \tmc requires much care during the sample cooling procedure. As it has been reported in the early study of this compound\cite{Takahashi82}, too rapid cooling (although then accidental). The   influence of the anion ordering on the nature of the \tmc ground state  is a good illustration for the famous talk pronounced by   Louis Pasteur  at the University at Lille  in 1854\cite{Pasteur54} \emph{"In the fields of observation, chance only favors the prepared mind".} 
The issue of the ordering of non-centrosymmetric anions challenged  Toshihiro Takahashi  who one day as he was trying at Orsay   to confirm  the magnetic nature of the strange phase observed at high fields above 6T or so  (now known as  the field induced SDW phase, \emph{see} Sec.~\ref{magnetoresistance oscillations}). When he  saw the disappearance of $^{77}$Se-NMR signal, he thought  he had got the evidence for the field induced phase. However, next morning, cooling  the sample again (probably too fast) to check its reproducibility, he failed to observe the same behavior and observed a magnetic ordering below $\approx$3.5 K nearly field-independent between 1 and 6.5T. This observation  has been subsequently  related to the anion ordering with a wave vector $q = (0, 1/2, 0)$ occurring at 24K by X-ray diffuse scattering experiments and providing the clue for the understanding of the dependence of the nature of the electronic  ground state on the  cooling rate dependence\cite{Pouget83}.

This effect called the  ``quenched" state of \tmc has been extensively used for the characterization of the superconducting phase as it enables a proper control of the electron elastic scattering in the superconducting phase  to be discussed below, see Sec.~\ref{Cooling rate-controlled superconductivity}.


 For other compounds with a non-centrosymetrical
anion like \re, the structural ordering is different and takes place at $q= (1/2, 1/2, 1/2)$; its impact on the electronic structure turns out to be more marked, since the anion potential at this wave vector creates a gap over the whole Fermi surface which is so large in amplitude (of the order of the Fermi energy) 
that it leads \emph{at variance} with \tmc to an insulating state at 200K in which electron-electron interactions probably play little role. The application of an hydrostatic pressure is then required to establish a less damaging anion ordering configuration\cite{Moret86}  and restore the metallic state with the possibility of long range ordering for the electronic degrees of freedom\cite{Parkin82,Tomic89a}.
\begin{figure}[h]
\includegraphics[width=1\hsize]{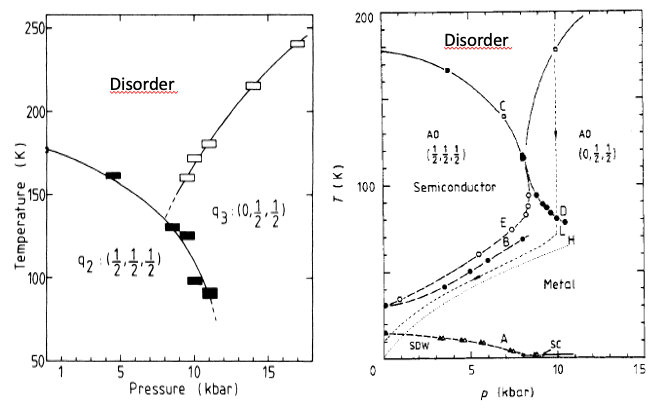}
\caption{\label{MoretTomic} (Left) T-P phase diagram  of \tmtsfreo4 showing the various phases with their anion ordering according to. (Right) Applying pressure above 200K, entering the  conducting (0,1/2,1/2) phase, cooling down to  72K and then releasing pressure and temperature above the melting line of helium (the small dotted line) allows to bypass the semiconducting (1/2,1/2,1/2) phase and keep the (0,1/2,1/2) ordering  in a metastable state to explore the electronic instabilities (SDW and SC) in the low temperature region, Sources: Fig. 1 [Ref\cite{Moret86}, p. 1916] (left), Fig. 1 [Ref\cite{Tomic89a}, p. 4453] (right).    }
\end{figure} 
The case of \tmtsfreo4 is particularly instructive since thanks to the possibility  to change the helium gas pressure at low temperature (although above the freezing point of helium gas) it is possible to bypass the phase (1/2,1/2,1/2) anion ordered and release the pressure at low temperature, \emph{see} Fig.~\ref{MoretTomic}. This experiment has shown that the (0,1/2,1/2) anion ordering does have a bad impact on the stability of SDW and SC phases at low temperature.

For (TMTSF)$_2$BF$_4$, having a tetrahedral anion as in (TMTSF)$_2$ReO$_4$, no structural analysis has been carried out so far; however, the very sharp transition at relatively high temperature (39K) compared with hexafluoro-compounds suggests a
transition similar to that of (TMTSF)$_2$ReO$_4$. Also, the observation of an ESR line
below the transition 
clearly shows that the insulating state is not magnetic. Although this anion ordered insulating phase can be removed by pressure  no superconductivity could be evidenced above 1.2K in (TMTSF)$_2$BF$_4$\cite{Parkin81a}. 
\subsection{A detailed investigation of the ambient pressure superconductor \tmc}

\subsubsection{Superconductivity under magnetic field, very anisotropic type II superconductors}
\label{very anisotropic type II}
The application of a magnetic field has led to important and unexpected results for the characterization of these anisotropic superconductors. 

The bulk nature of superconductivity  could be inferred from  diamagnetic shielding  (AC susceptibility) measurements   observed in the  early days  of this research, pointing to a type II superconductivity in \tmp6\cite{Ribault80}, but it has been confirmed by more detailed   Meissner effect experiments (i.e., a flux expulsion on cooling in a field)\cite{Andres80}  and subsequently in \tmc\cite{Gubser81,Schwenk84,Gubser82,Oh04}.
\begin{figure}[h]
\includegraphics[width=0.8\hsize]{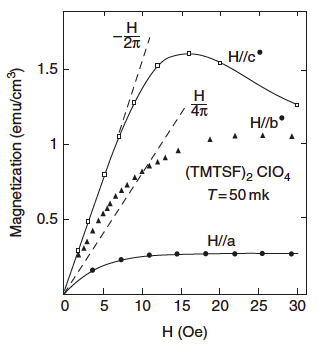}
\caption{\label{Mailly} Magnetization of \tmc at $T= 0.05$ K for magnetic fields along the three axes, Source: Fig. 3 [Ref\cite{Garoche82}, p. 714].  }
\end{figure} 
Regarding evidences for the Meissner expulsion, the lower critical field $H_{c1}$ is obtained from the magnetization curves of \tmc at 50 mK\cite{Mailly82}. 
According to Fig.~\ref{Mailly} the obtained values are 0.2, 1, and 10 Oe along the $a$, $b$, and $c^{\star}$ axes, respectively. 
\begin{figure}[h]
\includegraphics[width=1.05\hsize]{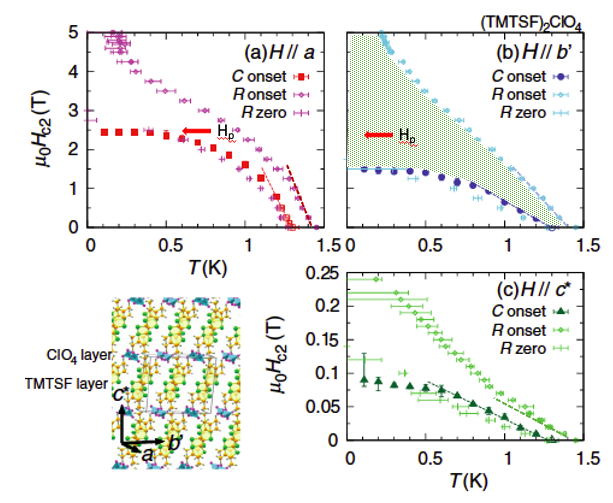}
\caption{\label{Hc2Cvrho} $H_{c2}-T$ phase diagram of \tmc obtained from specific heat and the $c^{\star}$ resistivity measurements   with the magnetic field aligned along the three crystal axes. The Clogston limit for spin-singlet pairing at $T=0$ is shown as the red arrow. The orbital limitation in field  applies to $H\parallel {b'}$ and  $H\parallel {c}$. The shaded region for $H\parallel {b'}$ corresponds to the HSC domain of superconductivity, Source: Fig. 5 [Ref\cite{Yonezawa12}, p. 3]. }
\end{figure} 
Following the values for the upper critical fields $H_{c2}$ derived either from the Meissner experiments and the knowledge of the thermodynamical field $H_{c}=44$ Oe\cite{Garoche82} or from a direct measurements of transport\cite{Yonezawa12}, it is found that superconductivity is in the extreme type-II limit. The Ginzburg–Landau parameter $\kappa$ can even overcome 1000 when the field is along the $a$ axis due to the weak interchain coupling, making field penetration very easy for this external-field configuration. An interpretation of the critical fields assuming the clean limit has been suggested in 1985\cite{Gorkov85}. The clean limit theory could be  questionable nowadays with the current knowledge of the evolution of \tc in the presence of controlled non magnetic impurities\cite{Yonezawa18}, but based on the microscopic expressions for the effective mass tensor in the Ginzburg–Landau equations for anisotropic superconductors near \tc lead, using the experimental slopes of the transport $H_{c2}$\cite{Yonezawa12,Gorkov85} (\emph{see} Fig.~\ref{Hc2Cvrho}),
$dH_{c2a}/dT = -67$kOe/K, $dH_{c2b'}/dT = -36$kOe/K, $dH_{c2c^{\star}}/dT = -1.5$kOe/K, to     
$t_{a}:t_{b'}:t_{c^{\star}}$ = 1200-818, 310-220 and 7-4.6 K, which are of the order  of the  tight binding band parameters\cite{Grant82,Ducasse85,Ducasse86,Ishiguro98}. From the experimental slope $dH_{c2}/dT$ in the vicinity of $T_c$, a derivation of the coherence length is possible for \tmc  and leads to 700, 330 and 20\AA\  along $a, b$ and $c^\star$ directions respectively\cite{Murata87}.

Figure.~\ref{Hc2Cvrho} reveals several important features. First, paying attention to the  behaviour of the transport $H_{c2}$ we notice a divergence decreasing temperature, in particular below 0.2K for both $a$ and $b'$ directions. Such a divergence had already been noticed in 1997 in the superconducting state of \tmp6\cite{Lee97,Lee00} with peculiarity that the $H_{c2}^{b'}$ becomes even larger than $H_{c2}^a$ above the field of order $2$ T. A few years later a similar situation has been encountered in \tmc\cite{Lee95} and from resistivity and torque measurements\cite{Oh04}, confirmed subsequently by the extensive measurements performed by the Kyoto group\cite{Yonezawa08a,Yonezawa08} showing that transport critical fields along $a$ and $b'$  largely exceed the orbital critical field of about $2.4$ T given by the Clogston limit\cite{Clogston62} $H_{p}(T=0)=1.84T_c$ for singlet superconductors. 

Second,  most papers following experiments  until 2004 were pointing toward triplet paring in the \tm2x series but following recent specific heat field-angle resolved measurements in the superconducting phase of \tmc\cite{Yonezawa12} a new $H_{c2}-T$ phase diagram has been drawn as also displayed on Fig.~\ref{Hc2Cvrho}, opening other possibilities.
The salient difference between resistivity and thermodynamic phase diagrams is most clearly seen on Fig.~\ref{Hc2Cvrho}.
The field at which the specific heat recovers its normal state value is much smaller than the field related to the onset of the resistivity drop in temperature and above this field superconductivity has a density of states nearly equal to that in the normal state\cite{Yonezawa12} and also Fig.~\ref{BrownShingo}, \emph{see} the forthcoming  discussion in Sec.~\ref{high field}.

\subsubsection{   Thermodynamics and NMR in the superconducting phase of \tmc}
\label{thermo}
The existence of superconductivity is also evidenced via the measurement of the thermodynamic properties.  But equally  important is the behaviour of the specific heat in superconducting phase which  provides crucial informations about the nature of single particle excitations  \textcolor{black}{ in the presence of a } superconducting gap. Whereas in the usual BCS case, excitations are created over the  finite gap with a probability e$^{-\Delta/k_{B}T}$, the specific heat of superconductors with nodes in their gap function exhibits  a power law behaviour in the limit  of $T\rightarrow0$\cite{Hirschfeld88}. 

 Although thermodynamic evidences for superconductivity in \tmc have appeared as early as 1982\cite{Garoche82}, we will focus this presentation on more recent data\cite{Jerome16}  and  will explain the reasons for this approach.
\begin{figure}[h]
\includegraphics[width=0.8\hsize]{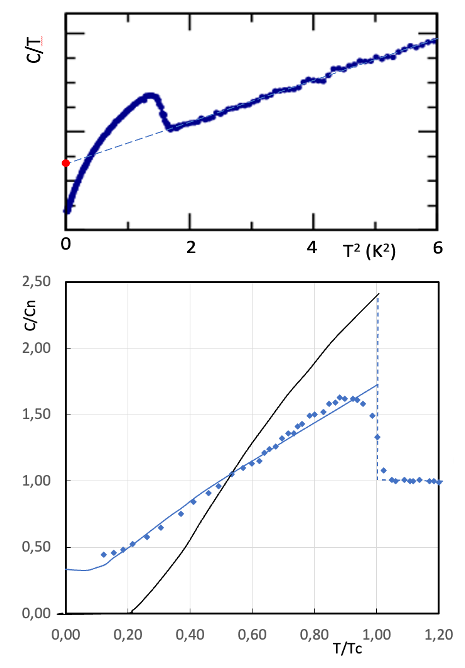}
\caption{\label{Cvsupra} (Top) Total specific heat for a \tmc single crystal of 0.1mg in the relaxed (R) state showing a  law $C/T = \gamma + \beta T^2$  at $T >1.3K$ for the electrons and phonons contributions  with the anomaly at the mid-transition \tc = 1.22K due to the superconducting transition. The red dot at $T=0$ is the value of the normal state electronic   specific heat, $\gamma=$10.6 mJ/K$^2$mol for \tmc.  (Bottom) Electronic contribution  normalized to the normal state value (blue dots) plotted as $C/C_n$ against $T/T_c$. The black continuous line is the behaviour expected for  a weak coupling fully gapped BCS superconductor while the blue continuous line is the behaviour expected for a polar superconducting state with  a ratio $\Delta C/C_{n}$ $\approx 0.73$
for $\Gamma /T_c$ =0.1 from ref\cite{Hirschfeld88}. This value of the scattering rate  for a pristine sample agrees with the measurements of a slowly cooled \tmc crystal\cite{Yonezawa18}. 
We thank Shingo Yonezawa for communicating the data used for this figure, also published in Ref\cite{Jerome16}.}
\end{figure} 
Fig.~\ref{Cvsupra} displays the specific heat of a \tmc single crystal weighing 0.3 mg versus temperature  behaving like  $C/T = \gamma + \beta T^2$ in the low temperature regime with $\gamma=$10.6 mJ/K$^2$mol and $\beta=$10 mJ/K$^4$mol. These values are in fairly good agreement with the previous data,\cite{Garoche82} but there are some notable differences which need serious comments. The two major differences between the data displayed on Fig.~\ref{Cvsupra} and those of 1982 are on the one hand the dependence of $C/T$ in temperature below \tc which reveals a striking difference with the usual BCS behaviour and on the other hand the jump of specific heat occurring at the transition even though P. Lee wrote in his article PRL 71, 1887, 1993 that "according to conventional wisdom, a single experiment showing activated behavior is sufficient to invalidate the d-wave hypothesis, we think that in the particular case of Q1D organic superconductivity there exists a considerable amount of recent experimental observations in favour of singlet d-wave which allows to take the conventional wisdom with a grain of salt!". While $\Delta C$/$\gamma$ \tc amounts to 1.67 in the data of ref\cite{Garoche82}, Fig.~\ref{Cvsupra} leads to a ratio smaller than unity, actually $\Delta C$/$\gamma$ \tc = 0.73 (which is quite reproducible between different samples)  fitting these data with the theory of polar superconductors\cite{Hirschfeld88}. Using the fit for the data on Fig.~\ref{Cvsupra}, the residual density of states at $T=0$ amounts to 35\% of the normal state value.
\begin{figure}[h]
\includegraphics[width=0.8\hsize]{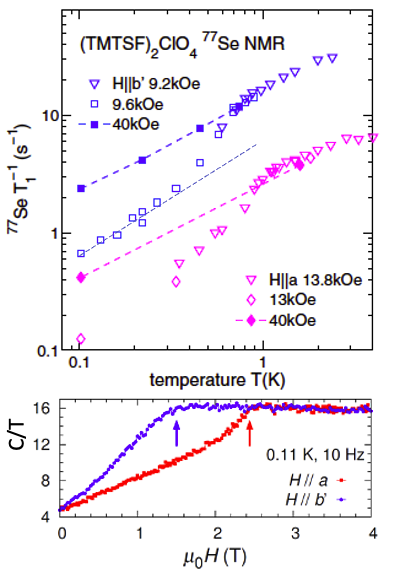}
\caption{\label{BrownShingo} (Top) $1/T_{1}$ versus $T$ for the $^{77}$Se spins of \tmc for $H\mid\mid b'$ and $H\mid\mid a$ according to Ref. The dotted blue  line at low temperature for the field $H\mid\mid b'=0.96T$  has a slope unity and suggests that the relaxation at low temperature is due to a residual density of states  provided by unpaired  single particles at the nodes of the SC gap. (Bottom) Field dependence of $C/T$ at $T=0.3$K, Sources: Fig. 2 [Ref\cite{Shinagawa07}, p. 3] (top), Fig. 1 (top) [Ref\cite{Yonezawa12}, p. 2] (bottom).
} 
\end{figure} 
The thermodynamic data suggest the existence of a finite residual density of states at Fermi level in the very low temperature regime even for a pristine sample. This experimental result is corroborated by the temperature dependence of  the  spin-lattice relaxation of $^{77}\mathrm{Se}$ nuclei   performed  in the superconducting phase of \tmc\cite{Shinagawa07} as single particle excitations are probed by $1/T_{1}T \propto \chi^2 (q=0,T) $ according to the celebrated Korringa law. Fig.~\ref{BrownShingo} shows that the Korringa law is fairly well obeyed above \tc (which amounts in the field of 0.96T~$ \| b'$ to 0.8K  and also below 0.3 K with a $T$-linear dependence of the relaxation rate with a prefactor in the field of 0.96T~$\| b'$  decreased by about 3.3 between high and low temperature regimes. The prefactor being proportional to the square of  the density of states, this result  shows that the density of states at Fermi level has decreased by a factor 1.8  from just above \tc  and low temperature in the SC state. 

In addition, since electronic specific heat measurements at low temperature have shown that the density of states measured by $C/T$  is $H$-field dependent, (\emph{see} bottom of Fig.~\ref{BrownShingo}) the density of states drops by another factor two from $0.96$ T down to zero field  leading to a final residual density of states at zero magnetic field which amounts to 28\% the normal state value. It is a value although different,  which compares favourably with the residual density of states given by the specific heat fit (35\%)  displayed on Fig.~\ref{Cvsupra}. 

The behaviour of the specific heat below \tc as shown on the latter figure  is very reminiscent of the numerous experimental studies performed on Uranium-based heavy fermions materials over the years 80 to 90's, systems in which non conventional pairing in the superconducting phase, in particular polar or axial gap functions must be considered\cite{Hirschfeld88,Flouquet19}. We intend to return to the comparison with heavy fermions when we look at the field-angular dependence of the specific heat.

One feature that the $^{77}\mathrm{Se}$ NMR experiments\cite{Shinagawa07} have provided is the existence of a residual density of states at $T=0$ in the superconducting state of \tmc but it is equally important to remark that between \tc and the the low temperature regime $1/T_1$ follows a $T^3$   law, the signature for lines of superconducting gap nodes. Actually, this possibility had been considered long ago  by Takigawa {\it et al.,} who reported a spin lattice relaxation for the protons of \tmc behaving like $1/T_{1} \propto T^{3}$ which goes  with the absence of the common enhancement of relaxation rate just below \tc in usual BCS superconductors\cite{Takigawa87}. This led Hasegawa and Fukuyama\cite{Hasegawa87} to propose  that superconductors in the \tmtsf2x family belong to a novel class with the order parameter exhibiting lines of nodes on the Fermi surface. However, both spin lattice relaxation and specific heat measurements suggesting nodal superconductivity were unable to discriminate between ($p$)-triplet  and ($d$)-singlet  pairing.

\begin{figure}[t]
\includegraphics[width=0.95\hsize]{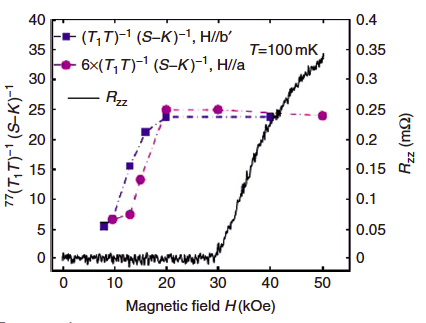}
\caption{\label{Relaxation} \tmc $^{77}$Se spin lattice relaxation for two orientation of the magnetic field showing the large difference in temperatures between the onset of the SC transition from relaxation and the zero of the resistive transition, Source: Fig. 3 [Ref\cite{Shinagawa07}, p. 3].}.
\end{figure}

In the early days of these experimental studies, the option of spin-triplet superconductivity has been claimed following a Knight shift experiment\cite{Lee02} performed on \tmp6 showing no change of the resonance shift through $T_c$. However, this conclusion could have been reached by a still poor understanding of the superconducting $H-T$ phase diagram at that time.
It can be noticed that the 2002 experiment in \tmp6 had been conducted under a relatively high magnetic field of $1.43$ T aligned along the most conducting $a$ axis. Given  NMR studies performed five years later on the superconducting state of the  sister compound \tmc, it was therefore quite likely taking into account the possible different $H-T$ superconducting phase diagrams between \tmp6 and \tmc, that the field of $1.43$T used in ref\cite{Lee02}   locates  the data of the \tmp6 sample in the high-field SC phase (so-called HSC, \emph{see} Fig.~\ref{Hc2Cvrho} for the case of \tmc) in which the spin susceptibility  does not reveal any noticeable change through the  resistive SC transition\cite{Shinagawa07} \emph{at variance} with  the low field state behaviour. 
 Different  thermodynamic and resistive SC transitions are also evidenced from NMR relaxation and resistive measurements, \emph{see} Fig.~\ref{Relaxation}.

\begin{figure}[t]
\includegraphics[width=0.75\hsize]{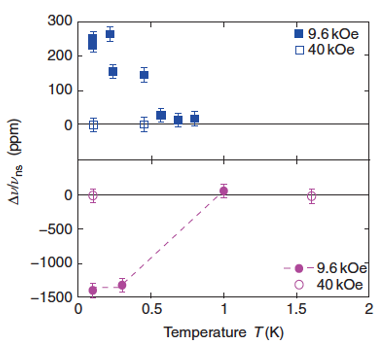}
\caption{\label{Knight}\tmc $^{77}$Se line shift versus temperature for  $H \parallel $ $b'$ (top) and $a$ (bottom)  with the zero shift value arbitrarily set to the normal state first moment position, Source: Fig. 1 (bottom) [Ref\cite{Shinagawa07}, p. 2]}.
\end{figure}

Therefore, it is the NMR result of 2007 which has really contributed to answer important questions regarding the nature of the spin pairing in the superconducting phase. As shown on Fig.~\ref{Knight}, in a field of $0.96$ T  parallel to  $b'$ or $a$ axes the drop of selenium Knight shift below \tc  is now providing a solid evidence in favour of spin-singlet pairing. It is unfortunate that the study of the superconducting state  as it was conducted on \tmc and to be presented in the following section, could not be performed on (TMTSF)$_2$PF$_6$. It does not seem that this crucial measurement should meet major technical challenges in the future.
\begin{figure}[h]
\includegraphics[width=0.95\hsize]{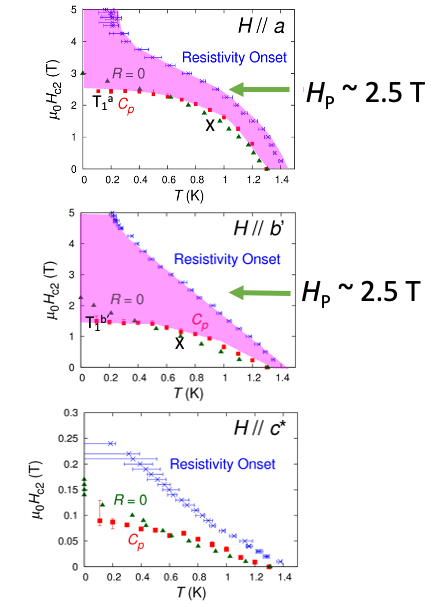}
\caption{\label{Locations} $H_{c2}-T$ phase diagram for \tmc, similar to Fig.~\ref{Hc2Cvrho} on which the location of NMR data points from ref.\cite{Shinagawa07} have been added. The high field phase is shown as the shaded area. ${\mathrm X}$ is related to the onset of the $1/T_{1}$ transition on cooling and  ${\mathrm T_1}$ marks the field location where the normal state value of  $1/T_{1}T$ is recovered at increasing field. Therefore, according to this diagram,  $0.96$ T$\| b'$  and $1.43$ T $\| b'$ are   in the LSC and HSC domains respectively. $b'$ is perpendicular to the $a-c^{\star}$ plane.}
\end{figure} 
In conclusion, although different symmetries for the order parameter in \tmc and \tmp6 cannot be totally ruled out as long as an  NMR investigation of the latter compound  is conducted at  fields below 1 T, we consider such a scenario as quite unlikely. In order to show  evidences for spin-singlet pairing, we have reported on the $H_{c2}-T$ phase diagram of (TMTSF)$_2$ClO$_4$, Fig.~\ref{Locations}, the locations of the various  NMR findings from Ref\cite{Shinagawa07}, (recovery of the normal state $1/T_{1}T$, onset of the $1/T_{1}T$$\propto$$\chi^{2}(T)$ SC transition and the $R_{zz}(H\parallel {a})=0$ point) showing that these data have been taken in the low field superconducting phase (LSC). As we can see, all these points are in good agreement with the diagram derived from  the zeroes of the resistivity or from the onset of the specific heat anomaly. Bulk superconductivity becomes Pauli limited when $H\parallel {a}$.  Consequently, it is now hard to dispute the experimental fact that the superconducting pairing in the \tms2x family is spin-singlet.

\subsubsection{The high magnetic field phase}
\label{high field}
The low field domain of superconductivity of \tm2x salts is fairly well understood in terms of a GL model for a 3D anisotropic type II superconductor as developed in Sec.~\ref{very anisotropic type II}.
However, as already mentioned in Sec.~\ref{thermo}, \tm2x salts are characterized by a divergent behavior  above the Pauli critical field at decreasing temperatures for the critical field $H_{c2}(T)$ determined from transport measurements. This feature is particularly clear on Fig.~\ref{Locations} for \tmc with the shaded areas,  but a similar behaviour had first  been reported in 2000 for \tmp6 under pressure in which the onset of superconductivity persists up to four times the Pauli limit when $H_{c2} \parallel b$\cite{Lee00}. These latter authors have considered the possibility of overcoming the Pauli effect thanks to the stabilization of  an inhomogeneous Larkin-Ovchinnikov-Fulde-Ferrell (LOFF) superconducting state\cite{Fulde64,Larkin65}, although they discarded such a scenario because of no evidence of first order transition between  homogeneous and inhomogeneous states, instead, they privileged the scenario of a triplet pairing in order to exceed the pair-breaking limit, a scenario that has been made obsolete after new NMR experiments\cite{Shinagawa07}.


Before looking at the SC in high magnetic field, let us concentrate on the normal state above \tc but in the 3D coherent regime, i.e. $T< t_{c}$,  ($T< 5$K or so). The role of the magnetic field perpendicular to the $c$ axis, suppressing the coherent hopping between adjacent $a-b$ planes thus resulting in a 2D confinement has been considered by Strong {\it et al.}\cite{Strong94} and its effect             
on the conductivity along $c$  has been treated  formally by the Kubo formula\cite{Joo06} but  a simple semiclassical treatment  gives rise, in real space, to a linear trajectory along the $a$ direction modulated by periodic oscillations along the $c$ axis with an
amplitude $\delta z$ = $4ct_{c}/ev_{F}H$ which decreases as the magnetic
field increases, making the electrons confined in the
($a-b$) plane\cite{Lebed86}. For a field strength of the order of $t_{c}$, the electrons
get confined in a single plane perpendicular to $c$. 
Finding $d\rho_{c}(T)/dT<0$  at fields above $H\parallel b'$$\approx$ 1T  is consistent with an inter-plane coupling of order of a few degrees\cite{Joo06}.

In this semi-classical picture, we do not expect a strong effect of the field on the conductivity along the $a$ direction. On
the contrary, the electron confinement in the ($a-b$) planes
should correspond to a field induced dimensional cross-over  (FIDCO) in the $c$
direction accompanied by a metal to insulator transition, as observed experimentally in \tmp6 under pressure\cite{Lee97} and also in \tmc at ambient pressure\cite{Lee95}. 

Lebed has shown that the FIDCO should suppress the orbital pair-breaking limitation leading in turn to the stabilization of superconductivity at fields higher than $H^{orb}_{c2}$.

A major breakthrough arose when the Kyoto group managed to perform a study of the high field SC phase under accurately aligned fields using a vector magnet\cite{Yonezawa08,Yonezawa08a}. For the sake of brevity, let us mention only the results  for $H$ perfectly aligned with the $a-b$ plane.

 In 2008, Yonezawa {\it et-al}\cite{Yonezawa08,Yonezawa08a} investigated the in-plane field-angle dependence of the onset temperature of superconductivity, $T_{co}$, based on the $c$-axis resistance measurements of \tmc single crystals\cite{Yonezawa08,Yonezawa08a}. 
 
\begin{figure}[h]
\includegraphics[width=0.7\hsize]{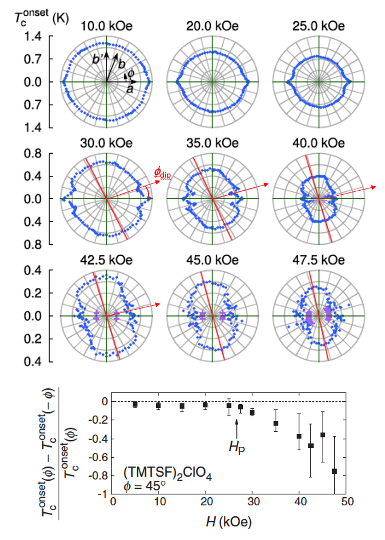}
\caption{\label{Tconset} (Top) Polar plot of the $\phi$ dependence of $T_{co}$ at several magnetic fields below and above the Pauli limit of 2.5 T. $\phi_{dip}=17^{\circ}$ under 30kOe corresponds to the angle above which  the 2D confinement begins when the field is rotated in the $a-b$ plane. The new principal axis $X$ is observed above 30 kOe. (Bottom) Field dependence of the relative  angular difference between positive and negative $\pm\phi=45^{\circ}$ showing that the stability of the high field phase is linked to the Pauli limiting field at 2.5 kOe, Sources: Fig. 3 [Ref\cite{Yonezawa08}, p. 3]. }
\end{figure}

  What was found is that for $H\parallel $ either  to  $b' $or to $a$, $T_{co}$ remains finite up to 5T at least, \emph{see} Fig.~\ref{Locations}. The behaviour for $H\parallel a$ resembles the theoretical treatment of a FFLO scenario\cite{Fuseya12,Miyawaki14} but the peculiar feature was found with  the maxima of the $T_{co}(\phi)$ curve,  located at $\phi$=0 deg for $H\parallel a$ and $\phi$=90 deg for $H\parallel b'$ at fields lower than 2.5 T shifting away from the crystalline $b'$ and $a$ axes at high fields, \emph{see} Fig.~\ref{Tconset}. A new principal axis related to the SC phase is emerging above 3 T (red line) and progressively moves back to $b$ at higher fields, after\cite{Yonezawa08,Yonezawa08a}.
   Although the existence of  high field SC phases have been suggested in heavy fermions superconductors (CeCoIn$_{5}$ in particular, although in this compound  antiferromagnetism may interfere with a text-book FFLO state)\cite{Matsuda07} and in several 2D organic superconductors according to magnetic, thermodynamic and NMR studies, \emph{see} Ref.\cite{Jerome16} for more details,   we choose to   focus in the present article on the possible FFLO state of \tmc because it can be considered as a textbook-like FFLO state supported by the numerous experimental studies performed on \tmc in the Kyoto laboratory\cite{Yonezawa08}. 

 What is remarkable in Fig.~\ref{Tconset} is first the sharp peaking of $T_{co}$ observed above 30 kOe at $\phi= 0^\circ$ together with dips at $\phi=17^{\circ}\pm1^\circ$. As shown by transport data under aligned fields\cite{Joo06}, the angle of $17^{\circ}$ corresponds also to the onset of the FIDCO. Since this angle shows a tendency to decrease at higher fields it has been inferred that these dips are related to an interplay between the FIDCO and the 3D GL theory.
 
 However, the most significant finding is that in magnetic fields above 30 kOe, the $b'$ axis is no longer a symmetry axis of $T_{co}(\phi)$ as it is under low fields, \emph{see} Fig.~\ref{Tconset}. The appearance of the  asymmetry in the rotation pattern is related to the field corresponding to the Pauli pair-breaking effect as shown on Fig.~\ref{Tconset}. 
 
 As seen on the rotation pattern, $T_{co}(\phi)$ is enhanced around $X$ \emph{at variance} with low fields where $T_{co}(\phi)$ exhibits a broad minimum around the $b'$ axis. In addition this $X$ axis shows a clear field dependence tendency  as its deviation from the $b'$ axis is reduced by 10$^{\circ}$ at 47.5 kOe.

\begin{figure}[h]
\includegraphics[width=0.6\hsize]{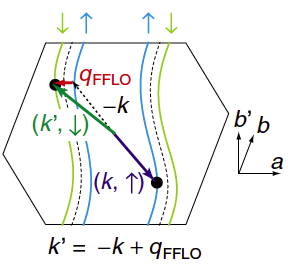}
\caption{\label{FFLO} Pair formation in a Q1D superconductor under magnetic field. Partners in a Cooper pair are the black dots, Source: Fig. 10 [Ref\cite{Jerome16}, p. 9].}
\end{figure}

 The FFLO state can be realized when spin-singlet Cooper pairs are formed between Zeeman-split Fermi surfaces in high magnetic fields. As a result of the Zeeman splitting, the Fermi wave number for the up-spin electron $\bf{k_{F}\uparrow}$
 and that for the down-spin electron $\bf{k_{F}\downarrow}$
 are non centro-symmetrical as shown on Fig.~\ref{FFLO} for Q1D superconductors.
 
 Thus, when a Cooper pair is formed between $\bf{k_{F}\uparrow}$ and $\bf{k_{F}\downarrow}$ electrons, the pair acquires the non-zero center-of-mass momentum $\bf{ q_{FFLO}}$= $\bf{k_{F}\uparrow}$-$\bf{k_{F}\downarrow}$. This momentum results in the spatial oscillation of the SC order parameter. Consequently, this non-zero center-of-mass momentum breaks the original translational symmetry.

 For Q1D systems, $\bf{ q_{FFLO}}$ should be nearly parallel to the $a$ axis, since the number of pairs can thus be maximized and the energy minimized if $\bf{ q_{FFLO}}$ matches the nesting vector between the spin-up and spin-down Fermi surfaces, which is nearly parallel to the $a$ axis, as schematically shown in Fig.~\ref{FFLO} providing the highest \tc under magnetic field\cite{Miyawaki14}. The evolution under magnetic field noticed in Fig.~\ref{Tconset} might be related to the influence of the magnetic field on the Q1D surface given the tight connexion existing between $\bf{ q_{FFLO}}$ and the warping of these  surfaces.
 
 The situation of a field parallel to the conducting axis  ($H\parallel a$) may be somewhat similar to spin-ladders\cite{Mayaffre98} where superconductivity was found experimentally to overcome the Pauli limit\cite{Braithwaite00,Nakanishi05} with the  possibility of a FFLO state, as  proposed by Roux\cite{Roux06}. 
 
\begin{figure}[h]
\includegraphics[width=0.7\hsize]{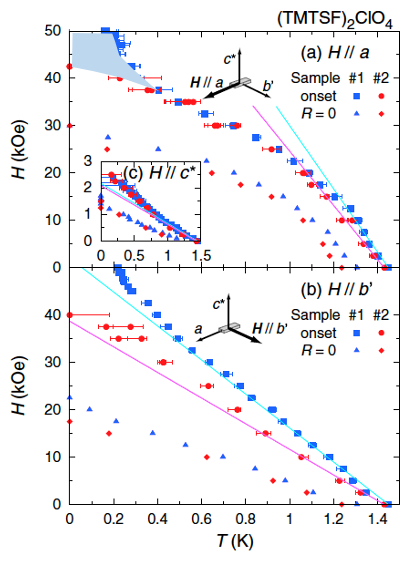}
\caption{\label{FFLO_impurities} Critical fields $H_{c2}$ from the onset of the resistive transition derived from two aligned  \tmc samples. The light blue shaded domain at low temperature and high fields illustrates the strong sensitivity of the FFLO state to impurities although \tc at $H$=0 is only weakly affected, Source: Fig. 4 [Ref\cite{Yonezawa08}, p. 5]. }
\end{figure}

 Another support to the existence of a FFLO phase for $H\parallel b'$ is provided by the experimental data showing that  the stability of the high field regime may be affected by impurities far more  than the superconductivity of the low field regime as predicted by  theories\cite{Agterberg01,Fuseya12}. While the critical field $H_{c2} \parallel b$  data reveal a clear up-turn going to low temperature in \tmp6\cite{Lee00}, the situation is more confused in \tmc. First, the distinct upturn of  $H_{c2} \parallel b$  noticed  in the first cool down of a sample  is no longer observed in subsequent cool downs\cite{Oh04} and second, study of different \tmc samples, Fig.~\ref{FFLO_impurities}, shows that the high-field phase may not be present in certain samples, suggesting that its stability is linked to sample purity\cite{Yonezawa08}. 
 
 The existence of the high field SC state is actually of significant importance for the theory of these superconductors, as it  as been ascribed to the signature of the interplay between magnetism and d-wave superconductivity in these Q1D conductors under magnetic field\cite{Fuseya12,Shahbazi17}.   
 
 Thermodynamics performed under oriented fields in \tmc  is also of great interest in the context of  possible FFLO phases. The recovery of the normal state density of states occurs at $H^{therm}_{c2}$ around 2.5 T (the Pauli limit) for $H\parallel a$\cite{Yonezawa12}, \emph{see} Fig.~\ref{BrownShingo}, i.e at a field much smaller than than the orbital critical field of about 7.7 T. Consequently, there exists a broad SC phase at low temperature above $H^{therm}_{c2}$  without any noticeable change for the density of states compared to the normal state in which transport exhibits a zero resistance. The nature of this state remains to be clarified. A fluctuating  order parameter might be a possibility\cite{Fuseya12},  as well as an inhomogeneous FFLO-like static order.
 
 With the physics of the superconducting phase of \tmc at high fields we are facing quite interesting questions; interplay between orbital pair-breaking
effect, the Pauli pair-breaking effect, the dimensionality
of the electronic system (FIDCO), and the  possible emergence of FFLO
states and their response to impurity scatterings or to the tilt
of the magnetic field away from the $a-b$ plane. Further investigations should be highly valuable 
 using for instance in case of \tmc  the cooling rate to monitor the amount of non-magnetic scatterers.
\subsubsection{Response of superconductivity to non-magnetic defects}
\label{non-magnetic defects}
 
A basic property of the \textit{s}-wave superconductivity proposed in the BCS theory is the isotropic ($k$-independent) gapping on the Fermi surface.  
Hence, no pair breaking is expected from the scattering of electrons against spinless impurities~\cite{Anderson59}, since such scatterings essentially just mix and average gaps at different $k$ positions. 
Experimentally, this property has been verified in non-magnetic dilute alloys of \textit{s}-wave superconductors and
provided a strong support to the BCS model of conventional \textit{s}-wave superconductors. 

However, the condition for an isotropic gap is no longer fulfilled for the case of non-$s$-wave pairing, in which the average of the gap $\varDelta(\bm{k})$ over the Fermi surface vanishes due to sign changes in $\varDelta(\bm{k})$, i.e. $\sum_{\text{FS}}\varDelta(\bm{k}) \sim 0$. 
Consequently, \tc
 for these superconductors should be strongly affected by any non-magnetic scattering, cancelling out positive and negative parts of the gap. 
Theories on effects of non-magnetic impurities on \tc
in such superconductors have been deduced by generalizing conventional pair-breaking theory of Abrikosov and Gorkov (A-G) for magnetic impurities in $s$-wave superconductors\cite{Abrikosov61}.
Then the famous relation,
\begin{eqnarray}
\ln\Big(\frac{T_{c}^0}{T_{c}}\Big)=\psi\Big(\frac{1}{2}+\frac{\alpha T_{c}^0}{2\pi T_{c}}\Big)-\psi\Big(\frac{1}{2}\Big),
\label{Digamma}
\end{eqnarray}
is obtained~\cite{Maki04,Larkin65}, with $\psi(x)$ being the Digamma function and $\alpha$ = $\hbar$ /2 $\tau k_{B}$ ${T_{c}^0}$, the depairing parameter related to the elastic scattering time $\tau$.

Experimentally, it has been found that this relation holds for non-$s$-wave superconductors such as $\mathrm{Sr_{2}RuO_{4}}$ (\tc = 1.5~K; most likely a $p$-wave spin-triplet superconductor)\cite{Mackenzie03,Maeno12}. 

It is also the remarkable sensitivity  of organic
superconductivity to irradiation detected in the early years \cite{Bouffard82,Choi82} that led Abrikosov to suggest the possibility of triplet pairing in these materials~\cite{Abrikosov83}.

A more recent investigation of the influence of non magnetic defects on organic superconductivity has been conducted following a procedure which rules out the addition of possible magnetic impurities, which is the case for X-ray irradiated samples~\cite{Miljak80}. Attempts to synthesize non-stoichiometric compounds have not been successful for these organic salts. However, what turned out to be feasible is an iso-electronic  anion solid solution keeping the charge transfer constant. One attempt has been to  create non-magnetic  disorder through the synthesis  of  solid solutions with centrosymetrical anions such as \as\ and \sb. This attempt turned out to be unsuccessful as the  effect of disorder happened to be very limited with only a minute effect on \tc\cite{Traettebergthesis}. 

Another scheme with which non-magnetic defects can be introduced in a controlled way for non-centro-symetrical anions in the \tms2x series is either by fast cooling preventing the complete ordering of the tetrahedral \C anions or by introducing \R
anions to the \C site by making the solid solution (TMTSF)$_2$(ClO$_4)_{(1-x)}$(ReO$_4)_{x}$. We will start with the study of the solids solutions \tmxns and consider the effect of quenching in a subsequent paragraph.

\begin{figure}[h]	
\includegraphics[width=0.85\hsize]{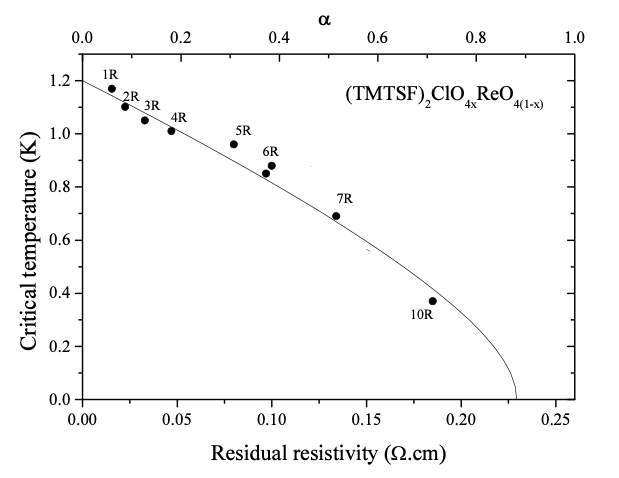}
\caption{Phase diagram of \tmxns, governed by non magnetic dilute disorder.
The data are obtained by newly analysing the temperature dependence of resistivity reported in Ref\cite{Joo05} (\emph{see} text). Data points with labels R refer to  very slowly cooled samples in the
R-state (the so-called relaxed state) with different \R  contents (\% \R). }
\label{TcvsRho0}  
\end{figure}

As displayed on Fig.~\ref{TcvsRho0},  superconductivity in the solid solution is  suppressed  and the  reduction in \tc is clearly related to the residual resistivity namely, the enhancement of the elastic scattering in the normal state. 
The data on Fig.~\ref{TcvsRho0} show that the relation \tc versus $\rho_0$ follows Eq.~\ref{Digamma} with a good accuracy using $T_{c}^{0}$=1.23~K. 

 At this stage, it is worth pointing out that the determination of the residual resistivity is not a trivial matter.  Various procedures have been used in the literature. First, the resistivity displays an usual quadratic temperature dependence both above the anion ordering temperature $T_{\rm AO}$= 24~K 
 and below down to approximately 10~K. Consequently, a first attempt to determine  $\rho_0$ was to extrapolate $\rho(T)$ down to zero temperature the $T+T^2$ behaviour observed between $T_{\rm AO}$ and 10~K. It turned out that $\rho_0$ is rather ill defined with this procedure (\emph{see} Ref.~\cite{Joo05}). Second, another procedure was to use a linear extrapolation of the temperature dependence below 10 K down to \tc, leading to lower values of $\rho_0$\cite{Joo04}. 

Furthermore, several recent re-analysis of the temperature dependence of the resistivity in the neighborhood of \tc in pure \tmcns~\cite{Doiron10} and in the alloy series \cite{Auban11} have emphasized the existence of two different regimes: a regime between 10 and 2K where the single particle scattering is dominated by antiferromagnetic fluctuations leading in turn to a linear dependence, and another regime between 2K and \tc where the downturn of the resistivity can be ascribed to a collective sliding of SDW waves  in the vicinity of an antiferromagnetic order.  This latter ordering is not accessible in \tmc since it would require  a negative hydrostatic pressure. However as shown by uniaxial elongation experiments along the $b$ axis~\cite{Kowada07} decreasing the interchain coupling, such a SDW state can be recovered  below 6K. The temperature domain between 2K and \tc will be more thoroughly presented in Sec.~\ref{conductingstate}.

The procedure used to derive $\rho_0$ in Fig.~\ref{TcvsRho0} was a linear extrapolation to zero temperature  of the linear regime between 2 and 10K where the scattering is dominated by  AF fluctuations (to be described in Sec.~\ref{conductingstate}).
This procedure should be rather accurate, in particular, in \tmcns. Beware that it is the quadratic fitting  procedure  which has been used for  the cooling rate dependence of $T_c$, \emph{see} also below.

Notice that it has been checked that the additional scattering cannot be ascribed to some magnetic scattering with the electron paramagnetic resonance (EPR) technique, which shows no additional traces of localized spins in the solid solution. Thus, the data in Fig.~\ref{TcvsRho0} 
 cannot be reconciled with the picture of a SC gap keeping a constant sign over the whole
$(\pm k_F)$ Fermi surface. They require a picture of pair breaking in a superconductor with an anisotropic gap
symmetry. 

It is interesting to compare the residual density of states predicted by theories with experimental data. 
Fig.~\ref{TcvsRho0} shows that the depairing  parameter of the pristine sample amounts to about 6.25\%  the critical value for the suppression of superconductivity. Given the ratio $\Gamma/\Gamma_0=0.0625$ for the pristine sample where $\Gamma$ is the scattering rate, the calculation of Sun and Maki~\cite{Sun95} performed for the unitarity scattering limit leads in turn to a residual density of states $N(0)$= 0.26$N_0$ which is fairly close to the residual density of states derived from specific heat experiments and from the NMR data,\cite{Shinagawa07} \emph{see} Section~\ref{thermo}. 

\subsubsection  {Cooling rate-controlled superconductivity}
\label{Cooling rate-controlled superconductivity}
A very peculiar property of the \tmc superconductor is the possibility to control \tc  without introducing any extrinsic impurities as achieved in the previous Section.  This property made it possible to examine the mechanism of superconductivity in \tmc and likely the same for all members of the \tm2x series.

It has long been known that the ground state of \tmc clearly depends on the rate at which the \clo4 anion ordering transition is crossed. The significant feature of \tmc
is actually related to the non-centrosymmetric nature of the
tetrahedral \clo4 anion located on the inversion centers of
the full structure. At high temperature, the thermal motion of
the \clo4 orientation makes it possible to preserve inversion
symmetry on average since \clo4 occupy randomly one or the
other inversion-symmetry-related orientations. The structural
disorder between two possible orientations no longer persists
at low temperatures below $T_{\rm AO}$ = 24 K, because an entropy
gain due to the reduction of degrees of freedom triggers the
anion ordering below this temperature.

Diffuse X-ray work has shown that, while
\clo4 anions adopt a uniform orientation along $a$ and $c$ axes,
they alternate along $b$\cite{Pouget83}. Furthermore, the ordering of these anions involves a slow dynamics. Unless  the compound is cooled slowly enough through the ordering transition, some anion disorder remains at low temperature ($T < 24$ K).  The two extreme situations, namely the relaxed state (slowly cooling) and the quenched state (fast cooling $\approx$ $dT/dt >$ 17K/mn) with SC and SDW ground states respectively, have been recognized from the beginning and are fairly well documented
. However, only limited studies have been published for the intermediate
cooling-rate regime, where superconductivity is moderately
suppressed\cite{Garoche82,Schwenk84,Matsunaga99,Joo05}. 
High-resolution X-ray investigations\cite{Kagoshima83,Moret85,Pouget90}
 have shown that samples in an intermediate cooling rate
regime exhibit a peculiar anion ordering, in which domains
with ordered anions of finite size are embedded in a disordered
background. In addition for the regime up to 5K/min, high-resolution
X-ray-diffraction measurements have determined both the volume
fraction of ordered anions and the average size of ordered
domains leading for the  relaxed state to more than 90$\%$ \clo4 ordered becoming  only $50\%$ at 5K/mn with  domains of size  (150$\times$150$\times$235) nm$^3$ and (30$\times$ 60$\times $ 50) nm$^3$  along ($a$-$b$-$c$) respectively\cite{Pouget90}.
\begin{figure}[t]	
\includegraphics[width=0.85\hsize]{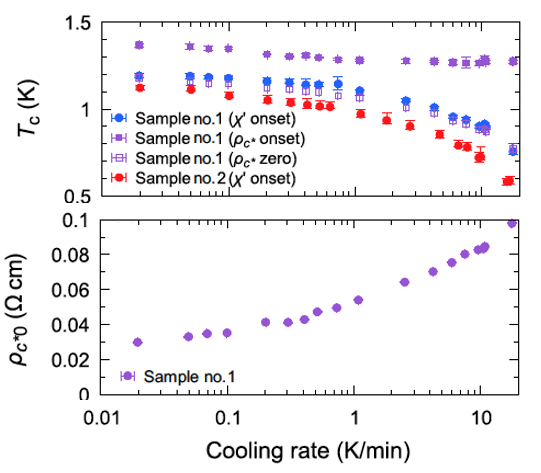}
\caption{ (Top) Cooling rate dependence of \tc in \tmc. Black dots   are the onset of the resistive transition. Lower blue and red marks are the temperature for zero resistivity, onset of  diamagnetic shielding and mid-transition for specific heat. (Bottom) Cooling-rate dependence of the residual resistivity $\rho_{c}$$_{\star0}$ derived from the polynomial fitting procedure explained in the text, Sources: Fig. 2bc [Ref\cite{Yonezawa18}, p. 4]. 
}
\label{coolingrate}  
\end{figure}

\begin{figure}[h]	
\includegraphics[width=0.9\hsize]{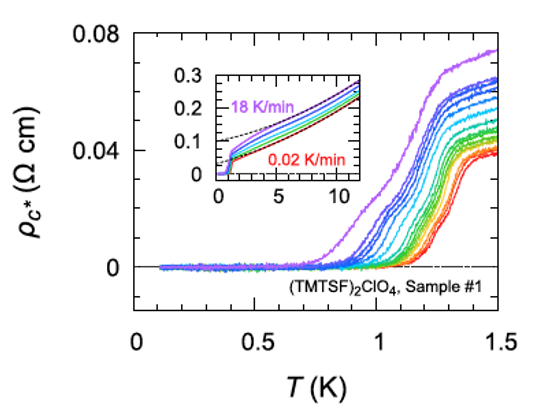}
\caption{
SC transition from   $\rho_{c}$$_{\star0}(T)$  after various cooling rates from 0.020K/mn up to 18K/mn. The inset
presents $\rho_{c}$$_{\star}(T)$ on a broader temperature range for 0.020, 0.52, 2.5,
7.6, and 18 K/min. For the 0.020 and 18K/min data, the results of
the fitting performed using the procedure explained in the text are presented with broken curves. The down turn of the resistivity mentioned in Sec.\ref{Precursor regime} in \tmc alloys and due to  fluctuating SDW's is clearly observed for the run  18K/mn, Source: Fig. 1d [Ref.cite{Yonezawa18}, p. 3]. }
\label{resistivity}  
\end{figure}

The influence of the cooling rate on the stability of the SC phase of \tmc has been recently revisited using transport along $c^{\star}$ and AC susceptibility measurements\cite{Yonezawa18} with a very careful control of the cooling speed from 0.02K/mn to 18K/mn through the anion ordering transition. The importance of this study is the control of the elastic electron mean free path in otherwise identical samples.
\begin{figure}[t]	
\includegraphics[width=0.85\hsize]{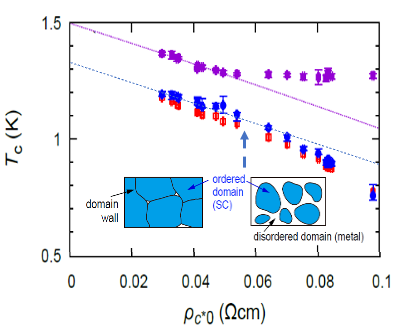}
\caption{ Cooling rate dependence of \tc in \tmc. black dots   are the onset of the resistive transition. Lower blue and red marks are the temperature for zero resistivity, onset of  diamagnetic shielding and middle of specific heat transition. The dashed blue arrow shows the cross-over between homogeneous SC below $\approx$ 1K/mn (at $\rho_{co}$=0.055 ($\Omega$.cm) and granular SC above, Source: Fig. 6 [Ref\cite{Yonezawa18}, p. 8].
}
\label{Yanov2}  
\end{figure}

The salient result is the finding of two regimes for the effect of the cooling rate as evidenced by a significant evolution of the  elastic mean free path dependence  on $T_c$, \emph{see} Fig.~\ref{Yanov2}. As far as the effects of scattering centers on SC are concerned,  the fast-cooled states cannot be described  by an average distance between disordered centers as it is the case for \C-\R alloys presented previously in Fig.~\ref{TcvsRho0},  evolving smoothly  as a function of the cooling speed, but instead as a state comprising a cooling-rate-dependent volume
fraction of well-ordered domains with the rest of the volume
occupied by disordered anions. There exists a crossover  around  1-2 K/mn where below this cooling rate the electron mean free path is given by the size of well \C ordered SC puddles and superconductivity propagating through the narrow disordered medium between puddles via a proximity effect\cite{Degennes,Deutscher80}, while above 1K/mn and up to 18 K/mn the system enters a regime of granular superconducting medium.  In order to get the residual elastic mean free path following the procedure developed in Ref.\cite{Yonezawa18} in which the residual resistivity is given by the zero temperature value of the  second order polynomial fit   for the normal state resistivity between 12 and 6K. 

\begin{figure}[t]
\begin{center}
\includegraphics[width=0.85\hsize]{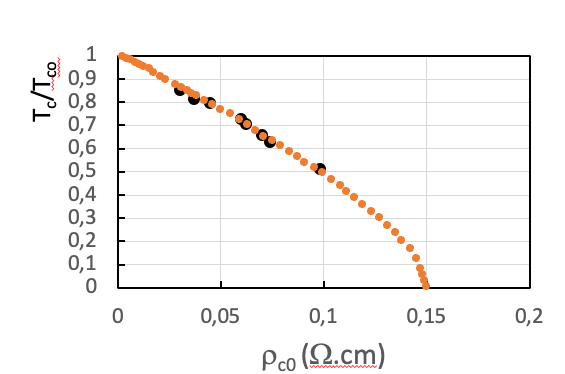}  
\caption{ Dependence of the normalized \tc for the half-height $C_{v}$ transition ($T_{c0}$= 1.32K) versus  residual resistivity displaying a very good agreement with the Abrikosov-Gorkov behaviour (yellow dots). Such a good agreement  cannot tell the difference between weak and strong scattering  limits, namely Born or unitarity limits. Actually, the strength of the electron scattering when the mean free path is governed by the size of the superconducting puddles  may be different from what it is when only residual chemical   impurities are at work.  In order to have access to the nature of the scattering leading to a decrease of \tc in fast cooled samples, the relevant measurement is  the evolution of the residual density of states as the mean free path of the electrons becomes controlled by the size of anion-ordered  superconducting domains in fast cooled samples. This on-going  experimental study will be the subject of a forthcoming publication\cite{Yonezawa24}.}

\label{mainfigure}
\end{center}
\end{figure}

As shown on Fig.~\ref{mainfigure},
the  fit of  \tc data for the specific heat transition  by the A-G theory extended to non-magnetic impurities in nodal superconductors\cite{Abrikosov61} provides a very good agreement with the theory  over the whole range of cooling rates. It is worth comparing mean free paths obtained from the A-G fit of the fast cooled data with  the determination of the domain sizes by high resolution X-ray scattering\cite{Pouget90}.    These features provide  additional supports in favour of a nodal $d$-wave superconducting coupling. The scattering rate $\Gamma_{0}$ of the slow cooled sample amounts  to $\Gamma_{c}$/5 according to Fig.~\ref{mainfigure}. Since $\Gamma_{c}$=0.88 $T_{c0}$\cite{Puchkaryov98}, a value of $\Gamma_{0}$= $3.4\times10^{10}$ s$^{-1}$ (or $\tau_{0}= 3\times 10^{-11}$s) is derived for the residual electronic  life time of a slowly-cooled pristine crystal. This would lead in turn to  a mean free path of about 5400 nm, given $v_{F}= 1.8\times10^{5}$m/s\cite{Yonezawa08a}, a value which is at least of the order of magnitude of  ordered domains size for pristine crystals\cite{Pouget90}.

However, from the plot on  Fig.~\ref{mainfigure} one cannot tell whether the scattering is   in the limit of strong scattering (unitarity)\cite{Sun95} or weak scattering (Born approximation)\cite{Suzumura89}, \emph{see} also Ref.\cite{Preosti94} for impurity-scattering effects in unconventional superconductors.

To conclude on the role of non-magnetic impurities,  we can anticipate that the influence of non-magnetic impurities on the SC phase  implies the existence of  positive as well as negative values for the SC order parameter on the Fermi surface.  The singlet nature is implied from the NMR data as shown in Sec.~\ref{thermo}  but NMR conclusions were  still unable to discriminate between the two possible options namely,  singlet-\textit{d} or \textit{g}~\cite{Nickel05}.

The main questions  still at hands now are to locate the $d$-wave nodes on the Fermi surface and be able to determine the strength  of the electron scattering in the superconducting state for  rapidely cooled samples affecting the ordering of  \clo4 anions below 24K. As far as the latter question is concerned,
 a recent investigation  of the  superconducting phase of \tmc under various cooling rates\cite{Yonezawa24}, has   inferred that when the size of the anion-ordered domains is governing the electron mean free path, the residual density of states stays constant at a finite value of about 25\% its normal state value, although  \tc can be suppressed down to $T_{c}/T_{c0}$=0.5. This residual DOS is actually due to preexisting chemical impurities in the pristine sample. This experiment suggests  that the borders of anion-ordered domains act as weak scatterers for the electrons (Born limit scattering)\cite{Puchkaryov98} unlike dilute  chemical defects which both suppress \tc and strongly increase the residual density of states (unitarity limit scattering)\cite{Sun95}. It would be very interesting to look at the residual DOS of slowly cooled alloyed samples since the behaviour which is displayed on Fig.~\ref{TcvsRho0} is unable to provide information regarding the strength  of the scattering. 

\subsubsection{(TMTSF)$_2$ClO$_4$, low field nodal and high field modulated superconductor}

As shown in the previous section, nodal superconductivity with spin singlet pairing is suspected from the NMR results  but these experiments were still unable to locate the nodes on the Fermi surface.

To reveal more precisely the gap structures and the location of nodes, one of the common technique is to look at the field-angle-dependent quasiparticle excitations through either specific heat or thermal conductivity measurements while rotating the magnetic field within a certain plane. These studies have been developed at the turn of 2000's in relation with quasi 2D heavy fermions\cite{Sakakibara07,An10}, quasi-2D organic\cite{Izawa02}, borocarbide\cite{Izawa01}and ruthenate\cite{Deguchi04} superconductors.

These experiments have made an extensive use of a property of type II superconductors with gap nodes where supercurrents surrounding vortices induce field-induced excitations of the quasi particles. Volovik made the remark that these quasi-particles  acquire an  energy shift (usually called Doppler shift), $\delta E = \bf{v}_{s}(r)\cdot v_{F}$, where $\bf{v_{s}(r)}$ is the superfluid velocity around a vortex and $\bf{v_{F}}$ the  Fermi velocity\cite{Volovik93,Volovik97a}.  Such field-induced excitations are now called the Volovik effect. These quasi-particle excitations are also field-direction dependent\cite{Vekhter99b}, because $\delta E$ is proportional to the scalar  product of the Fermi velocity $\bf{v_{F}}$ at the node and the superfluid velocity $\bf{v}_{s}$, the latter being in turn perpendicular to the applied field. Whenever  regions on the Fermi surface exist where  quasi particle excitations are larger than the superconducting gap, they will contribute to the density of states.

Thus, if one rotates the magnetic field within a certain plane, it is expected that the quasiparticle density of states will oscillate as a function of the field angle.  This effect becomes important at such angles where the
local energy gap becomes smaller than the Doppler shift
term, which can be realized in the case of superconductors with nodes\cite{Vekhter99b}. 
\begin{figure}[h]
\includegraphics[width=0.95\hsize]{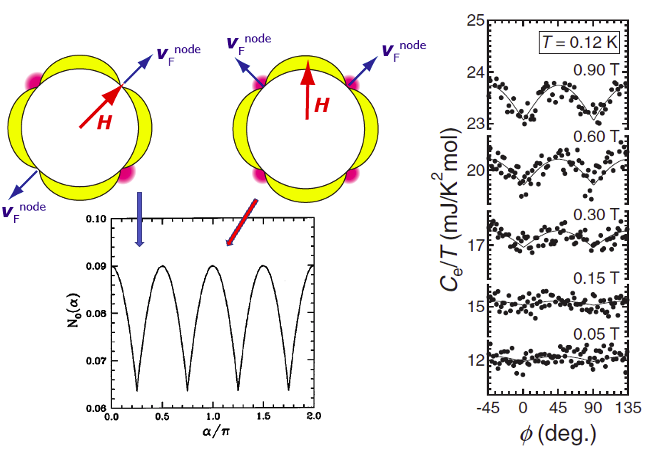}
\caption{\label{Sr2RuO4} Angular dependence under magnetic field of the electronic specific heat in  a 2D superconductor with a tetragonal symmetry.(Left) Theory for the gap function showing minima (maxima)  of the density of states for $\vec{H}$ parallel and (non parallel) to $\bf{v_{F}^{node}}$ respectively. (Right) Experimental data for $\mathrm{Sr_{2}Ru0_{4}}$  with the field rotated in the basal plane, Adapted from Fig. 1, 2 [Ref\cite{Vekhter99b}, p. 9024, 9025] and Fig. 3 left [Ref\cite{Deguchi04}, p. 2]. }
\end{figure}

 Hence, the density of states in the vortex state becomes angular-dependent, as found for instance in the compound $\mathrm{Sr_{2}Ru0_{4}}$, where the rotation pattern  displays a fourfold symmetry. Because the superfluid velocity  $\bf{v_{s}(r)}$ is perpendicular to  $\vec{H}$, when the field becomes parallel  to $\bf{v_{F}(k_n)}$ at a node $\bf{k_{n}}$  a minimum contribution of the QP's to the specific heat can in turn be expected in such conditions, \emph{see} Fig.~\ref{Sr2RuO4}.

\begin{figure}[h]
\includegraphics[width=0.8\hsize]{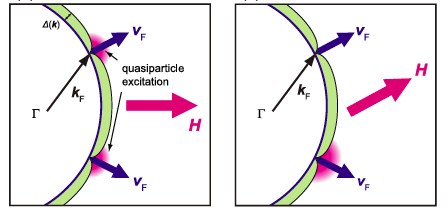}
\caption{\label{1Dnodal} A schematic representation of  1D Fermi surface showing the quasi particle excitations due to the Doppler shift shown in pink. Whenever the field is parallel to $\bf{v_{F}}$ at a node (right), the contribution of the excitations to the density of states is reduced, Source: Fig. 9bc [Ref\cite{Jerome16}, p. 365].  }
\end{figure}

For  2D conductors,
$\bf{v_{F}(k)}$ is usually collinear with $\bf{k}$. Therefore, the angular resolved
specific heat (or thermal conductivity) enables us to
reveal the positions of the gap nodes according to the angles
corresponding to minima of specific heat (thermal conductivity), as depicted in Fig.~\ref{Sr2RuO4}. However the situation is \emph{at variance} with the case of  a Q1D Fermi surface. 
More specifically, $\bf{v_F(}\bf{k})$, which is parallel to the gradient of the quasiparticle energy $\nabla\epsilon(\bf{k})$ in the reciprocal space and consequently perpendicular to the Fermi surface, is not always parallel to $\bf{k_F}$.
There, anomalies in the rotation pattern of $C_v/T$ are expected
at angles $\phi_n$ between $\vec{H}$ and $\bf{v_{F}(k_n)}$, where $\bf{k_{n}}$corresponds
to a nodal position on the Fermi surface, \emph{see} Fig.~\ref{1Dnodal}.
For such a situation  the interpretation of the angular dependence of the specific heat  requires  a model for the Fermi surface and also for the superconducting gap function. This research  has been performed very successfully by the Kyoto group  over the past fifteen years and reported in several publications\cite{Yonezawa12,Yonezawa13}. 
\begin{figure}[h]
\includegraphics[width=1\hsize]{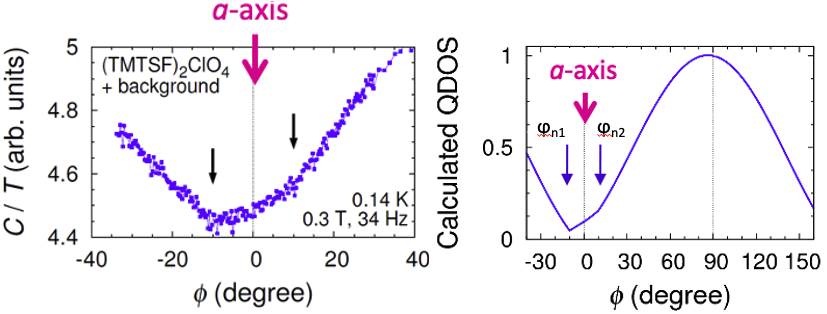}
\caption{\label{RotationCv} (Left) Observed in-plane field-angle dependence of the heat capacity of \tmc. What is remarkable is (i) the absence of symmetry of $C_v/T$ with respect to inversion around the  $a$ axis and (ii) the presence of two kinks in the rotation pattern at $\phi= \pm$ 10$^\circ$.(Right)  Simulated results by calculating the density of states based on a simple Doppler-shift model with nodes at $\phi= \pm$ 10$^\circ$,  Sources: Fig. 3a [Ref\cite{Yonezawa12}, p. 3] (left), Fig. 2g [Ref\cite{Yonezawa12}, p. 2] (right).  }
\end{figure} 

The extensive study of the angular dependence of  the specific heat in the low temperature and low field domain of the superconducting phase of \tmc, rotating the field in the basal plane $a-b'$. Detailed experimental results have been published in several articles\cite{Yonezawa12,Yonezawa13} and reviewed in\cite{Jerome16}. 

We will not present here on Fig.~\ref{RotationCv} the full treatment but only the most relevant results to characterize the nature of the superconducting coupling of these Q1D superconductors.
As far as Q1D superconductors are concerned, the angular dependence of $C_v/T$ can give only access to the direction of the Fermi velocity at nodes. To reveal the gap structure in $k$-space, one should know the band structure of the material. For these reasons, the gap-structure investigation of Q1D superconductors by the field-angle-induced quasiparticle excitation method had not been explored.

A simple model has been used to understand the data of
angular-resolved $C_v$ of \tmc of Fig.~\ref{RotationCv}. The rotation
pattern has been modeled by, 
\begin{eqnarray}
\label{N}
N(\phi) \propto (\Gamma^{2}\sin^{2}\phi + \cos^{2}\phi)^{1/4}\sqrt{\frac{H}{H_{c2}(0)}}\\\times\sum_{n,nodes}A_{n}\mid \sin (\phi-\phi_n) \mid\nonumber\end{eqnarray}
where the first factor reproduces the anisotropic character
of the critical field in the basal plane, while the weighted space summation over angles accounts for the existence of nodes at angles $\phi_n$ between the magnetic field and the special $\bf{k_{n}}$ points where the Fermi velocity is parallel to $\vec{H}$. 
\begin{figure}[t]
\includegraphics[width=0.6\hsize]{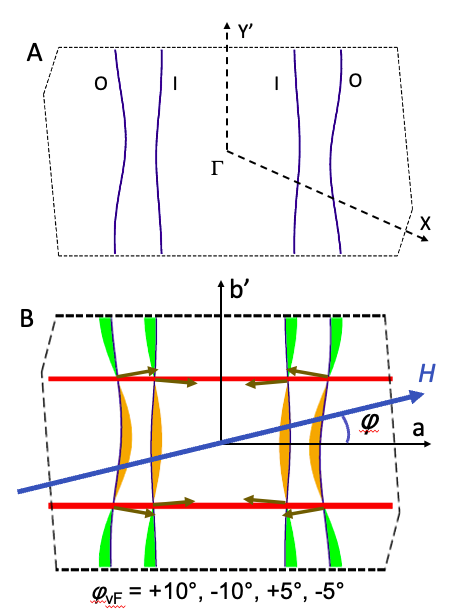}
\caption{\label{NodesClO4} (A) Doubling of the \tmc Fermi surface due to the anion-ordering, according to band structure calculations. The anion gap for this figure amounts to $\Delta$= 100 meV, Source: Fig. 8 [Ref\cite{Lepevelen01}, p. 361]. (B) The arrows indicate the $k$ points where the magnetic field is parallel  to $\bf{v_F(}\bf{k})$ leading to particular  angles $\phi$  at $\pm $10 and 5 degrees away from the $a$ axis which are located on the outer surfaces. Different colors indicate positive and negative signs for the superconducting gap function.}
\end{figure} 
The experimental observation of a rotation pattern at low field and low temperature which is non-symmetric with respect
to the inversion of $C_v$ around $a$  and the existence of kinks for $C_v/T$
at  $\pm$ 10$^\circ$ in Fig.~\ref{RotationCv} have been taken as the signature of line nodes along the $c^\star$ direction\cite{Yonezawa12}. In Eq.~\ref{N}, $\Gamma$ takes into account the in-plane anisotropy of the upper critical field $H_{c2}$ whereas the coefficients $A_{n}$ represent the contribution of every node to the field-angle dependent specific heat. Because of the triclinicity of the \tmc structure, $A_{n}$  depends on the details of the band structure and of the SC gap structure near the gap nodes. Regarding the overall anisotropy of the specific heat, a good fit of the data on Fig.~\ref{RotationCv} is obtained with the $H_{c2}$ anisotropy ratio $\Gamma$= 3.5 which compares favourably with a ratio of 5 announced in ref\cite{Gorkov85}. Furthermore, the  $A_n$ parameters  responsible for the $N(\phi)$  asymmetry around the $a$ axis  are such that $A_{n1}/A_{n2}$=3 on Fig.~\ref{RotationCv}.

In order to conclude about the  nodal nature of the superconducting gap, a comparison with a model calculation turned out to be indispensable\cite{Nagai11}. The observation of nodes at  $\phi_{ni} =\pm$ 10$^\circ$ and the asymmetry around the $a$ axis are strong constraints on the theory and consequently allow to discriminate between several possibilities of nodal superconductivity\cite{Yonezawa13}. The net result of this confrontation between experimental results and theory\cite{Nagai11} is that among the many possible configurations of the gap, the most plausible is the one which  locates the nodes of a d-wave pairing at $\bf{k_y}$$\sim \pm 0.25 b^{\star}$ as shown on Fig.~\ref{RotationCv} at the angles $\pm$ 10$^\circ$. Two other nodes are also expected at $\pm$ 5$^\circ$ but they contribute less to the rotation pattern of the specific heat\cite{Yonezawa13} and only a much improved sensitivity could make their detection possible.
\subsection{The conducting state above \tc: antiferromagnetic fluctuations  and their relation to transport and superconducting properties}
\label{conductingstate}
\subsubsection{Precursor regime from NMR data}
\label{precursors}
Interestingly, the metallic phase of \tmtsf2x in the 3D coherent \textcolor{black}{hopping} regime when pressure is located in the neighborhood of the critical pressure \pc, typically \tmp6 around 8 kbar or \tmc under ambient pressure or slightly higher  behave in a way far from what is expected for a Fermi liquid. This behavior indicates the dominance of quantum critical fluctuations  \textcolor{black}{emerging from the onset of superconductivity at the brink of antiferromagnetism} near $P_c$. Moreover, the close relation between the non-Fermi-liquid behavior and superconductivity has also been supported  by recent theoretical developments 
as described shortly  below.
\begin{figure}[t]
\includegraphics[width=0.8\hsize]{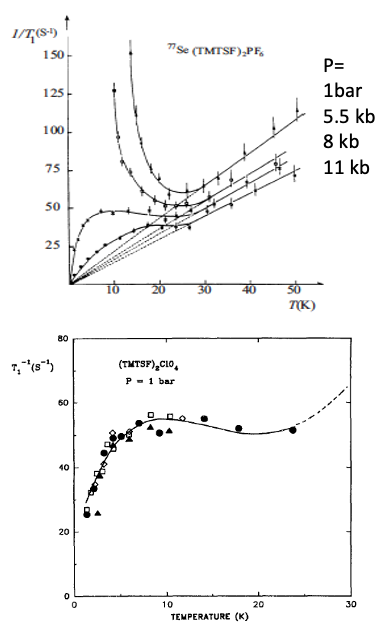}
\caption{\label{CreuzetWzietek} $^{77}$$\mathrm{Se (1/T_{1})}$ versus $T $ in \tmp6 at 1bar and 5.5 kbar (ground state SDW)  8 and 11 kbar (ground state SC). The lower part shows a similar behaviour for \tmc at ambient pressure showing how AF fluctuations are visible up to about 25 K, Sources: Fig. 1 [Ref\cite{Creuzet87b}, p. 278] (top), Fig. 11 [Ref\cite{Wzietek93}, p. 184] (bottom).
}
\end{figure} 

The existence of antiferromagnetic fluctuations below about 25 K have been discovered originally  by the NMR measurements of $1/T_1$\cite{Creuzet85e,Creuzet87b} . Its behaviour is such that the \textcolor{black}{temperature dependent} Korringa law, $1/T_{1}T \propto \chi^{2}(q=0,T)$, is well obeyed at high temperatures\cite{Wzietek93,Bourbon89}, say, above 25 K, but at the low-temperature its deviates strongly from the standard relaxation in paramagnetic metals. 

\begin{figure}[h]
\includegraphics[width=1\hsize]{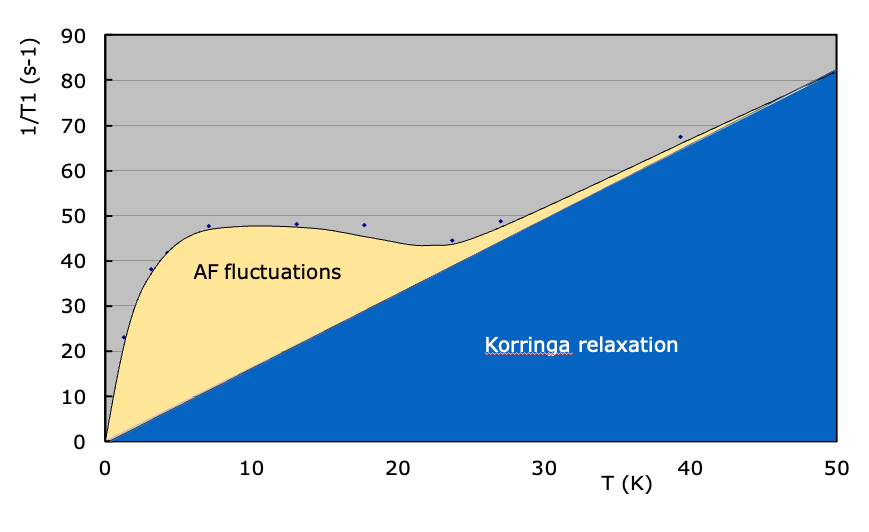}
\caption{\label{T1PF69kbar} The two different contributions to the spin lattice relaxation from the data on \tmp6 under 8 kbar\cite{Wzietek93}. }
\end{figure}

\begin{figure}[t]
\includegraphics[width=1\hsize]{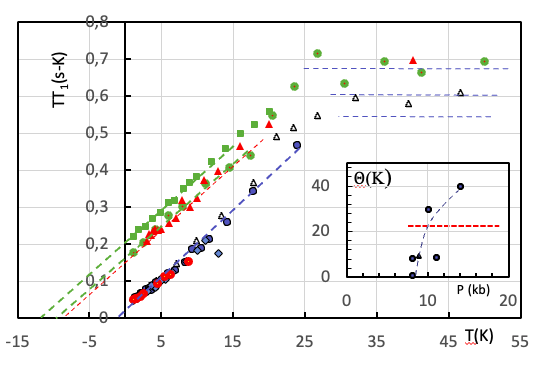}
\caption{\label{CW}Temperature dependence of the nuclear relaxation time multiplied by temperature versus temperature according to the data of Ref\cite{Creuzet87b}. A Korringa regime, $T_{1}T$= const is observed down to 25K. The 2D AF regime is observed below 15K and the small Curie-Weiss temperature of the 9 kbar run is the signature of the contribution of quantum critical fluctuations to the nuclear relaxation. The Curie-Weiss temperature becomes zero at the QCP (around  8 kbar for the present pressure scale). The dashed line in the inset sets the upper limit for the 2D AF regime. This figure is an other presentation for the data on Fig.~\ref{T1PF69kbar} .}
\end{figure}

As shown in Figs.~\ref{CreuzetWzietek} and~\ref{T1PF69kbar} an additional contribution to the relaxation rate emerges on top of the usual Korringa relaxation. 
This additional contribution rising at low temperatures, \emph{see} Fig.~\ref{T1PF69kbar}, has been attributed to the onset of antiferromagnetic fluctuations in the vicinity of \pc\cite{Bourbon84,Bourbonnais11}. In the lower-temperature regime, the relaxation rate follows a law such that $T_{1}T=C(T+\theta$), as shown in Fig.~\ref{CW}. This is the Curie-Weiss behavior for the relaxation which is to be observed in a 2D fluctuating antiferromagnet\cite{Wu05,Brown08,Shinagawa07,Moryia00,Bourbonnais09}. Similar behavior is also found in a $^{13}$C NMR study\cite{Kimura11}.
The positive Curie-Weiss temperature $\theta$, which provides the energy scale of the fluctuations, becomes zero when pressure is equal to \pc (the quantum critical conditions). When $\theta$ becomes large comparable to $T$, the standard relaxation mechanism is expected to recover down to low temperatures, in agreement 
with the observation at very high pressures\cite{Wzietek93}.

\begin{figure}[h]
\includegraphics[width=0.9\hsize]{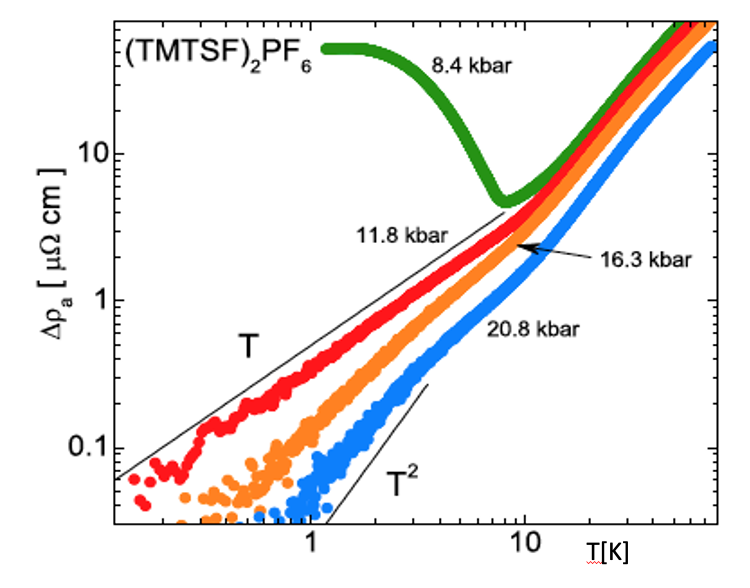}
\caption{\label{TandT2} Inelastic part $\Delta \rho_a(T)=\rho_a(T)-\rho_0$  of
the normal-state $a$-axis electrical resistivity of \tmp6 at 8.4,
11.8, 16.3, and 20.8 kbar. The lines represent $\Delta \rho(T)\propto T$  and $\Delta \rho(T)\propto T^2$ . Notice that for the 8.4 kbar run a sharp transition to the zero resistance state (not shown on the figure) is observed  around 0.9 K, Source: Fig. 4 (Top) [Ref\cite{Doiron09}], p. 3]..}
\end{figure}

These  fluctuations manifest themselves  also  as an anomalous (non Fermi liquid) behaviour  in transport at low temperature. At $P=P_c$, the inelastic scattering in transport reveals at once a strong linear term at low temperatures, most clearly seen in a log–log plot of the resistivity versus $T$, Fig.~\ref{TandT2}. This strongly linear behaviour, \textcolor{black}{more recently dubbed as a strange metal or Planckian dissipative behavior in other unconventional superconductors}, evolves to a quadratic behavior in the high-temperature regime. As pressure is increased away from $P_c$, the resistivity exhibits a general tendency to become quadratic at all temperatures, (\emph{see} Fig.~\ref{TandT2}). The existence of a linear temperature dependence of the resistivity is \emph{at variance} with the $T^2$ dependence expected from the ordinary electron–electron scattering in a conventional Fermi liquid, indicating that the dominant scattering would  involve spin fluctuations seen by NMR in the same temperature range. 

Notice that an early study\cite{Schulz81a} on \tmp6 also claimed a non-Fermi-liquid temperature dependence such as shown on Fig.~\ref{TandT2}. However, in the 1980's the research on organic superconductors has been over-influenced by the observation of a downward curvature of $\rho_{a} (T)$  close to \tc as shown on Fig.~\ref{Firstsupra}. Moreover, the theory in these years was very much influenced by the Landau-Ginzburg models of low-dimensional superconductors suggesting the possibility of non-Fermi-liquid behaviour with  superconducting fluctuations at temperatures well above the three-dimensional order\cite{Schulz81}. 

It is the extensive reinvestigation of \tmp6 and \tmc under pressure  performed in an experimental cooperation between Sherbrooke and Orsay from 2009 which  allowed to eliminate the interpretation of the low temperature transport in terms of paraconductivity of superconducting origin with the reservation as we will see below that  some paraconducting contribution is present in the very vicinity of the superconducting transition, although from the different  - SDW - origin.

\begin{figure}[h]
\includegraphics[width=0.9\hsize]{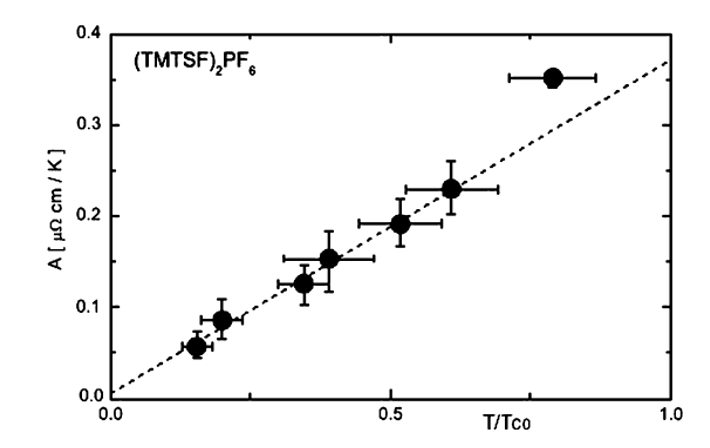} 
\caption{\label{ATc} Coefficient A of linear resistivity as a function of
normalized \tc (the maximum $T_{co}^{}$ =1.23 K occurs in the SDW/SC coexistence regime) for \tmp6, from a second order
polynomial fit over the range 0.1–4.0 K to all resistivity
curves at different pressure runs displayed on  Fig.~\ref{TandT2} between 11.8 and 20.8 kbar. The
vertical error bars show the variation of A when the upper limit of
the fit is changed by 1.0 K. \tc is defined as the midpoint of the
transition and the error bars come from the 10\% and 90\% points.
The dashed line is a linear fit to all the data points except that at
\tc=0.87 K, Source: Fig. 5 (Top) [Ref\cite{Doiron09}, p. 3]. } 
\end{figure}

In the previous paragraph, we mentioned the non-Fermi-liquid  behaviour which is manifested by the linear component of the $\rho(T)$  resistivity measured along the $a$ axis. However, comparing resistivity data of the Q1D superconductors \tmp6 and
\tmc along the least conducting $c$-axis and along the high conductivity $a$-axis as a
function of temperature and pressure, a low temperature 3D coherent  regime of the generic \tm2x phase diagram is observed in which a unique
scattering time governs the transport along both directions of these anisotropic conductors\cite{Auban11a}.
In this low temperature regime, both materials exhibit for $\rho_c$ an inelastic temperature dependence $\rho_{c}(T)$ $\sim AT + BT^2 $.
 The $T$-linear $\rho_c$ was also found to correlate with \tc in close analogy with the $\rho_a$ data displayed on Fig.~\ref{ATc}. 
\subsubsection{Precursor regime from DC transport data}
\label{Precursor regime}
\begin{figure}[h]
\includegraphics[width=0.9\hsize]{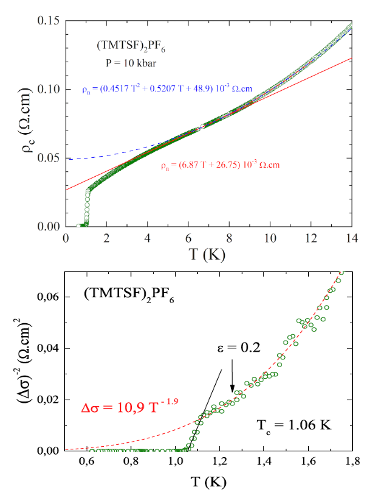} 
\caption{\label{rhocPF6v1}(Top) Temperature dependence of $\rho_c$ for \tmp6 at a pressure of 10 kbar. The dotted blue line is the second order polynomial fit between 6 and 12 K, while the continuous red line  is a linear fit between 4 and 8 K. These fits are supposed to be related to  the  single particle scattering. (Bottom) Plot of the temperature dependence of the additional conducting contribution for \tmp6 
as  $(\Delta\sigma)^{-2}$ versus $T$ using the linear fit for the single particle contribution. The dotted line is the power
law fit of the collective paraconduction after suppression of
superconductivity under a magnetic field of 0.0875 T. It is suggestive of a  contribution diverging like $1/T^{2}$ at very low temperature, Source: Fig. 1ab [Ref\cite{Auban11}, p. 2].
} 
\end{figure}

The data for $\rho(T)$ in  \tmtsf2x
reveal a particular sublinear behaviour  observed up to about two of three time \tc when the pressure is located in the vicinity of the critical pressure suppressing the insulating SDW phase, \emph{see} Fig.~\ref{rhocPF6v1}, such a behaviour being particularly visible on the $\rho_{c}$ component.

It is  instructive to analyze  the sub-linear  temperature regime  in the vicinity of \tc in terms of an additional conducting contribution, $\Delta\sigma$,  as performed on Fig.~\ref{rhocPF6v1} for \tmp6 under 10 kbar.  Then $\Delta\sigma$ is derived from the resistivity by,
\begin{eqnarray}
\label{paracon}
\Delta\sigma=\frac{\rho_{\rm n}-\rho_{\rm ex}}{\rho_{\rm n}\rho_{\rm ex}}
\end{eqnarray}
where $\rho_{\rm ex}$ is the measured resistivity while $\rho_{\rm n}$ is the behaviour of resistivity derived from the fit to a polynomial law of the measured  transport to a polynomial law  between 6 and 12 K. 

This figure reveals that the onset of superconductivity can be
characterized by two regimes, first the SC transition \emph{per se}
with a width in temperature of about $\Delta t \approx$ 0.1-0.2 K  and
second, a much broader temperature regime above \tc in which
the sublinear resistivity is observed.

The striking  feature of Fig.~\ref{rhocPF6v1} is the kink marking a cross over between the SC transition  regime and the precursor regime where the resistivity is sublinear. For the vicinity of the SC transition the plot such as $(\Delta\sigma)^{-2}=\alpha\epsilon$,  ($\epsilon$=$\frac{T-T_{c}}{T_{c}}$), could be related  to a 3D Azlamazov-Larkin regime\cite{Azlamazov68,Schulz81} for 3D superconducting fluctuations with \tc= 1.06K.  Although we may have some reservation about the actual existence of this AL regime,\tc of organics are known to be crucially affected by crystal defects.Therefore, the apparent AL law followed by the excess conductivity on the figure should be taken with a grain of salt as possibly not related to any sign of intrinsic paraconductivity. A support for this interpretation is provided by the extensive investigation undertaken in \tmc controlling the amount of \cl anions ordering  the broad temperature regime \textcolor{black}{may} reveal a different kind of contribution, namely a fluctuating AF contribution provided by the SDW channel\cite{Lee74}. 

\begin{figure}[h]
\includegraphics[width=0.9\hsize]{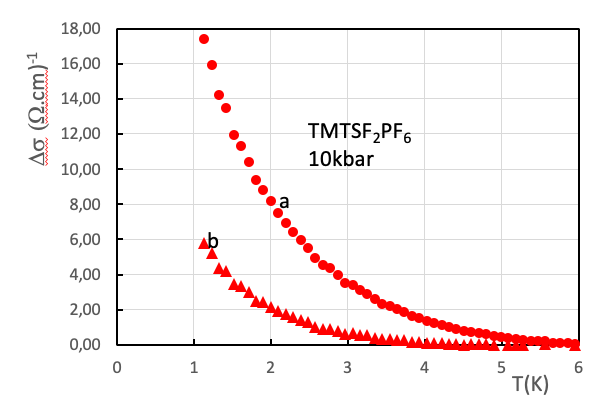} 
\caption{\label{PF6SDWpara} Derivation of the fluctuating SDW collective conduction in \tmp6 under 10 kbar  according to two different hypothesis for the single particle contribution, quadratic plot (a), linear plot (b). It is difficult to say which of the two hypotheses is the best, but given the predominance of the linear term in the single particle resistivity of (TMTSF)$_2$PF$_6$, \emph{see} Fig.~\ref{TandT2} and Ref.\cite{Auban11}, case (b) leading to a residual resistivity quite close to what is obtained in \tmc (\emph{vide infra})  may be more appropriate.  We thank P. Auban for the communication of the raw data of (TMTSF)$_2$PF$_6$.} 
\end{figure}

Such a contribution extending over a broad temperature regime has also been observed in \tmc\cite{Yonezawa18}. The collective magnetic channel can act in parallel with the single particle transport and leads to a experimentally  measured resistivity given by,  $1/\rho_{\rm ex}=\sigma_{\rm n}+\Delta\sigma_{\rm coll}$, where the collective contribution to the conductivity $\Delta\sigma_{\rm coll}$ according to Eq.~\ref{paracon} displayed on Fig.~\ref{PF6SDWpara}.

We can  estimate  the SDW paraconductive contribution using the value of $\Delta\sigma_{\rm coll}$ at $\epsilon = 0.2$ ($T$= 1.27K) \emph{see} Fig.~\ref{rhocPF6v1} leading in turn to $\Delta\sigma_{\rm coll}$ $\approx$  4 or 12.5 ($\Omega.{\rm cm})^{-1}$   for the linear or quadratic  single particle contribution respectively\cite{Auban11}.

The compound \tmc is quite interesting since the collective contribution is not only easily visible but it also can be controlled different ways, either through the cooling rate dependence of the elastic free  path  or by the addition of local impurities in alloys.  

\begin{figure}[h]
\includegraphics[width=0.9\hsize]{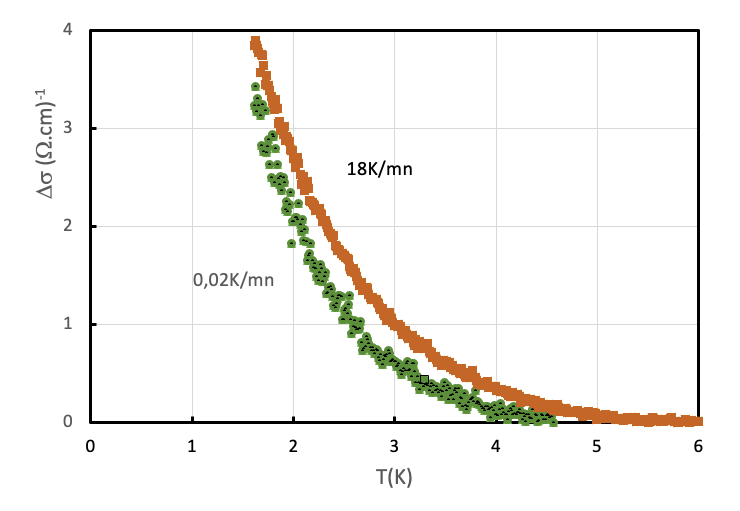} 
\caption{\label{SDWparacon} SDW paraconductive contribution to the conduction of \tmc cooled at 18 K/mn and 0.02 K/mn extracted from the original data of resistivity displayed on Fig.~\ref{resistivity}. We thank S. Yonezawa for communicating the full data sheets. The paraconductive contribution is quite similar for both cooling rates.  Notice the order of magnitude which is also similar to what is observed in \tmp6 under 10 kbar in the same temperature range, \emph{see} Fig.~\ref{rhocPF6v1}. This paraconductivity follows a power law $T^{-\alpha}$ with \tc =0 in this narrow  temperature range  and an exponent $\alpha$ in-between 3/2 and 2.
} 
\end{figure}

The study of transport properties under various cooling conditions\cite{Yonezawa18} has revealed a very strong down turn of the resistivity under fast-cooling as already displayed in the inset of Fig.~\ref{resistivity}. Interestingly, the SDW paraconductive contribution is  easily visible when the sample is mostly anion-disordered, i.e.  cooling rate of 18 K/mn, \emph{see }Fig.~\ref{resistivity} because of the large value of the normal state resistivity which enhances the contrast of resistivity. When the paraconductivity is extracted according to the procedure of  Eq.~\ref{paracon} a power law divergence is observed, Fig.~\ref{SDWparacon} with an exponent  1.5-2 similar to what has been obtained with (TMTSF)$_2$PF$_6$.

\begin{figure}[h]
\includegraphics[width=0.9\hsize]{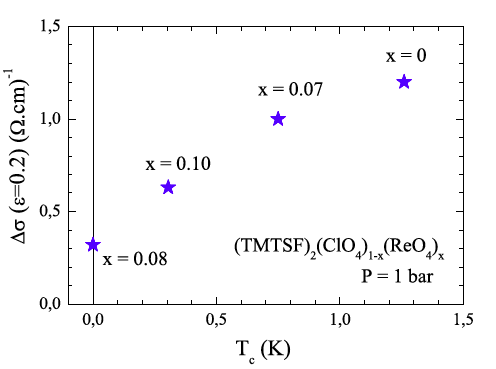} 
\caption{\label{SDWpinning} Paraconductive contribution characterised
by $\Delta\sigma$ at 1.2\tc in the precursor regime,
\emph{vs}. \tc in the alloy series of \cl-\re alloys
at 1 bar. For the sample not showing any finite \tc, the fluctuation
contribution has been derived from a linear extrapolation of
$\Delta\sigma$ to 0 K, Source: Fig. 4 [Ref\cite{Auban11}, p. 4].} 
\end{figure}

Although the incommensurate character allowing free sliding of the SDW  is established
by the observation of a double-horn shape of the $\mathrm{^{13}C}$-NMR
spectrum and the shape of the $^{1}H$ resonance line  in \tmp6\cite{Barthel93}, this excess of conduction can still be severely suppressed when the density of defects is increased. It is the  situation which is encountered in \cl-\re alloys\cite{Auban11} where \re anions play the role of active pinning centers, see Fig.~\ref{SDWpinning}.
Contribution between \tmp6 and \tmc. 
 When the mean distance between pinning centers
becomes smaller than the SDW coherence length, the
impurity pinning suppresses $\Delta\sigma$  heavily\cite{Jerome82}. 
 
Data for
the SDW coherence length support this interpretation as 
the zero-temperature longitudinal length $\xi_{oa}$ is of order
32nm in \tmp6\cite{Gruner94}. Consequently, it makes sense that
the SDW can be  strongly pinned by defects when the mean distance
between pinning centers, ($\approx 7.3$nm at 10\% \re doping)
is  four times smaller than the coherence length leading in
turn to a decrease of the fluctuation conduction independent
of the current direction. 

Given the size of the collective contribution, the sublinear behaviour for the resistivity requires a single particle conductivity small as  compared to the collective part. This is actually  the case for the $c$-axis component. Regarding the longitudinal component, the excess conduction should be more difficult to observe since the normal resistivity along this axis is several orders of magnitude smaller than along $\rho_{c}$. The first data for superconductivity in \tmp6 shown on Fig.\ref{Firstsupra} revealed a large temperature dependence of the resistivity in the helium temperature range which turned out to be confirmed by subsequent investigations but the claim of a downward curvature below 2K should be taken with caution since the resistivity was supposedly measured along the $a$-axis.

\subsubsection{A brief incursion into theory}
\label{Concluding experimental section}
To summarize, the investigation of both transport and superconductivity under pressure in \tmp6 has established a tight correlation between the amplitude of the linear temperature dependence of the resistivity and the value of $T_c$, as displayed in Fig.~\ref{ATc}. This is very suggestive of a common origin for the inelastic scattering of the metallic phase and pairing in the SC phase \tmp6 both rooted in
the low frequency antiferromagnetic fluctuations, as detected
by NMR experiments\cite{Wzietek93,Brown08,Bourbonnais93,Bourbonnais09} which we intend to summarize very briefly in the following. 

Such a correlation is not limited to 1D organics. It has also been observed in the pnictides \baas\cite{Doiron09}  {\color{black} and  electron-doped cuprates with  similar
phase diagrams}\cite{Fang09,Jin11} where a detailed temperature dependence of the
resistivity and the correlation with \tc strongly suggest that antiferromagnetic fluctuations play also  a  fundamental role in  pnictide superconductors and electron-doped cuprates, although their temperature scales $T_{\rm SDW}$ and \tc is  twenty times higher.

 Within the framework of a weak-coupling limit, the problem of the interplay between antiferromagnetism and superconductivity in the Bechgaard salts has been theoretically worked out using the renormalization group (RG) approach\cite{Bourbonnais09,Nickel06,Duprat01} as summarized very briefly below.  

The theory developed extensively by the Sherbrooke's school over several decades  takes into account a two dimensional  situation. 

The RG integration of high-energy electronic degrees of freedom was carried out down to the Fermi level, and leads to a renormalization of the couplings at the temperature $T$. The RG flow superposes the $2k_{F}$ electron–hole (density wave) and the electron-electron Cooper pairing many-body processes, which  \textcolor{black}{ become entangled  and interfere quantum mechanically at every order of perturbation}.
As a function of  pressure which is represented by the  parameter $t_{b}'$ \emph{i.e}. the \textcolor{black}{antinesting interchain coupling in a 2D tight binding spectrum \begin{eqnarray} 
\label{tightbinding}
E(k)=v_{F}(|k|-k_{F})-2t_{b}\cos k_{b}-2t'_{b}\cos 2k_{b},
\end{eqnarray}
 a singularity in the course of renormalization for the  scattering amplitudes signals an instability of the metallic state toward the formation of an ordered state at some characteristic temperature scale. }

\begin{figure}[h]
\includegraphics[width=0.9\hsize]{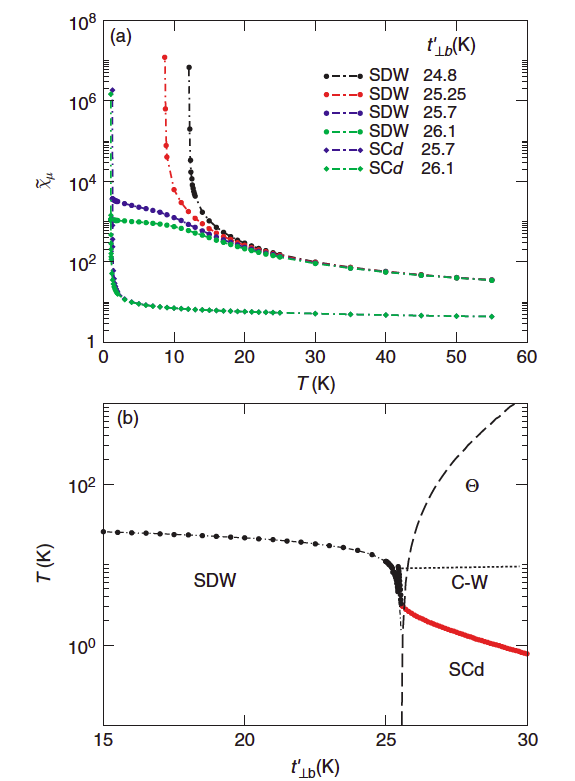}
\caption{\label{BourbonnaisSedeki} Renormalization group results of the Q1D electron gas model.  (a) Temperature dependence of the normalized static susceptibility $\chi_\mu$ of the $\mu={\rm SDW}$ and $\mu={\rm SCd}$ channels at various transverse coupling $t'_{b}$ on either side of the threshold value $t'^\star_{b}\simeq 25.4$K. (b) Phase diagram, Source: Fig. 1 [Ref\cite{Bourbonnais09}, p. 4]. 
}
\end{figure}

{\color{black} For typical input bare interactions and band parameters, the calculations show } that  at low $t_{b}'$ (small pressure) nesting is still  sufficiently good to induce a SDW instability in the temperature range of experimentally observed $T_{\rm SDW} \approx$ 10–20K as displayed, in Fig.~\ref{BourbonnaisSedeki}.
  When the antinesting parameter approaches the threshold coupling $t_{b}'^{\star}$ from below ($t_{b}'^{\star}$ $\sim$25.4K using the above parameters), $T_{\rm SDW}$ sharply decreases as a result of an interference between the Cooper and the Peierls channel (or SDW correlations). This situation leads in turn to an attractive pairing in the SC d-wave (SCd) channel. This gives rise to an instability of the normal state against SCd order at the temperature \tc with pairing coming from the exchange of antiferromagnetic spin fluctuations between carriers of neighboring chains\cite{Caron86,Bourbon88a}. Such a pairing model actually supports the conjecture of interchain pairing made by V. J. Emery in 1983 and 1986\cite{Emery83,Emery86} \textcolor{black}{and  taken up by others  \cite{Beal86,Scalapino86}, a mechanism  for  electrons to avoid the Coulomb repulsion}.

The calculated phase diagram shown in Fig.~\ref{BourbonnaisSedeki} is obtained with  a reasonable set of parameters  \textcolor{black}{ for a g-ology electron gas model introduced in Sec.~\ref{1DPhysics},  namely $g_1$=$g_{2}/2\approx  0.32\pi v_F$ for the backward and forward scattering amplitudes, respectively and $g_3=0.02\pi v_F$ for the longitudinal half-filling Umklapp scattering term for  weakly dimerized chains like for (TMTSF)$_2X$} \cite{Bourbonnais09,Bourbonnais11}. The model captures the essential features of the experimentally-determined phase diagram of \tmp6 presented in Fig.~\ref{Generic}.

Regarding the non-ordered \textcolor{black}{quantum critical phase,  Sedeki {\it et al.}\cite{Bourbonnais09,Sedeki09,Sedeki12} have shown the existence of Curie-Weiss (C-W) regime of SDW fluctuations above the critical $t_{b}'^{\star}$, where $\chi_{\rm SDW} \sim 1/(T+ \theta)$  is still enhanced according to Fig.~\ref{BourbonnaisSedeki}, consistently with the above NMR results. They also }proceeded to an evaluation of the imaginary part of the one-particle self-energy. In addition to the regular Fermi-liquid component whose scattering rate goes as $T^2$, low-frequency spin fluctuations yield $\tau^{-1}=aT\xi$, where $a$ is a constant and the antiferromagnetic correlation length $\xi(T)$ increases according to $\xi=c(T+\theta)^{-1/2}$ as $T{\rightarrow}$ \tc, 
  where $\theta$ is the Curie-Weiss temperature scale for spin fluctuations\cite{Bourbonnais09,Sedeki12}. It is then natural to expect the Umklapp resistivity to contain 
  (in the limit $T\ll$$\theta$ ) a linear term $AT$, whose
  magnitude would presumably be correlated with $T_c$, as both scattering and pairing are caused by the same antiferromagnetic correlations. 
  
  The observation of a $T$-linear law for the resistivity up to 8K in \tmp6 under a pressure of 11.8 kbar as displayed in Fig.~\ref{TandT2} is therefore consistent with the value of $\theta=8$K determined from NMR relaxation at 11kbar displayed in Fig.~\ref{CW}. 
  
  More recently, Bakrim {\it et al.}\cite{Bakrim14} studied
  the effects of electron–phonon interactions on the SC and SDW channels. Interestingly, it is revealed that intermolecular  electron–phonon coupling, \textcolor{black}{though much weaker in  amplitude than the Coulomb terms, enhances spin fluctuations, leading to unusual phenomena such as the prediction of a positive isotope effect.} \emph{"Unpublished experiments at Orsay on deuterated \tmc have shown a tendency to increase \tc. This is opposite to the usual isotope effect in superconductors but it may be due to an expansion of the deuterated volume cell unit."}

We add one comment that, in (TMTSF)$_2X$, \textcolor{black}{ the boundary between the SDW and SC phases very close to $P_c$  is observed to be weakly first-order\cite{Vuletic02,Lee02b}, indicating that the  quantum - second order - critical point within the pressure–temperature phase diagram is  actually unstable to first-order effects and the emergence of the superconducting dome}. However, it has been recently revealed that other typical “quantum critical” materials such as iron pnictides\cite{Goko09} indeed exhibit a first-order-like-behavior in the vicinity of the quantum critical point, evidenced by phase separation between magnetically ordered and paramagnetic phases detected by $\mu$SR studies\cite{Uemura15}. Thus, it is now getting clearer that the quantum criticality near a first-order transition observed in \tmtsf2x probably shares general and important physics with a broader class of materials.

A complete presentation of the theory related to all these experimental findings will be published in a forthcoming issue of Comptes Rendus Physique.

\subsection{A very brief overlook at the two dimensional organic superconductors}
\label{two dimensional organic superconductors}
The presentation of the \tm2x superconductors given in the previous sections has emphasized  the role of the  parent state, a `mother' insulating state neighbouring superconductivity. As an itinerant antiferromagnet (Slater-Overhauser phase), its stability is very sensitive to the intensity of the interchain coupling. The low temperature situation is quite different in two-dimensional organic solids.

The 2D compounds have provided not only higher values of \tc and a textbook
example for the  two dimensional fermiology  but also a  playground for the study of the 2D Mott transition as well as  the metallic
phase in its vicinity. A vast variety of quantum phenomena can be studied with these materials, based on the celebrated $\mathrm{BEDT-TTF}$ electron donating molecule and a variety of polymerized anions,  leading to strong coupling compounds in which    the mother state is a Mott insulating state. 
The  wealth of quantum phenomena exhibited by these 2D materials has been recently reviewed by Dressel and Tomic\cite{Dressel20}.

To complete this presentation it is useful to have a  brief look at the 2D family of organic superconductors whose physical properties are particularly interesting to compare with those of one-dimensional compounds.



\begin{figure}[h] 
\includegraphics[width=0.65\hsize]{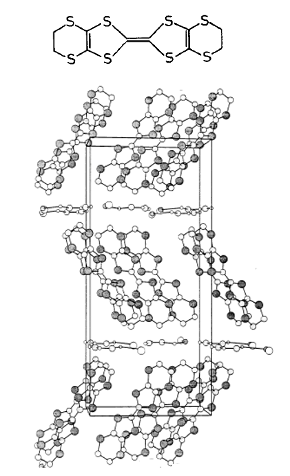} 
\caption{\label{ET2Xkappa} The unit cell of the \kappaet2cunbr   containing four ET dimers  arranged in a 2D structure separated by the polymeric anion layer.  The ET molecule is displayed at the top without their ethylene  end groups.  } 
\end{figure}

At the beginning of the 80's, a new organic cation radical of the fulvalene molecule has been proposed by G. Saito {\it et al.,}\cite{Saito82}which   paved the way for the synthesis of new  families of organic superconductors. This  molecule is  (BEDT-TTF)(bis(ethylenedithiolo)tetrathiafulvalene), an elongated version of \ttf  also called ET, Fig.~\ref{ET2Xkappa}.  
When complexed to the \cl anion, the material exhibits a metallic character down to 1.4K with pronounced 2D features (\emph{albeit }without traces of superconductivity)\cite{Saito82}. 

This molecule  has been the first to give a radical cation   salt with  \reo4 which becomes superconducting under pressure at a temperature of 2K, slightly higher temperature than the \tc of \tm2x\cite{Parkin83e}. However, the crystal structure  of \bedtttfreo4  revealed a pronounced 1D character with a molecular packing   bearing much similarity with the \tm2x series. Notice that the metal-insulator transition observed  around 80K under ambient pressure in this compound is very likely due to an anion ordering transition  similar to what has been encountered in \tmtsfreo4 and commented in Sec.~\ref{non-symmetrical anions}.

Furthermore, one year later, the same ET molecule  contributed to the elaboration of half-filled band organic conductors with quite  novel structures
namely, materials displaying a two dimensional (2D) conducting (or layer) structure. The first representatives of this 2D class of
organic materials have been the
$\beta-$(ET)$_{2}X$ compounds where $X$ is a linear triatomic anion such as I$_3$, IBr$_2$, AuI$_2$ etc,.. 

The salt with $X$= I$_3$ is of a 
particular interest since the so-called
$\beta_{H}$-phase, which can be stabilized at low temperature after a special pressure-temperature cycling,   has provided a large
increase of the superconducting \tc \,for organic superconductors  from 1-2K in the \tsx  series up to 8K\cite{Laukhin85,Murata85a,Creuzet85b}. 

When
$\mathrm{\beta-(ET)_{2}I_{3}}$ is cooled rapidly  without any pressure treatment, random disorder of the ethylene groups of the \et molecules
remains in the
$\beta_{L}$-phase and  strongly reduces the stability of the superconducting state with a
\tc dropping down to 1.4K\cite{Yagubskii84} in line with the role of non magnetic defects on  the superconductivity of 1D
organics\cite{Joo05} and discussed in Sec.~\ref{non-magnetic defects}. 

Besides, $\beta_{H}$-(ET)$_{2}$I$_{3}$  has provided a textbook example for  Shubnikov-de Haas  oscillations in a 2D conductor
with extraordinarily large oscillations of the magnetoresistance\cite{Kang89}. Such a study has also enabled the determination of the
overlap integral  between conducting planes of 0.5 meV, a clear-cut illustration for the pronounced 2D character and the existence
of angular magnetoresistance oscillations due to the $c$-axis warping of the Fermi surface in these
\et conductors\cite{Kartsovnik88,Yamaji89}. 

A further increase in the \tc above 10K has been accomplished in the 90's\,  with the discovery of the $\kappa$-phase salts namely $\mathrm{\kappa-(ET)_{2}X}$\,
with polymeric anions, $X$=
\cuncs at \tc=10.4K\cite{Urayama88}, in \cuncnbr ($\kappa$-Br)
at 11.6K  \cite{Kini90}, and  even up to 12.5K in \cuncncl  ($\kappa$-Cl)
  under a modest pressure of 0.3 kbar\cite{Williams90}.

\begin{figure}[t]
\includegraphics[width=1\hsize]{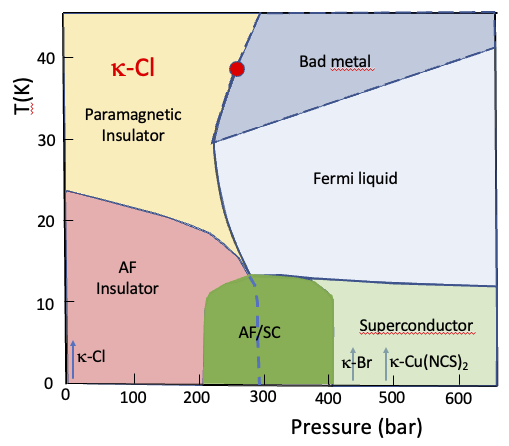}
\caption{\label{ET-Cldiagram}  T-P phase diagram of \cuncncl  ($\kappa$-Cl), Sources: Fig. 1 [Ref\cite{Lefebvre00}, p. 5421] and Fig. 1 [Ref\cite{Limelette03}, p. 1]. The arrows a and b are the  respective locations of compounds $\kappa-\mathrm{Br}$ and $\kappa-\mathrm{Cu(NCS)_{2}}$  
under ambient pressure. The dark green area AS/SC defines a region of coexistence between AFI and SC orders which  implies that the transition between the AF and SC phases (dashed blue line) is  first order with a finite  hysteresis under pressure, not shown on the figure. The first order transition between a paramagnetic Mott insulator and a Fermi liquid ends at the Mott critical point (red dot) above which the transition becomes second order. Under high pressures the material is crossing-over at the temperature $T^*$ (dashed line) from a low temperature Fermi liquid with a quadratic temperature dependence of the resistivity to a still metallic regime ($d\rho /dT<0$)  although exceeding the Mott-Ioffe-Regel criterion by several orders of magnitude, this regime is commonly called the bad metal regime\cite{Limelette03,Georges96}.  }
\end{figure}

Most $\kappa$-phase salts show an
isostructural face to face molecular dimers packing forming a 2D checker-board pattern. 
Very much like  the compounds belonging to the \tm2x family, most $\kappa$-phase   systems can be gathered together in the generic
Temperature-Pressure and Chemistry phase diagram displayed on Fig.~\ref{ET-Cldiagram}.

The compound $X$=\cuncnbr is thus of particular interest since 
superconductivity at
\tc = 11.6K occurs at ambient pressure making NMR studies in the superconducting state easily measurable over one decade in temperature. 
\begin{figure}[h] 
\includegraphics[width=0.85\hsize]{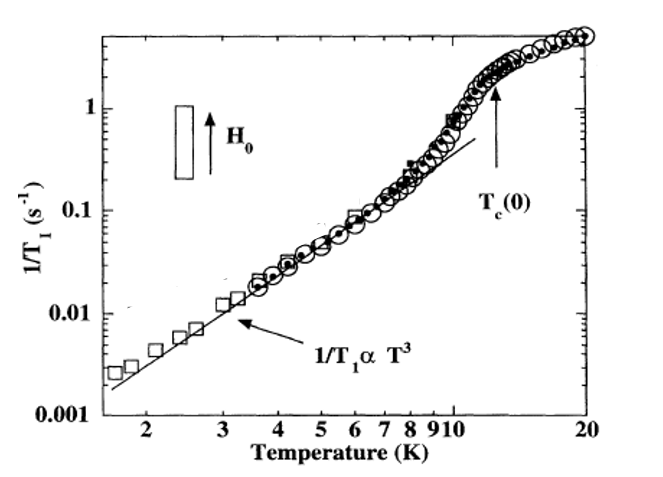} 

\caption{\label{ETBrT1}  $^{13}$C($1/T_{1})$ 
in the superconducting phase of  $\kappa  $-\cuncnbr in a parallel magnetic field showing the $T^{3}$ law  and the absence of any Hebel-Slichter peak at the transition as expected for d-wave superconductivity,\cite{Hasegawa87}, Source: Fig. 5 [Ref.\cite{Mayaffre95} p. 4124].
} 
\end{figure}
$^{13}$C Knight
shift and relaxation studies, \emph{see} Fig.~\ref{ETBrT1}, in a $\kappa$-(ET)$_2$\cuncnbr single crystal have shown conclusively that the pairing in  2D organic
superconductors must be spin singlet with the existence of nodes in the gap\cite{Mayaffre95,Kanoda96} in line with what has been observed in \tm2x superconductors, Sec.~\ref{thermo}.
 
  Although the the law $1/T_{1}\propto T^{3}$ which has been reported by several experimentalists  is very suggestive of a d-wave gap situation, it is still questionable to conclude for the existence of zeros on the Fermi surface of these 2D superconductors since it is not so easy to differentiate between a very anisotropic node-less  gap and a gap with nodes on the Fermi surface\cite{Mayaffre95}.  Investigations of anion alloying  performed in 
$\mathrm{\kappa-(ET)_{2}Cu(N(CN)_{2})Br_{x}Cl_{1-x}}$\, failed to report any significant effect of disorder on \tc $\approx$ 12K\cite{Yasin11,Bondarenko94} at variance with the findings in the $\beta$-type ET salts for the mixed anion $\mathrm{I_{3}-IBr_{2}}$ system\cite{Tokumoto87}.

Among all compounds,
$\kappa$-Cl is also particularly interesting as it is the prototype in the series showing under increasing pressure the complete sequence of states, namely, 
paramagnetic Mott insulating, antiferromagnetic, metallic and superconducting states within a few hundred bars which have been studied with the helium-gas pressure technique\cite{Lefebvre00}. Magnetic\cite{Lefebvre00}, acoustic\cite{Fournier03} and transport experiments\cite{Limelette03,Kagawa04} have shown with $\kappa$-Cl the existence of a first order transition line in the  T-P plane with a phase coexistence regime ending at a Mott critical point, \emph{see} Fig.~\ref{ET-Cldiagram}. 
\begin{figure}[h]
\includegraphics[width=0.8\hsize]{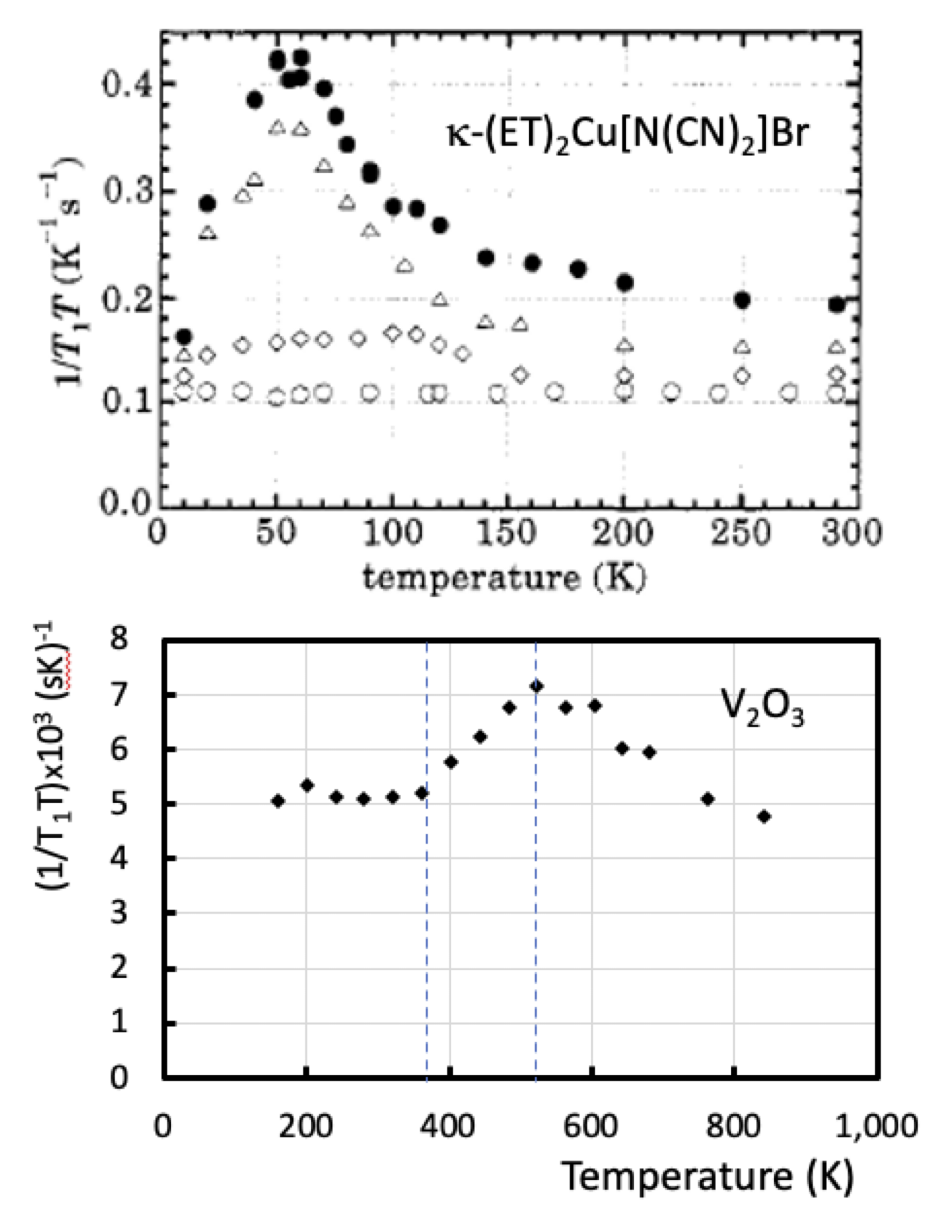}
\caption{\label{Relaxation_ETV2O3} (Top) $^{13}$C($1/T_{1}T)$ variation with temperature at various pressures from 1 bar up to 4 kbar, Source: Fig. 4 [Ref.\cite{Mayaffre94a} p. 208] 1 bar $(\bullet)$, 1.5 kbar $(\triangle)$, 3 kbar $(\diamond)$,4 kbar $(\circ)$. (Bottom) $^{51}$V($1/T_{1}T)$ variation with temperature in $\mathrm{V_{2}O_{3}}$ at ambient pressure, derived from Ref.\cite{Kerlin73}. }
\end{figure}
The recent reinvestigation of  $\kappa$-Br suggests that the critical regime around the critical end point
does not appear to obey the 2D Ising model universality class\cite{Kagawa05} \textit{at variance} with a similar study of the 3D Mott
transition performed in the vanadium sesquioxide\cite{Limelette03a}. 

A single crystal NMR relaxation study of the $\kappa$-Br compound\cite{Mayaffre94a} reveals a large enhancement of the relaxation rate $1/T_{1}T$ above the Korringa law ($1/T_{1}T\varpropto K^2$), Fig.\ref{Relaxation_ETV2O3}. The data on this figure have also been confirmed on a powder sample   by Kawamoto {et-al}\cite{Kawamoto95}.

\begin{figure}[t]
\includegraphics[width=0.8\hsize]{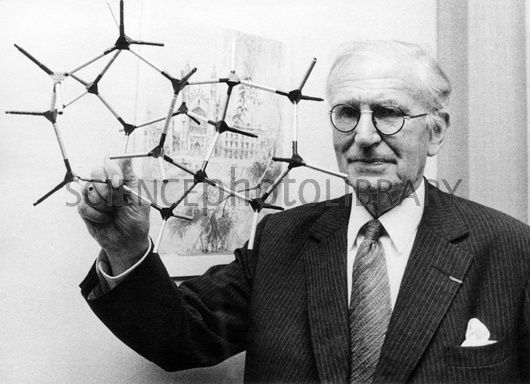}
\caption{\label{Mott} Sir Neville Mott (1905-1996), 1977 Nobel laureate in Physics. Sir Neville has been a pioneer in the physics of electrons in metals. The ideas presented in his textbook "Metal Insulator Transitions" have  influenced  the research on a large number of current materials\cite{Mott74}. We thank Jean Friedel for the communication of this photograph. }
\end{figure}

In close relation with these 2D organic compounds, it is illuminating to go back to the work of the 1970's particularly on vanadium sesquioxides\cite{McWhan70,Jayaraman70}, after Neville Mott, Fig.~\ref{Mott}, suggested in 1961\cite{Mott61,Mott71}  that by reducing the $U/W$ ratio between the Coulombic repulsion and the electron kinetic energy, a material  naturally insulating due to strong inter-electron correlations could undergo a metal to insulator transition. 

 NMR data displayed on Fig.~\ref{Relaxation_ETV2O3} show that for $\mathrm{(V_{0.99}Cr_{0.01})_2{O}_{3}}$  the relaxation rate is temperature independent in the  high temperature Mott insulator regime. A similar  behaviour is also observed for $\kappa$-(ET)$_2$\cuncnbr in the low pressure domain, \emph{see} Fig.~\ref{Relaxation_ETV2O3} (top) whereas a more common (Korringa) metallic behaviour  is recovered at higher pressures. 
 
 The large decrease of $1/T_{1}T$ observed below 50K for $\kappa$-(ET)$_2$\cuncnbr is the signature of  a magnetic pseudogap also visible in Knight shifts data\cite{Mayaffre94a,Kawamoto95} that disappears above 3 kbar or so. 
 
\begin{figure}[h]
\includegraphics[width=0.85\hsize]{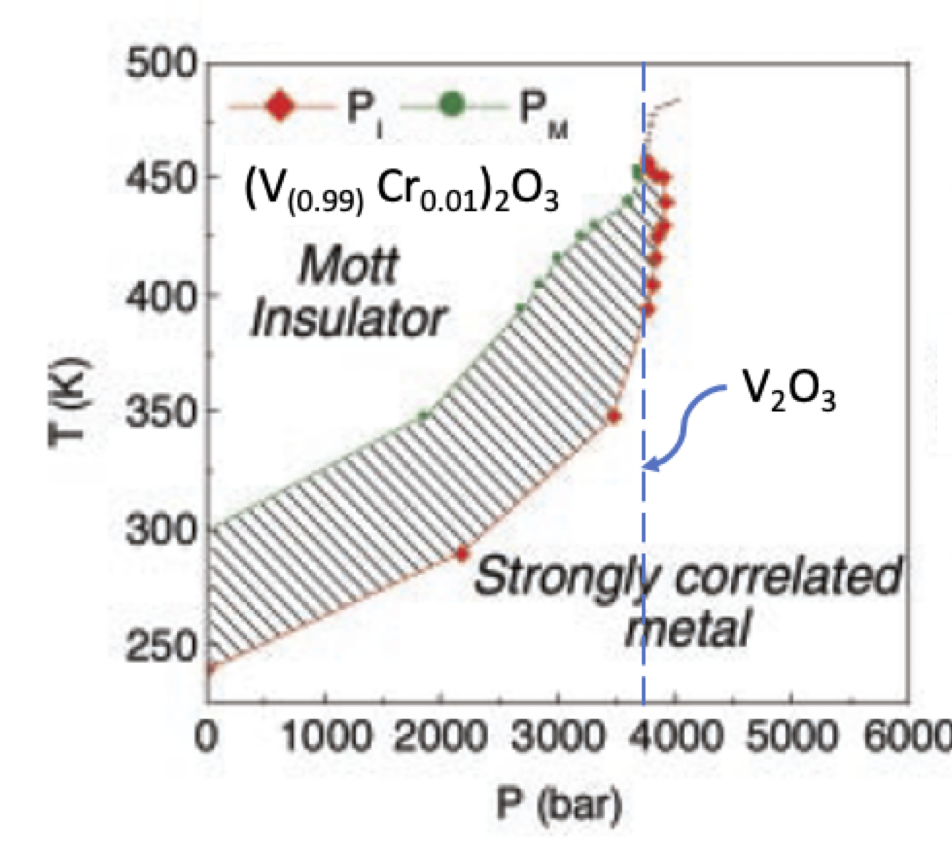}
\caption{\label{V203diagram} Limited phase diagram of $\mathrm{(V_{0.99}Cr_{0.01})_2{O}_{3}}$ as a function of temperature and pressure, Source: Fig. 1B [Ref.\cite{Limelette03a} p. 89]. The shaded phase coexistence  region is delimited by two spinodals and terminates at the critical point  3.7 kbar, 457K. Above the critical point, the Mott insulator is crossing over under pressure to a strongly correlated metal. The  dotted blue line shows the location of $\mathrm{V_{2}{O}_{3}}$ in this phase diagram. }
\end{figure}

 As far as $\mathrm{V_{2}{O}_{3}}$ is concerned, the decrease of $1/T_{1}T$ between the two blue dotted lines as shown on Fig.~\ref{Relaxation_ETV2O3} is the signature of a  phase coexistence regime with a progressive evolution between a high temperature Mott insulating regime  with $1/T_{1}$= const and the low temperature paramagnetic metal where $1/T_{1}\propto T$, \emph{see} also Fig.~\ref{V203diagram}.
 
 In conclusion, the comparative presentation of organics and oxides has shown that  in addition to being textbook models for one-dimensional physics, organic (2D) conductors are remarkable prototype compounds for studying the physics of the Mott transition.

Eventhough the stability of the low temperature antiferromagnetic state is known to be depressed by introduction of charge carriers in  $\mathrm{(V_{1-x}Cr_{x})_2{O}_{3}}$ alloys\cite{Limelette03},  an attempt to look for possible superconductivity in the alloy x=0.01   managed to  remove the AF insulating  state at helium temperature but failed to stabilize superconductivity\cite{Lesino,Limelette03}.

An other aspect of the 2D organics is the experimental realization
of the Jaccarino-Peter mechanism for superconductivity in systems comprising magnetic ions\cite{Jaccarino62}. This mechanism is probably
the one which allows the stabilization of superconductivity in
\betsfecl4 (where \bets is the selenide parent molecule of  ET ) under intense magnetic field\cite{Uji01,Balicas01} compensating the
exchange field of aligned Fe$^{3+}$ ions.

A review on organic conductors with much emphasis on 2D compounds has been  recently published in Ref.\cite{Naito21}.

\subsection{Charge density wave and superconductivity in 1D conductors}
\label{CDW-SC}
It is now important to dedicate a section of this overview to organic conductors that were among the first to be discovered and that have been studied exhaustively by several research teams in Portugal in cooperation with others, not only because these materials are interesting in their own right, but also for the contrast between their properties and those of   Bechgaard superconductors.

\begin{figure}[h]
\includegraphics[width=1\hsize]{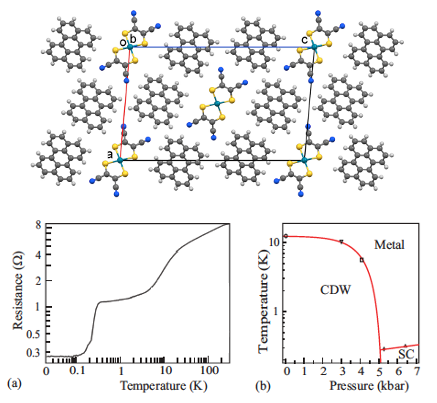}
\caption{\label{Per-Au} (Top) Room-temperature crystal structure of $\mathrm{(Per)_{2}Pt(mnt)_{2}}$ projected along the packing axis  $b$. (Bottom) (a) Temperature dependence of the resistance of  $\mathrm{(Per)_{2}Au(mnt)_{2}}$  under 5.3 kbar with \tc = 0.35K. (b) $T-P$ phase diagram. Adapted from Fig. 1 [Ref\cite{Alcacer80}, p. 946] (top),  Fig. 1, 5 [Ref\cite{Graf09}, p. 2, 4] (bottom). }
\end{figure}

In the search of highly conductive charge-transfer compounds of one-dimensional (1D)
character, perylene was one of the first donors used with bromine as anion exhibiting high conductivity but of non metallic character\cite{Akamatu54}. Then, in 1974, Alcacer reported electrically conducting metal dithiolate-perylene complexes in  with  the metal-maleonitriledithiolate ($\mathrm{M(mnt)^{2}}$) anion, ($\mathrm{D_{2}A}$-type salt)\cite{Alcacer74} giving a conductivity of 50 ($\Omega.${\rm cm})$^{-1}$ at room temperature but still behaving as  semiconductors at lower temperature where M is a transition metal atom.  A major progress has been accomplished with the synthesis of the 2:1 salts $\mathrm{(Per)_{2}Pt(mnt)_{2}}$. The single crystals conductivity of the M=Pt salt is large and metallic down to a metal insulator transition   at 15K\cite{Alcacer80}, \emph{see} structure on Fig.~\ref{Per-Au}.
The metallic properties are related to the existence of
stacks along the $b$ axis of cation molecules $\mathrm{Per}_{2}^{+}$ parallel to anions $\mathrm{M(mnt)}_{2}^{-}$.
The anion stacks  do not contribute to the electrical conductivity but play a key role in the magnetic properties. Whereas for M= Au the anion is diamagnetic and the metal insulator transition at 12K is of the Peierls nature, for  M=Pt the anions which are half filled bands   carry a magnetic moment  $(S= 1/2)$ and give rise to a Mott insulator leading in turn to a spin-Peierls transition for $\mathrm{(Per)_{2}Pt(mnt)_{2}}$  accompanied by a dimerization of the $\mathrm{Pt(mnt)_{2}}$ molecules and a concomitant  tetramerization of the interstack coupled perylene chain around 7.5K\cite{Bonfait91,Alcacer80}.

A $\mathrm{^{1}H}$-NMR investigation conducted on $\mathrm{(Per)_{2}Pt(mnt)_{2}}$ has confirmed the existence of localized spins on the $\mathrm{Pt(mnt)_{2}}$  chains and their
dominant influence on the relaxation through an interstack dipolar coupling\cite{Bourbonnais91a}. A scaling relation of the
form \textcolor{black}{$T_{1}^{-1} \propto T \chi_{s} (T)$} is found to exist in the entire temperature domain, including the regime of one dimensional
spin-Peierls $\mathrm{Pt(mnt)_{2}}$ lattice fluctuations observed by x-ray experiments below 30K\cite{Henriques84}. Contrasting with the previous compound, $\mathrm{Au(mnt)_{2}}$ there are no localized spins on the the dithiolate chains and the perylene proton relaxation is due to the fluctuating field provided by the mobile electrons of  perylene chains.

\begin{figure}[t] \includegraphics[width=1\hsize]{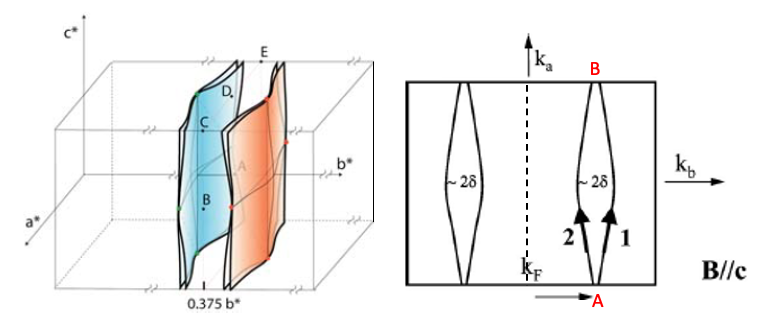} 
\caption{\label{CanadellFS} Schematic Fermi surface of $\mathrm{(Per)_{2}M(mnt)_{2}}$ according to\cite{Canadell04}  
showing one pair of the double-sheet surface intersecting the $b^*$ (stacking axis in reciprocal space) at $+\frac{3}{8}b^*$. The enlarged picture (after\cite{Graf07}) shows the   open electron trajectories  on the double sheet surface leading   to quantum interferences under magnetic field  at A and B. } 
\end{figure}

 Once the Peierls transition of $\mathrm{Au(mnt)_{2}}$ at 12K is suppressed by a moderate pressure of 5 kbar, a metallic state is stable at low temperature 	and the compound  becomes superconducting below 0.3K\cite{Graf09}, \emph{see} Fig.~\ref{Per-Au}. We may note that this metallic phase presents under magnetic field a phenomenon somewhat similar to what  will be detailed in Sec.~\ref{magnetoresistance oscillations} about \tm2x compounds namely, magnetoresistance oscillations due to quantum interferences between two open trajectories on open Fermi surfaces\cite{Graf07}. These oscillations are periodic in $1/B$ with a frequency of 19.5T which is determined  by the area enclosed by the two open trajectories in $k$ space, \emph{see} Fig.~\ref{CanadellFS}.  
 They can be related to the peculiar Fermi surface of  $\mathrm{(Per)_{2}M(mnt)_{2}}$  proposed by Canadell\cite{Canadell04} after the calculation of the electronic structure of these molecular conductors. 
 
 The unit cell containing four perylene units and two anions together with  a  pronounced 1D character resulting from a weak interstack coupling leads to the double sheet  Fermi surface displayed on Fig.~\ref{CanadellFS}. 
As this area is proportional to the magnitude of the transverse coupling, it is quite reasonable to find for the very 1D conductor $\mathrm{(Per)_{2}Au(mnt)_{2}}$ a value for the frequency much smaller than  what is observed in the more 2D coupled \tmtsf2x conductors, \emph{vide infra}.

 Two additional points on Fig.~\ref{Per-Au} are worth noticing. First, the resistance, which is weakly dependent on temperature below 10 K, \emph{see }Fig.~\ref{Per-Au}, is contrasting with the behaviour of the \tmtsf2x superconductors, Sec.~\ref{conductingstate}, for which the single particle scattering  is dominated by magnetism close to a quantum critical point. 
 
 Second, the $P-T$ phase diagram of 
 $\mathrm{(Per)_{2}Au(mnt)_{2}}$ reveals a tendency for \tc to increase under pressure which is also \emph{at variance} with the behaviour of \tc in the \tm2x family. Such a behaviour could be understood following the model developed by Bakrim and Bourbonnais\cite{Bakrim10} predicting a power law increase of the superconducting \tc$\approx\omega_{D}^{0.7}$ with the Debye frequency, from a quantum a quantum critical point of CDW order triggered by nesting alterations under pressure.

 There exists an other series of charge transfer conductors based on transition metal (M = Ni, Pd, Pt) complexes of dmit
(dmit = 1,3-dithia-2-thione-4,5-dithiolato)\cite{Bousseau86} which have attracted a great deal of attention after
the discovery of superconductivity in $\mathrm{TTF[Ni(dmit)_{2}]_2}$  at 1.6 K
under a pressure of 7 kbar\cite{Brossard86}. Two other  compounds of this family have also been found superconduting later,
$\mathrm{(CH_{3})_{4}N[Ni(dmit)_{2}]_{2}}$ (\tc = 5 K at 7 kbar)\cite{Kobayashi87} and $\alpha’-\mathrm{TTF[Pd(dmit)_{2}]_{2}}$ (\tc = 6.5 K at 20 kbar)\cite{Brossard89}. 

Moreover, an interesting feature of $\mathrm{TTF[M(dmit)_{2}]_{2}}$ (M = Ni,
Pd) is the occurrence of charge density wave (CDW) instabilities observed in X-ray diffuse
scattering studies at ambient pressure\cite{Ravy89}. Thus, these materials provide also an interesting
example of competition between CDW and superconductivity in molecular metals.
Unlike most molecular conductors presented up to now, the  $\mathrm{TTF[M(dmit)_{2}]_{2}}$ (M = Ni,
Pd) family is characterized by a small energy difference between  the $\mathrm{M(dmit)_{2}}$ highest occupied molecular
orbital and lowest unoccupied molecular orbitals
energy difference (HOMO-LUMO) which leads to a contribution at
the Fermi level of both LUMO and HOMO orbitals of the acceptor as evidenced by early band structure calculations\cite{Canadell89}. A recent the band structure and Fermi surface as a function
of temperature has been carried out on the basis of the centered
unit cell containing four TTF and eight $\mathrm{Ni(dmit)_{2}}$ molecules using a numerical atomic orbital density functional theory
(DFT) approach\cite{Kaddour14} which was developed for efficient
calculations in large systems and implemented in the SIESTA
code\cite{siestacode}.

 \begin{figure}[h]
\includegraphics[width=1\hsize]{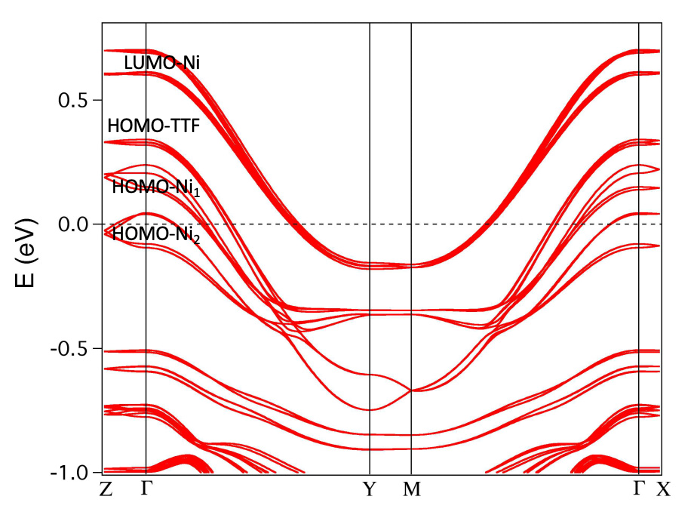}
\caption{\label{bandesNidmit} Most recent calculation of the band structure of  $\mathrm{TTF[Ni(dmit)_{2}]_{2}}$ at 150K, Source: Fig. 7 [Ref\cite{Kaddour14}, p. 4]. The Fermi level at zero energy is crossing the TTF HOMO band, the $\mathrm{Ni(dmit)_{2}}$ LUMO band and two low lying $\mathrm{Ni(dmit)_{2}}$ HOMO bands.  }
\end{figure}

As revealed  in the 1986 experiments\cite{Brossard86} $\mathrm{TTF[Ni(dmit)_{2}]_{2}}$ remains metallic under ambient pressure at low temperature with a conductivity of the order of $\mathrm{1.5\times10^{5}}$($\Omega.{\rm cm})^{-1}$ at 4K, but becomes a superconductor  at 1.62 K under 7 kbar. A subsequent investigation of this phase diagram under pressure revealed an unexpected complexity suggesting that superconductivity is coexisting with CDW instabilities at higher temperature\cite{Brossard90} as hinted by X-ray studies at ambient pressure\cite{Ravy89}. 
\begin{figure}[h]
\includegraphics[width=0.9\hsize]{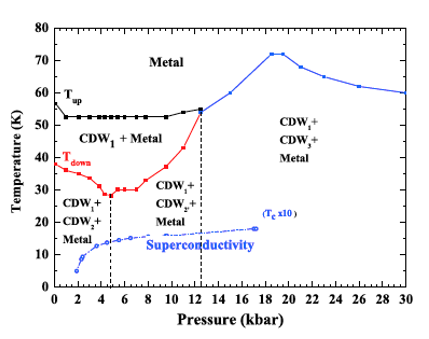}
\caption{\label{TPNidmit} A recent determination of the phase diagram of  $\mathrm{TTF[Ni(dmit)_{2}]_{2}}$ from transport properties under pressure, Source: Fig. 10 [Ref\cite{Kaddour14}, p. 6]. Onset of CDW orderings have been detected by small anomalies in susceptibility and sharp anomalies on transverse transport (only). The superconducting \tc  is multiplied by ten on this diagram.}
\end{figure}
This phase diagram   re-investigated 14 years  later using  transport properties under pressure has led  to a better knowledge of this complex system\cite{Kaddour14}. The most recent determination of the T-P phase diagram is displayed on Fig.~\ref{TPNidmit}. This diagram is confirming previous experimental results in particular, $^{13}$C NMR Knight shift data showing that even the 3D ordered CDW below 60 K does not open a gap throughout the entire Fermi surface and instead decreases the density of states at Fermi level\cite{Vainrub90}. 
 \begin{figure}[h]
\includegraphics[width=0.9\hsize]{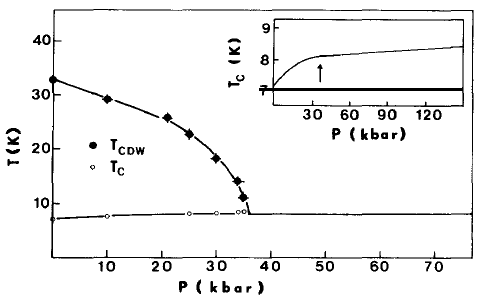}
\caption{\label{PTNbse2} T-P phase diagram of 2H-$\mathrm{NbSe_{2}}$ determined on the same sample batch by transport and NMR\cite{Berthier76}, Source: Fig. 2 [Ref\cite{Berthier76c}, p. 1394]. \emph{see also} Fig.~\ref{Claude-Berthier}.  }
\end{figure}

As a result of the complex situation of this  multiband system a superconducting state can be stabilized at low temperature even in
the presence of several CDW instabilities but when the consequence of distortions on the density of states is reduced under pressure, $N(E_{F})$ increases and \tc is favored. The situation where  a competition arises between superconductivity and CDW is reminiscent of   what has been revealed long ago in the study of  
transition-metal dichalcogenides\cite{Molinie74}, \emph{see} Fig.~\ref{PTNbse2}. 
\begin{figure}[h]
\includegraphics[width=0.4\hsize]{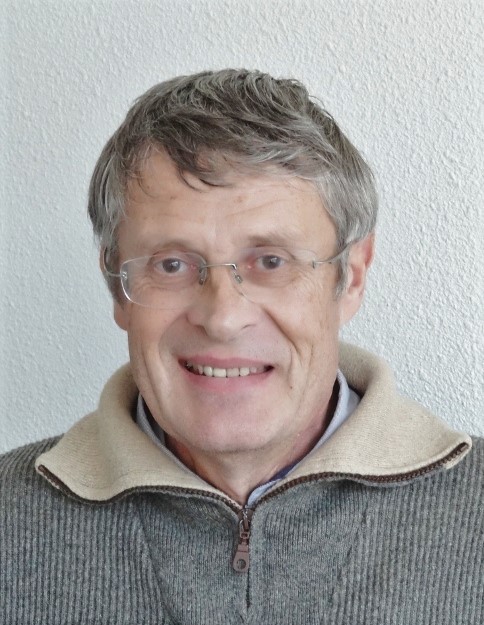}
\caption{\label{Claude-Berthier} Claude Berthier (1946-2018) after a PhD in  NMR in solids at Grenoble University, did a postdoctoral stay at Orsay in 1975. He has been the first to carry out very smart  NMR studies on small single crystals of transition metal dichgalcogenides and organic conductors. Back to Grenoble, he became a leading scientist at the High Magnetic Fields Laboratory and contributed to major advances in the physics of one  and two dimensional compounds.}
\end{figure}
According to a simple argument proposed by Friedel in 1975\cite{Friedel75}, as far as 2D chalcogenides are concerned the creation of a superlattice lowers \tc because of a concomitant lowering of $N(E_{F})$. Consequently, suppressing the distortion under pressure results in   an enhancement of  $T_c$. If the same argument  is applied to $\mathrm{TTF[Ni(dmit)_{2}]_{2}}$  it must be applied with caution since  unlike  di-chalcogenides the temperature-pressure phase diagram on Fig.~\ref{TPNidmit} does not provide a complete suppression of the distortions. However this interpretation may be more relevant   for the superconducting phase of $\mathrm{TTF[Pd(dmit)_{2}]_{2}}$ which is stabilized at 5.5K under a pressure of 24 kbar instead of an insulating  phase at lower pressures\cite{Brossard89}. 

\subsection{\textcolor{black}{Single-component molecular superconductors}}
\textcolor{black}{
Until now, all the molecular conductors mentioned in this review have two components, either charge transfer compounds such as TTF-TCNQ, or  organic cationic salts like Bechgaard-Fabre salts, (TMTSF)$_2X$. We have  presented in Sec.~\ref{CDW-SC} the case  of different molecular metals  in which  some highest occupied molecular orbitals  (HOMO)  belonging to the   acceptor molecule being only slightly  depressed  in energy below the lowest unoccupied orbitals (LUMO)  contribute both  to the density of states at the Fermi level (together with HOMO levels from the TTF donor molecule) as shown on Fig.~\ref{bandesNidmit}. }
\textcolor{black}{These compounds are based on transition metal complex molecules such as $\mathrm{Ni(dmit)_{2}}$ which have been at the origin  of the development of new molecular conductors and superconductors composed of only one chemical specie at  Nihon University Tokyo\cite{Akiko04}.}

\textcolor{black}{The design of a single-component molecular metal is difficult because most molecules have an even number of electrons and their HOMO is usually doubly occupied with a  HOMO-LUMO gap larger than the intermolecular interaction leading to a finite bandwidth in the solid state. Therefore, in order to achieve metallicity  with partially filled bands,  the energy separation between HOMO and the lowest unoccupied molecular LUMO  orbitals should be small enough to make the HOMO and LUMO bands overlap each other as a result of two-dimensional or three-dimensional intermolecular interactions in order to form.}

\begin{figure}[h]
\includegraphics[width=0.8\hsize]{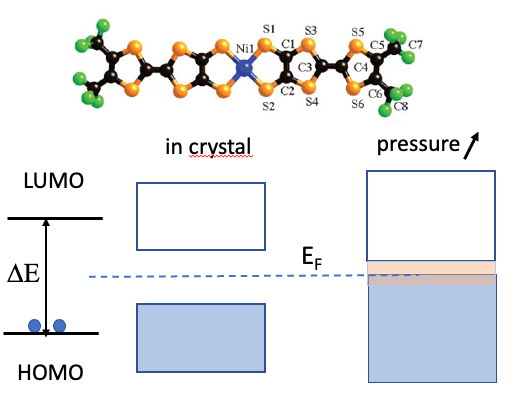}
\caption{\label{HOMOLUMO} Molecular structure of the transition metal complex $\mathrm{[Ni(hfdt)_{2}]}$ which is the unique ingredient in the single component molecular superconductor, Source [Ref\cite{Sasa05}]. Notice the bulky fluorinated terminal groups  which enhance the coupling between the $a-b$ layers \emph{at variance} with other  transition metal  single-component molecular conductors.       Schematic diagram showing how a single component molecular semi-metal such as $\mathrm{[Ni(hfdt)_{2}]}$ can form if the HOMO-LUMO gap $\Delta E$  is not too large compared to  individual bandwidths in the solid. The compound is a classical band insulator under ambient and low pressure. In the  case of  $\mathrm{[Ni(tmdt)_{2}]}$ the overlap between valence and conduction bands exists already under ambient pressure leading in turn to a metallic behaviour\cite{Tanaka01}.  However, for $\mathrm{[Ni(hfdt)_{2}]}$ it is the  high pressure increasing the intermolecular interaction  which makes both bands overlapping, giving rise to a semi-metallic band structure with a superconducting phase\cite{Cui14}.}
\end{figure}

\textcolor{black}{To obtain a molecule with a very small HOMO-LUMO gap, 
 metal complexes with
extended-TTF dithiolate ligands were adopted because on the one hand the HOMO of the metal dithiolate complex can be roughly expressed as a bonding combination of left and right ligand orbitals while the LUMO is roughly the anti-bonding combination of them, including very small ${d_{xz}}$ orbital of the central metal atom with appropriate symmetry to mix in the LUMO leading to a relatively weak interaction between left and right ligands through sulfur atoms\cite{Kobayashi21}, \emph{see} Fig.~\ref{HOMOLUMO}. }

\textcolor{black}{These HOMO and LUMO orbitals suggest the possibility of a small
HOMO-LUMO gap due to the relatively weak interaction between left and right ligands through ligand sulfur atoms.
Furthermore, strong intermolecular interactions through S-S atoms in extended-TTF ligands play a very important role in the enlargement of the bandwidth. Consequently, as sketched on Fig.~\ref{HOMOLUMO}, the small HOMO-LUMO gap and strong intermolecular interaction through S-S atoms produce HOMO band and LUMO band overlaps around  $E_F$, and electron and hole Fermi surfaces are produced.}

 \begin{figure}[h]
\includegraphics[width=0.8\hsize]{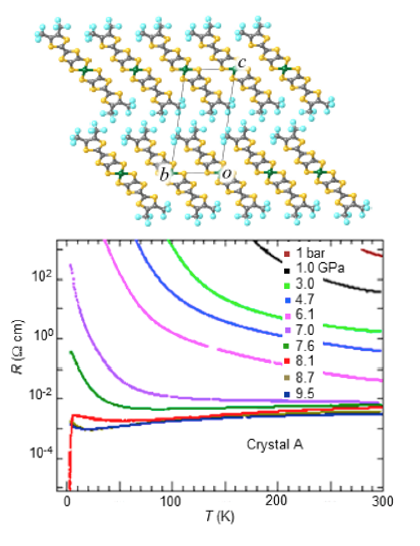}
\caption{\label{hfdt} \textcolor{black}{Molecular packing of $\mathrm{[Ni(hfdt)_{2}]}$. Temperature  dependence of the resistivity up to the pressure of 9.5 Gpa showing the superconducting transition at $\sim 5.5$ K in a narrow pressure window around 8 Gpa, Source: Fig. 3a [Ref\cite{Cui14}, p. 7620].}  }
\end{figure}

\textcolor{black}{A lot of transition metal single-component compounds have been synthesized by the Nihon group and their physical properties studied in temperature and very high pressure\cite{Kobayashi01,Kobayashi21}. Among them, let us mention $\mathrm{[Ni(tmdt)_{2}]}$ (tmdt = trimethylenetetrathiafulvalenedithiolate),  which reveals a metallic resistivity behaviour down to 0.6 K and a semi-metallic electronic structure consistent with extended H\"uckel band calculations   of a 3D crystal and a HOMO-LUMO gap of the order of 0.1 eV\cite{Tanaka01}.}

\textcolor{black}{ The compound $\mathrm{[Ni(hfdt)_{2}]}$ \hbox{(hfdt~=~bis(trifluoromethyl} \hbox{-tetrathiafulvalenedithiolate)}  is a semiconductor under ambient pressure with an activated resistivity, (\hbox{$E_{a}\sim 0.14$}  eV), in fair agreement with the LDA band structure calculation\cite{Sasa05}. According to the band structure  performed with  lattice parameters  under pressure derived by first principle calculation, the band gap disappears around 6 GPa in agreement with the observed stabilization of a two dimensional metallic state leading to superconductivity below $\sim 5.5$ K in a very narrow pressure range (7.5-8.6) GPa\cite{Cui14}. However, what is remarkable about the  stabilization of superconductivity in this  single-component molecular conductor  is the absence of lattice or magnetic precursors  in the high temperature metal phase.
It can thus be concluded that the physics of this one-component superconductor is likely to be closer to that of classical band semiconductors than to that of low-dimensional conductors presented above.}


\subsection{Magnetic field-induced SDW phases in 1D conductors and magnetoresistance oscillations }
\label{magnetoresistance oscillations}
Quasi one-dimensional conductors pertaining to the \tm2x family exhibit under magnetic field properties that are unique in solid state physics. These properties are the magnetic  Field Induced Spin Density Wave (FISDW) phases on the brink of superconductivity, which  have contributed to another success story mixing theory and experiments (with the  quantized nesting model and a nice application  of the magnetic breakdown theory). Even if the link between FISDW and superconductivity is not conclusively established, it would be a shame to conclude this overview without mentioning even very briefly these phenomena under magnetic field. 

It is remarkable to note that this research domain was developed very actively   at the beginning of 1983 simultaneously    by two groups on both sides of the Atlantic and at the Landau Institute of Moscow. While we  tend to focus on the historical development of the field, the interested reader can find more details in  several articles published in the  textbook edited by A. Lebed\cite{Lebedbook}.
 \begin{figure}[h]
\includegraphics[width=0.8\hsize]{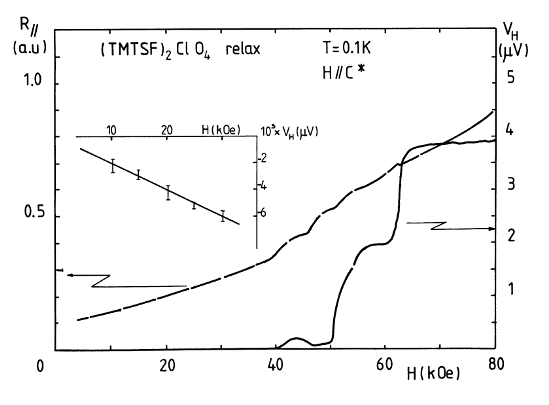}
\caption{\label{QuantizedHall} Magnetic field dependence of the Hall voltage in slowly-cooled \tmc at 0.1K. The inset shows the low field regular hole-like part  before the onset of the FISDW phases, Source: Fig. 2b [Ref\cite{Ribault83}, p. 956].  }
\end{figure}

  \begin{figure}[h]
\includegraphics[width=0.9\hsize]{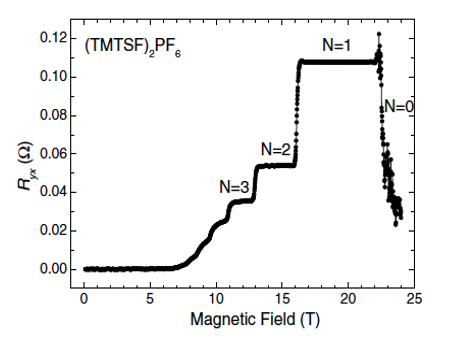}
\caption{\label{Kang}  Magnetic field dependence of the transverse resistance of \tmp6 under 10 kbar,  Source: Fig. 3b [Ref\cite{Kang04}, p. 265]. These results have been chosen among many others for their high quality. They even enable a quantum measurement of the sample thickness namely, 0.29 mm. }
\end{figure}

 \begin{figure}[h]
\includegraphics[width=0.6\hsize]{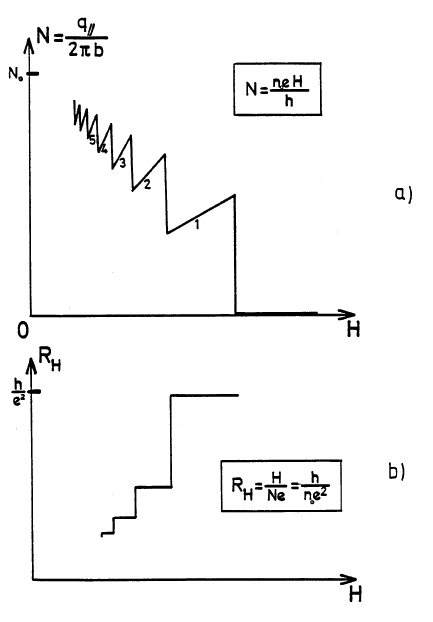}
\caption{\label{theseGM} a) N is the number of unpaired carriers  in the pocket formed by the Fermi surface  and the translated surface by $\mathbf{Q}(H)$. This vector varies with the field so as to maintain in the pocket an integer number n of completely full Landau levels, a condition for the minimum of diamagnetic energy. b) The Hall resistance (without spin degeneracy) $\mathrm{R_{H}=\frac{H}{Ne}}$ is quantized  within each phases with n filled Landau levels since $\mathrm{N=\frac{neH}{h}}$ according to a). This is in perfect agreement with the experimental data displayed on Fig.~\ref{Balicasrhoxy}. This figure is borrowed from the thesis of G. Montambaux\cite{Montambaux85c}. }
\end{figure}
 \begin{figure}[h]
\includegraphics[width=0.9\hsize]{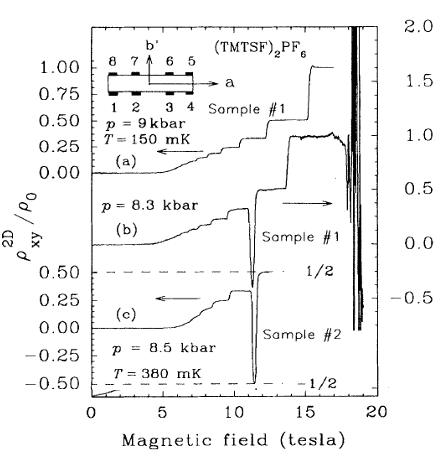}
\caption{\label{Balicasrhoxy}  Transverse resistance $\rho^{2D}_{xy}$ for two samples at pressures less than 10 kbar,  multiplied by the number of conducting layers to obtain the single layer resistance and normalized to the resistance quantum $\rho_{0}$ which amounts to 12906 $\Omega$ including spin. Negative Hall conductivity becomes visible in the pressure range  $p\leq$ 8.5 kbar, (N=-2 signal in sample \#1 at 8.5 kbar for instance), Source: Fig. 1 [Ref\cite{Balicas95}, p. 2001]. These data corroborate  previously published results,\cite{Cooper89,Hannahs89} but they are displayed here because of their higher quality.  }
\end{figure}

The new physics concerning the conductive phase of \tm2x compounds under magnetic field began with the experimental observation that in magnetic fields of a few Tesla applied along the lowest conductivity axis, novel phases appear in (TMTSF)$_2$ClO$_4$, of semi-metallic nature\cite{Garoche82}, characterized by a magnetic order of the SDW nature\cite{Takahashi82} and an unexpected step-like behaviour of the Hall voltage\cite{Ribault83}, \emph{see} Fig.~\ref{QuantizedHall}. Subsequent investigations revealed other unexpected phenomena such as the existence of Hall plateaus alternating in sign at increasing magnetic field the so-called  ``Ribault anomaly"\cite{Ribault85}. Negative plateaus have been also observed and extensively studied in \tmp6 under pressure\cite{Piveteau86,Balicas95,Cho99}.
 
Shortly after publications of these experiments, 
 Gorkov and Lebed proposed a model\cite{Gorkov85} in which the orbital effect of a magnetic field aligned perpendicular to the $a-b$ plane of the Q1D conductor with a finite transverse coupling along $b$ tends to restrict the excursion along the $b$ direction  for an electron moving along the $a$ axis. 
   The effect of the magnetic field  amounts  to  increasing the one dimensionality as  in the case of open orbits the transverse excursion along $b$ is inversely proportional to the magnetic field\cite{Lebed86}. Hence, going to low temperature, the Q1D electron gas is likely to undergo an instability towards a semi-metallic SDW phase with a wave vector $\mathbf{Q}$= (2$k_{F}, \pi/b^{*}, \pi/c^{*})$ .
The  Gorkov and Lebed’s argument on the effect of a magnetic field on nesting has been improved by the Orsay group, \textcolor{black}{H\'eritier, Montambaux and Ledderer}\cite{Heritier84}. G.~Montambaux using Gorkov-Lebed's formalism in his PhD thesis work, derived the staggered static 2D spin susceptibility $\chi_{0}(\mathbf{Q},H, T)$ of the noninteracting electron gas with open Fermi surface \cite{Montambaux85,Montambaux86}. Starting from  the condition of minimum diamagnetic energy, allowing the wave vector $\mathbf{Q}(H)$
to adapt to the field so that the nesting becomes quantized, it was predicted that the field induces a succession of SDW
phases with $\mathbf{Q}(H)$ varying linearly, separated by first order transitions with  $\mathbf{Q}(H)$ discontinuities and Hall voltage plateaus within each phases. This theory is  known as the Quantized Nesting Model, \emph{see} Fig.~\ref{theseGM}. 
 
The model was even improved taking into account slight modifications of the dispersion relation in the metallic phase\cite{Montambaux85a} to explain the existence of negative Hall voltage plateaus as displayed on Fig.~\ref{Balicasrhoxy} for \tmp6 although first evidenced in \tmc\cite{Ribault85}. 
 \begin{figure}[h]
\includegraphics[width=0.8\hsize]{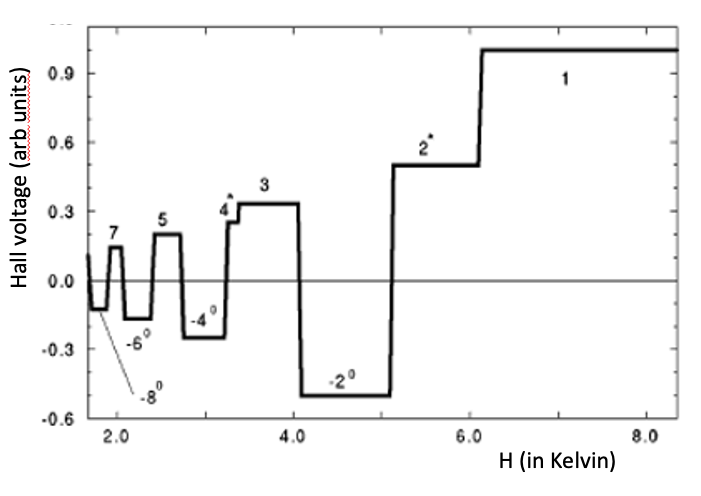}
\caption{\label{Zanchi} Theory of the Hall voltage versus magnetic field (1K=0.5T), Source: Fig. 3b [Ref\cite{Zanchi96}, p. 369]. }
\end{figure}
The sequence of negative plateaus appearing in \tmp6\cite{Balicas95,Cho99} below 9 kbar  with  even quantum
numbers   have been understood by a slight modification of the  strongly pressure sensitive dispersion relation of the metallic phase\cite{Zanchi96}, \emph{see} Fig.~\ref{Zanchi}. As far as \tmc is concerned the existence of the  negative plateaus  is very dependent on the ordering of the \cl anions\cite{Ribault85} which is closely related to the peculiar  granular structure  likely to develop  upon fast cooling, \emph{see} Sec.~\ref{Cooling rate-controlled superconductivity}. 
Taking into account the third direction along $c$, Montambaux\cite{Montambaux86} noticed that \emph{at variance} with 2D case given by  $T_{c 2D}(H)$ ,   there exists  at $T=0$ K in the 3D situation a finite  threshold field $H_{t}$ for the stability of the SDW phase such as, $T_{c 2D}(H_{t})= t_{c}'$ where $t_{c}'$ is the measure of the deviation from perfect nesting along the third direction.

 Furthermore, a wealth of fascinating phenomena have been observed  regarding the  magnetoresistance of these model materials, either at fixed orientations of the field with respect to  crystals axes or rotating a fixed field in well defined planes. This area of research has been very productive and has given rise to many publications.  In the present article we give only a short overview and refer the interested reader to the various contributions published in  textbooks\cite{Ishiguro,Lebedbook}. A recent account for part of these questions has been published in Comptes Rendus Physique\cite{Montambaux16}.

 First, it is well established in low dimensional organic conductors  that  the Landau quantization of closed electronic orbits originating from pockets of unpaired carriers left after  density waves instabilities of the 1D electron gas (CDW or SDW) give rise to magnetic oscillations of transport properties.  The quantization of the electronic motion along these closed pockets leads to Shubnikov-de Haas (SdH) oscillations  (periodic in $1/B$) the period of which is proportional to the size of the pocket.  The typical field $B_{f}$ characteristic of the oscillations is proportional to the area ${\cal A}$ of the closed orbits in reciprocal space. 
The characteristic energy of deviation from perfect nesting, named $t'_b$ in Eq.~\ref{tightbinding}, is usually of order of $10$-$30$\,K, so that ${\cal A}   \propto {  t'_b \over \hbar   v_F b}$ and the typical field $B_{f} \propto {t'_b \over e   v_F b}$ is of order of a few dozen Tesla. In Bechgaard salts, the competition between Spin Density Wave (SDW) ordering and the quantization due the  magnetic field   leads to the cascade of SDW sub-phases in which the  Hall effect is quantized as presented above. 

\begin{figure}[h]
\includegraphics[width=0.7\hsize]{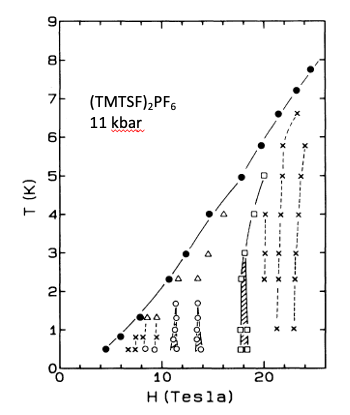}
\caption{\label{TMTSF2PF6Rapidopscillations} T-H phase diagram of \tmp6 under pressure displaying the FISDW phases. Shaded areas denote hysteresis between Hall plateaus. The high field state arises above 18T and the crosses represent peaks in $\rho_{xx}$, Source: Fig. 4 [Ref\cite{Cooper89}, p. 1986].  }  
\end{figure}
\begin{figure}[b]
\includegraphics[width=1.05\hsize]{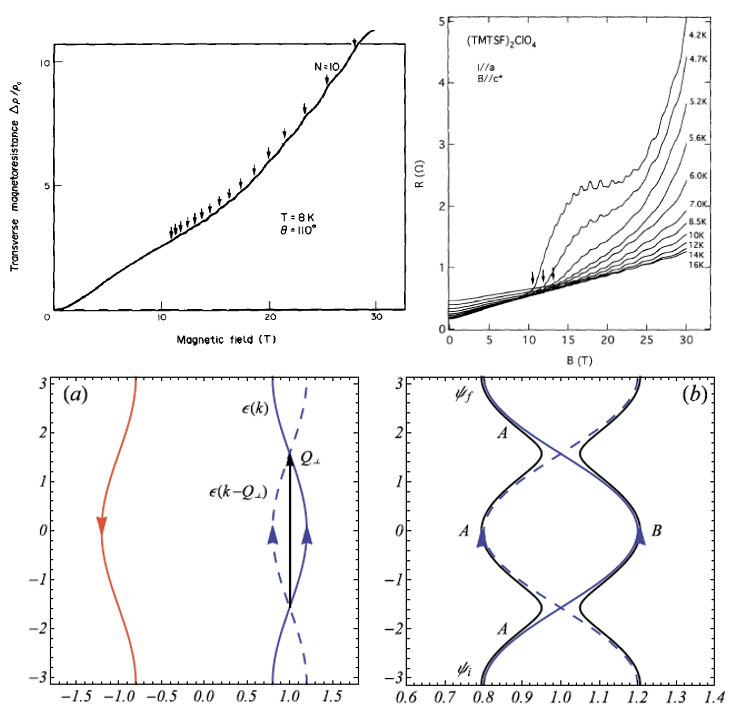}
\caption{\label{RO} (Top left) Observation of rapid oscillations in the metallic phase of R-\tmc ($a$ axis magnetoresistance with the field   perpendicular to $a$ but 20 degrees off $c*$ axis, $T=8$K). (Top right) Magnetoresistance in the metallic phase of R-\tmc  from $T= 16$K down to the FISDW phase
. (Bottom) (a) A schematic picture of the \tmc Fermi surface  in the presence of a modulation with transverse wave vector  $\bold{Q}_{\perp}= (0, \pi/b$) which couples states on the same side of the Fermi surface. The dashed blue electron trajectory under magnetic field $\| c*$ is the result of the FS folding at $\pm\pi/2b$ due to the anionic external potential creating two open warped sheets AAA and BBB on (b). An electron starting on sheet A may travel either along two interfering paths  either AAA  or  ABA trajectories through a magnetic breakdown, leading in turn to magnetic oscillations in the conductance with the frequency $\propto t_{b}$, according to , Sources: Fig. 3 [Ref\cite{Uji97}, p. 548] (Top left), Fig. 2 [Ref\cite{Ulmet84}, p. 389] (Top right), Fig. 1 [Ref\cite{Montambaux16}, p. 377] (Bottom). }  
\end{figure}


What is more intriguing 
are the so-called rapid oscillations of the magnetoresistance, characterized by a much larger characteristic field of the order of a few hundred Tesla\cite{Ulmet84}. These rapid oscillations have been observed in the  FISDW phases of \tmp6 under pressure (11 kbar at room temperature) measuring $\rho_{xx}$ above 18 T with a period of 286T\cite{Cooper89}, \emph{see} Fig.~\ref{TMTSF2PF6Rapidopscillations}, and reexamined  in more details  by Kornilov \emph{et-al}\cite{Kornilov07b} 18 years later. As noticed in Refs.\cite{Cooper89} and \cite{Kornilov07b}, the amplitude of the ``fast" oscillations periodic in 1/H with the frequency of 286 T do not follow the classical temperature dependence of Shubnikov oscillations. Instead their amplitude progressively diminishes below 2 K.

 Rapid oscillations had also been observed in the metallic phase at temperatures larger than those of the FISDW instabilities. This is for instance the case for the  slowly-cooled phase of \tmc (R-\tmc)\cite{Ulmet84} or also in \tmre under pressure\cite{Kang91b}.

Rapid oscillations are  observed as well in the fast-cooled phase of \tmc (Q-\tmc), but only   in the presence of a FISDW instability\cite{Brooks99}. At \emph{variance} with the slow oscillations, the  frequency of these rapid oscillations is related to  interchain coupling $t_b$, that is to the warping of the open Fermi surface much larger than the unnesting contribution $t'_{b}$ in Eq.~\ref{tightbinding}.

Since an overview of all situations where rapid oscillations arise in these materials  with an unified theoretical model based on the experimental results has been published recently in Ref.~\cite{Montambaux16}  we only wish to illustrate here the concept of open orbits, magnetic breakdown  and Stark interference with the simple situation of the  anion-ordered \tmc. 
 
As displayed on Fig.~\ref{RO}, under a magnetic field perpendicular to the $a-b$ plane, the electrons travel on both FS sheets along the same direction, and may experience a tunneling through magnetic breakdown from one sheet to the other. Hence, the   two different paths AAA and ABA on Fig.~\ref{RO}  may interfere and lead to  St\"uckelberg oscillations of the conductivity\cite{Stuck32}, 
 which are proportional to cos $(2\pi \frac{B_{f}}{B}$)  where  the frequency becomes now $B_{f}$   $\propto{t_b \over e   v_F b}$ \emph{i.e}. a much larger frequency ($\approx $260T)  compare to the slow oscillations ($\approx $ 30T). A detailed review covering all cases of rapid oscillations in 1D conductors is proposed in the  reference\cite{Montambaux16} on open access.

\begin{figure}[h]
\includegraphics[width=0.75\hsize]{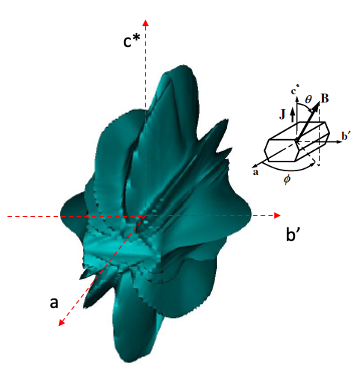}
\caption{\label{Stereo_exp}  A 3-D representation of the measured interlayer conductivity $\sigma_{zz}$ in polar coordinates of \tmp6 at high pressure  under a magnetic field of fixed amplitude with respect to the   field orientation defined  in an orthogonal frame  by angles $\theta$ and $\phi$. The radial distance from the origin is proportional to $\log\sigma_{zz}(\theta,\phi)$, Source: Fig. 1 [Ref\cite{Kang07}, p. 1]. We thank Wuon Kang  for allowing us to use the figure   for an adaptation. }  
\end{figure}
 
In addition,  a wealth of  phenomena have emerged from the magnetoresistance measured   when a constant  magnetic field is rotated in a frame characterized by angles  $\theta$ and $\phi$  (the angles  between $H$ and $a'$ in the  $a-b'$ plane   and between $H$ and $c'$ in the $c^{*}-b'$ planes respectively).  

In 2007  W. Kang and his colleagues\cite{Kang07} \emph{"The nice view presented in this figure is a result which has been made possible thanks to the construction of a two-axis rotator for the pressurized sample under high magnetic field".} have revisited this topic and  summarized all previous experimental works   providing a 3-D representation of the angular dependence of transport over 4$\pi$ steradian  in the case of the  simplest 1D conductor namely, \tmp6 under pressure, \emph{see} Fig.~\ref{Stereo_exp}. The figure  shows that  the interlayer conductivity ($\sigma_{zz}\approx 1/\rho_{zz}$) behaves  in a way far from the  smooth expectation of the Boltzmann transport theory   applied to an anisotropic conductor, peaks and dips of magneto-conductivity at well established angles  are superimposed to the semiclassical Boltzmann transport.

 The field of angular magnetoresistance in Q1D conductors  began after Lebed\cite{Lebed86a} predicted that tilting the field away from the $c*$ direction in the $c^{*}-b'$ plane should decrease the threshold field $H_{t}$ requested for the stabilization of the FISDW phase at $T=0$ K
 \emph{see}, Sec.~\ref{magnetoresistance oscillations} recovering the 2D limit. For these special orientations of the field  the electron motion would become commensurate along the $c$ and $b$ directions, \emph{i.e } frequencies  to cross the Brillouin zone along $b$ and $c$ directions become commensurate. Subsequently Lebed and Bak\cite{Lebed89}  dealing with transport in the metallic phase,  derived from the calculation of the scattering rate of carriers moving along the 1D axis the existence of anomalies of the magnetoresistance (peaks) departing from the Boltzmann calculation, at well identified  orientation (so-called magic angles) when the magnetic field is  rotated in the $b^{*}-c^{*}$ plane. Crudely speaking these magic angles are given for an orthorhombic approximation by,
 \begin{eqnarray}
\label{magicangles}
\mathrm{tan}\phi=\frac{m}{n}\frac{b}{c}
\end{eqnarray}
where $\phi$ is the angle for $H$ in the $b^{*}-c^{*}$ plane   from the $c^*$ axis and $n$ and $m$ integers.  Under magnetic field in general,  the 3D electron motion along the 1D axis   follows a complex trajectory  winding around the 1D axis with  incommensurate periods along $b$ and $c$  unless the field is tilted from the $c^*$ by an angle satisfying Eq.~\ref{magicangles}. When the two trajectories become commensurate, anomalies of magnetoresistance are expected at magic angles  according to Eq.~\ref{magicangles}  for example at $29.6^\circ$ and  $48.7^\circ$ for $m/n$ = 1 or 2 respectively with the \tmc lattice parameters at low temperature\cite{Gallois83}.

\begin{figure}[h]
\includegraphics[width=0.8\hsize]{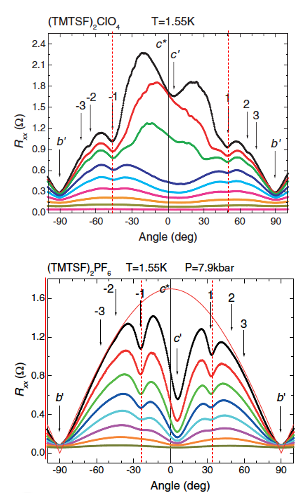}
\caption{\label{KangClO4andPF6Rxx} Angular magnetoresistance $\rho_{xx}$ of \tmc (top) and \tmp6  under 7.9 kbar(bottom), Source [Ref\cite{Kang04}]. Arrows indicate the position of magic angles  proposed by  Lebed for the respective compounds, Source [Ref\cite{Lebed86}]. Magnetic fields start from 1T and increase  by steps of 1T. }  
\end{figure}

The first experimental confirmation of these angular resonances came in 1991 with the observation of  fine structures superimposed on the angular dependence of magnetoresistance in the metallic  phase  of \tmc\cite{Osada91,Naughton91}. Weak dip structures are clearly observed around $\pm$ 48$^\circ$ and $\pm$ 65$^\circ$   in \tmc\cite{Osada91} with similar results in Ref.\cite{Naughton91} at magnetic fields lower than any threshold field for the stabilization of the FISDW phase\cite{Osada91}. This hypothesis is consistent with the fact that the unit cell of \tmc is doubled along $b$ at low temperature after  the ordering of the \cl anions. More recent results have been published by  W. Kang and colleagues, see Fig.~\ref{KangClO4andPF6Rxx} where the difference between the position of the oscillations in  \tmc and \tmp6 is made clear.

Although the position of the  angular anomalies agrees with the theory\cite{Lebed89}, there is still a contradiction with their sign since the experiments  show a suppression of the resistivity instead of the enhancement predicted by Lebed's theory\cite{Lebed86a} although   dips  derived from Osada's theory\cite{Osada92}.

\begin{figure}[h]
\includegraphics[width=0.9\hsize]{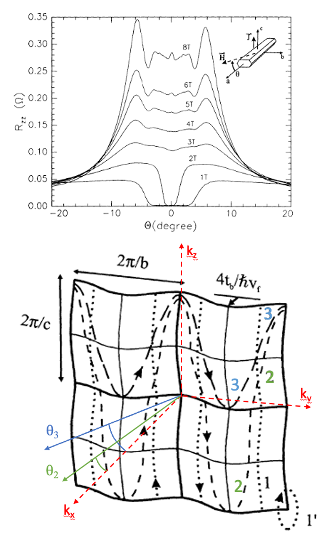}
\caption{\label{DannerChaikin} (Top) Angular dependence of the $c$ axis resistance for a rotation in the $a-c^*$ plane at a temperature of 0.5 K. Above 4T the angular dependence is no longer influenced by superconductivity according to the data in Fig.~\ref{Hc2Cvrho}.  Superimposed to the semi-classical Boltzmann angular dependence, a primary spike of resistance is observed at $\pm$ 6$^\circ$ which, is related to the transverse coupling $t_{b}$ and  possibly a secondary spike around $\pm$ 2$^\circ$  developing at 8T. The small spike   visible at $
\theta =\pm 0^\circ$ enables the measurement of the tunneling coupling $t_{c}$ between $a-b$ planes. (Bottom) Some possible orbits according to the field  orientation  in the $a-c^*$ plane, \emph{see} text for details, Sources: Fig. 1, 2 [Ref\cite{Danner94}, p. 3715]. }  
\end{figure}

Until now the Lebed's angular oscillations depend only on the geometry of the Q1D lattice and do not provide any information about the band structure  of the materials. However, the band  structure parameters become relevant for an other kind of oscillations, when the magnetic field is rotated  from $a$ to $c^*$ in the  plane parallel to $a-c^*$ as shown in Fig.~\ref{DannerChaikin}. 

In such a situation, the resistance presents a major peak when the field is slightly tilted off the $a$ axis. This is known as the Danner-Kang-Chaikin (DKC) oscillation (DKC)\cite{Danner94} as shown on Fig.~\ref{DannerChaikin}.

Using the equation of motion for an electron in a uniform magnetic field, Danner {\textit{et-al}}\cite{Danner94} managed to derive the time dependence of the velocity  component $v_{z} (t)$  along $z$. Depending on the orientation  of the magnetic field with respect to  the quasi planar Fermi surface, the averaging of  $v_{z} (t)$ over time is more or less efficient.  When the field is parallel to the   $a$ axis, orbits traversing the Fermi surface along $c$ average $v_{z} (t)$ to zero (such as orbits 1 on Fig.~\ref{DannerChaikin}) and therefore do not contribute to the magneto-resistivity. When the magnetic field is tilted by a small angle $\theta$ the orbits follow  a restricted path along $c$ but move to infinity along the   $b$  direction (orbit 2 on Fig.~\ref{DannerChaikin}). However, the averaging of  $v_{z} (t)$ to zero is still particularly efficient when the orbit sweeps periodically an integral number of lattice points along $c$ as shown by the path 3 on Fig.~\ref{DannerChaikin}  which sweeps one Brillouin zone  leading in turn to a strong spike of angular magnetoresistance at an angle  corresponding to $\theta_{3}$ in the figure. Path 3 can be approximated by the cross section of the Fermi surface by a plane tilted  $\theta_{3}$ degrees from the normal. Therefore,  tan $\theta_{3}$ amounts to $\approx (4t_{b}/\hbar v_{F})/(2\pi/c)$ and provides information about $t_{b}/v_{F}$. Given the existence of a magnetoresistance peak around $\pm$ 6$^\circ$, a value of  $t_{b}$= 12 meV is then derived (with $v_{F}=1.8\times10^{5}$m/s). A much weaker peak is also visible around $\pm$ 2$^\circ$ under the highest field. This result is in agreement with the more quantitative treatment according to which anomalies are given by Bessel function zeros\cite{Danner94}. In addition the small peak observed at $\theta$ =0$^\circ$ has been taken as an  evidence for a coherent transport along the least conductive direction in line with similar findings in 2D organic compounds when a magnetic field is aligned parallel to the 2D layers\cite{Singleton02}. 

The DKC oscillations can therefore provide informations on the parameters of the electronic structure namely $t_{b}$= 12 meV and $t_{c}$= 0.8 meV  for \tmc\cite{Danner94} and $t_{b}$= 32.5 meV for \tmp6\cite{Danner95}. The difference by a factor of  about 2 between $b$ axis couplings of \tmc and \tmp6 has to be considered in the context of the \cl ordering reducing the coupling $\thicksim1/2$ from its room temperature value. Similar magnetoresistance studies along the $a-c^*$ plane on \tmp6 under pressure  adding a small field along the $b$ direction have suggested that  this compound, unlike (TMTSF)$_2$ClO$_4$,  could be a marginal Fermi liquid\cite{Danner95}.

\begin{figure}[h]
\includegraphics[width=0.9\hsize]{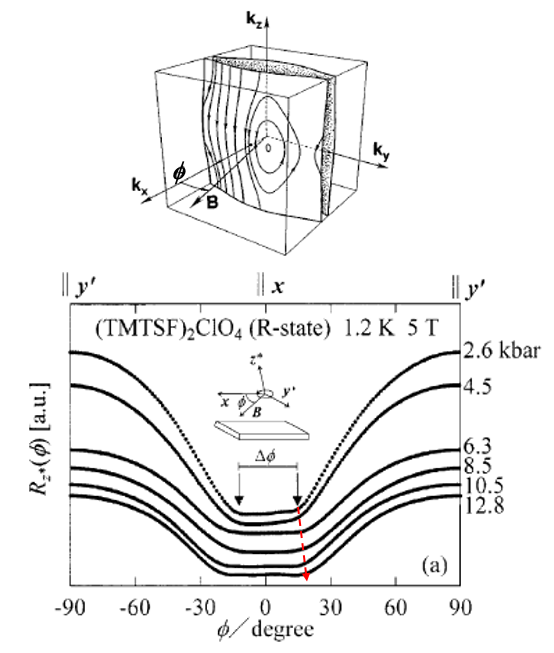}
\caption{\label{Thirdoscill} (Top) Semiclassical electron orbits on the warped Fermi surface under magnetic field tilted by an angle $\phi$ from the normal to the surface.
(Bottom) Angular dependence of $\rho_{zz}$ in slowly cooled \tmc at 1.2 K. The dashed red line shows how the ratio $t_{b}/t_{a}$ is increasing under pressure, Sources: (Top) [Fig.2 [Ref\cite{Osada96}, p. 5263], (Bottom) Fig. 1a [Ref\cite{Yoshino03}, p. 56].}  
\end{figure}

The section about angular magnetoresistance cannot be concluded without mentioning the anomaly of magnetoresistance when the magnetic field is rotated in the  most conducting $a-b$  plane  (the so-called third angular oscillation)\cite{Osada96}. Under such conditions, $\rho_{zz}$ reveals a single kink behaviour at an angle measured from the $a$ axis  in the vicinity of 15$^{\rm o}$. This phenomenon has been related to the vanishing above a critical tilt angle of the closed orbits    existing  on the sides of the Q1D surface with an incomplete cancellation of the $v_{z}$ carrier velocity when the field is weakly  tilted  from the normal axis, \emph{see} top of Fig.~\ref{Thirdoscill}. The value of the critical angle being a measure of the ratio $t_{b}/t_{a}$ is shown in Fig.~\ref{Thirdoscill} to increase under pressure as the conductor becomes more 2D coupled. A similar behaviour  has been reported in \tmp6\cite{Yoshino01} and in other Q1D conductors\cite{Yoshino95}.

These studies of transport as a function of a controlled orientation of the magnetic field have provided an overview of   the generality of the method and its usefulness in gaining knowledge of the band parameters.






\section{General Conclusion}
This short retrospective for the search  of superconductivity in organic matter has been based on an historical approach since we feel  
it is  instructive to see how the whole story has developed from novel, stimulating and sometime provocative theoretical statements leading finally to the
discovery of the first organic superconductor thanks to a very productive work between Chemistry and Physics communities  over the past 50 years. The results described in this review which have enabled a comparison with theoretical models  could not have been obtained without a sustained effort in the synthesis and growth of single crystals of very good quality.




Thanks to the determining role of high pressure governing their physical properties, organic conductors (and superconductors) have shown
their interest compared to high \tc cuprates in terms of purity and  variety of phenomena which can be studied in a single
system keeping both  structure and  chemical purity constant. 

 It has been claimed in 2000 that    the  physics of organics
including superconductivity could be reproduced without the burden of chemistry and  high pressure techniques by simply  tuning the carrier concentration using a
field effect technique. We all wished  that this could be true, but it turned out one year later that the claims were fraudulent. However, when the storm calmed down, more fruitful research began a few years later  trying to fabricate organic electronic devices. Some of them will be reported shortly in the next section of this overview.

As early as 1974 Aviram and Ratner had considered the use of molecules as components of electronic circuitry and suggested according to a calculation the feasibility of such molecular devices\cite{Aviram74}. It is not the goal of the present article to cover the vast domain of organics electronics which is exhaustively presented in the recent texbook by Alcacer\cite{Alcacer23}. However, we wish to emphasize  only in a few words in this conclusion about  the fabrication of  prototype organic field effect transistors because they are directly inspired by  the diversified physical properties of the 2D organics  presented in Sec.~\ref{two dimensional organic superconductors}. 

Taking advantage of the proximity between  insulating and superconducting phases   which is evidenced by the phase diagram of $\mathrm{\kappa-(ET)_{2}X}$ compounds\cite{Lefebvre00}, the control of the ground state  of organic materials via the application of  a gate voltage turned out to be  successful\cite{Suda14,Yamamoto13}.

The low temperature state of the device depends on both  chemical nature of the bulk and  substrate (in addition to  cooling rate) since what matters is the differential contraction between bulk and substrate\cite{Kawasugi11,Kawaguchi19}, \emph{see} Fig.~\ref{ET-Cldiagram}.  The use of the various physical properties of these organic conductors appears therefore very promising for the fabrication of future devices.

\begin{figure}[t]
\includegraphics[width=1.05\hsize]{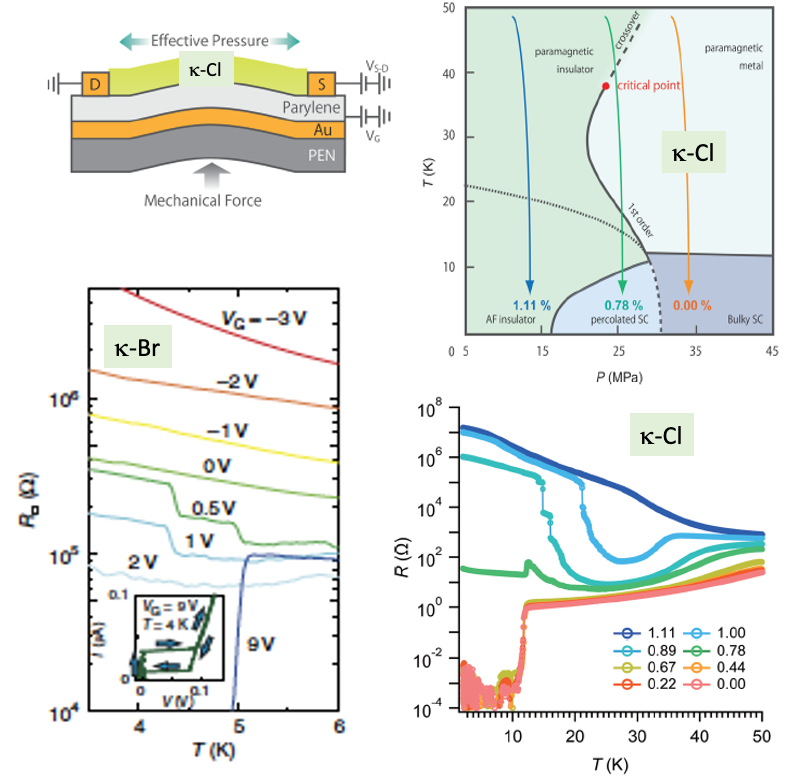}
\caption{\label{organicFET} Organic FET devices  based on $\kappa$-$\mathrm{(ET)_{2}X}$.(Top,left) A thin layer of  
$\mathrm{\kappa-(ET)_{2}Cl}$
 is deposited  on a substrate of polyethylene naphtalate sheet with thermally evaporated Au-gate electrodes and parylene gate-dielectric. Given the differential thermal contraction between the bulk sample  and the substrate, cooling the device  without any strain (0.00\%), drives it  into the SC state at low temperature (Bottom, right),  while the bulk sample SC phase is stable only above 30 Mpa (= 300 bar)\cite{Lefebvre00}. Increasing the strain  (up to 1.11\%) at high temperature and cooling down (Top, right), moves  the device towards the AF insulating state,  Sources: Fig. 1a, 3ac [Ref\cite{Suda14}]. 
 An other FET device using $\kappa$-$\mathrm{(ET)_{2}Br}$ on top of a metallic Nb-doped $\mathrm{SrTiO_{3}}$ substrate covered with a 28 nm of $\mathrm{Al_{2}O_{3}}$ dielectric layer grown by atomic  deposition showing the stabilization of the SC state under a gate voltage of 9V and switching  to an insulating behaviour at negative gate voltages (Bottom, left), Source: Fig. 2 [Ref\cite{Yamamoto13}, p. 4].}
\end{figure}


A perfect illustration for the scientific interest of organic conductors is provided by the amount of knowledge and new physical concepts which came out of the studies of organic superconductors. 

 To summmarize, both experimental and theoretical results point to the contribution of electron correlations to the SC pairing problem. The extensive experimental evidence in favour of the emergence of superconductivity in the \tm2x family next to the stability pressure threshold for the Slater-Overhauser antiferromagnetism has shown the need for a unified description of all electronic excitations that lies at the core of both density-wave and SC correlations. In this matter, the recent progresses of the renormalization group method for the 1D–2D electron gas model have resulted in predictions about the possible symmetries of the SC order parameter when a purely electronic mechanism is involved, predictions that often differ from phenomenologically based approaches to superconductivity but are in fair agreement with recent experimental findings.

Firstly, the SC order parameter is displaying lines of nodes that are governing the stability against impurity and the thermodynamics of the SC phase. Important constraints on the nodal position have been obtained by the field angular dependence of the specific heat. 

Secondly, electron scattering in the metallic phase above \tc suggests the existence of strong antiferromagnetic fluctuations leading to the possibility of a spin-mediated pairing in the SC phase. The pairing mechanism behind organic superconductivity is likely different from the proposal made by Little, but it is nevertheless a phonon-less mechanism, at least in \tm2xsuperconductors.

What is also emerging from the work on these prototype 1D organic superconductors is their very simple electronic structure with only a single band at the Fermi level, no prominent spin orbit coupling and an extremely high chemical purity and stability. 

They should be considered in several respects as very simple model systems to inspire the physics of the more complex high \tc superconductors, especially for pnictides and electron-doped cuprates. 

Most concepts discovered in these simple low-dimensional conductors may also become of interest for the study of other 1D or Q1D systems such as carbon nanotubes, artificial 1D structures, the black bronze superconductor $\mathrm{Li_{0.9}Mo_{6}O_{17}}$ with Mo-O chains\cite{Greenblatt84}, the newly-discovered telluride superconductor $\mathrm{Ta_{4}Pd_{3}Te_{16}}$ with Ta-Pd chains\cite{Jiao14}, and the recently discovered $\mathrm{A_{2}Cr_{3}As_{3}} $(A= K, Rb, Cs) materials comprising $\mathrm{[(Cr_{3}As_{3})^{-2}}]_{\infty}$ chains\cite{Tang15}. It should be noted that the electronic anisotropy of the latter two classes of compounds seems to be weaker than originally expected and much weaker than those of the Bechgaard superconductors. Nevertheless, unconventional behaviours, such as possible nodal superconductivity in $\mathrm{Ta_{4}Pd_{3}Te_{16}}$\cite{Pan15} and unusually large $\mathrm{H_{c2}}$ in $\mathrm{Li_{0.9}Mo_{6}O_{17}}$\cite{Mercure12} and $\mathrm{A_{2}Cr_{3}As_{3}}$\cite{Tang15,Bao15,Kong15}, resemble those observed in \tm2x and thus it is interesting to explore the common nature of Q1D superconductivity amongst a wide class of materials. 

In addition, it might be profitable to reconsider the properties of copper-free superconducting perovskites such as the $\mathrm{Sr_{2}RuO_{4}}$ type\cite{Maeno94}  in the light of the properties of quasi-1D organic superconductors presented in this overview.  Although the electronic structure of  these compounds are two dimensional with three bands  crossing the Fermi level instead of a single band in organics, many of their electronic properties\cite{Mackenzie03} are very similar to those observed in the far more simple Q1D conductors, e.g. temperature dependence of  transport along the stacking axis, specific heat in the superconducting state, the sensitivity of \tc to impurities\cite{Mackenzie19} or the behaviour of \tc  under pressure\cite{Shirakawa97}, the existence of a FFLO state\cite{Kinjo22}, not to mention recent NMR results which are favoring singlet over triplet superconducting pairing\cite{Pustogow19,Ishida20,Chronister20}. A new visit to $\mathrm{Sr_{2}RuO_{4}}$ taking into account experimental results on Q1D superconductors and the theories that explain them successfully, would now be very valuable.

Two dimensional organic superconductors although not extensively reviewed in the present article  have been very popular due to the possibility to stabilize  superconductivity   up to 10 K even under ambient pressure and also to the great diversity of stable phases which can be achieved  depending on the chemical components. It has been shown that the optimized location for  the superconducting  \tc is the with  a Mott localization resulting from a compromise between the interchain coupling and the strength of the correlations.  They also revealed  the likeliness  of an unconventional   pairing mechanism which makes them model materials for 2D superconductors.


And last but not least, the lesson to be learned  from these 1D and 2D organic superconductors is that the term unconventional superconductivity currently used  should be taken with a grain of salt and possibly no longer be  used. The proper denomination should be more in line with the results of research in 1D and 2D organics as well as in cuprates, etc... and qualify superconducting materials according to their Cooper pairs binding mechanism  carried out either via phonons or via an exchange of magnetic fluctuations.

Concluding this  retrospective overview, it seems appropriate to quote the words of V. L. Ginzburg in 1989\cite{Ginzburg89}. Talking about organic superconductors and intercalated layered superconductors, he wrote:
\emph{``I believe that it was undoubtedly the discussion of the possible exciton mechanism of superconductivity that stimulated the
search for  such superconductors and studies of them"}.
In addition,  the other  statement of Ginzburg: \emph{``the
organic superconductors (produced for the first time in 1980) are clearly interesting by themselves or, to be more  precise, irrespective of the high \tc problem"} cannot be more appropriate. 
 
 The reader may realize that a lot has been accomplished since 1991 when P. W. Anderson declared  in a conversation with  R. Schrieffer reported by Physics Today\cite{Anderson91} 
 \emph{``Organic superconductors are still almost a complete mystery}".

It is also a pleasure to  agree with  Paul Chaikin and cite  his  presentation celebrating the fiftieth  anniversary  of the BCS theory  in 2007 at University of Illinois  \emph{``Quasi 1D organic \tmp6 superconductor is  the most interesting material ever discovered''}.
\section{Bibliography (not appearing as articles in scientific journals)}
\label{Bibliography}
-Organic Semiconductors, F. Gutman and L. E. Lyons Wiley, New York, 1967\\
-Low Dimensional Cooperative Phenomena, edited by H. J. Keller, Plenum Press, New York, 1975\\
-Chemistry and Physics of One-Dimensional Metals, edited by H. J. Keller, Plenum, Press, New York, 1976\\
-Electron-Phonon Interactions and Phase Transitions, edited by T. Riste, Nato Advanced Study Institute, Plenum Press, 1977\\
-Organic Conductors and Semiconductors, edited by L. P\'al, G. Gr\"uner, A. Janossy and J. Solyom, Lecture Notes in Physics 65, Springer, Berlin, 1997\\
-Quasi One-Dimensional Conductors I and II, edited by S. Barisic, A. Bjelis, J. R. Cooper and B. Leontic, Lecture Notes in Physics, 95 and 96, Springer, Berlin, 1978\\
-Molecular Metals, edited by W. E. Hatfield, Plenum Press, New York, 1979\\
-Highly Conducting One-Dimensional Solids, edited by J. T. Devreese, R. P. Evrard and V. E. van Doren, Plenum Press, New York, 1979\\
-The Physics and Chemistry of Low Dimensional Solids, edited by L. Alcacer, Proceedings of the Nato Advanced Institute, D. Reidel Publishing Company, Dordrecht, 1980\\
-Physics in One Dimension, edited by J. Bernasconi and T. Schneider, Springer, 1981\\ 
-One Dimensional Conductors, S. Kagoshima, T. Sambongi and H. Nagasawa, Tokyo Syokabo in Japanese, 1982\\
-Extended Linear Chain Compounds, vol 1 to 4, edited by J. S. Miller, Plenum Press, New York, 1982\\
-Charge Density Waves in Solids, edited by Gy. Hutiray and J. Solyom, Lecture Notes in Physics, 217, Springer, 1985\\
-Introduction to Synthetic Electrical Conductors, J. R. Ferraro and J. W. Williams, Academic Press, London, 1987\\
-Novel Superconductivity, edited by S. E. Wolf and V. Z. Kresin, Plenum Press, New York, 1987\\
-Low Dimensional Conductors and Superconductors, edited by D. Jerome and L. G. Caron, NATO ASI Series, vol 155, Plenum Press, New York, 1987\\
-Introduction to Synthetic Electrical Conductors, J. R. Ferraro and J. W. Williams, Academic Press, London, 1987\\
-One-Dimensional Conductors, S. Kagoshima, H. Nagasawa and T. Sambongi, Springer, Berlun, 1988\\
-Lower-Dimensional Systems and Molecular Electronics, edited by R. M. Metzger, P. Day and G. C. Papavassiliou, Proceedings of Nato Advanced Study Institute, Plenum Press, New York, 1989\\
-The Physics and Chemistry of Organic Superconductors, edited by G. Saito and S. Kagoshima, Proceedings of the International ISSP Symposium, Springer, Berlin, 1989\\
-Organic Superconductivity, edited by V. Z. Kresin and W. A. Little, Plenum Press, New York, 1990\\
-Low Dimensional Organic Conductors, A. Graja, World Scientific, Singapore, 1992\\
-Organic Superconductors (including fullerenes), J. M. Williams \emph{et-al},Prentice Hall, Englewood Cliffs, New Jersey, 1992\\
-Organic Conductors (fundamental and applications), edited by J. P. Farges, M. Dekker, New York, 1994\\
-Organic Superconductors, T. Ishiguro, K. Yamaji and G. Saito, Springer Series in Solid-Sciences, n. 88, Springer, Berlin, 1998\\
-Advances in Synthetic Metals, Twenty Years of Progress in Science and Technology, edited by P. Bernier, S. Lefrant, and G. Bidan, Elsevier, Amsterdam, 1999\\
-Quantum Physics in One Dimension, T. Giamarchi, Oxford Science Publications, Clarendon Press, Oxford, 2004\\
-The Physics of Organic Superconductors and Conductors, edited by A. Lebed, Springer Series in Materials Science, n.110, Springer, Berlin, 2008\\
-Advances in Organic Conductors and Superconductors, edited by M. Dressel, Special Issue in Crystals, MDPI, 2018\\
-The Physics of Organic Electronics, L. Alcacer, IOP Publishing, Bristol, 2022\\
\begin{acknowledgements}


The research activity on low dimensional conductors has involved researchers from all continents over a period of more than 50 years. Within the framework of this worldwide cooperation, the Laboratoire de Physique des Solides de  l'Université de  Paris-Sud à Orsay and the HC \O ersted Institute at Copenhagen together with the Département de  Physique de l'Université de Sherbrooke   have played a major role that we wish to underline and acknowledge.

CNRS, French Ministry of Research, Université Paris-Sud and the European Commission are acknowledged for the support of this research.

The particular characteristic of the Orsay laboratory, of which Jacques Friedel was one of the co-founders in 1959, was its multidisciplinary character, bringing together physical measurements (electronic and structural properties of solids, condensed matter theory and chemistry of materials with an efficient internal cooperation. As the partnerships of the  Orsay  laboratory have been so numerous, we will rather mention  countries and institutions involved.

To be noticed is the openness of the Orsay  laboratory to numerous  national cooperations (Montpellier, Rennes, Nantes and Angers for chemistry, Marseilles for tunnel junctions measurements,  Grenoble and Toulouse for high magnetic fields physics),   as well as international cooperations (Great Britain with the Cavendish laboratory at Cambridge, Switzerland with the Institut de Physique at Geneva and Ecole Polytechnique Fédérale at Lausanne, Germany with the University at Stuttgart and the Max Planck Institut at Heidelberg,  Croatia with the Institute of Physics at Zagreb, Tunisia with the University at Tunis, Japan with the Universities of Kyoto, Tsukuba Research Centers  and Osaka and Gakushuin University at Tokyo, Russia with the Landau Institute at Moscow, Korea with the Ewha Women University at Seoul, Israel with the Hebrew University,  the United States with the University of California at Los Angeles and the National High Magnetic Fields Laboratory at Tallahassee, Portugal with the Department of Chemistry at Sacavem and Instituto Superior Tecnico at Lisbon, Spain with  the University  at Barcelona).  These cooperations have enabled  to welcome a large number of brilliant thesis students and international post-doctoral visitors who, once back in their countries, have successfully pursued their activities in the field.

We wish to make a special mention to our colleagues who have contributed in a decisive way to the progress of the field and who sadly passed away much  too soon; Gen Soda who died accidentally  at the Institute for Molecular Science, Okazaki in 1987, Klaus Bechgaard (1945-2017) without whom this field would not have been able to develop and Heinz Schulz (1954-1998) one of the  major  contributors to theoretical low dimensional physics and François Creuzet (1957-2020) whose NMR contribution is fundamental to the understanding of these materials and recently Robert Comés (1937-2022) who has been the first with his collaborators  to use diffuse X-ray scattering techniques to reveal charge density wave fluctuations in the one-dimensional compound, platinum chains, KCP. It is this result obtained in 1973 at Orsay which triggered the beginning of the study of one-dimensional conductors, rapidly extended to organic conductors.


\end{acknowledgements}
\bibliographystyle{unsrt}
\bibliography{biblioClO4v3.bib,AjoutsRefsCB.bib}

\providecommand{\noopsort}[1]{}\providecommand{\singleletter}[1]{#1}%
\begin{thebibliography}{100}

\bibitem{Onnes11a}
H.~K. Onnes.
\newblock Further experiments with liquid helium. on the change of electrical
  resistance of pure metals at very low temperatures, etc. the disappearance of
  the resistance of mercury.
\newblock In KNAW, editor, {\em KNAW, Proceedings}, volume 14 I, pages
  113--115, 1911.

\bibitem{Onnes11}
H.~K. Onnes.
\newblock Further experiments with liquid helium. on the electrical resistance
  of pure metals etc. on the sudden change in the rate at which the resistance
  of mercury disappears.
\newblock In KNAW, editor, {\em KNAW Proceedings}, volume 14 II, pages
  818--821, 1911-1912.

\bibitem{Meijer94}
P.~H.~E. Meijer.
\newblock Kamerlingh {O}nnes and the discovery of superconductivity.
\newblock {\em American Journal of Physics}, 62:1105, 1994.

\bibitem{Bardeen57}
L.~N.~Cooper J.~Bardeen and J.~R. Schrieffer.
\newblock Microscopic theory of superconductivity.
\newblock {\em Physical Review}, 106:162, 1957.

\bibitem{Bardeen57a}
J.~Bardeen, L.N. Cooper, and J.R. Schrieffer.
\newblock Theory of superconductivity.
\newblock {\em Phys. Rev.}, 108:1175, 1957.

\bibitem{McCoy11}
H.N. McCoy and W.C. Moore.
\newblock Organic amalgams: substances with metallic properties composed in
  part of non-metallic elements.
\newblock {\em J. Am. Chem. Soc.}, 33:273, 1911.

\bibitem{London35}
F.~London and H.~London.
\newblock Supraleitung und diamagnetismus.
\newblock {\em Physica}, 2:341, 1935.

\bibitem{Pauling36}
L.~Pauling.
\newblock The diamagnetic anisotropy of aromatic molecules.
\newblock {\em Jour of Chemical Physics}, 4:673, 1936.

\bibitem{London37}
F.~London.
\newblock Theorie quantique des courants interatomiques dans les combinaisons
  aromatiques.
\newblock {\em Journal de Physique}, 8(10):397, 1937.

\bibitem{London37a}
F.~London.
\newblock Supraconductivity in {A}romatic {C}ompounds.
\newblock {\em Jour of Chemical Physics. Chem. Physics}, 5:837, 1937.

\bibitem{Bouchiat89}
H.~Bouchiat and G.~Montambaux.
\newblock Persistent currents in mesoscopic rings : ensemble averages and
  half-flux-quantum periodicity.
\newblock {\em J. Phys. (France)}, 50:2695, 1989.

\bibitem{Keller74}
H.~J. Keller, editor.
\newblock {\em Low-Dimensional Cooperative Phenomena}.
\newblock Plenum Press, 1975.
\newblock Lectures presented at the 1974 Nato Advanced Study Institute,
  Low-Dimensional Cooperative Phenomena and the Possibility of High Temperature
  Superconductivity, held in Starnberg, Germany, September 3-15, 1974.

\bibitem{Eley51}
M.~J.~Perry D.~D.~Eley, G . D.~Parfitt and D.~H. Taysum.
\newblock The semiconductivity of organic substances.
\newblock {\em Transactions of the Faraday Society}, page~79, 1951.

\bibitem{Akamatu50}
H.~Akamatu and H.~Inokuchi.
\newblock On the electrical conductivity of violanthrene, iso-violanthrene, and
  pyranthrene.
\newblock {\em The Journal of Chemical Physics}, 18(6):810, 1950.

\bibitem{Inokuchi55}
H.~Inokuchi.
\newblock The effect of pressure on the semi-conductivity of isoviolanthrene.
\newblock {\em The Bulletin of the Chemical Society of Japan}, 28(8):570, 1955.

\bibitem{Akamatu54}
H.~Akamatu, H.~Inokuchi, and Y.~Matsunaga.
\newblock Electrical conductivity of the perylene-bromine complex.
\newblock {\em Nature}, 173:168, 1954.

\bibitem{Acker60}
D.~S. Acker, R.~J. Harder, W.~R. Hertler, W.~Mahler, L.~R. Melby, R.~E. Benson,
  and W.~E. Mochel.
\newblock 7,7,8,8-tetracyanoquinodimethane and its electrically conducting
  anion-radical derivatives.
\newblock {\em Jour. Am. Chem. Soc}, 82:6408, 1960.

\bibitem{Andre76}
J.J. Andr\'e, A.~Bieber, and F.~Gautier.
\newblock Physical properties of highly anisotropic systems: Radical-ion salts
  and charge transfer complexes radical-ion salts and charge transfer
  complexes.
\newblock {\em Annales de Physique}, I:145, 1976.

\bibitem{Melby62a}
L.~R. Melby, R.~J. Harder, W.R. Hertler, W.~Mahler, R.~E. Benson, and W.~E.
  Mochel.
\newblock Substituted quinodimethans. anion-radical derivatives and complexes
  of 7,7,8,8-tetracyanoquinodimethan.
\newblock {\em Jour. Am. Chem. Soc}, 84:3374, 1962.

\bibitem{Shchegolev68}
I.~F. Shchegolev, L.~I Burarov, A.~V. Zvarykina, and R.~B. Lyubovskii.
\newblock Conduction mechanism of highly-conducting organic complexes based on
  \tq.
\newblock {\em JETP Letters}, 8:218, 1968.

\bibitem{Shchegolev72}
I.F.Shchegolev.
\newblock Electric and magnetic properties of linear conducting chains.
\newblock {\em Phys. stat. sol. (a)}, 12:9, 1972.

\bibitem{Ginzburg68}
V.~L. Ginzburg.
\newblock The problem of high temperature superconductivity.
\newblock {\em Contemp. Phys.}, 9(4):355--374, 1968.

\bibitem{Melby65}
L.~R. Melby.
\newblock Salts derived from the 7,7,8,8-tetracyanoquinodimethans.
  anion-radical and benzologues of quaternary pyrazinium cations.
\newblock {\em Canadian Journal of Chemistry}, 43:1448, 1965.

\bibitem{Epstein71}
A.J. Epstein, A.F. Garito, S.~Etemad, and A.J. Heeger.
\newblock Metal---insulator transition in an organic solid: Experimental
  realization of the one-dimensional hubbard model.
\newblock {\em Solid State Comm}, 9:1803, 1971.

\bibitem{Devreux78}
F.~Devreux and M.~Nechtchein.
\newblock Nuclear relaxation in 1d conductors.
\newblock In {\em Proc. of the International Conference of
  Quasi-One-Dimensional Conductors I}, page 145, New York, 1978. Edited by S.
  Barisic, A. Bjeli\'s, J. R. Cooper and B. Leonti\'c, Springer, Lectures Notes
  in Physics, Vol. 95.

\bibitem{Flandrois77}
S.~Flandrois and D.~Chasseau.
\newblock Longueurs de liaison et transfert de charge dans les sels du
  tetracyanoquinodim{\'e}thane.
\newblock {\em Acta Cryst}, B 33:2744, 1977.

\bibitem{Comes73}
R.~Com\`es, M.~Lambert, H.~Launois, and H.~R. Zeller.
\newblock Evidence for a {P}eierls distortion or a {K}ohn anomaly in
  one-dimensional conductors of the type $\mathrm{K_{2}Pt(CN)_{4}Br_{0.30}
  xH_{2}O}$.
\newblock {\em Phys. Rev. B}, 8:571, 1973.

\bibitem{Denoyer75}
F.~Denoyer, R.~Com\`es, A.F. Garito, and A.H. Heeger.
\newblock X-ray diffuse sattering evidence for a phase transition in
  {TTF-TCNQ}.
\newblock {\em Phys. Rev. Lett.}, 35:445, 1975.

\bibitem{Pouget82}
J.P. Pouget, R.~Com\`es, A.J. Epstein, and J.S. Miller.
\newblock X-ray observation of cross over of $\mathrm{2k_{F}}$ to
  $\mathrm{4k_{F}}$ scattering in $\mathrm{NMP)_{x}Phen_{1-x}(TCNQ)}$
  $\mathrm{0.5\leq 1\leq 1}$.
\newblock {\em Mol. Cryst. Liq. Cryst.}, 85:203, 1982.

\bibitem{Hardy53}
G.~Hardy and J.K. Hulm.
\newblock Superconducting silicides and germanides.
\newblock {\em Phys. Rev.}, 87:884, 1953.

\bibitem{Matthias54}
B.T. Matthias, T.H. Geballe, S.~Geller, and E.~Corenwitz.
\newblock Superconductivity of $\mathrm{Nb_{3}Sn}$.
\newblock {\em Phys. Rev}, 95:1435, 1954.

\bibitem{Gavaler73}
J.~R. Gavaler.
\newblock Superconductivity in {Nb-Ge} films above 22 {K}.
\newblock {\em Appl. Phys. Lett}, 23:480, 1973.

\bibitem{Matthias71}
B.T. Matthias.
\newblock The search for high-temperature superconductors.
\newblock {\em Physics Today}, August:23, 1971.
\newblock (page 27).

\bibitem{Weger64}
M.~Weger.
\newblock The electronic band structure of $\mathrm{V_{3}Si}$ and
  $\mathrm{V_{3}Ga}$.
\newblock {\em Reviews of Modern Physics}, 36:175, 1964.

\bibitem{Labbe66}
J.Labb\'e and J.Friedel.
\newblock Instabilit{\'e} {\'e}lectronique et changement de phase cristalline
  des compos{\'e}s du type $\mathrm{V_{3}Si}$ {\`a} basse temp{\'e}rature.
\newblock {\em J.Physique}, 27:153, 1966.

\bibitem{Labbe67}
J.~Labb\'e, S.~Barisic, and J.~Friedel.
\newblock Strong-coupling superconductivity in $\mathrm{V_{3}X}$ type of
  compounds.
\newblock {\em Physical Review Letters}, 19(18):1039, 1967.

\bibitem{Eliashberg60}
G.~M. Eliashberg.
\newblock Interactions between electrons and lattice vibrations in a
  superconductor.
\newblock {\em Sov.Phys.JETP.}, 11(3):696--702, 1960.

\bibitem{Kohn65}
W.~Kohn and J.M. Luttinger.
\newblock New mechanism for superconductivity.
\newblock {\em Phys. Rev. Lett}, 15:524, 1965.

\bibitem{Friedel58}
J.~Friedel.
\newblock Metallic alloys.
\newblock {\em Nuovo Cimento Suppl}, 7(2):287--311, 1958.

\bibitem{Little64}
W.A. Little.
\newblock Possibility of synthesizing an organic superconductor.
\newblock {\em Phys. Rev. A}, 134:1416, 1964.

\bibitem{Jerome08}
D.~Jerome.
\newblock Historical approach to organic superconductivity.
\newblock In A.~Lebed, editor, {\em The Physics of Organic Superconductors and
  Conductors}, volume Vol 110 of {\em Springer Series in Materials Science}.
  Springer, Berlin, Heidelberg, 2008.

\bibitem{Little65}
W.~A. Little.
\newblock Superconductivity at room temperature.
\newblock {\em Scientific American}, 212(2):21--27, 1965.

\bibitem{Ginzburg70a}
V.L. Ginzburg.
\newblock High temperature superconductivity.
\newblock {\em J. Polymer. Sci.C.}, 29:3, 1970.

\bibitem{Ladik69}
J.~Ladik, G.~Bicz\'o, and J.~Redly.
\newblock Possibility of superconductive-type enhanced conductivity in {DNA} at
  room temperature.
\newblock {\em Physical Review}, 188:710, 1969.

\bibitem{1969LIttleconf}
W.~A. Little.
\newblock In W.~A. Little, editor, {\em International symposium on the physical
  and chemical problems of possible organic superconductors}, University of
  Hawai, Honolulu, 1969. Physics Department, Stanford.

\bibitem{Little70}
W.~A. Little.
\newblock The exciton mechanism in superconductivity.
\newblock In Wiley~Online Library, editor, {\em Journal of Polymer Science Part
  C: Polymer Symposia}, volume~9, page~17, 1970.

\bibitem{Byschkov66}
L.~P.~Gorkov Yu. A.~Byschkov and I.~E. Dzyaloshinskii.
\newblock Possibility of superconductivity type phenomena in a one-dimensional
  system.
\newblock {\em Sov Phys JETP}, 23:489, 1966.

\bibitem{Landau59}
L.~D. Landau and E.~M. Lifshitz.
\newblock {\em Statistical Physics}.
\newblock Pergamon, London, 1959.

\bibitem{Mermin66}
N.~D. Mermin and H.~Wagner.
\newblock Absence of ferromagnetism or antiferromagnetism in one or
  two-dimensional isotropic heisenberg models.
\newblock {\em Phys. Rev. Lett}, 17:1133, 1966.

\bibitem{Wudl70}
F.~Wudl, G.~M. Smith, and E.~J. Hufnagel.
\newblock Bis- 1,3=dithiolium chloride: an unusually stable organic radical
  cation.
\newblock {\em Chemical Communications}, page 1453, 1970.

\bibitem{Wudl72}
F.~Wudl, D.~Wobschall, and E.J. Hufnagel.
\newblock Electrical conductivity by the bis-l,3-dithiole-bis-l,3-dithiolium
  system.
\newblock {\em Journal of the American Chemical Society}, page 670, 1972.

\bibitem{Bryce99}
M.~R. Bryce.
\newblock Tetrathiafulvalenes as pi-electron donors for intramolecular
  charge-transfer materials.
\newblock {\em Advanced Materials}, 1:11, 199.

\bibitem{Aviram74}
A.~Aviram and M.~A. Ratner.
\newblock Molecular rectifiers.
\newblock {\em Chemical Physics Letters}, 29:277, 1974.

\bibitem{Wudl04}
F.~Wudl M.~Bendikov and D.~M. Perepichka.
\newblock Tetrathiafulvalenes, oligoacenenes, and their buckminsterfullerene
  derivatives: The brick and mortar of organic electronics.
\newblock {\em Chemical Reviews}, 104:4891--4945, 2004.

\bibitem{Laplaca75}
S.~J.~La Placa, P.~W.~R. Corfield, R.~Thomas, and B.~A. Scott.
\newblock Non-integral charge transfer in an organic metal: The structure and
  stability range of $\mathrm{(TTF)Br_{x}}$.
\newblock {\em Solid State Comm}, 17:635--638, 1975.

\bibitem{Chaikin80}
P.~M. Chaikin, R.~A Craven, and S.~Etemad et~al.
\newblock Commensurate {P}eierls transition in a quasi-one-dimensional
  compound: The bromide salt of tetrathiafulvalene {(TTF)}.
\newblock {\em Phys Rev B}, 22:5599, 1980.

\bibitem{Coleman73}
L.B. Coleman et~al.
\newblock Superconducting fluctuations and the peierls instability in an
  organic solid.
\newblock {\em Solid State Comm}, 12:1125, 1973.

\bibitem{Ferraris73}
J.~Ferraris, D.O. Cowan, W.~Walatka, and J.H. Perlstein.
\newblock Electron transfer in a new highly conducting donor-acceptor complex.
\newblock {\em J. Am. Chem. Soc}, 95:948, 1973.

\bibitem{Bright73}
A.~A. Bright, A.~F. Garito, and A.~J. Heeger.
\newblock Optical properties of \tq in the visible and infrared.
\newblock {\em Solid Sate Comm}, 13:943--948, 1973.

\bibitem{Kistenmacher74}
T.~J. Kistenmacher, T~E. Phillips, and D.~O. Cowan.
\newblock The crystal structure of \tq.
\newblock {\em Acta Cryst B}, 30:763, 1974.

\bibitem{Bechgaard80}
K.~Bechgaard, C.~S. Jacobsen, K.~Mortensen, H.~J. Pedersen, and N.~Thorup.
\newblock The properties of five highly conducting salts derived from {TMTSF}.
\newblock {\em Solid State Comm.}, 33:1119--1125, 1980.

\bibitem{Vanduyne86}
R.~P.~Van Duyne, T.~W. Cape, M.~R. Suchanski, and A.~R. Seidle.
\newblock Determination of the extent of charge transfer in partially oxidized
  derivatives of tetrathiafulvalene and tetracyanoquinodimethan by resonance
  raman spectroscopy.
\newblock {\em J. Phys. Chem}, 90:739--743, 1986.

\bibitem{Anzai76}
H.~Anzai.
\newblock Growth of large-crystals of charge-transfer complex,
  tetrathiofulvalene-tetracyanoquinodimethane {(TTF-TCNQ)}.
\newblock {\em Journal of Crystal Growth}, 33:185--187, 1976.

\bibitem{Chemrev04}
P.~Batail, editor.
\newblock {\em Molecular Conductors}, volume 104.
\newblock American Chemical Society, 2004.

\bibitem{Greene76}
R.L. Greene, J.J. Mayerle, R.~Schumaker, G.~Castro, P.~Chaikin, S.~Etemad, and
  S.J.~La Placa.
\newblock The structure, conductivity, and thermopower of {HMTTF-TCNQ}.
\newblock {\em Solid State Comm.}, 20:943, 1976.

\bibitem{Bright74}
A.~A. Bright, A.~F. Garito, and A.~J. Heeger.
\newblock Opticai conductivity studies in a one-dimensional organic metal:
  Tetrathiofulvalene tetracyanoquinodimethan (\tq).
\newblock {\em Phys Rev B}, 10:1328, 1974.

\bibitem{Calas75}
P.~Calas, J.~M. Fabre, E.~Toreilles, and L.~Giral.
\newblock Synth\'ese de d\'eriv\'es du t{\'e}trathiofulval{\`e}ne.
\newblock {\em C.R. Acad. Sc (Paris) s\'erie C}, 280:901--903, 1975.

\bibitem{Brun77}
G.~Brun, S.~Peytavin, B.~Liautard, E.~Toreilles M.~Maurin, J.M. Fabre, and
  L.~Giral.
\newblock Sur quelques nouveaux complexes organiques {\`a} propri{\'e}t{\'e}s
  anisotropes.
\newblock {\em C.R. Acad. Sc. (Paris)}, 284 C:211, 1977.

\bibitem{Scott74}
J.~C. Scott, A.~F Garito, and A.~J. Heeger.
\newblock Magnetic susceptibility studies of
  tetrathiofulvalene-tetracyanoquinodimethan \tq and related organic metals.
\newblock {\em Phys Rev B}, 10:3131, 1974.

\bibitem{Bechgaard74}
K~.Bechgaard, D.~O. Cowan, and A.~N. Bloch.
\newblock Synthesis of the organic conductor tetramethyltetraselenofulvalenium.
\newblock {\em J. C. S. Chem. Comm}, page 937, 1974.

\bibitem{Engler74}
E.~M. Engler.
\newblock Structure control in organic metals. synthesis of
  tetraselenofulvalene and its charge transfer salt with tetracyano-p
  -quinodimethane.
\newblock {\em J. Am. Chem. Soc}, 96:7376, 1974.

\bibitem{Bechgaard76}
K.~Bechgaard, D.~O. Cowan, and A.~N. Bloch.
\newblock Stabilization of the organic metallic state.
\newblock {\em Mol. Crystals. Liq. Crystals Crystals}, 49(32):227, 1976.

\bibitem{Wudl76}
F.~Wudl D.~E. Schafer and W.~M.~Walsh et~al.
\newblock A systematic study of an isomorphous series of organic solid state
  conductors based on tetrathiafulvalene.
\newblock {\em J. Chem. Phys}, 66:377, 1977.

\bibitem{Strzelecka77}
H.~Strzelecka, L.~Giral, J-M. Fabre, E.~Torreilles, and G.~Brun.
\newblock Nouvelle voie d'acc{\`e}s {\`a} des conducteurs organiques.
\newblock {\em C. R. Acad. Sc. Paris. S{\'e}rie C}, 284:463--465, 1977.

\bibitem{Delhaes79}
P.~Delhaes, J.~Amiell C.~Coulon, E.~Toreilles S.~Flandrois, J.M. Fabre, and
  L.~Giral.
\newblock Physical properties of one dimensional conductors.
\newblock {\em Mol. Cryst. Liq. Cryst.}, 50:43, 1979.

\bibitem{Edelsack87}
A.~A. Edelsack, D.~U. Gubser, and S.~A. Wolf.
\newblock The rocky road to high temperature superconductivity.
\newblock In S.~A. Wolf and V.~Z. Kresin, editors, {\em Novel
  Supercondcutivity}, volume~12, page~1. Plenum Press, New Ork, 1987.

\bibitem{Bednorz86}
J.~Bednorz and K.A. M\"uller.
\newblock Possible high {T}c superconductivity in the {BaLaCu0} system.
\newblock {\em Z. Phys. B - Condensed Matter}, 64:189, 1986.

\bibitem{KB}
Klaus Bechgaard,~{\it Wikipedia},\\ https://en.wikipedia.org/wiki/Bechgaard.

\bibitem{Jacobsen78}
C.S. Jacobsen, K.~Mortensen, J.R. Andersen, and K.~Bechgaard.
\newblock Transport properties of some derivatives of
  tetrathiafulvalene-tetracyano-p-quinodimethane (\tq).
\newblock {\em Phys. Rev. B}, 18:905, 1978.

\bibitem{Laukhin78}
V.~N.~Laukhin et~al.
\newblock Metallic high pressure phase in tetraselenotetracene chloride
  $\mathrm{(TSeT)_{2}Cl}$.
\newblock {\em JETP Letters}, 28(5):284--287, 1978.

\bibitem{Bechgaard82}
K.~Bechgaard.
\newblock $\mathrm{(TMTSF)_{2}X}$ salts: {P}reparation, {S}tructure and effect
  of the anions.
\newblock In {\em Proc International Conference on Low-Dimensional Conductors},
  volume~79, pages 357--369. Gordon and Breach, Science Publisher, 1982.

\bibitem{Batail98}
P.~Batail, K.~Boubekeur, M.~Fourmigu\'e, and J.~C.~P. Gabriel.
\newblock Electrocrystallization, an invaluable tool for the construction of
  ordered, electroactive molecular solids.
\newblock {\em Chem. Mater}, 10:3005--3015, 1998.

\bibitem{videocrystals}
K.~Bechgaard, D.~Jerome, and A.~Moradpour.
\newblock Electrocrystallisation of organic superconductors. {V}ideo, {Z}enodo,
  2023, on line at https://doi.org/10.5281/zenodo.8064104.

\bibitem{Garito74}
A.~F. Garito and A.~J. Heeger.
\newblock The design and synthesis of organic metals.
\newblock {\em Accounts of Chemical Research}, 7:232, 1974.

\bibitem{Berlinsky74}
A.J. Berlinsky, J.F. Carolan, and L.~Weiler.
\newblock Band structure parameters for solid \tq.
\newblock {\em Solid State Comm}, 15:795, 1974.

\bibitem{Jerome82}
D.~Jerome and H.J. Schulz.
\newblock Organic conductors and superconductors.
\newblock {\em Adv in Physics}, 31:299--479, 1982.
\newblock DOI: 10.1080/00018738200101398.

\bibitem{Shitzkovsky78}
S.~Shitzkovsky, M.~Weger, and H.~Gutfreund.
\newblock Band structure of {TTF-TCNQ}.
\newblock {\em Journal de Physique}, 39:711, 1978.

\bibitem{Jerome77b}
D.~Jerome and M.~Weger.
\newblock Electronic properties of organic conductors: pressure effects.
\newblock In H.~J. Keller, editor, {\em Chemistry and Physics of
  One-Dimensional Metals}, chapter Electronic properties of organic conductors
  : pressure effects, page 341. Plenum Press (New York), 1977.

\bibitem{Soda76b}
G.~Soda, D.~Jerome, M.~Weger, K.~Bechgaard, and E.~Pedersen.
\newblock Spin relaxation and magnetic susceptibility studies of {HMTSF-TCNQ}.
\newblock {\em Solid Sate Comm}, 20:107--113, 1976.

\bibitem{Peierls55}
R.E. Peierls.
\newblock {\em in Quantum Theory of Solids}.
\newblock Oxford University Press, London, 1955.

\bibitem{Heeger}
Alan Heeger is a co-laureate of the 2000 Nobel Prize in Chemistry together with
  Alan Mac Diarmid and Hideki Shirakawa for the discovery organic conductive
  polymers.

\bibitem{Lubkin73}
G.~Lubkin.
\newblock Superconducting fluctuations at $\mathrm{60K}$?
\newblock {\em Physics Today}, May:17, 1973.

\bibitem{Patton71}
B.~R. Patton.
\newblock Fluctuation theory of the superconducting transition in restricted
  dimensionality.
\newblock {\em Phys. Rev. Lett}, 27(19):1273, 1971.

\bibitem{Etemad75}
S.~Etemad, T.~Penney, E.M. Engler, B.A. Scott, and P.E. Seiden.
\newblock $\mathrm{DC}$ conductivity in an isostructural family of organic
  metals.
\newblock {\em Phys. Rev. Lett.}, 34:741, 1975.

\bibitem{Barisic72a}
S.~Barisic.
\newblock Rigid-atom electron-phonon coupling in the tight-binding
  approximation.
\newblock {\em Physical Review B}, 5(3):932, 1972.

\bibitem{Schafer74}
D.~E. Schafer, F.~Wudl, G.~A. Thomas, J.~P. Ferraris, and D.~O. Cowan.
\newblock Apparent giant conductivity peaks in an anisotropic medium: \tq.
\newblock {\em Solid. State. Comm}, 14:347--351, 1974.

\bibitem{Thomas76}
G.A.Thomas {\textit{with 29 coauthors}}.
\newblock Electrical conductivity of
  tetrathiofulvalenium-tetracyanoquinodimethanide {(TTF-TCNQ)}.
\newblock {\em Phys.Rev.B}, 13(11):5105, 1976.

\bibitem{Cooper78}
J.~R.~Cooper et~al.
\newblock Isotope effect on the {P}eierls transition temperature of {TTF-TCNQ}.
\newblock {\em Solid State Comm}, 25:699, 1978.

\bibitem{Friend78b}
R.~H. Friend, M.~Miljak, and D.~Jerome.
\newblock Pressure dependence of the phase transitions in
  tetrathiafulvalene-tetracyanoquinodimethane: Evidence for a longitudinal
  lockin at 20 kbar.
\newblock {\em Phys. Rev. Lett.}, 40:1048, 1978.

\bibitem{Soda77}
G.~Soda, D.~Jerome, M.~Weger, J.~Alizon, J.~Gallice, H.~Robert, J.~M. Fabre,
  and L.~Giral.
\newblock Electronic properties of $\mathrm{TTF-TCNQ}$ : An {NMR} approach.
\newblock {\em J. Phys. (Paris)}, 38:931, 1977.
\newblock Open archive on HAL http://hal.archives-ouvertes.fr/.

\bibitem{Cooper77a}
J.~R. Cooper, D.~Jerome, S.~D. Etemad, and E.~M. Engler.
\newblock On the behaviour of {TSF-TCNQ} under pressure.
\newblock {\em Solid Sate Comm}, 22:257--263, 1977.

\bibitem{Grant73}
P.M. Grant, R.L. Greene, G.C. Wrighton, and G.~Castro.
\newblock Temperature dependence of the near-infrared optical properties of
  tetrathiofulvalinium tetracyanoquinodimethane \tq.
\newblock {\em Phys. Rev. Lett.}, 31:1311, 1973.

\bibitem{Ong77}
N.~P. Ong and A.~M. Portis.
\newblock Microwave {H}all effect in a quasi-one-dimensional system:
  Tetrathiafulvalenium-tetracyanoquinodimethanide \tq.
\newblock {\em Phys Rev B}, 15:1782, 1977.

\bibitem{Cohen74}
M.~J. Cohen, L.~B. Coleman, A.~F. Garito, and A.~J. Heeger.
\newblock Electrical conductivity of tetrathiofulvalinium
  tetracyanoquinodimethan \tq.
\newblock {\em Phys. Rev. B.}, 10(4):1298, 1974.

\bibitem{Jerome79}
D.~Jerome.
\newblock Fluctuating collective conductivity and single particle conductivity
  in 1-{D} organic conductors.
\newblock In L.~Alcacer, editor, {\em The Physics and Chemistry of Low
  Dimensional Solids}, pages 123--142. D. Reidel Publishing Company, 1980.

\bibitem{Bardeen73}
J.~Bardeen.
\newblock Superconducting fluctuations in one-dimensional organic solids.
\newblock {\em Solid State Comm}, 13:357, 1973.

\bibitem{Allender74}
D.~Allender, J.~W. Bray, and J.~Bardeen.
\newblock Theory of fluctuation superconductivity from electron-phonon
  interactions in pseudo-one-dimensional systems.
\newblock {\em Physical Review B}, 9(1):119, 1974.

\bibitem{Froehlich54}
H.~Fr\"ohlich.
\newblock On the theory of superconductivity: the one-dimensional case.
\newblock {\em Proc. R. Soc. London}, A 223:296, 1954.

\bibitem{Patton74}
B.~R. Patton and L.~J. Sham.
\newblock Fluctuation conductivity in the incommensurate {P}eierls system.
\newblock {\em Phys Rev Letters}, 35(11):638, 1974.

\bibitem{Kagoshima75}
S.~Kagoshima, H.~Anzai, K.~Kajimura, and T.~Ishiguro.
\newblock Observation of the {K}ohn anomaly and the {P}eierls transition in \tq
  by {X}-ray scattering.
\newblock {\em J. Phys. Soc. Jap.}, 39(4):1143, 1975.

\bibitem{Comes75}
R.~Com\'es, S.M. Shapiro, G.~Shirane, A.F. Garito, and A.J. Heeger.
\newblock Neutron-scattering study of the 38 and 54{K} phase transitions in
  deuterated tetrathiafulvalene-tetracyanoquinodimethane \tq.
\newblock {\em Phys. Rev. Lett.}, 35(22):1518--1521, 1975.

\bibitem{Takahashi84a}
T.~Takahashi, D.~Jerome, F~Masin, J.~M. Fabre, and L~Giral.
\newblock $\mathrm{^{13}C}$ {NMR} studies of $\mathrm{TTF(^{13}C)-TCNQ}$.
\newblock {\em J. Phys.C. Solid State Phys}, 17:3777--3792, 1984.

\bibitem{Rybaczewski76}
E.~F. Rybaczewski, S.~Smith, A.~F. Garito, A.~J. Heeger, and B.~G. Silbernagel.
\newblock $\mathrm{^{13} C}$ {K}night shift in \tq$\mathrm{^{13} C}$:
  Determination of the local susceptibility.
\newblock {\em Physical Review B}, 14(7):2746, 1976.

\bibitem{Tomkiewicz77}
Y.~Tomkiewicz, A.~R. Taranko, and J.~B. Torrance.
\newblock Spin susceptibility of tetrathiafulvalene tetracyanofluinodimethane,
  \tq, in the semiconducting regime: Comparison with conductivity.
\newblock {\em Physical Review B}, 15(2):1017, 1977.

\bibitem{Bak76}
P.~Bak and V.~J.Emery.
\newblock Theory of the structural phase transformations in
  tetrathiafulvalene-tetracyanoquinodimethane \tq.
\newblock {\em Phys. Rev. Letters}, 36:978, 1976.

\bibitem{Nishiguchi98}
T.~Nishiguchi, M.~Kageshima, N.~Ara-Kato, and A.~Kawazu.
\newblock Behavior of charge density waves in a one-dimensional organic
  conductor visualized by scanning tunneling microscopy.
\newblock {\em Phys Rev Letters}, 81(15):3187, 1998.

\bibitem{Wang03}
Z.Z. Wang, J.C. Girard, C.~Pasquier, D.~Jerome, and K.~Bechgaard.
\newblock Scanning tunneling microscopy in ttf-tcnq: Phase and amplitude
  modulated charge density waves.
\newblock {\em Phys. Rev. B.}, 67:R--121401, 2003.

\bibitem{Kato96}
Norihiko~A Kato, Masahiko Hara, Hiroyuki Sasabe, and Wolfgang Knoll.
\newblock An interpretation for the {STM} imaging of an organic molecule,
  tetrathiafulvalene-- tetracyanoquinodimethane (\tq).
\newblock {\em Nanotechnology}, 7:122--127, 1996.

\bibitem{Abrahams77}
E.~Abrahams, J.~Solyom, and F.~Woynarovich.
\newblock The landau theory of phase transitions in \tq.
\newblock {\em Physical Review B}, 16:5238, 1977.

\bibitem{Bjelis76}
A.Bjeli\'s and S.Barisi\'c.
\newblock Commensurate ordering in tetrathiafulvalene-tetracyanoquinodimethane.
\newblock {\em Phys.Rev.Lett}, 37:1517, 1976.

\bibitem{Shirane79}
R.~Com\'es and G.~Shirane.
\newblock X-ray scattering and neuron scattering from one-dimensional
  conductors.
\newblock In J.~T. Devreese, editor, {\em Highly Conducting One-Dimensional
  Solids}, page~17. Plenum Press, New York, 1979.

\bibitem{Pouget16}
J.~P. Pouget.
\newblock The {P}eierls instability and charge density wave in one-dimensional
  electronic conductors.
\newblock {\em Comptes Rendus Physique}, 17:332--356, 2016.

\bibitem{Pouget76}
J.~P. Pouget, S.~K. Khanna, F.~Denoyer, R.~Com\`es, A.~F. Garito, and A.~J.
  Heeger.
\newblock X ray observation of $\mathrm{2k_{F}}$ and $\mathrm{4k_{F}}$
  scatterings in tetrathiafulvalene- tetracyanoquinodimethane (\tq).
\newblock {\em Phys. Rev. Lett.}, 37:437, 1976.

\bibitem{Khanna77}
S.~K. Khanna, J.~P. Pouget, R.~Comes, A.~F. Garito, and A.~J. Heeger.
\newblock X-ray studies of $\mathrm{2k_{F}}$ and $\mathrm{4k_{F}}$ anomalies in
  tetrathiafulvalene-tetracyanoquinodimethane ({TTF-TCNQ}).
\newblock {\em Phys Rev B}, 16(4):1468, 1977.

\bibitem{Emery76}
V.~J. Emery.
\newblock New mechanism for a phonon anomaly and lattice distortion in quasi
  one-dimensional conductors.
\newblock {\em Phys Rev Letters}, 37:107, 1976.

\bibitem{Voit88}
J.~Voit and H.J. Schulz.
\newblock Electron-phonon interaction and phonon dynamics in one-dimensional
  conductors.
\newblock {\em Phys. Rev. B}, 37:10068, 1988.

\bibitem{Basista90}
H.~Basista, D.A. Bonn, T.~Timusk, J.~Voit, D.~Jerome, and K.~Bechgaard.
\newblock Far-infrared optical properties of
  tetrathiofulvalene-tetracyanoquinodimethane (\tq).
\newblock {\em Phys. Rev. B}, 42:4088, 1990.

\bibitem{Pouget88}
J.P. Pouget.
\newblock {\em Semiconductors and Semimetals}, volume~27.
\newblock E. Conwell editor, Academic Press, New York, 1988.

\bibitem{Bernasconi75}
J.~Bernasconi, M.~J. Rice, W.~R. Schneider, and S.~Str\"asseler.
\newblock Peierls transition in the strong-coupling hubbard chain.
\newblock {\em Phys. Rev. B}, 12:1090, 1975.

\bibitem{Malfait69}
G.~Malfait and D.~Jerome.
\newblock Techniques de hautes pressions {\`a} tr{\'e}s basses
  temp{\'e}ratures.
\newblock {\em Revue de Physqiue Appliqu{\'e}e}, 4:467, 1969.

\bibitem{Delplanque70}
G.~Delpanque et~al.
\newblock Appareil de pression hysdrostatique pour mesures {\'e}l{\'e}ctriques
  jusqu'{\`a} 17 kbar {\`a} tr{\`e}s basse temp{\'e}rature.
\newblock {\em Revue de Physique Appliqu{\'e}e}, 5:731, 1970.

\bibitem{Andrieux79f}
A.~Andrieux, H.~J. Schulz, D.~Jerome, and K.~Bechgaard.
\newblock Conductivity of the one-dimensional conductor
  tetrathiafulvalene-tetracyanoquinodimethane {(TTF-TCNQ)} near
  commensurability.
\newblock {\em Phys Rev Letters}, 43(3):227, 1979.

\bibitem{Andrieux79g}
A.~Andrieux, H.~J. Schulz, D.~Jerome, and K.~Bechgaard.
\newblock Fluctuation conductivity in 1-d conductor
  tetrathiafulvalene-tetracyanoquinodimethane (\tq).
\newblock {\em Jour. Physique.Lett}, 40:L--385, 1979.

\bibitem{Megtert81}
S.~Megtert, R.~Com\'es, C.~Vettier, R.~Pynn, and A.~Garito.
\newblock Structural evidence of $\mathrm{2k_{F}}$ commensurablity in \tq under
  pressure.
\newblock {\em Solid. State. Comm.}, 37:875--877, 1981.

\bibitem{Jerome82a}
D.~Jerome and H.J. Schulz.
\newblock Quasi-one-dimensional conductors:{T}he {P}eierls instability,
  pressure and fluctuations effects.
\newblock In J.~S. Miller, editor, {\em Extended Linear Chain Compounds},
  volume~2, chapter~4, page p. 159. Plenum Press, New York, 1982.

\bibitem{Welber78}
B.~Welber, P.E. Seiden, and P.M. Grant.
\newblock Pressure dependence of the {D}rude optical edge of
  tetrathiafulvalenium ({TTF}) and tetraselenafulvalenium ({TSF})
  tetracyanoquinodimethanide ({TCNQ}).
\newblock {\em Phys. Rev. B}, 118:2692, 1978.

\bibitem{Bouffard81}
S.~Bouffard, R.~Chipaux, D.~Jerome, and K.Bechgaard.
\newblock Pinning of charge density waves in irradiated \tq.
\newblock {\em Solid State Comm}, 37:405, 1981.

\bibitem{Tanner81}
D.~B. Tanner, K.~D. Cummings, and C.~S. Jacobsen.
\newblock Far-infrared study of the charge density wave in tetrathiafulvalene
  tetracyanoquinodimethane (\tq).
\newblock {\em Phys Rev Letters}, 47:597--600, 1981.

\bibitem{Friend78a}
R.~H.~Friend et~al.
\newblock Linear temperature dependence of the constant volume resistivity of
  \tq.
\newblock {\em Jour Physique Lettres}, 39:134, 1978.

\bibitem{Herman77}
F.~Herman.
\newblock Scattered wave calculations of monomers and dimers of
  {T}etraselenafulvalene.
\newblock {\em Physica Scripta}, 16:303--306, 1977.

\bibitem{Herman77a}
F.~Herman, D.~R. Salahub, and R.~P. Messmer.
\newblock Xa scattered-wave calculations for dimers and trimers of
  tetrathiafulvalene ({TTF}) and tetracyanoquinodimethane ({TCNQ}).
\newblock {\em Phys Rev B}, 16:2453, 1977.

\bibitem{Monceau}
G.~Gr\"uner and P.~Monceau.
\newblock {\em in Charge Density Waves in Solids}.
\newblock L.V. Gorkov and G. Gr\"uner editors, p 137 North Holland, Amsterdam,
  1989.

\bibitem{Forro87}
L.~Forro, R.~Lacoe, S.~Bouffard, and D.~Jerome.
\newblock Defect-concentration dependence of the charge-density-wave transport
  in tetrathiafulvalene tetracyanoquinodimethane.
\newblock {\em Phys. Rev. B}, 35:5884, 1987.

\bibitem{Lacoe85}
R.~C. Lacoe, H.~J. Schulz, D.~Jerome, K.~Bechgaard, and I.~Johannsen.
\newblock Observation of nonlinear electrical transport at the onset of a
  {P}eierls transition in an organic conductor.
\newblock {\em Phys Rev Letters}, 55(21):2351, 1985.

\bibitem{Lacoe87}
R.~C. Lacoe, J.~R. Cooper, D.~Jerome, F.~Creuzet, K.~Bechgaard, and
  I.~Johannsen.
\newblock Nonlinear electrical transport effects in
  tetrathiafulvalene-tetracyanoquinodimethane as driven through
  {C}harge-{D}ensity-{W}ave commensurability.
\newblock {\em Phys Rev Letters}, 58(3):262, 1987.

\bibitem{Gruner89}
G.~Gr\"uner and P.~Monceau.
\newblock Dynamical properties of charge density waves.
\newblock In L.~V. Gorkov and G.~Gr\"uner, editors, {\em Charge Density Waves
  in Solids}, page 137. North-Holland, 1989.

\bibitem{Yasuzuka07}
S.~Yasuzuka, K.~Murata, T.~Arimoto, and R.~Kato.
\newblock Temperature-pressure phase diagram in \tq: Strong suppression of
  charge-density-wave state under extremely high pressure.
\newblock {\em Jour. Phys. Soc. Japan}, 76(3):033701, 2007.

\bibitem{Horovitz75}
B.~Horovitz, H.~Gutfreund, and M.~Weger.
\newblock Interchain coupling and the {P}eierls transition in linear-chain
  systems.
\newblock {\em Phys. Rev. B}, 12:3174, 1975.

\bibitem{Engler77}
E.M.~Engler et~al.
\newblock Organic alloys: Synthesis and properties of solid solutions of
  tetraselenafulvalene- tetracyano-p-quinodimethane {(TSF-TCNQ)} and
  tetrathiafulvalene-tetracyano-p-quinodimethane (\tq).
\newblock {\em Journal of the American Chemical Society}, 99:5909, 1977.

\bibitem{Pouget79}
J.P. Pouget, S.~Megtert, and R.~Com\`es.
\newblock X ray diffuse scattering study of {1D} organic conductors: \tq and
  its family.
\newblock {\em Lecture Notes in Physics}, 95:14, 1979.

\bibitem{Chaikin76}
P.M. Chaikin, R.L.Greene, S.~Etemad, and E.M. Engler.
\newblock Thermopower of an isostructural series of organic conductors.
\newblock {\em Phys. Rev. B}, 13:1627, 1976.

\bibitem{Weyl76}
C.~Weyl, E.M. Engler, K.~Bechgaard, G.~Jehanno, and S.~Etemad.
\newblock Diffuse {X}-ray scattering in the metallic state of {TSF-TCNQ} and
  {HMTSF-TCNQ}.
\newblock {\em Solid State Comm}, 19:925, 1976.

\bibitem{Etemad76}
S.~Etemad.
\newblock Systematic study of the transitions in
  tetrathiafulvalene-tetracyanoquinodimethane (\tq) and its selenium analogs.
\newblock {\em Phys. Rev. B}, 13(6):2254, 1976.

\bibitem{Scott78}
J.~C. Scott, S.~Etemad, and E.~M. Engler.
\newblock Magnetic susceptibility of {TSF-TCNQ}
  (tetraselenafulvalene-tetracyanoquinodimethane) and its alloys with \tq.
\newblock {\em Phys Rev B}, 17:2269, 1978.

\bibitem{Bates81}
F.~E. Bates, J.~E. Eldridge, and M.~R. Bryce.
\newblock High-resolution polarized far-infrared vibrational spectra of
  semiconducting {TTF-TCNQ} and {TSF-TCNQ}.
\newblock {\em Can. J. Phys}, 59:339, 1981.

\bibitem{Kagoshima79}
S.~Kagoshima, T.~Ishiguro, T.~D. Schultz, and Y.~Tomkiewicz.
\newblock Peierls transition and short range order of charge-density waves in
  {TSeF-TCNQ} - an x ray study.
\newblock {\em Lecture Notes in Physics}, 95:28, 1979.

\bibitem{Thomas82}
J.F. Thomas.
\newblock Commensurability channels in \tsq.
\newblock {\em Solid. State. Comm}, 42:587, 1982.

\bibitem{Megtert79}
S.~Megtert, J.P. Pouget, and R.~Com\`es.
\newblock {\em in Molecular Metals}.
\newblock W.E. Hatfield editor, Plenum Press, New York, 1979.

\bibitem{Thomas80}
J.F. Thomas and D.~Jerome.
\newblock Commensurability and fluctuating conductivity in the organic
  conductor \tsq.
\newblock {\em Solid State Comm}, 36:813, 1980.

\bibitem{Beni74}
G.~Beni.
\newblock Peierls transition in a quasi---one dimensionalsystem.
\newblock {\em Solid State Comm}, 15:269--272, 1974.

\bibitem{Bloch75}
A.N. Bloch, D.O. Cowan, K.~Bechgaard, R~E. Pyle, R.H. Banks, and T.O. Poehler.
\newblock Low-temperature metallic behavior and resistance minimum in a new
  quasi one-dimensional organic conductor.
\newblock {\em Phys. Rev. Lett.}, 34:1561, 1975.

\bibitem{Friend78}
R.~H.~Friend et~al.
\newblock Stabilisation of the metallic state at low temperatures in
  {HMTTF-TCNQ} under pressure.
\newblock {\em J. Phys. C. Solid State Phys}, 11:263, 1978.

\bibitem{Phillips76}
T.~E. Phillips, T.~J. Kistenmacher, A.~N. Bloch, and D.~O. Cowan.
\newblock X-ray crystal structure of the organic conductor from 2,2' - bi -
  (2,4 - diselenabicyclo[3.3.0]octylidene) and 7,7,8,8-tetracyano - p
  -quinodimethane ({HMTSF-TCNQ}).
\newblock {\em J. Chem. Soc. Chem. Commun}, pages 334--335, 1976.

\bibitem{Korin81}
B.~Korin, J.R. Cooper, M.~Miljak, A.~Hamzic, and K.~Bechgaard.
\newblock Magnetoresistance studies oif $\mathrm{HMTSF-TCNQ}$.
\newblock {\em Chemica Scripta}, 17:45, 1981.

\bibitem{Cooper76b}
R.J. Cooper, M.~Weger, D.~Jerome, D.~Lefur, K.~Bechgaard, A.N. Bloch, and D.O.
  Cowan.
\newblock Semi-metallic behaviour of {HMTSF-TCNQ} at low temperatures under
  pressure.
\newblock {\em Solid State Comm}, 19:749--754, 1976.

\bibitem{Weger76}
M.~Weger.
\newblock A model for the electronic band structure of {HMTSeF-TCNQ}.
\newblock {\em Solid State Comm}, 19:1149--1155, 1976.

\bibitem{Cooper76d}
J.~R. Cooper, M.~Weger, G.~Delplanque, D.~Jerome, and K.~Bechgaard.
\newblock The {H}all effect in {HMTSF-TCNQ}.
\newblock {\em Jour Physique Lettres}, 37:L--349, 1976.

\bibitem{Miljak78}
M.~Miljak, A.~Andrieux, R.H. Friend, G.~Malfait, D.~Jerome, and K.~Bechgaard.
\newblock Observation of de {H}aas-{S}hubnikov oscillations in an organic
  metal, {HMTSF-TCNQ}.
\newblock {\em Solid. State. Comm.}, 26:969--971, 1978.

\bibitem{Murata14}
K.~Murata \textit{}et al.
\newblock Magnetic-field-induced phase transitions in the quasi-one-dimensional
  organic conductor {HMTSF--TCNQ}.
\newblock {\em Low {t}emperature {p}hysics}, 40(4, a Memorial Issue for the
  60th Year Anniversary of {L}ifschitz-{K}osevich theory):371, 2014.
\newblock http://dx.doi.org/10.1063/1.4869591.

\bibitem{Lebed09}
A.~Lebed, editor.
\newblock {\em The {P}hysics of {O}rganic {S}uperconductors and {C}onductors}.
\newblock Springer Verlag, 2009.
\newblock Several references about FICDW phases in low dimensional conductors
  are to be found in this book.

\bibitem{Murata14a}
K.~Murata \textit{et al}.
\newblock Magnetic-field-induced phase transition and a possible quantum hall
  effect in the quasi-one-dimensional {CDW} organic conductor {HMTSF-TCNQ}.
\newblock {\em Journal of Modern Physics}, 5:673--679, 2014.
\newblock http://dx.doi.org/10.4236/jmp.2014.58078.

\bibitem{Tomkiewicz78}
Y.~Tomkiewicz, J.~R. Andersen, and A.~R. Taranko.
\newblock Relative stability of donor and acceptor stacks against {P}eierls
  distortion in the tetrathia- and
  tetraselenafulvalene-tetracyanoquinodimethane family of organic metals.
\newblock {\em Phys. Rev. B}, 17:1579, 1978.

\bibitem{Pouget81}
J.P. Pouget.
\newblock Diffuse {X}-ray scattering studies of one-dimensional organic metals.
\newblock {\em Chemica Scripta}, 55:85--91, 1981.

\bibitem{Andrieux79b}
A.~Andrieux, P.M. Chaikin, C.~Duroure, D.~Jerome, C.~Weyl, K.~Bechgaard, and
  J.R. Andersen.
\newblock Transport properties of the metallic state of {TMTSF-DMTCNQ}.
\newblock {\em J. Physique. Paris}, 40:1199, 1979.

\bibitem{Bouffard83}
S.~Bouffard, M.~Ribault, D.~Jerome, and K.~Bechgaard.
\newblock {S}hubnikov-de {H}aas oscillations in an organic conductor,
  {TMTSF-DMTCNQ}.
\newblock {\em J. Physique Lett}, 44:L 285--293, 1983.

\bibitem{Roth66}
L.~M. Roth and P.~N. Argyres.
\newblock Semiconductors and semimetals.
\newblock volume~1, page 159. Academic Press, 1966.

\bibitem{Hardebusch79}
U.~Hardebusch et~al.
\newblock The magnetic susceptibility of {TMTSF-DMTCNQ} under pressure.
\newblock {\em Solid State Comm}, 32:1151--1154, 1979.

\bibitem{Schulz81a}
H.J. Schulz, D.~Jerome, A.~Mazaud, M.~Ribault, and K.~Bechgaard.
\newblock Possibility of superconducting precursor effects in
  quasi-one-dimensional organic conductors : theory and experiments.
\newblock {\em J.Physique}, 42:991--1002, 1981.

\bibitem{Friedel82}
J.~Friedel and D.~Jerome.
\newblock Organic superconductors: the $\mathrm{(TMTSF)_{2}X}$ family.
\newblock {\em Contemp. Phys}, 23(6):583--624, 1982.

\bibitem{Galigne78}
J.L. Galign\'e, B.~Liautard, S.~Peytavin, G.~Brun, , J.-M. Fabre,
  E.~Torreilles, and L.~Giral.
\newblock Etude structurale du bromure de
  t{\'e}tram{\'e}thylt{\'e}trathiafulval{\`e}ne \tmttfbr.
\newblock {\em Acta Cryst.}, B 34:620, 1978.

\bibitem{Jerome12}
D.~Jerome.
\newblock Organic superconductors: When correlations and magnetism walk in.
\newblock {\em Jour Supercond Nov Magn}, 25:633--655, 2012.

\bibitem{Drozdov15}
A.~P. Drozdov, M.~I. Eremets, I.~A. Troyan, V.~Ksenofontov, and S.~I. Shylin.
\newblock Conventional superconductivity at 203 {K}elvin at high pressures in
  the sulfur hydride system.
\newblock {\em Nature}, 525:73, 2015.

\bibitem{Andersen78}
J.R. Andersen, K.~Bechgaard, C.S. Jacobsen, G.~Rindorf, H.~Soling, and
  N.~Thorup.
\newblock The {C}rystal and {M}olecular {S}tructure of the {O}rganic
  {C}onductor 2,3,6,7-tetramethyl 1,4,5,8-tetraselenafulvalenium 2,5- dimethyl-
  7,7,8,8-tetracyano-p-quinodimethanide {(TMTSF-DMTCNQ)}.
\newblock {\em Acta Cryst.B.}, 34:1901, 1978.

\bibitem{Andrieux79c}
A.~Andrieux, C.~Duroure, D.~Jerome, and K.~Bechgaard.
\newblock The metallic state of the organic conductor $\mathrm{TMTSF-DMTCNQ}$
  at low temperature under pressure.
\newblock {\em J. Phys. (Paris) Lett.}, 40:381, 1979.

\bibitem{Schegolev79}
I.~F. Schegolev and R.~B. Lubovskii.
\newblock Properties of the quasi-one-dimensional organic metal \tset2cl.
\newblock In {\em Quasi One-Dimensional Conductors I}, volume~95, page~39.
  Springer-Verlag, 1979.

\bibitem{Laukhin80}
V.N. Laukhin and I.~F. Schegolev.
\newblock Study of phase transitions in \tset2cl under pressure at low
  temperatures.
\newblock {\em JETP}, 51:1170, 1980.

\bibitem{Jerome10}
D.~Jerome.
\newblock {\em Superconductivity in New Materials}, volume~04, chapter Organic
  Superconductivity: A Mouse may be of Service to a Lion, page 149.
\newblock Elsevier, 2011.

\bibitem{Bechgaard80c}
K.~Bechgaard and J.~R. Andersen.
\newblock {\em The {P}hysics of {O}rganic {S}uperconductors and {C}onductors},
  page 247.
\newblock D. Reidel, Heidelberg, 1980.

\bibitem{Jerome80}
D.~Jerome, A.~Mazaud, M.~Ribault, and K.~Bechgaard.
\newblock Superconductivity in a synthetic organic conductor \tmp6.
\newblock {\em J. Phys. (Paris) Lett.}, 41:L95, 1980.
\newblock Open archive https://hal.archives-ouvertes.fr/jpa-00231730.

\bibitem{Pouget97}
J.~P. Pouget and S.~Ravy.
\newblock X-ray evidence of charge density wave modulations in the magnetic
  phases of \tmp6 and \tmttfbr.
\newblock {\em Synthetic Metals}, 85:1523, 1997.

\bibitem{Andres80}
K.~Andres, F.~Wudl, D.B. McWhan, G.A. Thomas, D.~Nalewajek, and A.L. Stevens.
\newblock Observation of the {M}eissener effect in an {O}rganic
  {S}uperconductor.
\newblock {\em Phys. Rev. Lett.}, 45:1449, 1980.

\bibitem{Scott80}
J.~C. Scott, H.~J. Pedersen, and K.~Bechgaard.
\newblock Magnetic properties of the organic conductor \tmp6: a new phase
  transition.
\newblock {\em Phys Rev Letters}, 45:2125, 1980.

\bibitem{Walsh80}
W.~M.~Walsh et~al.
\newblock Restoration of {M}etallic {B}ehavior in organic {C}onductors by small
  electric fields.
\newblock {\em Phys Rev Letters}, 45:829, 1980.

\bibitem{Andrieux81}
A.~Andrieux, D.~Jerome, and K.~Bechgaard.
\newblock Spin-density wave ground state in the one-dimensional conductor \tmp6
  : microscopic evidence from $^{77}${S}e and $^{1}${H} {NMR} experiments.
\newblock {\em J. Phys. Lett. Paris}, 42:L87, 1981.
\newblock Open archive on HAL http://hal.archives-ouvertes.fr/jpa-00231880.

\bibitem{Mortensen81}
K.~Mortensen, Y.~Tomkiewicz, T.~D. Schultz, and E.~M. Engler.
\newblock Antiferromagnetic ordering in the organic conductor
  his-tetramethyltetraselenafulvalene-hexafluorophosphate \tmp6.
\newblock {\em Phys Rev Letters}, 46(18):1234--1237, 1981.

\bibitem{Slater51}
J.~C. Slater.
\newblock Magnetic {E}ffects and the {H}artree-{F}ock {E}quation.
\newblock {\em Phys Rev}, 82:538, 1951.

\bibitem{Lomer62}
W.~M. Lomer.
\newblock Electronic {S}tructure of {C}hromium {G}roup {M}etals.
\newblock {\em Proc. Phys. Soc.A}, 80:489, 1962.

\bibitem{Overhauser62}
A.~W. Overhauser.
\newblock Spin {D}ensity {W}aves in an {E}lectron {G}as.
\newblock {\em Phys Rev}, 128:1437, 1962.

\bibitem{Friend79}
R.~H. Friend and D.~Jerome.
\newblock Periodic lattice distortions and charge density waves in one-and
  two-dimensional metals.
\newblock {\em J. Phys. C. Solid State}, 12:1441, 1979.

\bibitem{Molinie74}
P.~Molini\'e, D.~Jerome, and A.~J. Grant.
\newblock Pressure enhanced superconductivity and superlattice structures in
  transition metal dichalcogenide layer crystals.
\newblock {\em Phil. Mag}, 30:1091--1103, 1974.

\bibitem{Wilson75}
J.~A. Wilson, F.~J.~Di Salvo, and S.~Mahajan.
\newblock Charge-density waves and superlattices in the metallic layered
  transition metal dichalcogenides.
\newblock {\em Adv in Physics}, 24:117, 1975.

\bibitem{Friedel75}
J.~Friedel.
\newblock On the pressure {D}ependence of {S}uperconductivity in transition
  metal dichalcogenide layer crystals.
\newblock {\em J. Phys. Lett}, 36:L279, 1975.
\newblock Open archive on HAL: http://hal.archives-ouvertes.fr/.

\bibitem{Greene80}
R.L. Greene and E.M. Engler.
\newblock Pressure dependence of superconductivity in a organic superconductor
  \tmp6.
\newblock {\em Phys. Rev. Lett.}, 45:1587, 1980.

\bibitem{Jerome91}
D.~Jerome.
\newblock The physics of organic superconductors.
\newblock {\em Science}, 252:1509, 1991.

\bibitem{Schulz81}
H.J. Schulz, D.~Jerome, M.~Ribault, A.~Mazaud, and K.~Bechgaard.
\newblock Pressure dependence of the organic superconductivity in \tmp6.
\newblock {\em J.Physique. Lett}, 42:L--51, 1981.
\newblock Open archive on HAL http://hal.archives-ouvertes.fr/.

\bibitem{Brusetti82a}
R.~Brusetti, M.~Ribault, and D.~Jerome.
\newblock Insulating, conducting and superconducting states of \tmtsfasf6 under
  pressure and magnetic field.
\newblock {\em J. Physique}, 43:801, 1982.

\bibitem{Jerome82b}
D.~Jerome.
\newblock {O}rganic {S}uperconductors: A survey of low dimensional phenomena.
\newblock {\em Mol. Cryst.Liq. Cryst}, 79:155--182, 1982.

\bibitem{Jaccard01}
D~Jaccard, H~Wilhelm, D~Jerome, J~Moser, C~Carcel, , and J~M Fabre.
\newblock From spin-{P}eierls to superconductivity: \tmps under high pressure.
\newblock {\em J. Phys. Cond. Matter}, 13:L89, 2001.

\bibitem{Itoi07}
M.~Itoi {\textit{et al}}.
\newblock Pressure-induced superconductivity in the quasi-one-dimensional
  organic conductor $\mathrm{(TMTTF)_{2}AsF_{6}}$.
\newblock {\em Journal of the Physical Society of Japan}, 76(5):053703, 2007.
\newblock https://doi.org/10.1143/JPSJ.76.053703.

\bibitem{Itoi08}
M.~Itoi {\textit{et al}}.
\newblock Anomalously wide superconducting phase of one-dimensional organic
  conductor $\mathrm{(TMTTF)_{2}SbF_{6}}$.
\newblock {\em Journal of the Physical Society of Japan}, 77(2), 2008.
\newblock https://doi.org/10.1143/JPSJ.77.023701.

\bibitem{Parkin81a}
S.~S.~P. Parkin, M~Ribault, D.~Jerome, and K~Bechgaard.
\newblock Superconductivity in the family of organic salts based on the
  tetramethyltetraselenafulvalene {(TMTSF)} molecule.
\newblock {\em J. Phys. C. Solid State}, 14:5305, 1981.

\bibitem{Parkin82a}
S.S.P. Parkin, F.~Creuzet, M.~Ribault, D.~J{\'e}rome, K.~Bechgaard, and J.~M.
  Fabre.
\newblock Superconductivity in the organic charge transfer salts:
  $\mathrm{(TMTTF)_{2}X}$ and $\mathrm{(TMTSF)_{2}X}$.
\newblock {\em Molecular Crystals and Liquid Crystals}, 79:605--615, 1982.

\bibitem{Balicas94}
L.~Balicas, K.~Behnia, W.~Kang, E.~Canadell, P.~Auban-Senzier, D.~Jerome,
  M.~Ribault, and J.M. Fabre.
\newblock Superconductivity and magnetic field induced spin density waves in
  the \tmttf2x family.
\newblock {\em J. Phys. I (France)}, 4:1539, 1994.

\bibitem{Tomic89a}
S.~Tomi\'c and D.~Jerome.
\newblock A hidden low-temperature phase in the organic conductor \tmtsfreo4.
\newblock {\em J.Phys.Cond.Matter.Letters.}, 1:4451, 1989.

\bibitem{Wilhelm01}
H.~Wilhelm et~al.
\newblock The case for universality of the phase diagram of the {F}abre and
  {B}echgaard salts.
\newblock {\em Eur.Phys.Jour.B}, 21:175, 2001.

\bibitem{Adachi00}
T.~Adachi, E.~Ojima, K.~Kato, and H.~Kobayashi.
\newblock Superconducting transition of \tmps above 50 kbar.
\newblock {\em Jour. Am. Chem. Soc.}, 122:3238, 2000.

\bibitem{Emery79}
V.~J. Emery.
\newblock Theory of the one-dimensional electron gas.
\newblock In J.~T. Devreese, R.~E. Evrard, and V.~E. van Doren, editors, {\em
  Highly Conducting One-Dimensional Solids}, page 247. Plenum Press, New York,
  1979.

\bibitem{Solyom79}
J.~Solyom.
\newblock The {F}ermi gas model of one-dimensional conductors.
\newblock {\em Adv. Phys.}, 28:201, 1979.

\bibitem{Schulz95}
H.~J. Schulz.
\newblock Fermi liquids and non-{F}ermi liquids.
\newblock In E.~Akkermans, G.~Montambaux, J.~Pichard, and J.~Zinn-Justin,
  editors, {\em Proceedings of Les Houches Summer School LXI}, page 533.
  Elsevier, Amsterdam, 1995.
\newblock https://doi.org/10.48550/arXiv.cond-mat/9503150.

\bibitem{Voit95}
J.~Voit.
\newblock One-dimensional {F}ermi liquids.
\newblock {\em Rep. Prog. Phys.}, 58:977, 1995.

\bibitem{Giamarchi04}
T.~Giamarchi.
\newblock {\em Quantum Physics in One-Dimension}.
\newblock Clarendon Press, Oxford, 2004.

\bibitem{Emery82}
V.~J. Emery, R.~Bruinsma, and S.~Barisi\'c.
\newblock Electron-{E}lectron {U}mklapp scattering in organic superconductors.
\newblock {\em Phys. Rev. Lett.}, 48:1039, 1982.

\bibitem{Barisic81}
S.~Barisi\'c and S.~Brazovskii.
\newblock In J.~T. Devreese, editor, {\em Recent Developments in Condensed
  Matter Physics}, volume~1, page 327. Plenum Press, New York, 1981.

\bibitem{Penc94b}
K.~Penc and F.~Mila.
\newblock Charge gap in the one-dimensional dimerized {H}ubbard model at
  quarter-filling.
\newblock {\em Phys Rev B}, 50:11429, 1994.

\bibitem{Grant82}
P.~M. Grant.
\newblock Band-structure parameters of a series of
  tetramethyltetraselenafulvalene \tmtsf2x compounds.
\newblock {\em Physical Review B}, 26:6888, 1982.

\bibitem{Coulon82}
C.~Coulon, P.~Delhaes, S.~Flandrois, R.~Lagnier, E.~Bonjour, and J.M. Fabre.
\newblock Effect of doping \tmc with {TMTTF} {I}. ambient pressure results : a
  competition between the different possible ground states.
\newblock {\em J. Phys. (Paris)}, 43:1059, 1982.

\bibitem{Caron86}
L.~G. Caron and C.~Bourbonnais.
\newblock Importance of one-dimensional correlations in the phase diagram of
  the $\mathrm{(TMTTF)_{2}X-(TMTSF)_{2}X}$ salts.
\newblock {\em Physica B}, 143B:453, 1986.

\bibitem{Brazo85}
S.~Brazovskii and Y.~Yakovenko.
\newblock On the theory of organic superconducting materials.
\newblock {\em Sov. Phys. JETP}, 62:1340, 1985.

\bibitem{Giamarchi91}
T.~Giamarchi.
\newblock Umklapp process and resistivity in one-dimensional fermion systems.
\newblock {\em Phys. Rev. B}, 44:2905, 1991.

\bibitem{Shahbazi15}
M.~Shahbazi and C.~Bourbonnais.
\newblock Electrical transport near quantum criticality in low dimensional
  organic superconductors.
\newblock {\em Phys. Rev. B}, 92:195141, 2015.

\bibitem{Giamarchi97}
T.~Giamarchi.
\newblock Mott transition in one dimension.
\newblock {\em Physica}, B 230-232:975, 1997.

\bibitem{Mila93}
F.~Mila and X.~Zotos.
\newblock Phase diagram of the one-dimensional extended {H}ubbard model at
  quarter-filling.
\newblock {\em Europhys. Lett.}, 24:133, 1993.

\bibitem{Tsuchiizu01}
M.~Tsuchiizu, H.~Yoshika, and Y.~Suzumura.
\newblock Crossover from quarter-filling to half-filling in a one-dimensional
  electron system with a dimerized and quarter-filled band.
\newblock {\em Jour. Phys. Soc. Japan}, 70:1460, 2001.

\bibitem{Giamarchi04a}
T.~Giamarchi.
\newblock Theoretical framework for quasi-one dimensional systems.
\newblock {\em Chem. Rev.}, 104:5037, 2004.

\bibitem{Ruetschi09}
A.~S. R\"uetschi and D.~Jaccard.
\newblock High pressure study of the organic compound
  $\mathrm{(TMTTF)_{2}BF_{4}}$.
\newblock {\em Eur. Phys. J. B}, 67:43, 2009.

\bibitem{Bourbonnais98}
C.~Bourbonnais and D.~Jerome.
\newblock Electronic confinement in organic metals.
\newblock {\em Science}, 281:1156, 1998.

\bibitem{Ducasse86}
L.~Ducasse, A.~Abderraba, J.~Hoarau, M.~Pesquer, B.~Gallois, and J.~Gaultier.
\newblock Temperature dependence of the transfer integrals in the \tmtsf2x and
  \tmttf2x families.
\newblock {\em J. Phys. C: Solid State}, 19:3805, 1986.

\bibitem{Fabre04}
J.~M. Fabre.
\newblock Synthesis strategies and chemistry of nonsymmetrically substituted
  tetrachalcogenafulvalenes.
\newblock {\em Chemical Reviews}, 104:5133, 2004.

\bibitem{Creuzet87a}
F.~Creuzet, C.~Bourbonnais, L.~G. Caron, D.~Jerome, and K.~Bechgaard.
\newblock A $^{13}${C NMR} study of the interplay between the spin-{P}eierls
  and antiferromagnetic ground states in \tmps under pressure.
\newblock {\em Synthetic. Metals}, 19:289, 1987.

\bibitem{Bourbon89}
C.~Bourbonnais, P.~Wzietek, D.~Jerome F.~Creuzet, K.~Bechgaard, and P.~Batail.
\newblock Scaling relation between nuclear relaxation and magnetic
  susceptibility in organic conductors: Evidence for {1D} paramagnon effects.
\newblock {\em Phys. Rev. Lett.}, 62:1532, 1989.

\bibitem{Wzietek93}
P.~Wzietek, F.~Creuzet, C.~Bourbonnais, D.~Jerome, K.~Bechgaard, and P.~Batail.
\newblock Nuclear relaxation and electronic correlations in
  quasi-one-dimensional organic conductors. {II}. experiments.
\newblock {\em J. Phys. I (France)}, 3:171, 1993.
\newblock Open archive on HAL http://hal.archives-ouvertes.fr/.

\bibitem{Pouget82a}
J.P. Pouget, R.~Moret, R.~Com\`es, K.~Bechgaard, J.-M. Fabre, and L.~Giral.
\newblock X-ray diffuse scattering study of some {(TMTSF)2X} and
  {(TMTTF)2X}salts.
\newblock {\em Mol. Cryst. Liq. Cryst.}, 79:129, 1982.

\bibitem{Auban04}
P.~Auban-Senzier, D.~Jerome, C.~Carcel, and J.M Fabre.
\newblock Longitudinal and transverse transport of the quasi-one dimensional
  organic conductor \tmps studied under high pressure.
\newblock {\em J. Phy. IV France}, 114:41, 2004.
\newblock Open archive on HAL http://hal.archives-ouvertes.fr/.

\bibitem{Bourbonnais88}
C.~Bourbonnais and L.G. Caron.
\newblock New mechanisms for phase transitions in quasi- one-dimensional
  conductors.
\newblock {\em Europhys. Lett.}, 5:209, 1988.

\bibitem{Grant83}
P.~M. Grant.
\newblock Electronic structure of the 2 : 1 charge transfer salts of {TMTCF}.
\newblock {\em J. Phys. (Paris) Coll.}, 44:847, 1983.

\bibitem{Jerome04}
D.~Jerome.
\newblock Organic conductors: From charge density wave \tq to superconducting
  \tmp6.
\newblock {\em Chem. Rev}, 104:5565, 2004.

\bibitem{Schwartz98}
A.~Schwartz, M.~Dressel, G.~Gr\"uner, V.~Vescoli, L.~Degiorgi, and
  T.~Giamarchi.
\newblock On-chain electrodynamics of metallic \tm2x salts: Observation of
  {T}omonaga-{L}uttinger liquid response.
\newblock {\em Phys. Rev. B.}, 58:1261, 1998.

\bibitem{Vescoli98}
V.~Vescoli, L.~Degiorgi, W.~Henderson, G.~Gr\"uner, K.~P. Starkey, and L.K.
  Montgomery.
\newblock Dimensionality-driven insulator-to-metal transition in the
  {B}echgaard salts.
\newblock {\em Science}, 281:1181, 1998.

\bibitem{Degiorgi06}
L.~Degiorgi and D.~Jerome.
\newblock Transport and optics in quasi-one-dimensional organic conductors.
\newblock {\em Jour. Phys. Soc. Japan}, 75:051004, 2006.

\bibitem{Biermann01}
S.Biermann, A.Georges, A.Lichtenstein, and T.Giamarchi.
\newblock Deconfinement transition and {L}uttinger to {F}ermi liquid crossover
  in quasi-one-dimensional systems.
\newblock {\em Phys. Rev. Lett.}, 87:276405, 2001.

\bibitem{Moser98}
J.~Moser, M.~Gabay, P.~Auban-Senzier, D.~Jerome, K.~Bechgaard, and J.~M. Fabre.
\newblock Transverse transport in \tm2x organic conductors: possible evidence
  for a {L}uttinger liquid.
\newblock {\em Eur. Phys. Jour. B}, 1:39, 1998.

\bibitem{Creuzet87b}
F.~Creuzet, C.Bourbonnais, L.G. Caron, D.~Jerome, and A.~Moradpour.
\newblock $^{77}${S}e {NMR} spin-lattice relaxation rate properties in the
  \tmtsf2x series under pressure : Cooperative phenomena and {SDW} transition.
\newblock {\em Synthetic Metals}, 19:277, 1987.

\bibitem{Gallois87}
B.~Gallois et~al.
\newblock Cristallographic structures of \tmp6 under constraints:evidence of a
  change in the electronic structure.
\newblock {\em Synth. Metals}, 19:321--326, 1987.

\bibitem{constantvolume}
Using the helium gas pressure technique it has been established that the
  pressure coefficient of the longitudinal conductivity $\sigma_{a}$ is
  temperature independent between 80 and 300K, namely,
  $(\sigma_{a}(P)-\sigma_{a}(0)/\sigma_{a}(0)$=+25\%/$kbar^{-1}$.

\bibitem{Bourbonnais99}
C.~Bourbonnais and D.~Jerome.
\newblock The normal phase of quasi-one-dimensional organic superconductors.
\newblock In P.~Bernier, S.~Lefrant, and G.Bidan, editors, {\em Adv. in Synth.
  Metals}, pages p.206--261. Elsevier, 1999.

\bibitem{Auban99}
P.~Auban-Senzier, D.~Jerome, and J.~Moser.
\newblock Non fermi-liquid features in \tm2x 1-{D} conductors from transport
  properties.
\newblock In R.~Schrieffer Z.~Fisk, L.~Gorkov, editor, {\em Physical Phenomena
  at High Magnetic Fields}, page 211. World Scientific, Singapore, 1999.

\bibitem{Chow00}
D.S. Chow, F.~Zamborsky, B.~Alavi, D.J. Tantillo, A.~Baur, C.A. Merli\'c, and
  S.E. Brown.
\newblock Charge ordering in the {TMTTF} family of molecular conductors.
\newblock {\em Phys. Rev. Lett.}, 85:1698, 2000.

\bibitem{Nad00}
F.~Nad, P.~Monceau, C.~Carcel, and J.~M. Fabre.
\newblock Dielectric response of the charge-induced correlated state in the
  quasi-one-dimensional conductor \tmps.
\newblock {\em Phys.Rev.B}, 62:1753, 2000.

\bibitem{Monceau01}
P.~Monceau, F.~Nad, and S.Brazovskii.
\newblock Ferroelectric {M}ott-{H}ubbard phase of organic \tmttf2x conductors.
\newblock {\em Phys.Rev.Lett.}, 86:4080, 2001.

\bibitem{Bourbonnais84}
C.~Bourbonnais, F.~Creuzet, D.~Jerome, K.~Bechgaard, and A.~Moradpour.
\newblock Cooperative phenomena in \tmc : an {NMR} evidence.
\newblock {\em J. Phys. (Paris) Lett.}, 45:L755, 1984.

\bibitem{Tomic91}
S.~Tomi\'c, J.~R. Cooper, W.~Kang, D.~Jerome, and K.~Maki.
\newblock The influence of chemical impurities and {X}-ray induced defects on
  the single-particle and spin-density wave conductivity in the {B}echgaard
  salts.
\newblock {\em J. Phys. I (France)}, 1:1603, 1991.
\newblock Open archive on HAL http://hal.archives-ouvertes.fr/.

\bibitem{Cao96}
N.~Cao~T. Timusk and K.~Bechgaard.
\newblock Unconventional electrodynamic response of the quasi one dimensional
  organic conductor \tmc.
\newblock {\em J. Physique (France)}, 6:1719, 1996.
\newblock Open archive on HAL http://hal.archives-ouvertes.fr/.

\bibitem{Henderson99}
W.~Henderson, V.~Vescoli, P.~Tran, L.~Degiorgi, and G.~Gr\"uner.
\newblock Anisotropic electrodynamics of low dimensional metals: Optical
  studies of \tmc.
\newblock {\em Eur. Phys. J. B}, 11:365, 1999.

\bibitem{Clarke95}
D.~G. Clarke, S.~P. Strong, and P.~W. Andersot.
\newblock Conductivity between {L}uttinger liquids in the confinement regime
  and c-axis conductivity in the cuprate superconductors.
\newblock {\em Phys Rev Letters}, 74:4499, 1995.

\bibitem{Dressel96}
M.~Dressel, A.~Schwartz, G.~Gr\"uner, , and L.~De Giorgi.
\newblock Deviations from {D}rude response in low-dimensional metals:
  Electrodynamics of the metallic state of \tmp6.
\newblock {\em Phys. Rev. Lett.}, 77:398, 1996.

\bibitem{Jacobsen81}
C.~S. Jacobsen, D.~B. Tanner, and K.~Bechgaard.
\newblock Dimensionality crossover in the organic superconductor
  tetramethyltetraselenafulvalene hexafluorophosphate \tmp6.
\newblock {\em Phys. Rev. Letts}, 46:1142, 1981.

\bibitem{Dressel12}
M.~Dressel.
\newblock Electrodynamics of {B}echgaard salts: Optical properties of
  one-dimensional metals.
\newblock {\em International Scholarly Research Network Condensed Matter
  Physics}, 2012, 2012.
\newblock doi:10.5402/2012/732973.

\bibitem{Georges00}
A.~Georges, T.~Giamarchi, and N.~Sandler.
\newblock Interchain conductivity of coupled {L}uttinger liquids and organic
  conductors.
\newblock {\em Phys. Rev. B.}, 61:16393, 2000.

\bibitem{Heuze03}
K.~Heuz\'e, M.~Fourmigu\'e, P.~Batail, C.~Coulon, R.~Cl\'erac, E.~Canadell,
  P.~Auban-Senzier, R.~Ravy, and D.~Jerome.
\newblock A genuine quarter-filled band {M}ott insulator, \edtasf6:where the
  chemistry and physics of weak intermolecular interactions act in unison.
\newblock {\em Adv.Materials}, 15, 2003.

\bibitem{Chow98a}
D.~S.~Chow {\textit{et al}}.
\newblock Singular behavior in the pressure-tuned competition between
  spin-{P}eierls and antiferromagnetic ground states of \tmps.
\newblock {\em Phys Rev Letters}, 81:3984, 1998.

\bibitem{Zorina09}
L.~Zorina, S.~Simonov, C.~M\'{e}zi\`{e}re, E.~Canadell, S.~Suh, S.~E. Brown,
  P.~Foury-Leylekian, P.~Fertey, J-P. Pouget, and P.~Batail.
\newblock Charge ordering, symmetry and electronic structure issues and
  {W}igner crystal structure of the quarter-filled band {M}ott insulators and
  high pressure metals \edtx, {X}= {B}r and \as.
\newblock {\em Journal of Materials Chemistry}, 19:6980, 2009.

\bibitem{Auban09a}
P.~Auban-Senzier, C.~R. Pasquier, D.~Jerome, S.~Suh, S.~E. Brown, C.~Meziere,
  and P.~Batail.
\newblock Phase diagram of quarter-filled band organic salts \edtx, {X} = \as
  and {B}r.
\newblock {\em Phys. Rev. Lett}, 102:255001, 2009.

\bibitem{Bechgaard81}
K.~Bechgaard, M.~Carneiro, M.~Olsen, F.B. Rasmussen, and C.~S. Jacobsen.
\newblock Zero-pressure organic superconductor \tmc.
\newblock {\em Phys. Rev. Lett.}, 46:852, 1981.

\bibitem{Bechgaard81a}
K.~Bechgaard, K.~Carneiro, M.Olsen, F.~B. Rasmussen, and M.~Olsen.
\newblock Superconductivity in an organic solid. synthesis, structure, and
  conductivity of bis(tetramethyltetraselenafulva1enium) perchlorate, \tmc.
\newblock {\em J. Am. Chem. Soc}, 103:2440--2442, 1981.

\bibitem{Thorup81}
N.~Thorup, G.~Rindorf, H.~Soling, and K.~Bechgaard.
\newblock The structure of \tmp6, the first superconducting organic solid.
\newblock {\em Acta Cryst B}, 37:1236--1240, 1981.

\bibitem{Kowada07}
H.~Kowada, R.~Kondo, and S.~Kagoshima.
\newblock Development of uniaxial elongation method and its application to low
  dimensional conductors.
\newblock {\em Jour. Phys.Soc. Japan}, 76:114710, 2007.

\bibitem{Ducasse85}
L~Ducasse, M~Abderrabba, and B~Gallois.
\newblock Temperature dependence of the {F}ermi surface topography in the
  \tmtsf2x and \tmttf2x families.
\newblock {\em J.Phys C:Solid State}, 18:L947, 1985.

\bibitem{Bruinsma83}
R.~Bruinsma and V.~J. Emery.
\newblock Theory of anion ordering in {TMTSF} compounds.
\newblock {\em J. Phys. (Paris) Coll.}, 44-C3:1115, 1983.

\bibitem{Emery83}
V.~J. Emery.
\newblock Some basic questions in organic superconductivity.
\newblock {\em J. Phys. (Paris) Coll.}, 44:C3--977, 1983.

\bibitem{Pouget83}
J.~P. Pouget, G.~Shirane, K.~Bechgaard, and J.~M. Fabre.
\newblock X-ray evidence of a structural transition in \tmc, pristine and
  slightly doped.
\newblock {\em Phys. Rev. B}, 27:5203, 1983.

\bibitem{Takahashi82}
T.~Takahashi, K.~Bechgaard, and D.~Jerome.
\newblock Observation of a magnetic state in the organic superconductor \tmc :
  influence of the cooling rate.
\newblock {\em J. Physique. Lett}, 43:L565, 1982.
\newblock Open archive on HAL http://hal.archives-ouvertes.fr/.

\bibitem{Pasteur54}
L.~Pasteur.
\newblock Talk at the {U}niversity of {L}ille.
\newblock 1854.
\newblock
  https://innovationetserendipite.files.wordpress.com/2011/01/discours-de-louis-pasteur.pdf.

\bibitem{Moret86}
R.~Moret, S.~Ravy, J.~P. Pouget, R.~Com\'es, and K.~Bechgaard.
\newblock Anion-ordering phase diagram of di(tetramethyltetraselenafulvalenium)
  perrhenate,\tmtsfreo4.
\newblock {\em Phys. Rev. Lett.}, 57:1915, 1986.

\bibitem{Parkin82}
S.~S.~P. Parkin, D.~Jerome, and K.~Bechgaard.
\newblock Pressure dependence of the metal-insulator and superconducting phase
  transitions in {(TMTSF)2Re04}.
\newblock {\em Mol. Cryst. Liq. Cryst}, 79:213, 1982.

\bibitem{Ribault80}
M.~Ribault, G.~Benedek, D.~Jerome, and K.~Bechgaard.
\newblock Diamagnetic {AC} susceptibility in the quasi-one dimensional organic
  conductor : \tmp6.
\newblock {\em J. Phys. Lett. Paris}, 41:L--397, 1980.
\newblock Open archive https://hal.archives-ouvertes.fr/jpa-00231806.

\bibitem{Gubser81}
D.~U. Gubser, W.~W. Fuller, T.~O. Poehler, D.~O. Cowan, M.~Lee, M.~Lee, R.~S.
  Potemberg, L.~Y. Chiang, and A.~N. Bloch.
\newblock Magnetic susceptibility and resistive transitions of superconducting
  \tmc : critical fields.
\newblock {\em Phys.Rev.B}, 24:478, 1981.

\bibitem{Schwenk84}
H.Schwenk, K.Andres, and F.Wudl.
\newblock Resistivity of the organic superconductor \tmc in its relaxed,
  quenched, and intermediate state.
\newblock {\em Phys. Rev. B.}, 29:500, 1984.

\bibitem{Gubser82}
D.U. Gubser, W.W. Fuller, T.O. Poehler, J.~Stokes, D.O. Cowan, M.~Lee, and A.N.
  Bloch.
\newblock Resistive and magnetic susceptibility transitions in superconducting
  \tmc.
\newblock {\em Mol. Cryst. Liq. Cryst.}, 79:225, 1982.

\bibitem{Oh04}
J.~I. Oh and M.~J. Naughton.
\newblock Magnetic determination of {H}c2 under accurate alignment in \tmc.
\newblock {\em Phys Rev Letters}, 92:067001, 2004.

\bibitem{Garoche82}
P.~Garoche, R.~Brusetti, D.~Jerome, and K.~Bechgaard.
\newblock Specific heat measurements of organic superconductivity in \tmc.
\newblock {\em J. Phys. Lett. Paris}, 43:L--147, 1982.
\newblock https://hal.archives-ouvertes.fr/jpa-00232023/document.

\bibitem{Mailly82}
D.~Mailly, M.~Ribault, K.~Bechgaard, J.~M. Fabre, and L.~Giral.
\newblock Anisotropy of the {M}eissner effect and the diamagnetic shielding in
  \tmc.
\newblock {\em J. Physique. Lett}, 43:L--711, 1982.

\bibitem{Yonezawa12}
S.~Yonezawa, Y.~Maeno, K.~Bechgaard, and D.~Jerome.
\newblock Nodal superconducting order parameter and thermodynamic phase diagram
  of \tmc.
\newblock {\em Phys. Rev. B}, 85:140502R, 2012.
\newblock arxiv.org/abs/1112.5974.

\bibitem{Gorkov85}
L.P. Gorkov and D.~Jerome.
\newblock Back to the problem of the upper critical fields in organic
  superconductors.
\newblock {\em J. Phys. Lett.}, 46:L--643, 1985.

\bibitem{Yonezawa18}
S.~Yonezawa, C.~A. Marrache-Kikuchi, K.~Bechgaard, and D.~Jerome.
\newblock Crossover from impurity-controlled to granular superconductivity in
  \tmc.
\newblock {\em Phys Rev B}, 97:014521, 2018.

\bibitem{Ishiguro98}
T.~Ishiguro, K.~Yamaji, and G.~Saito.
\newblock {\em Organic Superconductors}.
\newblock Springer, Heidelberg, 1998.

\bibitem{Murata87}
K.~Murata, M.~Tokumoto, H.~Anzai, K.Kajimura, and T.~Ishiguro.
\newblock Upper critical field of the anisotropic organic superconductors,
  \tmc.
\newblock {\em Japanese Jour. of Appl. Physics}, Suppl. 26-3:1367, 1987.

\bibitem{Lee97}
I.~J. Lee, M.~J. Naughton, G.~M. Danner, and P.~M. Chaikin.
\newblock Anisotropy of the upper critical field in \tmp6.
\newblock {\em Phys. Rev. Lett.}, 78:3555, 1997.

\bibitem{Lee00}
I.J. Lee, P.M. Chaikin, and M.J. Naughton.
\newblock Exceeding the {P}auli paramagnetic limit in the critical field of
  \tmp6.
\newblock {\em Phys. Rev. B}, 62:R14669, 2000.

\bibitem{Lee95}
I.~J. Lee, A.~P. Hope, M.~J. Leone, and M.~J. Naughton.
\newblock Revisiting the superconducting phase diagram of \tmc.
\newblock {\em Synthetic Metals}, 70:747--750, 1995.

\bibitem{Yonezawa08a}
S.~Yonezawa, S.~Kusaba, Y.~Maeno, P.~Auban-Senzier, C.~Pasquier, and D.~Jerome.
\newblock Magnetic-field variations of the pair-breaking effects of
  superconductivity in \tmc.
\newblock {\em J. Phys. Soc. Japan}, 77:054712, 2008.

\bibitem{Yonezawa08}
S.~Yonezawa, S.~Kusaba, Y.~Maeno, P.~Auban-Senzier, C.~Pasquier, K.~Bechgaard,
  and D.~Jerome.
\newblock Anomalous in-plane anisotropy of the onset of superconductivity in
  \tmc.
\newblock {\em Phys. Rev. Lett.}, 100:117002, 2008.

\bibitem{Clogston62}
A.~M. Clogston.
\newblock Upper limit for the critical field in hard superconductors.
\newblock {\em Phys Rev Letters}, 9:266, 1962.

\bibitem{Hirschfeld88}
P.~J. Hirschfeld, P.~Wolfe, and D.~Einzel.
\newblock Consequences of resonant impurity scattering in anisotropic
  superconductors: Thermal and spin relaxation properties.
\newblock {\em Phys Rev B}, 37:83, 1988.

\bibitem{Jerome16}
D.~Jerome and S.~Yonezawa.
\newblock Novel superconducting phenomena in quasi-one-dimensional {B}echgaard
  salts.
\newblock {\em C. R. Physique}, 17:357, 2016.
\newblock Open Access Sciencedirect.

\bibitem{Shinagawa07}
J.~Shinagawa, Y.~Kurosaki, F.~Zhang, C.~Parker, S.~E. Brown, and D.~Jerome.
\newblock Superconducting state of the organic conductor \tmc.
\newblock {\em Phys.Rev.Lett.}, 98:147002, 2007.

\bibitem{Flouquet19}
D.~Aoki, K.Ishida, and J.~Flouquet.
\newblock Review of {U}-based ferromagnetic superconductors: comparaison betwen
  $\mathrm{UGe_{2}}$, $\mathrm{URhGe}$ and $\mathrm{UCoGe}$.
\newblock {\em Jour. Phys. Soc. Japan}, 88:022001, 2018.

\bibitem{Takigawa87}
M.~Takigawa, H.~Yasuoka, and G.~Saito.
\newblock Proton spin lattice relaxation in the superconducting state of \tmc.
\newblock {\em J. Phys. Soc. Japan}, 56:873, 1987.

\bibitem{Hasegawa87}
Y.~Hasegawa and H.~Fukuyama.
\newblock {NMR} rexation time of anisotropic superconding state in quasi-one
  dimensional systems.
\newblock {\em Jour. Phys.Soc. Japan}, 56:877, 1987.

\bibitem{Lee02}
I.J. Lee, M.J. Naughton, and P.M. Chaikin.
\newblock Triplet superconductivity in an organic superconductor probed by
  {NMR} {K}night shift.
\newblock {\em Phys. Rev. Lett.}, 88:207002, 2002.

\bibitem{Fulde64}
P.~Fulde and R.~A. Ferrell.
\newblock Superconductivity in a strong spin-exchange field.
\newblock {\em Phys. Rev.}, 135:A550, 1964.

\bibitem{Larkin65}
A.~I. Larkin and Y.~N. Ovchinnikov.
\newblock Nonuniform state of superconductors.
\newblock {\em Sov. Phys. JETP.}, 20:762, 1965.

\bibitem{Strong94}
S.P Strong, D.~G. Clarke, and P.~W. Anderson.
\newblock Magnetic field induced confinement in strongly correlated anisotropic
  materials.
\newblock {\em Phys Rev Letters}, 73:1007, 1994.

\bibitem{Joo06}
N.~Joo, P.~Auban-Senzier, C.R. Pasquier, S.~Yonezawa, R.~Higashinaka, Y.~Maeno,
  S.~Haddad, S.~Charfi-Kaddour, M.~H\'eritier, and D.~Jerome.
\newblock Field-induced confinement in \tmc under accurately aligned magnetic
  fields.
\newblock {\em Eur. Phys. B}, 52:337, 2006.

\bibitem{Lebed86}
A.~Lebed.
\newblock Reversible nature of the orbital mechanism for the suppresion of
  superconductivity.
\newblock {\em JETP.Letters}, 44:114, 1986.

\bibitem{Fuseya12}
Y.~Fuseya, C.~Bourbonnais, and K.~Miyake.
\newblock On the origin of the anomalous upper critical field in quasi-
  one-dimensional superconductors.
\newblock {\em Europhysics Letters}, 100:57008, 2012.

\bibitem{Miyawaki14}
N.~Miyawaki an~S.~Shimahara.
\newblock Dimensional crossover of the fulde--ferrell--larkin--ovchinnikov
  state in strongly {P}auli-limited quasi-one-dimensional superconductor.
\newblock {\em J. Phys.Soc.Japan}, 83:024703, 2014.

\bibitem{Matsuda07}
Y.~Matsuda and H.~Shimahara.
\newblock {F}ulde--{F}errell--{L}arkin--{O}vchinnikov state in heavy fermion
  superconductors.
\newblock {\em J. Phys. Soc. Japan}, 76:051005, 2007.

\bibitem{Mayaffre98}
H.~Mayaffre, P.~Auban-Senzier, M.~Nardone, D.~Jerome, D.~Poilblanc,
  C.~Bourbonnais, U.~Ammerrahl, G.~Dhalenne, and A.~Revcolevschi.
\newblock Absence of a spin gap in the superconducting ladder compound
  \srcacuo.
\newblock {\em Science}, 279:345, 1998.

\bibitem{Braithwaite00}
D.~Braithwaite et~al.
\newblock Upper critical field of the spin ladder system \srcacuo.
\newblock {\em Solid.State.Com}, 114:533, 2000.

\bibitem{Nakanishi05}
T.~Nakanishi et~al.
\newblock Magnetic field effect on the pressure-induced superconducting state
  in the hole-doped two-leg ladder compound \srcacuo.
\newblock {\em Phys Rev B}, 72:054520, 2005.

\bibitem{Roux06}
G.~Roux, S.~R. White, S.~Capponi, and D.~Poilblanc.
\newblock Zeeman effect in superconducting two-leg ladders: Irrational
  magnetization plateaus and exceeding the pauli limit.
\newblock {\em Phys Rev Letters}, 97:087207, 2006.

\bibitem{Agterberg01}
D.~F. Agterberg and K.~Yang.
\newblock The effect of impurities on
  {F}ulde--{F}errell--{L}arkin--{O}vchinnikov superconductors.
\newblock {\em J. Phys. Condensed. Matter}, 13:9259, 2001.

\bibitem{Shahbazi17}
M.~Shahbazi, Y.~Fuseya, H.~Bakrim, A.~Sedeki, and C.~Bourbonnais.
\newblock Superconducting and density-wave instabilities of low-dimensional
  conductors with a zeeman coupling to a magnetic field.
\newblock {\em Phys. Rev. B}, 95:165111, Apr 2017.

\bibitem{Anderson59}
P.~W. Anderson.
\newblock Theory of dirty superconductors.
\newblock {\em J. Phys. Chem. Solids}, 11:26--30, 1959.

\bibitem{Abrikosov61}
A.~A. Abrikosov and L.~D. Gorkov.
\newblock Contribution to the theory of superconducting alloys with
  paramagnetic impurities.
\newblock {\em Sov. Phys. JETP}, 12:1243, 1961.

\bibitem{Maki04}
K.~Maki, H.~Won, and S.~Haas.
\newblock Quasiparticle spectrum of the hybrid s+g-wave superconductors
  $\mathrm{YNi_{2}B_{2}C}$ and $\mathrm{LuNi_{2}B_{2}C}$.
\newblock {\em Phys. Rev. B}, 69:012502, 2004.

\bibitem{Mackenzie03}
A.~P. MacKenzie and Y.~Maeno.
\newblock The superconductivity of $\mathrm{{S}r_{2}{R}u{O}_{4}}$ and the
  physics of spin-triplet pairing.
\newblock {\em Reviews of Modern Physics}, 75:657, 2003.

\bibitem{Maeno12}
Y.~Maeno et~al.
\newblock Evaluation of spin-triplet superconductivity in
  $\mathrm{{S}r_{2}{R}u{O}_{4}}$.
\newblock {\em Jour of the Physical Society of Japan}, 81:011009, 2012.

\bibitem{Bouffard82}
S.~Bouffard, M.~Ribault, R.~Brusetti, D.~Jerome, and K.~Bechgaard.
\newblock Low-temperature metallic state and superconductivity in
  quasi-one-dimensional organic conductors: pressure and irradiation
  investigations.
\newblock {\em J. Phys. C}, 15:295, 1982.

\bibitem{Choi82}
M.Y. Choi, P.M. Chaikin, S.Z. Huang, P.~Haen, E.M. Engler, and R.L. Greene.
\newblock Effect of radiation damage on the metal-insulator transition and low
  temperature transport in the \tmp6 salt.
\newblock {\em Phys. Rev. B.}, 25:6208, 1982.

\bibitem{Abrikosov83}
A.A. Abrikosov.
\newblock Superconductivity in a quasi-one-dimensional metal with impurities.
\newblock {\em Jour.Low.Temp.Physics}, 53:359, 1983.

\bibitem{Miljak80}
M.~Miljak, B.~Korin, J.~R. Cooper, K.~Holczer, and A.~J{\'a}nossy.
\newblock Low temperature magnetic susceptibility of quasi one-dimensional
  conductors.
\newblock {\em J. Physique}, 41:639, 1980.

\bibitem{Traettebergthesis}
O.~Traetteberg.
\newblock {\em The spin density wave collective mode in
  $\mathrm{(TMTSF)_{2}AsF_{6(1-x)}SbF_{6x}}$}.
\newblock PhD thesis, Orsay University, 1993.

\bibitem{Joo05}
N.~Joo, P.~Auban-Senzier, C.~Pasquier, D.~Jerome, and K.~Bechgaard.
\newblock Impurity-controlled superconductivity spin density wave interplay in
  the organic superconductor: \tmc.
\newblock {\em Eur.Phys.Lett}, 72:645, 2005.

\bibitem{Joo04}
N.~Joo, P.~Auban-Senzier, C.R. Pasquier, P.Monod, D.~Jerome, and K.~Bechgaard.
\newblock Suppression of superconductivity by non-magnetic disorder in the
  organic superconductor \tmxns.
\newblock {\em Eur.Phys.J. B}, 40:43--48, 2004.

\bibitem{Doiron10}
N.~Doiron-Leyraud, P.~Auban-Senzier, S.~Ren\'e de~Cotret, C.~Bourbonnais,
  D.~Jerome, K.~Bechgaard, and L.~Taillefer.
\newblock Linear-{T} scattering and pairing from antiferromagnetic fluctuations
  in the \tm2x organic superconductors.
\newblock {\em Eur. Phys. Jour.B}, 78:23, 2010.
\newblock DOI 10.1140/epjb/e2010-10571-4.

\bibitem{Auban11}
P.~Auban-Senzier et~al.
\newblock Fluctuating spin density wave conduction in \tmx organic
  superconductors.
\newblock {\em Europhysics Letters}, 94:17002, 2011.

\bibitem{Sun95}
Y.~Sun and K.~Maki.
\newblock Impurity effects in d-wave superconductors.
\newblock {\em Phys. Rev. B}, 51:6059, 1995.

\bibitem{Matsunaga99}
N.~Matsunaga et~al.
\newblock Anion disorder and two-dimensionality in the superconducting and sdw
  states of \tmc.
\newblock {\em Jour of Low Temp. Physics}, 117:1735, 1999.

\bibitem{Kagoshima83}
S.~Kagoshima {\textit{et al}}.
\newblock Quenching effect on the anion ordering in the organic superconductor
  \tmc.
\newblock {\em Solid. State. Comm}, 46:867, 1983.

\bibitem{Moret85}
R.~Moret, J.~P. Pouget, R.~Com\`es, and K.~Bechgaard.
\newblock X-ray study of the anion ordering transition in \tmc: quenching and
  irradiation effects.
\newblock {\em J. Phys. (France)}, 46:1521, 1985.
\newblock Open archive on HAL
  https://hal.archives-ouvertes.fr/jpa-00210098/document.

\bibitem{Pouget90}
Jean-Paul Pouget, Seiichi Kagoshima, Tsuyoshi Tamegai, Yoshio Nogami, Koichi
  Kubo, Tetsuo Nakajima, and Klaus Bechgaard.
\newblock High resolution x-ray scattering study of the anion ordering phase
  transition of \tmc.
\newblock {\em Journal of the Physical Society of Japan}, 59(6):2036--2053,
  1990.

\bibitem{Degennes}
P.G~De Gennes.
\newblock {\em Superconductivity of metals and alloys}.
\newblock Benjamin-New York, 1966.

\bibitem{Deutscher80}
G.~Deutscher, O.~Entin-Wohlman, S.~Fishman, and Y.~Shapira.
\newblock Percolation description of granular superconductors.
\newblock {\em Phys Rev B}, 21:5041, 1980.

\bibitem{Puchkaryov98}
E.~Puchkaryov and K.~Maki.
\newblock Impurity scattering in d-wave superconductivity. {U}nitarity limit
  versus {B}orn limit.
\newblock {\em Eur. Phys. J. B}, 4:191, 1998.

\bibitem{Suzumura89}
Y.~Suzumura and H.~J. Schulz.
\newblock Thermodynamic properties of impure anisotropic quasi-one-dimensional
  superconductors.
\newblock {\em Phys. Rev. B}, 39:11398, 1989.

\bibitem{Preosti94}
G.~Preosti, H.~Kim, and P.~Muskar.
\newblock Density of states in unconventional superconductors:
  {I}mpurity-scattering effects.
\newblock {\em Phys Rev B}, 50:1259, 1994.

\bibitem{Yonezawa24}
S.~Yonezawa \textit{et al}.
\newblock Born-limit for non-magnetic scattering in the organic superconductor
  \tmc.
\newblock 2024.
\newblock to be published.

\bibitem{Nickel05}
J.~C. Nickel, R.~Duprat, C.~Bourbonnais, and N.~Dupuis.
\newblock Triplet superconducting pairing and density-wave instabilities in
  organic conductors.
\newblock {\em Phys. Rev. Lett.}, 95:247001, 2005.

\bibitem{Sakakibara07}
T.~Sakakibara et~al.
\newblock Nodal structures of heavy fermion superconductors probed by the
  specific-heat measurements in magnetic fields.
\newblock {\em J. Phys. Soc. Japan}, 76:051004, 2007.

\bibitem{An10}
K.~An et~al.
\newblock Sign reversal of field-angle resolved heat capacity oscillations in a
  heavy fermion superconductor $\mathrm{CeCoIn_{5}}$ and d(x2-y2) pairing
  symmetry.
\newblock {\em Phys Rev Letters}, 104:037002, 2010.

\bibitem{Izawa02}
K.~Izawa, H.~Yamaguchi, T.~Sasaki, and Y~Matsuda.
\newblock Superconducting gap structure of
  $\kappa$-$\mathrm{(BEDT-TTF)_{2}Cu(NCS)_{2}}$ probed by thermal conductivity
  tensor.
\newblock {\em Phys Rev Letters}, 88:027002, 2002.

\bibitem{Izawa01}
K.~Izawa et~al.
\newblock Low energy quasiparticle excitation in the vortex state of
  borocarbide superconductor $\mathrm{YNi_{2}B_{2}C}$.
\newblock {\em Phys Rev Letters}, 86:1327, 2001.

\bibitem{Deguchi04}
K.~Deguchi, Z.Q. Mao, H.~Yaguchi, and Y.~Maeno.
\newblock Gap structure of the spin-triplet superconductor {Sr2RuO4} determined
  from the field orientation dependence of the specific heat.
\newblock {\em Phys Rev Letters}, 92:047002, 2004.

\bibitem{Volovik93}
G.~E. Volovik.
\newblock Superconductivityu with lines of gap nodes:density of states in the
  vortex.
\newblock {\em JETP Letters}, 58:469, 1993.

\bibitem{Volovik97a}
G.~E. Volovik.
\newblock Fermionic entropy of the vortex state in d-wave superconductors.
\newblock {\em JETP Letters}, 65:491, 1997.

\bibitem{Vekhter99b}
I.~Vekhter, P.~J. Hirschfeld, J.~P. Carbotte, and E.~J. Nicol.
\newblock Anisotropic thermodynamics of d-wave superconductors in the vortex
  state.
\newblock {\em Phys Rev B}, 59:R9023, 1999.

\bibitem{Yonezawa13}
S.~Yonezawa, Y.~Maeno, and D.~Jerome.
\newblock Extended analysis of the field-angle-dependent heat capacity of \tmc
  toward identification of the superconducting gap structure.
\newblock {\em Journal of Physics: Conference Series}, 449:012032, 1993.

\bibitem{Lepevelen01}
D.~Le Pevelen et~al.
\newblock Temperature and pressure dependencies of the crystal structure of the
  organic superconductor \tmc.
\newblock {\em Eur. Phys. J. B}, 19:363, 2001.

\bibitem{Nagai11}
Y.~Nagai, H.~Nakamura, and M.~Machida.
\newblock Superconducting gap function in an organic superconductor \tmc with
  anion ordering; first-principles calculations and quasi-classical analyses
  for angle-resolved heat capacity.
\newblock {\em Phys Rev B}, 83:104523, 2011.

\bibitem{Creuzet85e}
F.~Creuzet, C.~Bourbonnais, D.~Jerome, and K.~Bechgaard.
\newblock Cooperative phenommena in \tmc {NMR} relaxation.
\newblock {\em Mol. Cryst. Liq. Cryst}, 119:45--51, 1985.

\bibitem{Bourbon84}
C.~Bourbonnais, F.~Creuzet, D.~Jerome, K.~Bechgaard, and A.~Moradpour.
\newblock Cooperative phenomena in \tmc : an {NMR} evidence.
\newblock {\em J. Phys. (Paris) Lett.}, 45:L755, 1984.

\bibitem{Bourbonnais11}
C.~Bourbonnais and A.~Sedeki.
\newblock Superconductivity and antiferromagnetism as interfering orders in
  organic conductors.
\newblock {\em Comptes Rendus Physique}, 12:532--541, 2011.

\bibitem{Wu05}
W.~Wu et~al.
\newblock $^{77}${S}e {NMR} probe of magnetic excitations of the magic angle
  effect in \tmp6.
\newblock {\em Phys. Rev. Lett}, 94:097004, 2005.

\bibitem{Brown08}
S.E. Brown, P.M. Chaikin, and M.J. Naughton.
\newblock La tour des sels de {B}echgaard.
\newblock In A.~Lebed, editor, {\em The Physics of Organic Superconductors and
  Conductors}, page~49. Springer, Heidelberg, 2008.

\bibitem{Moryia00}
T.~Moryia and K.~Ueda.
\newblock Spin fluctuations and high temperature superconductivity.
\newblock {\em Advances in Physics}, 49:555, 2000.

\bibitem{Bourbonnais09}
C.~Bourbonnais and A.~Sedeki.
\newblock Link between antiferromagnetism and superconductivity probed by
  nuclear spin relaxation in organic conductors.
\newblock {\em Phys. Rev. B}, 80:085105, 2009.
\newblock arXiv.org:0904.2858.

\bibitem{Kimura11}
Y.~Kimura, M.~Misawa, and A.~Kawamoto.
\newblock Correlation between non-{F}ermi-liquid behavior and antiferromagnetic
  fluctuations in \tmp6 observed using $^{13}${C} {NMR} spectroscopy.
\newblock {\em Phys Rev B}, 84(045123), 2011.

\bibitem{Doiron09}
N.~Doiron-Leyraud, P.~Auban-Senzier, S.~Ren\'e de~Cotret, C.~Bourbonnais,
  D.~Jerome, K.~Bechgaard, and L.~Taillefer.
\newblock Correlation between linear resistivity and \tc in the {B}echgaard
  salts and the pnictide superconductor \baas.
\newblock {\em Phys. Rev. B}, 80:214531, 2009.
\newblock arXiv:0905.0964.

\bibitem{Auban11a}
P.~Auban-Senzier, D~J\'{e}rome, N~Doiron-Leyraud, S~{Ren\'{e} de Cotret},
  A~Sedeki, C~Bourbonnais, L~Taillefer, P~Alemany, E~Canadell, and K~Bechgaard.
\newblock The metallic transport of \tm2x organic conductors close to the
  superconducting phase.
\newblock {\em J. Phys.: Condens. Matter}, 23:345702, 2011.

\bibitem{Azlamazov68}
L.~G. Azlamazov and A.~I. Larkin.
\newblock Effect of fluctuations on the properties of a superconductor above
  the critical temperature.
\newblock {\em Sov. Phys. Solid State}, 10:875, 1968.

\bibitem{Lee74}
P.A. Lee, T.M. Rice, and P.W. Anderson.
\newblock Conductivity from charge or spin density waves.
\newblock {\em Solid State Comm.}, 14:703, 1974.

\bibitem{Barthel93}
E.~Barthel {\textit{et al}}.
\newblock {NMR} incommensurate and incommesurate spin density waves.
\newblock {\em Europhys. Lett}, 21:87, 1993.

\bibitem{Gruner94}
G.~Gruner.
\newblock The dynamics of spin density waves.
\newblock {\em Rev. Mod. Phys.}, 66:1--24, 1994.

\bibitem{Bourbonnais93}
C.~Bourbonnais.
\newblock Nuclear relaxation and electronic correlations in quasi-one
  dimensional organic conductors. i. {S}caling theory.
\newblock {\em J. Phys. I (France)}, 3:143, 1993.

\bibitem{Fang09}
L.Fang et~al.
\newblock Roles of multiband effects and electron hole asymmetry in the
  superconductivity and normal properties of \baas.
\newblock {\em Phys. Rev. B}, 80:140508(R), 2009.
\newblock arXiv.org:0903.2418.

\bibitem{Jin11}
K.~Jin, N.~P. Butch, K.~Kirshenbaum, J.~Paglione, and R.~L. Greene.
\newblock Link between spin fluctuations and electron pairing in copper oxide
  superconductors.
\newblock {\em Nature}, 476:73, 2011.

\bibitem{Nickel06}
J.~C. Nickel, R.~Duprat, C.~Bourbonnais, and N.~Dupuis.
\newblock Superconducting pairing and density-wave instabilities in
  quasi-one-dimensional conductors.
\newblock {\em Phys. Rev. B}, 73:165126, 2006.

\bibitem{Duprat01}
R.~Duprat and C.~Bourbonnais.
\newblock Interplay between spin-density-wave and superconducting states in
  quasi-one-dimensional conductors.
\newblock {\em Eur. Phys. J. B}, 21:219, 2001.

\bibitem{Bourbon88a}
C.~Bourbonnais and L.~G. Caron.
\newblock New mechanisms for phase transitions in quasi-one-dimensional
  conductors.
\newblock {\em Europhys. Lett.}, 5:209, 1988.

\bibitem{Emery86}
V.~J. Emery.
\newblock The mechanisms of organic superconductivity.
\newblock {\em Synthetic. Met}, 13:21, 1986.

\bibitem{Beal86}
M.T. B\'eal-Monod, C.~Bourbonnais, and V.~J. Emery.
\newblock Possible superconductivity in nearly antiferromagnetic itinerant
  fermion systems.
\newblock {\em Phys. Rev. B}, 34:7716, 1986.

\bibitem{Scalapino86}
D.~J. Scalapino, E.~Loh, and J.~E. Hirsch.
\newblock d-wave pairing near a spin-density-wave instability.
\newblock {\em Phys. Rev. B}, 34:R8190, 1986.

\bibitem{Sedeki09}
A.~Sedeki, D.~Bergeron, and C.~Bourbonnais.
\newblock Interfering antiferromagnetism and superconductivity in
  quasi-one-dimensional organic conductors.
\newblock {\em Physica B}, 405:S 89, 2009.

\bibitem{Sedeki12}
A.~Sedeki, D.~Bergeron, and C.~Bourbonnais.
\newblock Extended quantum criticality of low-dimensional superconductors near
  a spin-density-wave instability.
\newblock {\em Phys. Rev. B}, 85:165129, 2012.

\bibitem{Bakrim14}
H.~Bakrim and C.~Bourbonnais.
\newblock Role of electron-phonon interaction in a magnetically driven
  mechanism for superconductivity.
\newblock {\em Physical Review B}, 90:125119, 2014.

\bibitem{Vuletic02}
T.~Vuleti\'c, P.~Auban-Senzier, C.~Pasquier, S.~Tomi\u c, D.~Jerome,
  M.~H\'eritier, and K.~Bechgaard.
\newblock Coexistence of superconductivity and spin density wave orderings in
  the organic superconductor \tmp6.
\newblock {\em Eur. Phys. J. B}, 25:319, 2002.

\bibitem{Lee02b}
I.~J. Lee, P.M. Chaikin, and M.~J. Naughton.
\newblock Critical field enhancement near a superconductor-insulator
  transition.
\newblock {\em Phys Rev Letters}, 88:207002, 2002.

\bibitem{Goko09}
T.~Goko et~al.
\newblock Superconductivity coexisting with phase-separated static magnetic
  order in \bafeas, \srfeas and \cafeas.
\newblock {\em Phys Rev B}, 80:024508, 2009.

\bibitem{Uemura15}
Y.~J. Uemura.
\newblock {\em Strongly Correlated Systems}, volume 180 of {\em {S}olid {S}tate
  {S}ciences}, chapter Muon spin relaxation studies of unconventional
  superconductors: first-order behavior and comparable spin-charge energy
  scales, pages 237--267.
\newblock Springer, 2015.

\bibitem{Dressel20}
M.~Dressel and S.~Tomic.
\newblock Molecular quantum materials: electronic phases and charge dynamics in
  two-dimensional organic solids.
\newblock {\em Adv in Physics}, 69:1, 2020.

\bibitem{Saito82}
G.~Saito, T.~Enoki, K.~Toriumi, and H.~Inokuchi.
\newblock Two-dimensionality and suppression of metal-semiconductor transition
  in a new organic metal with alkylthio substituted {TTF} and perchlorate.
\newblock {\em Solid Sate Comm}, 42(8):557--560, 1983.

\bibitem{Parkin83e}
S.~S. P.~Parkin et~al.
\newblock Superconductivity in a {N}ew {F}amily of {O}rganic {C}onductors.
\newblock {\em Phys Rev Letters}, 50(4):270, 1983.

\bibitem{Laukhin85}
V.N. Laukhin, E.E. Kostyuchenko, Yu. Sushko, I.F. Shchegolev, and E.B.
  Yagubskii.
\newblock Effect of pressure on the superconductivity of
  $\beta$-$\mathrm{(ET)_{2}I_{3}}$.
\newblock {\em JETP.Lett.}, 41:81, 1985.

\bibitem{Murata85a}
K.Murata, M.Tokumoto, H.~Anzai, H.~Bando, G.~Saito, K.~Kajimura, and
  T.~Ishiguro.
\newblock Pressure phase diagram of the organic supercon,ductor
  $\beta$-\bedtttf2i3.
\newblock {\em J.Phys.Soc.Jpn.}, 54:2084, 1985.

\bibitem{Creuzet85b}
F.Creuzet, G.Creuzet, D.Jerome, D.Schweitzer, and H.J. Keller.
\newblock Homogeneous superconducting state at 8.1 k under ambient pressure ion
  the organic conductor $\beta$-\bedtttf2i3.
\newblock {\em J.Physique.Lett.}, 46:L--1079, 1988.

\bibitem{Yagubskii84}
E.B. Yagubskii, I.F. Shchegolev, V.N. Laukhin, P.A. Kononovich, M.V.
  Kartsovnik, A.V. Zvarykina, and L.I. Bubarov.
\newblock Superconducting properties of the orthorhombic phase of bis-
  (ethylenedithiolo)tetrathiofulvalene triiodide.
\newblock {\em JETP.Lett.}, 39:12, 1984.

\bibitem{Kang89}
W.Kang, G.~Montambaux, J.R. Cooper, D.~Jerome, P.~Batail, and C.~Lenoir.
\newblock Observation of giant magnetoresistance oscillations in the high-\tc
  phase of the two-dimensional organic conductor $\beta$-\bedtttf2i3.
\newblock {\em Phys. Rev. Lett}, 62:2559, 1989.

\bibitem{Kartsovnik88}
M.V. Kartsovnik, P.A. Kononovich, V.N. Laukhin, and I.F. Shchegolev.
\newblock Anisoropy of magnetoresitance and {S}hubnikov-de {H}aas oscillations
  in the organic meta $\beta$-$\mathrm{(ET)_{2}IBr_{2}}$.
\newblock {\em Anisotropy of Magnetoresistance and {S}hubnikov-de{H}aas
  oscillations in the organic metal $\beta$-\bedtttf2i3}, 48:541, 1988.

\bibitem{Yamaji89}
K.~Yamaji.
\newblock On the angle dependence of the magnetoresistance in
  quasi-two-dimensional organic superconductors.
\newblock {\em J.Phys.Soc.Jpn.}, 58:1520, 1989.

\bibitem{Urayama88}
H.~Urayama, H.~Yamochi, G.~Saito, K.~Nozawa, T.~Sugano, M.~Kinoshita, S.~Sato,
  K.~Oshima, A.Kawamoto, and J.~Tanaka.
\newblock A new ambient pressure organic superconductor based on {BEDT-TTF}
  with \tc higher than {10K} (\tc={10.4K}).
\newblock {\em Chem. Lett}, 55, 1988.

\bibitem{Kini90}
A.M.~Kini {\textit{et al}}.
\newblock A new ambient-pressure organic superconductor, \kappaet2cunbr, with
  the highest transition temperature yet observed (inductive onset \tc = 11.6
  {K}, resistive onset = 12.5 {K}.
\newblock {\em Inorg.Chem.}, 29:2555, 1990.

\bibitem{Williams90}
J.M.~Williams {\textit{et al}}.
\newblock From semiconductor-semiconductor transition {(42 K)} to the highest-
  \tc organic superconductor, \kappacl (\tc = {12.5 K}).
\newblock {\em Inorg.Chem.}, 29:3272--3274, 1990.

\bibitem{Lefebvre00}
S.~Lefebvre, P.~Wzietek, S.~Brown, Bourbonnais, D.~Jerome, C.~M\'ezi\`ere,
  M.~Fourmigu\'e, and P.~Batail.
\newblock Mott transition, antiferromagnetism, and unconventional
  superconductivity in layered organic superconductors.
\newblock {\em Phys. Rev. Lett.}, 85(25):5420--5423, Dec 2000.

\bibitem{Limelette03}
P.~Limelette, P.~Wzietek, S.Florens, A.~Georges, T.~A. Costi, P.~Pasquier,
  D.~Jerome, C.~M\'eziere, and P.~Batail.
\newblock Mott transition and transport crossovers in the organic compound
  \kappacl.
\newblock {\em Phys. Rev. Lett}, 91:016401, 2003.

\bibitem{Georges96}
A.Georges, G.~Kotliar, W.~Krauth, and M.~Rosenberg.
\newblock Dynamical mean-field theory of strongly correlated fermion systems
  and the limit of infinite dimensions.
\newblock {\em Rev.Mod.Phys.}, 68:13, 1996.

\bibitem{Mayaffre95}
H.~Mayaffre, P.~Wzietek~D. Jerome, C.~Lenoir, and P.~Batail.
\newblock Superconducting state of \kappaet2cunbr studied by 13{C NMR}:
  Evidence for vortex-core-induced nuclear relaxation and unconventional
  pairing.
\newblock {\em Phys.Rev.Lett}, 75, 1995.

\bibitem{Kanoda96}
K.~Kanoda, K.~Miyagawa, A.~Kawamoto, and Y.~Nakazawa.
\newblock {NMR} relaxation rate in the superconducting state of the organic
  conductor nmr relaxation rate in the superconducting state of the organic
  conductor \kappaet2cunbr.
\newblock {\em Phys. Rev. B.}, 54, 1996.

\bibitem{Yasin11}
S.~Yasin {\textit{et al}}.
\newblock Transport studies at the {M}ott transition of the two-dimensional
  organic metal \kappabrcl.
\newblock {\em Eur. Phys. J. B}, 79:383, 2011.

\bibitem{Bondarenko94}
V.~A.~Bondarenko {\textit{et al}}.
\newblock Transport properties of \kappabrcl (0 $\leq$ x $\leq$ 1) organic
  superconductors.
\newblock {\em Physica C}, 235:2467, 1994.

\bibitem{Tokumoto87}
M.~Tokumoto, H.~Anzai, K.~Kajimura, and T.~Ishiguro.
\newblock High \tc superconducting states in organic metals
  $\beta$-$\mathrm{(ET)_{2}X}$.
\newblock {\em Japanese Jour of Appl. Physics}, 26-3:1977, 1987.

\bibitem{Fournier03}
D.~Fournier, M.~Poirier, M.~Castonguay, and M.~Truong.
\newblock Mott transition, compressibility divergence and the {P-T} phase
  diagram of layered organic superconductors: An ultrasonic investigation.
\newblock {\em Phys.Rev. Lett}, 90:127002, 2003.

\bibitem{Kagawa04}
F.~Kagawa, T.~Itou, K.~Miyagawa, and K.~Kanoda.
\newblock Transport criticality of the first-order {M}ott transition in the
  quasi-two-dimensional organic conductor \kappacl.
\newblock {\em Phys.Rev.}, B 69:064511, 2004.

\bibitem{Mayaffre94a}
H.~Mayaffre, P.~Wzietek, C.~Lenoir, D.~Jerome, and P.~Batail.
\newblock $^{13}${C NMR} study of a quasi-two dimensional organic
  superconductor \kappaet2cunbr.
\newblock {\em Europhysics Letters}, 28:205--210, 1994.

\bibitem{Kerlin73}
A.~L. Kerlin, H.~Nagasawa, and D.~Jerome.
\newblock {M}etal-{I}nsulator transition investigation in $\mathrm{V_{2}O_{3}}$
  by nuclear magnetic resonance and relaxation.
\newblock {\em Solid State Communications}, 13:1125--1129, 1973.

\bibitem{Kagawa05}
F.~Kagawa, K.~Miyagawa, and K.~Kanoda.
\newblock Unconventional critical behaviour in a quasi-twodimensional organic
  conductor.
\newblock {\em Nature}, 436:534, 2005.

\bibitem{Limelette03a}
P.~Limelette, A.~Georges, D.~Jerome, P.Wzietek, P.~Metcalf, and J.M. Honig.
\newblock Universality and critical behavior at the {M}ott transition.
\newblock {\em Science}, 302:89, 2003.

\bibitem{Kawamoto95}
A.~Kamawamoto \textit {et al}.
\newblock $^{13}$ {NMR} study of layered organic superconductors based on
  {BEDT-TTF} molecules.
\newblock {\em Physical Review Letters}, 74:3455, 1995.

\bibitem{McWhan70}
D.~B. McWhan and J.~P. Remeika.
\newblock Metal-insulator transition in \vxcrx.
\newblock {\em Phys. Rev. B.}, 2(9):3734, 1970.

\bibitem{Jayaraman70}
A.~Jayaraman, D.~B. McWhan, J.~P. Remeika, and P.~D. Dernier.
\newblock Critical behavior of the {M}ott transition in {C}r-doped \v2o3.
\newblock {\em Phys. Rev. B}, 2(9):3751, 1970.

\bibitem{Mott61}
N.~F. Mott.
\newblock The transition to the metallic state.
\newblock {\em Phil. Mag}, 6:287, 1961.

\bibitem{Mott71}
N.~F. Mott.
\newblock The {M}etal-non-{M}etal {T}ransition.
\newblock {\em Jour. Physique. Colloques}, 32:C1--14, 1971.

\bibitem{Lesino}
High pressure investigation of $\mathrm{V_{2}O_{3}}$ performed in the thesis
  work of G. Lesino at Orsay in 1972, unpublished.

\bibitem{Jaccarino62}
V.~Jaccarino and M.~Peter.
\newblock Ultra-high-field superconductivity.
\newblock {\em Phys.Rev.Lett}, 9:290, 1962.

\bibitem{Uji01}
S.Uji, H.~Shinagawa, C.~Terakura, T.~Terashima, T.~Yakabe, Y.~Terai,
  M.~Tokumoto, A.Kobayashi, H.~Tanaka, and H.~Kobayashi.
\newblock Magnetic-field-induced superconductivity in a two-dimensional organic
  conductor.
\newblock {\em Nature}, 410:908, 2001.

\bibitem{Balicas01}
L.Balicas, J.S. Brooks, K.~Storr, S.Uji, M.~Tokumoto, K.Kobayashi,
  A.~Kobayashi, H.~Tanaka, V.~Barzykin, and L.P. Gorkov.
\newblock Superconductivity in an organic insulator at very high magnetic
  fields.
\newblock {\em Phys. Rev. Lett.}, 87:067002, 2001.

\bibitem{Naito21}
T.~Naito.
\newblock Modern history of organic conductors: An overview.
\newblock {\em Crystals}, 11:838, 2021.
\newblock https://doi.org/10.3390/cryst11070838.

\bibitem{Mott74}
N.F. Mott.
\newblock {\em Metal-Insulator Transitions}.
\newblock Taylor and Francis, London, 1974.

\bibitem{Alcacer74}
L.~Alcacer and A.~H. Maki.
\newblock Electrically conducting metal dithiolate-perylene complexes.
\newblock {\em Jour of Physical Chemistry}, 78(3):1974, 1974.

\bibitem{Alcacer80}
L.~Alcacer {\textit{et al}}.
\newblock Synthesis, structure and preliminary results on electrical and
  magnetic properties of $\mathrm{(Perylene)_{2} [Pt(mnt)_{2}]}$.
\newblock {\em Solid State Comm}, 35:945, 1980.

\bibitem{Graf09}
D.~Graf {\textit{et al}}.
\newblock Evolution of superconductivity from a charge-density-wave ground
  state in pressurized $\mathrm{(Per)_{2}Au(mnt)_{2}}$.
\newblock {\em Eur. Phys. Lett}, 85:27009, 2009.

\bibitem{Bonfait91}
G.~Bonfait {\textit{et al}}.
\newblock Magnetic field dependence of the metal-insulator transition in
  $\mathrm{(Per)_{2}Pt(mnt)_{2}}$ and $\mathrm{(Per)_{2}Au(mnt)_{2}}$.
\newblock {\em Solid State Comm.}, 6:391--394, 1991.

\bibitem{Bourbonnais91a}
C.~Bourbonnais {\textit{et al}}.
\newblock Nuclear and electronic resonance approaches to magnetic and lattice
  fluctuations in the two-chain family of organic compounds
  $\mathrm{(Per)_{2}M(mnt)_{2}}$ ($\mathrm{M =Pt, Au)}$.
\newblock {\em Phys Rev B}, 44:641, 1991.

\bibitem{Henriques84}
R.~T. Henriques, L.~Alcacer, J.~P. Pouget, and D.~Jerome.
\newblock Electrical conductivity and x-ray diffuse scattering study of the
  family of organic conductors $\mathrm{(Per)_{2}M(mnt)_{2}}$, $\mathrm{M= Pt,
  Pd, Au)}$.
\newblock {\em J. Phys. C Solid State}, 17:5197, 1984.

\bibitem{Graf07}
D.~Graf {\textit{et al}}.
\newblock Quantum interference in the quasi-one-dimensional organic conductor
  $\mathrm{(Per)_{2}Pt(mnt)_{2}}$.
\newblock {\em Phys. Rev. B}, 75:255101, 1997.

\bibitem{Canadell04}
E.~Canadell, M.~Almeida, and J.~Brooks.
\newblock Electronic band structure of. $\alpha$-$\mathrm{(Per)_{2}M(mnt)_{2}}$
  compounds.
\newblock {\em Eur. Phys. J. B}, 42:453, 2004.

\bibitem{Bakrim10}
H.~Bakrim and C.~Bourbonnais.
\newblock Superconductivity close to the charge-density-wave instability.
\newblock {\em Eur. Phys. J. B}, 90:27001, 2010.

\bibitem{Bousseau86}
M.~Bousseau {\textit{et al}}.
\newblock Highly conducting charge-transfer compounds of tetrathiafulvalene and
  transition metal-dmit-complexes.
\newblock {\em J. Am. Chem. Soc}, 108:1908--1916, 1986.

\bibitem{Brossard86}
L.~Brossard, M.~Ribault, M.~Bousseau, L.~Valade, and P.~Cassoux.
\newblock Un nouveau type de supraconducteur mol{\'e}culaire:
  $\mathrm{TTF[Ni(dmit)_{2}]_{2}}$.
\newblock {\em Comptes Rendus Acad. Sc. Paris}, 302:205, 1986.

\bibitem{Kobayashi87}
A.Kobayashi, R.Kato, H.Kobayashi, S.Moriyama, Y.Nishino, K.Kajita, and
  W.Sasaki.
\newblock Anion arrangement in a new molecular superconductor.
\newblock {\em Chem. Lett.}, pages 2017--2020, 1986.

\bibitem{Brossard89}
L.~Brossard, M.~Ribault, L.~Valade, and P.~Cassoux.
\newblock Pressure induced superconductivity in molecular
  $\mathrm{TTF[Pd(dmit)_{2}]_{2}}$.
\newblock {\em J. Phys. France}, 50:1521--1534, 1989.

\bibitem{Ravy89}
S.~Ravy, J.~P. Pouget, L.~Valade, and J.~P. Legros.
\newblock Structural evidence of charge density waves in the series of
  molecular conductors and superconductors: $\mathrm{TTF[M(dmit)_{2}]_{2}}$
  $\mathrm{(M = Pd, Ni)}$.
\newblock {\em Europhysics Letters}, 9:391--396, 1989.

\bibitem{Canadell89}
E.~Canadell {\textit{et al}}.
\newblock On the band electronic structure of $\mathrm{X [M (dmit)_{2}]_{2}}$
  ({X = TTF}, $\mathrm{(CH_{3})_{4}N}$; {M = Ni, Pd)} molecular conductors and
  superconductors.
\newblock {\em J. Phys. (France)}, 50:2967, 1989.

\bibitem{Kaddour14}
W.~Kaddour {\textit{et al}}.
\newblock Charge density wave and metallic state coexistence in the multiband
  conductor $\mathrm{TTF[Ni(dmit)_{2}]2}$.
\newblock {\em Phys Rev B}, 90:205132, 2014.

\bibitem{siestacode}
{SIESTA} code.
\newblock For more information on the SIESTA code visit:http://
  icmab.cat/leem/siesta/.

\bibitem{Brossard90}
L.~Brossard, M.~Ribault, L.~Valade, and P.~Cassoux.
\newblock Simultaneous competition and coexistence between charge density waves
  and reentrant superconductivity in the pressure-temperature phase diagram of
  the molecular conductor $\mathrm{TTF[Ni(dmit)_{2}]_{2}}$.
\newblock {\em Phys Rev B}, 42:3935, 1990.

\bibitem{Vainrub90}
A.~Vainrub, D.~Jerome, M.~F. Bruniquel, and P.~Cassoux.
\newblock Coexistence of metallic character and charge density wave on
  $\mathrm{Ni(dmit)_{2}}$ stacks in $\mathrm{TTF[Ni(dmit)_{2}]_{2}}$.
\newblock {\em Europhysics Letters}, 12:267, 1990.

\bibitem{Berthier76}
C.~Berthier, D.~Jerome, P.~Molini{\'e}, and J.~Rouxel.
\newblock Charge density waves in layer structures: A {NMR} study on a
  {2H}-$\mathrm{NbSe_{2}}$ single crystal.
\newblock {\em Solid State Comm.}, 19:131--135, 1976.

\bibitem{Berthier76c}
C.~Berthier, P.~MoIini{\'e}, and D.~Jerome.
\newblock Evidence for a connection between charge density waves and the
  pressure enhancement of superconductivity in {2H}---$\mathrm{NbSe_{2}}$.
\newblock {\em Solid State Comm.}, 18:1393, 1976.

\bibitem{Akiko04}
A.~Kobayashi, E.~Fujiwara, and H.~Kobayashi.
\newblock Single-component molecular metals with extended-{TTF} dithiolate
  ligands.
\newblock In P.~Batail, editor, {\em Molecular Conductors}, volume 104, pages
  5243--5264. American Chemical Society, 2004.

\bibitem{Kobayashi21}
A.~Kobayashi {\textit{et al}}.
\newblock Single-component molecular conductors.
\newblock {\em Bull.Chem. Soc. Jpn}, 94:2540--2562, 2021.

\bibitem{Sasa05}
M.~Sasa {\textit{et al}}.
\newblock Crystal structures and physical properties of single-component
  molecular conductors consisting of nickel and gold complexes with
  bis(trifluoromethyl)tetrathiafulvalenedithiolate ligands.
\newblock {\em J. Mater. Chem}, 15:155--163, 2005.

\bibitem{Tanaka01}
H.~Tanaka {\textit{et al}}.
\newblock A three-dimensional synthetic metallic crystal composed of
  single-component molecules.
\newblock {\em Science}, 291:285, 2001.

\bibitem{Cui14}
H.~Cui {\textit{et al}}.
\newblock A single-component molecular superconductor.
\newblock {\em J. Am. Chem. Soc}, 136:7619--7622, 2014.

\bibitem{Kobayashi01}
A.~Kobayashi, H.~Tanaka, and H.~Kobayashi.
\newblock Molecular design and development of single-component molecular
  metals.
\newblock {\em J. Mater. Chem}, 11:2078--2088, 2001.

\bibitem{Lebedbook}
A.~Lebed, editor.
\newblock {\em The Physics of Organic Superconductors and Conductors}.
\newblock Number 110 in Materials Science. Springer-Verlag, Berlin, 2008.

\bibitem{Ribault83}
M.~Ribault, D.~Jerome, J.~Tuchendler, C.~Weyl, and K.~Bechgaard.
\newblock Low-field and anomalous high-field hall effect in \tmc.
\newblock {\em J. Physique Lett}, 44:L--953, 1982.

\bibitem{Kang04}
W.~Kang.
\newblock Magnetoresistance study of low-dimensional electrons in the
  {B}echgaard salts.
\newblock {\em Current Applied Physics}, 4:263--266, 2004.

\bibitem{Montambaux85c}
G.~Montambaux.
\newblock {\em Contribution {\`a} l'{\'e}tude des conducteurs
  quasi-unidimensionnels sous champ magn{\'e}tique}.
\newblock PhD thesis, Universit{\'e} Paris-Sud, Orsay, December 1985.

\bibitem{Balicas95}
L.~Balicas, G.~Kriza, and F.~I.~B. Williams.
\newblock Sign reversal of the quantum {H}all number in \tmp6.
\newblock {\em Phys Rev Letters}, 75(10):2000, 1995.

\bibitem{Cooper89}
J.R. Cooper, W.~Kang, P.~Auban, G.~Montambaux, D.~Jerome, and K.~Bechgaard.
\newblock Quantized {H}all effect and a new field-induced phase transition in
  the organic superconductor \tmp6.
\newblock {\em Phys. Rev. Lett}, 65:1984, 1989.

\bibitem{Hannahs89}
S.T. Hannahs, J.S. Brooks, W.~Kang, L.Y. Chiang, and P.M. Chaikin.
\newblock Quantum {H}all effect in a bulk crysal.
\newblock {\em Phys.Rev.Lett.}, 63:1988, 1989.

\bibitem{Ribault85}
M.~Ribault.
\newblock Electronic states below 5{K} in \tmc.
\newblock {\em Mol. Cryst. Liq. Cryst}, 119:91--65, 1985.

\bibitem{Piveteau86}
B.~Piveteau {\textit{et al}}.
\newblock Hall effect study of the field-induced instabilities in \tmp6 under
  pressure.
\newblock {\em J. Phys. C. Solid State}, 19:4483, 1986.

\bibitem{Cho99}
H.~Cho and W.~Kang.
\newblock Negative {H}all plateaus and quantum {H}all effect in \tmp6.
\newblock {\em Phys. Rev. B.}, 59:9814, 1999.

\bibitem{Heritier84}
M.~H\'eritier, G.~Montambaux, and P.~Lederer.
\newblock Stability of the spin density wave phases in \tmc : quantized nesting
  effect.
\newblock {\em J. Phys. (Paris) Lett.}, 45:L943, 1984.

\bibitem{Montambaux85}
G.~Montambaux, M.~H\'eritier, and P.~Lederer.
\newblock Spin susceptibility of the two-dimensional electron gas with open
  fermi surface under magnetic field.
\newblock {\em Phys Rev Letters}, 55:2078, 1985.

\bibitem{Montambaux86}
G.~Montambaux.
\newblock Susceptibility and instability of the {Q1D} electron gas under
  magnetic field.
\newblock In D.~Jerome and L.~G. Caron, editors, {\em Low dimensional
  Conductors and Superconductors}, page 233. Plenum Press (New York), 1986.

\bibitem{Montambaux85a}
G.~Montambaux and D.~Zanchi.
\newblock The quantum {H}all effect in {Q1D} conductors.
\newblock {\em Synthetic Metals}, 86:2235--2240, 1997.

\bibitem{Zanchi96}
D.~Zanchi and G.~Montambaux.
\newblock Sign reversals of the quantum {H}all effect in quasi-1d conductors.
\newblock {\em Phys. Rev. Lett}, 77:366, 1996.

\bibitem{Ishiguro}
T.~Ishiguro, K.~Yamaji, and G.~Saito.
\newblock {\em Organic Superconductors}.
\newblock Number~88 in Springer Series in Solid State Sciences. Springer,
  Heidelberg, 2008.

\bibitem{Montambaux16}
G.~Montambaux and D.~Jerome.
\newblock Rapid magnetic oscillations and magnetic breakdown in quasi-{1D}
  conductors.
\newblock {\em Comptes Rendus Physique}, 17:376--388, 2016.
\newblock http://dx.doi.org/10.1016/j.crhy.2015.11.007.

\bibitem{Uji97}
S.~Uji, J.~S. Brooks, S.~Takasaki, J.~Yamada, and H.~Anzai.
\newblock Origin of rapid oscillations in the metallic phase for the organic
  conductor \tmc.
\newblock {\em Solid State Comm}, 103:387, 1997.

\bibitem{Ulmet84}
J.P. Ulmet, P.~Auban, and S.~Askenazy.
\newblock High field {S}hubnikov-de-{H}aas effect and magnetoresistance in the
  organic metal \tmc.
\newblock {\em Solid State Comm}, 52:547--549, 1984.

\bibitem{Kornilov07b}
A.~V.~Kornilov {\textit{et al}}.
\newblock Rapid oscillations in \tmp6.
\newblock {\em Phys Rev B}, 76:045109, 2007.

\bibitem{Kang91b}
W.~Kang, J.~R. Cooper, and D.~Jerome.
\newblock Quantized {H}all effect in the organic superconductor \tmre.
\newblock {\em Phys. Rev. B}, 43:11467, 1991.

\bibitem{Brooks99}
J.~S.~Brooks {\textit{et al}}.
\newblock Quantum oscillations in quasi-one-dimensional metals with
  spin-density-wave ground states.
\newblock {\em Phys. Rev. B.}, 59:2604, 1999.

\bibitem{Stuck32}
E.~C.~G. Stuckelberg.
\newblock Theory of inelastic collisions between atoms.
\newblock {\em Helv.Phys.Acta}, 5:369, 1932.

\bibitem{Kang07}
W.~Kang, T.~Osada, Y.~J. Jo, and H.~Kang.
\newblock Interlayer magnetoresistance of quasi-one-dimensional layered organic
  conductors.
\newblock {\em Phys Rev Letters}, 99:017002, 2007.

\bibitem{Lebed86a}
A.G. Lebed.
\newblock Anisotropy of an instability for a spin density wave induced by a
  magnetic field in a {Q1D} conductor.
\newblock {\em JETP. Letters}, 43:174, 1986.

\bibitem{Lebed89}
A.~Lebed and P.~Bak.
\newblock Theory of unusual anisotropy of magnetoresistance in organic
  superconductors.
\newblock {\em Phys Rev Letters}, 63:1315, 1989.

\bibitem{Gallois83}
B.~Gallois {\textit{et al}}.
\newblock \tmc: Meanstructure at 7{K} comparative studt with 300{K} and 125{K}
  data.
\newblock {\em J. Phys. (Paris) Coll.}, 44:C3--1071, 1983.

\bibitem{Osada91}
T.~Osada, A.~Kawasumi, S.~Kagoshima, N.~Miura, and G.~Saito.
\newblock Commensurability effect of magnetoresistance anisotropy in the
  quasi-one-dimensional conductor tetramethyltetraselenafulvalenium
  perchlorate, \tmc.
\newblock {\em Phys.Rev.Lett.}, 66:1525, 1991.

\bibitem{Naughton91}
M.~J.~Naughton {\textit{et al}}.
\newblock Commensurate fine structure in angular-dependent studies of \tmc.
\newblock {\em Phys Rev Letters}, 67:3712, 1991.

\bibitem{Osada92}
T.~Osada, S.~Kagoshima, and N.~Miura.
\newblock Resonance effect in magnetotransport anisotropy of
  quasi-one-dimensional conductors.
\newblock {\em Phys. Rev. B}, 46:1812, 1992.

\bibitem{Danner94}
G.~M. Danner, W.~Kang, and P.~M. Chaikin.
\newblock Measuring the {F}ermi surface of quasi-one-dimensional metals.
\newblock {\em Phys Rev Letters}, 72:3714, 1994.

\bibitem{Singleton02}
J.~Singleton {\textit{et al}}.
\newblock Test for interlayer coherence in a quasi-two-dimensional
  superconductor.
\newblock {\em Phys Rev Letters}, 88:037001, 2002.

\bibitem{Danner95}
G.~M. Danner and P.~M. Chaikin.
\newblock Non {F}ermi liquid transport in \tmp6.
\newblock {\em Phys. Rev. Lett}, 75:4690, 1995.

\bibitem{Osada96}
T.~Osada, S.~Kagoshima, and N.~Miura.
\newblock Third angular effect of magnetoresistance in quasi-one-dimensional
  conductors.
\newblock {\em Phys. Rev. Lett}, 77:5261, 1996.

\bibitem{Yoshino01}
H.~Yoshino {\textit{et al}}.
\newblock Direct evidence of dimensionality enhancement of {Q1D} {TMTSF} and
  {DMET} salts.
\newblock {\em Synthetic Metals}, 120:885--886, 2001.

\bibitem{Yoshino95}
H.~Yoshino {\textit{et al}}.
\newblock Anomalous angular dependence of magnetoresistance of an organic
  superconductor, $\mathrm{(DMET)_{2}I_{3}}$.
\newblock {\em J. Phys. Soc.Japan}, 64:2307, 1995.

\bibitem{Yoshino03}
H.~Yoshino, S.~Shodai, and K.~Murata.
\newblock Third angular effect of \tmc in {R}- and {Q}-states under pressure.
\newblock {\em Synthetic. Metals}, 133-134:55--56, 2003.

\bibitem{Alcacer23}
L.~Alcacer.
\newblock {\em The {P}hysics of organic electronics}.
\newblock IOP Publishing, 2023.

\bibitem{Suda14}
M.~Suda {\textit{}}.
\newblock Strain-tunable superconducting field-effect transistor with an
  organic strongly-correlated electron system.
\newblock {\em Adv. Materials}, 26:3490--3495, 2014.

\bibitem{Yamamoto13}
H.~M.~Yamamoto {\textit{et al}}.
\newblock A strained organic field-effect transistor with a gate-tunable
  superconducting channel.
\newblock {\em Nature Comm}, 4:1--7, 2013.

\bibitem{Kawasugi11}
Y.~Kawasugi {\textit{}}.
\newblock Electric-field-induced {M}ott transition in an organic molecular
  crystal.
\newblock {\em Phys. Rev. B .}, 84:125129, 2011.

\bibitem{Kawaguchi19}
G.~Kawaguchi and H.~M. Yamamoto.
\newblock Control of organic superconducting field effect transistor by cooling
  rate.
\newblock {\em Crystals}, 9:605, 2019.

\bibitem{Greenblatt84}
M.~Greenblatt {\textit{et al}}.
\newblock Quasi two-dinensional electronic properties of the lithium molybdenum
  bronze,$\mathrm{Li_{0.9}Mo_{6}O_{17}}$.
\newblock {\em Solid State Commun.}, 51:671, 1984.

\bibitem{Jiao14}
W.~H.~Jiao {\textit{et al}}.
\newblock Superconductivity in a layered $\mathrm{Ta_{4}Pd_{3}Te_{16}}$ with
  $\mathrm{PdTe_{2}}$ chains.
\newblock {\em J. Am. Chem. Soc}, 136:1284--1287, 2014.

\bibitem{Tang15}
Z-Tu~Tang {\textit{et al}}.
\newblock Unconventional superconductivity in quasi-one-dimensional
  $\mathrm{A_{2}Cr_{3}As_{3}} $.
\newblock {\em Phys Rev B}, 91:020506, 2015.

\bibitem{Pan15}
J.~Pan {\textit{et al}}.
\newblock Nodal superconductivity and superconducting dome in new layered
  superconductor $\mathrm{Ta_{4}Pd_{3}Te_{16}}$.
\newblock {\em Phys Rev B}, 92:180505, 2015.

\bibitem{Mercure12}
J.~F.~Mercure {\textit{et al}}.
\newblock Upper critical magnetic field far above the paramagnetic
  pair-breaking limit of superconducting one-dimensional
  $\mathrm{Li_{0.9}Mo_{6}O_{17}}$ single crystals.
\newblock {\em Phys. Rev. Lett}, 108:187003, 2012.

\bibitem{Bao15}
J.~K.~Bao {\textit{et al}}.
\newblock Superconductivity in quasi-one-dimensional
  $\mathrm{K_{2}Cr_{3}As_{3}}$ with significant electron correlations.
\newblock {\em Phys Rev X}, 5:011013, 2015.

\bibitem{Kong15}
T.~Kong, S.~L. Bud'ko, and P.~C. Canfield.
\newblock Anisotropic $\mathrm{H_{c2}}$, thermodynamic and transport
  measurements, and pressure dependence of $\mathrm{T_{c}}$ in
  $\mathrm{K_{2}Cr_{3}As_{3}}$ single crystals.
\newblock {\em Phys Rev B}, 91:020507, 2015.

\bibitem{Maeno94}
Y.~Maeno, H.~Hashimoto, K.~Yoshida, S.~Nishizaki, T.~Fujita, J.~G. Bednorz, and
  F.~Lichtenberg.
\newblock Superconducvity in a layered perovskite whitout copper.
\newblock {\em Nature}, 372:532, 1994.

\bibitem{Mackenzie19}
A.~P. Mackenzie.
\newblock A personal perspective on the unconventional superconductivity of
  $\mathrm{Sr_{2}RuO_{4}}$.
\newblock {\em Journal of Superconductivity and Novel Magnetism}, 33:177--182,
  2019.
\newblock https://doi.org/10.1007/s10948-019-05312-4.

\bibitem{Shirakawa97}
N.~Shirakawa {\textit{et al}}.
\newblock Pressure dependence of superconducting critical temperature of
  $\mathrm{Sr_{2}RuO_{4}}$.
\newblock {\em Phys Rev B}, 56(13):7890, 1997.

\bibitem{Kinjo22}
K.~Kinjo {\textit{et al}}.
\newblock Superconducting spin smecticity evidencing the {F}ulde-{F}errell-
  {L}arkin-{O}vchinnikov state in $\mathrm{Sr_{2}RuO_{4}}$.
\newblock {\em Science}, 376(6591):397--400, 2022.

\bibitem{Pustogow19}
A.~Pustogow {\textit{et al}}.
\newblock Constraints on the superconducting order parameter in
  $\mathrm{Sr_{2}RuO_{4}}$ from oxygen-{17} nuclear magnetic resonance.
\newblock {\em Nature}, 574:72, 2019.
\newblock https://doi.org/10.1038/s41586-019-1596-2.

\bibitem{Ishida20}
K.~Ishida, M.~Manago, K.~Kinjo, and Y.~Maeno.
\newblock Reduction of the $\mathrm{^{17} {O}}$ {K}night shift in the
  superconducting state and the heat-up effect by {NMR} pulses on
  $\mathrm{Sr_{2}RuO_{4}}$.
\newblock {\em J. Phys. Soc. Japan}, 89:034712, 2020.

\bibitem{Chronister20}
A.~Chronister {\textit{et al}}.
\newblock Evidence for even parity unconventional superconductivity in
  $\mathrm{Sr_{2}RuO_{4}}$.
\newblock {\em PNAS}, 118(25), 2020.
\newblock https://doi.org/10.1073/pnas.2025313118.

\bibitem{Ginzburg89}
V.L. Ginzburg.
\newblock High-temperature superconductivity: some remarks.
\newblock {\em Progress in Low Temperature Physics}, 12:1--44, 1989.

\bibitem{Anderson91}
P.~W. Anderson and J.~R. Schrieffer.
\newblock A dialogue on the theory of {H}igh \tc.
\newblock {\em Physics Today}, June:54--61, 1991.

\end{thebibliography}
\end{document}